\documentclass[a4paper,numbers=noenddot,12pt]{book}
\pdfoutput=1
\usepackage[utf8]{inputenc}

\usepackage{amsmath,amssymb,amsfonts,array,marvosym,dsfont}
\usepackage[pdftex]{color,graphicx}
\graphicspath{{.}{Images_NMSSM/}{Images_CKM/}{Images_U2SUSY/}{Images_U2EFT/}{Images_Introduction/}}
\usepackage{color}
\usepackage{slashed,fancybox,empheq,accents}
\usepackage[compress]{cite}
\usepackage{enumerate}
\usepackage{subfigure}
\usepackage[english]{babel}
\usepackage{bbm}
\usepackage{slashed}
\usepackage{multirow}
\usepackage{fancybox}

\usepackage[dvipsnames]{xcolor}
\definecolor{darkblue}{rgb}{0,0.2,0.6}
\definecolor{darkgreen}{rgb}{0,0.4,0}

\usepackage[linktoc=page,bookmarks=false,colorlinks=false,
linkbordercolor=RoyalBlue,citebordercolor=ForestGreen,urlbordercolor=CornflowerBlue]{hyperref}

\usepackage[normalem]{ulem}
\allowdisplaybreaks

\def\L{\mathcal{L}}
\def\Q{\mathcal{Q}}

\def\T{\mathcal{T}}

\def\D{\mathcal{D}}

\DeclareFontFamily{OT1}{pzc}{}
\DeclareFontShape{OT1}{pzc}{m}{it}{<-> s * [1.10] pzcmi7t}{}
\DeclareMathAlphabet{\mathpzc}{OT1}{pzc}{m}{it}
\DeclareMathOperator{\tr}{tr} 

\definecolor{rosso}{cmyk}{0,1,1,0.4}

\newcommand{\be}{\begin{equation}}
\newcommand{\ee}{\end{equation}}
\newcommand{\bea}{\begin{eqnarray}}
\newcommand{\eea}{\end{eqnarray}}
\newcommand{\beq}{\begin{equation}}
\newcommand{\eeq}{\end{equation}}
\newcommand{\beqa}{\begin{eqnarray}}
\newcommand{\eeqa}{\end{eqnarray}}
\newcommand{\ba}{\begin{array}}
\newcommand{\ea}{\end{array}}

\newcommand{\tsum}{{\textstyle\sum}}

\newcommand{\GeV}{\,{\rm GeV}}

\newcommand{\ELqL}{\boldsymbol{q}_{\boldsymbol{L}}}
\newcommand{\ELuR}{\boldsymbol{u}_{\boldsymbol{R}}}
\newcommand{\ELdR}{\boldsymbol{d}_{\boldsymbol{R}}}
\newcommand{\EHqL}{q_{3L}}
\newcommand{\EHuR}{t_{R}}

\newcommand{\ELqLbar}{\boldsymbol{\bar q}_{\boldsymbol{L}}}

\newcommand{\EHqLbar}{\bar q_{3L}}

\newcommand{\CLquR}{\boldsymbol{Q_R^u}}

\newcommand{\CHquR}{Q_{3R}^u}

\newcommand{\CLuLbar}{\boldsymbol{\bar U_L}}

\newcommand{\CHuLbar}{\bar T_{L}}

\newcommand{\ELeR}{\boldsymbol{e}_{\boldsymbol{R}}}

\newcommand{\EHeR}{\tau_{R}}
\newcommand{\ELlLbar}{\boldsymbol{\bar l}_{\boldsymbol{L}}}

\newcommand{\EHlLbar}{\bar l_{3L}}

\newcommand{\CLlR}{\boldsymbol{L_R}}

\newcommand{\CHlR}{L_{3R}}

\newcommand{\CLeLbar}{\boldsymbol{\bar E_L}}

\newcommand{\CHeLbar}{\bar{\mathcal{T}}_{L}}

\newcommand{\V}{\boldsymbol{V}}
\newcommand{\Ve}{\boldsymbol{V_e}}
\newcommand{\Vu}{\boldsymbol{V_u}}
\newcommand{\Vd}{\boldsymbol{V_d}}

\def\I{\mathcal{I}}

\newcommand{\qL}{\boldsymbol{q}_{\boldsymbol{L}}}
\newcommand{\uR}{\boldsymbol{u}_{\boldsymbol{R}}}
\newcommand{\dR}{\boldsymbol{d}_{\boldsymbol{R}}}

\newcommand{\qLbar}{\boldsymbol{\bar q}_{\boldsymbol{L}}}
\newcommand{\uRbar}{\boldsymbol{\bar u}_{\boldsymbol{R}}}

\newcommand{\UtreLC}{$U(3)^3_\mathrm{LC}$}
\newcommand{\UtreRC}{$U(3)^3_\mathrm{RC}$}
\newcommand{\UdueLC}{$U(2)^3_\mathrm{LC}$}
\newcommand{\UdueRC}{$U(2)^3_\mathrm{RC}$}
\newcommand{\Utre}{$U(3)^3$}
\newcommand{\Udue}{$U(2)^3$}

\begin{document}

\begin{titlepage}
\begin{center}
\begin{center}
\includegraphics[width=.3\textwidth]{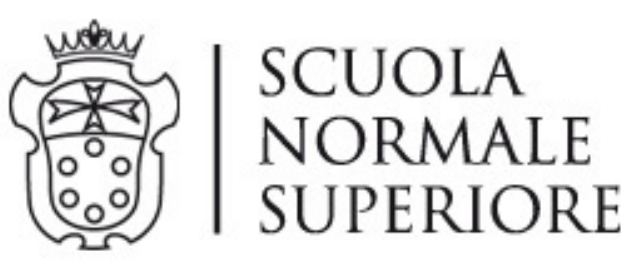}
\end{center}
\vspace{0.4cm}
\textsc{\large Classe di Scienze}\\ \vspace{0.2cm}
\large{Corso di Perfezionamento in Fisica}\\ \vspace{1 cm}
\textsc{\large PhD Thesis}\\ \vspace{2.5 cm}
\textsc{\Huge \bf Flavour and Higgs physics\\ \vspace{.4 cm} near the Fermi scale}\\
\vspace{4 cm}
\begin{minipage}{0.4\linewidth}
\flushleft  {\large Candidate}\\ \vspace{0.2cm} \textbf{\large Filippo Sala} \\
\end{minipage}
\hspace{.05\linewidth}
\begin{minipage}{.5\linewidth}
\flushright {\large Supervisor}\\ \vspace{0.2cm}
{\large \bf Prof. Riccardo Barbieri} \\
\end{minipage}
\vspace{3 cm}
\textsc{\Large }\\ \large{November 2013}\\
\end{center}

\newpage
\thispagestyle{empty} \phantom{}
\end{titlepage}

\frontmatter

\chapter*{\centering \begin{normalsize}Abstract\end{normalsize}}
\begin{quotation}
\noindent

After a discussion of the hierarchy problem of the Fermi scale, we study the flavour and Higgs boson(s) phenomenology of new physics close to it, both in general and in the specific cases of Supersymmetry and composite Higgs models.
First, we promote the approximate $U(2)^3$ symmetry exhibited by the quark sector of the Standard Model to be a more fundamental symmetry of Nature, and explore its phenomenological consequences from an effective field theory point of view. 
We also study the embedding of natural Supersymmetry within the $U(2)^3$ framework, 
focusing in particular on the pattern of flavour and CP violation in $B$ decays.
Then we consider the CP-even scalar sector of the next-to-minimal supersymmetric Standard Model. We quantify the impact of current and foreseen measurements of the Higgs boson signal strengths on the physical parameters, deriving analytical relations for this purpose, and we outline a possible overall strategy to search for the other Higgs scalars. Finally we analyze the constraints from flavour and electroweak precison tests on composite Higgs models, for different representations of the composite fermions, and comparing the case of an anarchic flavour structure to models with a $U(3)^3$ and a $U(2)^3$ flavour symmetry.
\end{quotation}
\clearpage

\tableofcontents

\mainmatter

\setcounter{secnumdepth}{-1}
\chapter{Invitation}

The Standard Model of particle physics was proposed more than fourty years ago, and has been crowned last year with the Higgs boson discovery at the LHC. Its predictions have been tested to a remarkable level of accuracy, from flavour and CP violation, to electroweak precision tests (EWPT) and, recently, to the couplings of the Higgs boson. Interestingly, the measured value of its mass makes even possible to extrapolate the Standard Model (SM) up to the Planck scale, where the effects of gravity become important and a new theory is needed, without running into inconsistencies.

While representing an impressive success of our understanding of Nature, the SM leaves some fundamental problems unanswered. On the experimental side, it lacks of a suitable candidate for Dark Matter (and of an explanation of neutrino masses and oscillations) and it does not reproduce the observed matter/antimatter asymmetry of the Universe. On the theoretical side, reasons to be unsatisfied are for example the absence of an explanation of the masses and mixings of quarks and leptons, the lack of a unified description of gauge interactions, and the so-called hierarchy problem of the Fermi scale.
Radiative corrections make in fact its value sensitive to higher energy scales, and pose the question of why it is so small, $\sim 250$ GeV, if compared to the Planck scale, $\sim 10^{19}$ GeV. A reasonable ``natural'' solution to this problem is the presence of some new physics (NP) close to the Fermi scale, that screens its sensitivity to higher energies. The two main candidates for this purpose are Supersymmetry and composite Higgs models. This attitude, which has steered most of the theoretical speculations in the last thirty years or so, has provided the main motivation to expect some NP to show up at experiments, either directly or indirectly.

 The experimental strive of the last decades has been focused to test the SM in a huge amount of independent ways, in the hope to falsify it, even in a single prediction, and get in this way a clue to solve some of its problems. However, this has not happened yet. This has induced some physicists to dedicate their energies in the exploration of other solutions to the hierarchy problem, that do not require the appearence of NP close to the Fermi scale. The fact that it is possible to find solutions to the experimental problems of the SM, without spoiling its consistency up to the Planck scale nor implying deviations from it at foreseen experiments, has further corroborated this theoretical exploration.

However the last word on the validity of naturalness as a guiding principle has not been cast yet.
Near future experiments will provide determinant (but not definitive) pieces of information about this issue.
Eventually, abandoning naturalness as a guiding principle would have far reaching consequences for the way we investigate Nature at a fundamental level. With the only exception of the cosmological constant, naturalness has always provided a good criterion for doing physics. It is in fact intimately tied with the assumption that physics at small energy scales is not substantially influenced by physics at much higher scales. This is the very base of reductionism as an approach to make progresses in fundamental physics. 
Reviewing naturalness might call for a change in this approach. Also, in the absence of physics close to the Fermi scale, it appears more difficult to imagine ways of testing NP theories, necessary for example to solve other SM problems. Currently, it seems that a definitive change of attitude towards the hierarchy problem would somehow lower our expectations about the falsifiability of theories.

For these reasons, it is of great importance both to make further steps in the exploration of other paradigms, as well as to thoroughly probe the Fermi scale and its immediate proximities. Before possibly considering naturalness dead, one would like to have looked for it at the best of its possibilities.
During my three years of PhD studies, I have been mainly focused on the latter program, in the research papers  \cite{Barbieri:2011fc,Barbieri:2012uh,Barbieri:2012bh,Barbieri:2012tu,Barbieri:2013hxa,Barbieri:2013nka}, and along the lines exposed in the next paragraph. They will be the main subject of this thesis.
Recently I have also contributed to the former program in \cite{Buttazzo:2013uya}, where the running of the SM parameters is computed with the highest precision to date, and their values at the high scale are interpreted in terms of some criteria alternative to naturalness. We will mention \cite{Buttazzo:2013uya} only to the extent that it is related to the discussion of the hierarchy problem, in Section~\ref{sec:hierarchy_problem}.

This thesis is organized as follows.
In Chapter~\ref{cha:chapter1} we give a general overview of the Standard Model of particle physics, with
the purposes to fix some conventions and to introduce its problems. Among those, we critically discuss the hierarchy problem and its possible solutions. 
The main body of the exposition is then divided in three parts. The subject of the first one is flavour physics.
In Chapter~\ref{cha:SMCKM} we discuss in more detail the SM description of flavour and CP violation in the quark sector, with emphasis on the relation between data and the Cabibbo Kobayashi Maskawa (CKM) picture. Then, in Chapter~\ref{cha:EFT}, we motivate the introduction of a $U(2)^3$ symmetry in the quark sector, 
and discuss in details its phenomenology. We also attempt an extension to the lepton sector.
The second part of the thesis concentrates instead on Supersymmetry (SUSY). In Chapter~\ref{cha:SUSYintro} we introduce it, and put the focus first on its relation with the hierarchy problem, then on the SUSY flavour and CP problems. We also introduce the next-to-minimal supersymmetric Standard Model (NMSSM) as a most natural theory. The supersymmetric realization of $U(2)^3$ is discussed in Chapter~\ref{cha:U2_SUSY}, where also some implications that go beyond flavour and CP phenomenology are emphasised. In Chapter~\ref{cha:SUSY_Higgs} we explore the phenomenology of the NMSSM CP-even scalar sector, both in light of the measurement of the Higgs signal strenghts, and of direct searches of the new scalar states. For comparison, we carry out the same analysis also for the MSSM case. We finally present an explicit NMSSM model.
In the third and last part we discuss composite Higgs models (CHM). After a brief introduction, in Chapter~\ref{cha:CHM} we analyze how some simple CHM realizations deal with flavour and electroweak precision constraints, and determine those that allow for relatively light composite fermions, and for other distinctive experimental signatures. Detailed summaries and conclusions are reported at the end of Chapters \ref{cha:EFT},\ref{cha:U2_SUSY}, \ref{cha:SUSY_Higgs} and \ref{cha:CHM}.

\setcounter{secnumdepth}{2}

\chapter{Problems of interest and strategy}
\label{cha:chapter1}

\section{The Standard Model}
\label{sec:TheSM}

The present understanding of Nature at a fundamental level can be briefly sketched as follows.
At the shortest distances accessible by current experiments, the observed behaviour of elementary particles can be consistently described in the language of a quantum field theory on a flat space, in what is known as the \textit{Standard Model of particle physics}. The gravitational interaction is instead described by General Relativity, which can be seen as a geometrical theory of spacetime. 
The Standard Model of particle physics plus General Relativity constitute the established knowledge of Nature: their predictions have been confirmed to a high level of accuracy by all experiments to date, with a very few exceptions, some of them we will discuss in the next section. They are both necessary ingredients in the standard cosmological model, known as $\Lambda$CDM, which is today the reference for describing the history of the Universe we live in, back to a few minutes (time of nucleosynthesis) after its origin. 

We focus here on the Standard Model of particle physics, from now on we refer to it as the Standard Model (SM) for brevity\footnote{For a quite recent, compact presentation we refer the reader to \cite{Barbieri:2007gi}.}. A model of elementary particles in the language of a quantum field theory (QFT) is unambiguously defined by 
\begin{itemize}
 \item the gauge symmetry of the Lagrangian, which also fixes the spin-1 field content of the model, the gauge bosons, and their self-interactions,
 \item the eventual breaking pattern of the symmetry,
 \item the fermionic (spin-1/2) and bosonic (spin-0) field content, together with its representation under the above symmetry group, which fixes their interactions with the gauge boson.
\end{itemize}
The recipe for obtaining a model with the use of these ingredients is writing the most general renormalizable Lagrangian invariant under the gauge symmetry at the quantum level.
Starting from this Lagrangian one can derive the Green functions of the theory, and make in this way predictions to be compared with experiments.\\
In this terms, the building blocks defining the Standard Model are:
\begin{enumerate}

 \item \textbf{The Gauge sector}\\
 The symmetry group of the Standard Model is
\begin{equation}
 G_{\text{SM}} = SU(3)_c \times SU(2)_L \times U(1)_Y\,,
\end{equation}
we denote the associated gauge couplings with $g_s$, $g$ and $g_Y$ respectively, and introduce the weak angle by $\theta_w = \arctan{(g_Y/g)}$. The \textit{colour} symmetry $SU(3)_c$ implies the existence of eight massless spin-one fields, the gluons, and governs the so-called ``strong dynamics'' of the Standard Model. Despite its very simple foundations, this dynamics is extremely rich from the phenomenological point of view, and it is in excellent agreement with experiments, the interested reader can find a review in \cite{Beringer:1900zz}.
The \textit{electroweak} (EW) symmetry $SU(2)_L \times U(1)_Y$ calls for four extra massless spin-one fields $W_{1,2,3}^\mu$ and $B^\mu$, but only a combination of them, the photon, is observed in Nature. 

 \item \textbf{The Electroweak Symmetry Breaking (EWSB) sector}\\
  The only known viable 
  way to give mass to the gauge bosons, compatibly with $G_{\text{SM}}$, is via spontaneous symmetry breaking: the vacuum state of the theory does not respect the symmetry which is instead manifest in the Lagrangian. The massless Goldstone bosons associated with the spontaneous breaking are commonly said to be ``eaten'' by the gauge bosons, in what is known in the literature as the Brout-Englert-Higgs (BEH) mechanism. The choice that reproduces the observed vectors masses is the breaking of the electroweak symmetry $SU(2)_L \times U(1)_Y$ down to the electromagnetic one
\begin{equation}
 SU(2)_L \times U(1)_Y \to U(1)_{\text{em}}\, .
\end{equation}
One ends up in this way with a massless gauge boson, the photon (associated with the unbroken symmetry $U(1)_{\text{em}}$), and with two gauge bosons $W$ and $Z$ with a mass fixed in terms of the gauge couplings and of the vacuum expectation value (vev) of the theory $v/\sqrt{2}$. 

In the Standard Model EWSB is realized in a minimal way, by adding an extra complex scalar field
\begin{equation}
\phi \sim (1,2)_{\frac{1}{2}}\, ,
\end{equation}
with a potential whose minimum is invariant under the electromagnetic symmetry but not under the electroweak one. This requirement, via use of the recipe defined at the beginning of this section, univocally fixes the potential to the form
\begin{equation}
 V_h(\phi) = -m^2 |\phi|^2 + \lambda |\phi|^4,
 \label{eq:Higgs_Pot}
\end{equation}
 so that at tree level $v^2 = m^2/\lambda$. Three out of the four real components of $\phi$ are the massless Goldstone bosons associated with the breaking of the symmetry. The fourth one is a neutral scalar field $h$, known as the Standard Model ``Higgs boson'', with a physical mass $m_h^2 = 2 \lambda v^2 = 2 m^2$. The choice $Y_{\phi} = 1/2$ is necessary to allow its interactions with the SM fermionic fields, and the charge convention $Q = T_{3L} + Y$ forces the vev to lie in the lower $SU(2)_L$-component of $\phi$. We stress that the BEH mechanism that gives mass to the gauge bosons and the existence of a Higgs boson are logically independent, in particular the former does not imply the latter. 
One can in fact think of other ways to realize the BEH mechanisms, even without the association of any fluctuation (either elementary or composite field) to $v$. However, since LEP2 they were generically disfavoured by electroweak precision tests (EWPT), which on the contrary provided indirect indications in favour of the existence of a SM-like Higgs boson with a mass $114 < m_h < 175$ GeV \cite{Beringer:1900zz}, the lower bound being the direct $95 \%$ confidence level (C.L.) exclusion \cite{Barate:2003sz}. The issue has been settled in 2012 by the LHC discovery of a Higgs boson \cite{:2012gu,:2012gk} of mass $m_h \simeq 126$ GeV, whose measured couplings are now remarkably close to those of a SM one \cite{CMS,ATLAS,tevatron:2013} (univocally predicted as a function of $m_h$, $v$ and the masses of the SM fields), even if they still leave space for sizeable deviations from this minimal picture.

Note that the potential $V_h$ is invariant under a $SO(4)$ symmetry acting on the four real components of $\phi$. This symmetry is called ``custodial symmetry'' because it is responsible for the tree-level relation
\begin{equation}
\rho \equiv \frac{m_W^2}{m_Z^2 \cos^2\theta_w} = 1\,,
 \label{eq:rho_SM}
\end{equation}
as we will now see. 
It is convenient to discuss it exploiting the fact that $SO(4)$ is isomorphic to an $SU(2)_L \times SU(2)_R$ algebra. After rewriting the Higgs field as $\mathcal{H}~=~(i \sigma_2 \phi^*, \phi)$, this symmetry acts on it like $\mathcal{H} \to e^{i \theta_L^i \sigma_i/2} \mathcal{H} e^{-i \theta_R^i \sigma_i/2}$, where $\theta_{L,R}$ are three-dimensional vectors parametrizing the two independent $SU(2)$ rotations, and $\sigma_i$ are the Pauli matrices.
Note that the action of $SU(2)_L$ coincides with the one of the $SU(2)_L$ gauge group, while the one of $U(1)_Y$ is equivalent to a $e^{-i \theta_R^3 \sigma_3/2}$ acting on the right-hand side of $\mathcal{H}$. Thus the gauging of $SU(2)_L$ respects the full $SO(4)$ symmetry, the one of $U(1)_Y$ breaks it. The vacuum configuration $\langle \mathcal{H} \rangle = v \mathds{1}$ is left invariant by a diagonal $SU(2)_V$ transformation with $\theta_L = \theta_R$, which explains why the ratio of the $W$ and $Z$ boson masses is proportional to $g_Y$ (i.e. why $\rho = 1$).

The Higgs mass, together with the vev $v$ and with the $SU(2)_L$ and $U(1)_Y$ gauge couplings $g$ and $g_Y$, completes the list of the four parameters of the EW SM Lagrangian\footnote{The most precise measurements that are usually considered to define the electroweak sector (i.e. to determine also $v$, $g$ and $g_Y$) are the Fermi constant $G_\mu$ (from the muon lifetime measured at PSI \cite{Webber:2010zf}) the Z mass $M_Z$ measured at LEP2 \cite{ALEPH:2005ab}, and the fine-structure constant $\alpha_{\text{em}}$ (from rubidum recoils \cite{Bouchendira:2010es} and/or the electron anomalous magnetic moment \cite{Hanneke:2008tm}). The determination of $g_s$, the only other fundamental parameter introduced till now, is more involved: it depends on world averages of both lattice simulations and experimental data, the interested reader is referred to \cite{Beringer:1900zz}.}. Its many predictions are in excellent agreement with experimental data, see for example \cite{Baak:2013ppa} for an updated discussion. 
Note 
that $m_h$ can be traded for the Higgs quartic coupling $\lambda$, leaving only one massive parameter, $v \simeq 246$ GeV, that sets the so-called ``Fermi scale'' which is currently being explored by the LHC.

 \item \textbf{The Yukawa sector}\\
 The fermionic field content of the SM consists in three copies (``flavours'') of the following left-handed Weyl fermions
\begin{equation}
q_L \sim (3,2)_{\frac{1}{6}},\; u_R^\dagger \sim (\bar{3},1)_{-\frac{2}{3}},\; d_R^\dagger \sim (\bar{3},1)_{\frac{1}{3}},\; \ell_L \sim (1,2)_{-\frac{1}{2}},\; e_R^\dagger \sim (1,1)_{1}\,,
\label{SMfermions}
\end{equation}
where in brackets we have denoted the field representation under $SU(3)_c$ and $SU(2)_L$ respectively, and in the pedix the $U(1)_Y$ charge (hypercharge). The most general, gauge-invariant, renormalizable terms one can add to the Lagrangian using these fields are the so-called ``Yukawa terms''
\begin{equation}
u_R^\dagger Y_u^\dagger \phi q_L\, +\, d_R^\dagger Y_d^\dagger \phi^c q_L \, + \, e_R^\dagger Y_e^\dagger \phi^c \ell_L\,+ h.c.\,,
\label{SMYuk}
\end{equation}
where $\phi^c_i = \epsilon_{ij} \phi^*_j$ (with $i$, $j$ $SU(2)_L$ indices) and $Y_u$, $Y_d$ and $Y_e$ are arbitrary complex matrices in flavour space. They can be written as $Y_u = L_u  Y_u^{\text{diag}}  R_u$, $Y_d = L_d Y_d^{\text{diag}} R_d$ and $Y_e = L_e Y_e^{\text{diag}} R_e$, where $L_{u,d,e}$ and $R_{u,d,e}$ are unitary matrices. To avoid redundancy in the parameters, five of the six unitary matrices above can be reabsorbed in redefinitions of the fields \eqref{SMfermions}. Then, once the Higgs field takes vev, in terms of the $SU(2)_L$ components of $q_L = (u_L, d_L)$ and $\ell_L = (\nu_L, e_L)$ the interactions \eqref{SMYuk} generate the mass terms
\begin{equation}
\frac{v}{\sqrt{2}} \big( \bar{u}_L V^\dagger Y_u^\text{diag} u_R\,+\, \bar{d}_L Y_d^\text{diag} d_R \,+\, \bar{e}_L Y_e^\text{diag} e_R\,+ \,h.c.\big)\,,
\label{SMmass}
\end{equation}
where we have moved to Dirac notation for the fermions (we will keep it for the rest of the exposition), and we have introduced the so-called Cabibbo Kobayashi Maskawa matrix $V = L_u^\dagger L_d$.
To go to the mass-diagonal basis (i.e. to remove $V$) the field $u_L$ has to be rotated independently of its $SU(2)_L$-partner $d_L$. The only term that feels this rotation in the SM Lagrangian is the charged current interaction of quarks with the $SU(2)_L$ gauge bosons $W^\pm_\mu$, which in the physical basis reads
\begin{equation}
 \mathcal{L}_{\text{ch.c.}} = \frac{g}{\sqrt{2}} \big( \bar{u}_L V \gamma^\mu d_L\, W^+_\mu\, + \, h.c.\big)\,.
 \label{SMChargedCurrent}
 \end{equation}
A discussion of the physical parameters added by the Yukawa sector to the SM can be useful here. Starting from the definitions $Y_X = L_X  Y_X^{\text{diag}}  R_X$, with $X=u,d,e$, the $3 \times 3$ Yukawa matrices can be viewed as containing 9 real and 9 complex parameters each one. Concerning the quarks, the three field redefinitions we performed allow to remove $3 \times 3$ real and $(3 \times 6) - 1$ complex parameters, where the $- 1$ accounts for an over-all $U(1)_\text{B}$, acting on all the quarks irrespectevely of their flavour, which therefore does not affect the parameter counting we are performing. One is left with 6 real ``yukawas'', each one fixed by the relative quark mass, and 3 real and 1 complex parameters appearing in the CKM matrix.
Concerning the leptons, the analogous two field redefinitions allow to remove $2 \times 3$ real and $(2 \times 6) - 3$ complex parameters, where the $- 3$ accounts for three over-all $U(1)_{e,\mu,\tau}$, acting independenlty on all the leptons of a given flavour, which are ineffective here\footnote{These $U(1)_{e,\mu,\tau}$, together with $U(1)_\text{B}$, are ``accidental'' symmetries of the Standard Model, in the sense that they are obtained without having been imposed. They are associated respectively with individual lepton number and baryon number conservation.}. What remains are just three real yukawas, which are fixed in terms of the lepton masses.
 
It is important to notice that $V$ encodes all the flavour and CP violation (FV and CPV) in the Standard Model, which is then very predictive for this kind of observables, and which have shown up to now a good agreement with data. Due to the importance of this statement for the subjects discussed in the first part of this thesis, we will explore it thoroughly in Chapter \ref{cha:SMCKM}.

\end{enumerate}
The Standard Model makes testable predictions, functions of the 18 ``input'' parameters, that have been confirmed by all high energy experiments to date, very often with an impressive level of accuracy.
Despite that, there are many reasons not to be satisfied with it, and why it cannot be the end of the story for particle physics. They will be the subject of the next section.\\

\textit{A digression: neutrinos}\\
The Standard Model as we have built it predicts the neutrinos $\nu_\ell$ to be massles and the lepton sector to be flavour and CP conserving. Both these statements are contradicted by experiments: at least two neutrinos have mass, and they are observed to ``oscillate'' in flavour space, i.e. the mass eigenstates and the weak interaction eigenstates do not coincide. One could account for this by introducing three copies of a new Weyl fermion 
\begin{equation}
N \sim (1,1)_0\, ,
\end{equation}
and applying the recipe defined at the beginning of this section to obtain all the new terms in the Lagrangian of the model
. This procedure introduces new parameters, some of which have not been measured yet (mean value of the left-handed neutrino masses, CP violating phases, \dots), making this a not-(yet)-established sector of the current model of particle physics. Since we will not be interested in neutrinos in this thesis, we will not discuss these issues further, and refer the reader to \cite{Barbieri:2007gi} for an introduction and to \cite{Strumia:2006db} for a comprehensive review.

\section{The hierarchy problem as a guideline}
\label{sec:hierarchy_problem}

In this section we give an overview of the problems of the Standard Model of particle physics. Since the hierarchy problem has provided a guide for most of the theoretical speculations in the last thirty years, we discuss it in more details with respect to the other ones.
In particular we try to summarize the possible attitudes towards it. The choice of one with respect to the others have strong implications for the possible solutions to many of the other SM problems. 


\subsection{Problems of the Standard Model}

From an \textit{experimental} point of view, the Standard Model does not account for:
\begin{itemize}
 \item Dark matter\\
 Many independent astrophysical and cosmological observations imply the existence of a new kind of matter, which is not accounted for by any SM particle.
 The ``standard'' mechanism for the production of the new particles in the early universe requires them to be weakly interacting, and with a mass in the TeV range, thus setting a possible energy scale at which NP could manifest itself. However, other viable production mechanisms exist, which do not point clearly to a mass scale for the Dark Matter (DM) particle(s).
 \item Baryon asymmetry\\
 This is the only established proof against the CKM picture of CP violation, in the sense that it implies new sources of CPV to exist beyond the CKM phase $\delta_{\text{CKM}}$ (CP violation is a necessary condition for baryogenesis to be possible). This is because $\delta_{\text{CKM}}$ alone, combined with the Universe history, predicts the present baryon number density to be many orders of magnitude below the observed value. At which scale the NP responsible for this new source of CPV should show up? The answer to this question depends on the model of baryogenesis, the relevant scale can easily be much higher than a TeV (e.g. in models of baryogenesis via leptogenesis), even if models where this is not the case do exist (e.g. electroweak baryogenesis).

\end{itemize}

From a \textit{theoretical} point of view, it is \textit{highly desirable} to address the following issues:
\begin{itemize}
\item Standard Model flavour puzzle\\
 The SM parameters of the Yukawa sector are hierarchical in magnitude, and the majority of them is very small:
 \begin{align}
  &y_t \sim 1\,, \qquad & y_c \sim 10^{-2}\,, \qquad &  \quad y_u \sim 10^{-5}\,,\\
  &y_b \sim 10^{-2}\,,  & y_s \sim 10^{-3}\,, \qquad &  \quad y_d \sim 10^{-4}\,,\\
  &y_\tau \sim 10^{-2}\,, & y_\mu \sim 10^{-3}\,, \qquad & \quad y_e \sim 10^{-6}\,,\\
  &|V_{us}| \sim 0.2\,,  & |V_{cb}| \sim 0.04\,, \qquad & \quad |V_{ub}| \sim 0.004\,,
 \end{align}
where we have shown the order of magnitude of the yukawas of the SM fermionic fields, and chose three entries of the CKM matrix as representative of the orders of magnitude of the three real independent parameters contained in $V$. 
In the SM the understanding of the flavour sector is merely parametrical, far from the elegance of the gauge principle that allows to describe the strong and electromagnetic interactions. The number of generations of matter is also chosen by hand.
Why three? Why does the flavour parameters display such a clear hierarchy? Are there any underlying symmetries or/and dynamical features that can provide an answer to these questions?
 \item Strong CP problem\\
 The recipe we defined for building the Lagrangian of a given model implies the presence of an additional CP violating term in the SM lagrangian, proportional to the adimensional parameter $\theta_{\text{QCD}}$. This term induces an electric dipole moment of the neutron which, for $\theta_{\text{QCD}} \sim O(1)$, exceed the experimental bound by roughly ten orders of magnitude: what forces $\theta_{\text{QCD}}$ to be so small?
 \item Charge quantization\\
 What forces the fermion $Y$-charges to take the specific values listed in \eqref{SMfermions}?
 One could argue that the requirement of anomaly cancellation (i.e. gauge invariance at the quantum level) fixes them to their values. A reason to be nonetheless unsatisfied is that, in general, adding fields introduces some arbitrariness in their determination, and it is hard to imagine a solution to all the above problems (plus neutrino masses and oscillations) without the addition of new fields. Also, \eqref{SMfermions} is not the minimal matter content compatible with anomaly cancellation.
 \item Unification of forces\\
 It would be extremely appealing to reach a unified description of all the forces existing in Nature. Gravity and the Higgs interactions left aside, this could amount in identifying a unique gauge group in which $G_{\text{SM}}$ could be embedded (that could at the same time solve the charge quantization problem). The running of the three SM gauge couplings in the absence of NP give reasons to hope that this could actually be the case, their values approximately crossing at energies of $10^{13\div16}$ GeV when $g_Y$ is properly normalized for the embedding.
 \item The hierarchy problem(s)\footnote{The identification of a father for this idea is not an immediate task. The first clear formulation of it, together with the proposal of the ``naturalness criterion'', was put forward by 't Hooft in 1979\cite{'tHooft:1979bh}.
 However the issue was known also before, at least in the context of Unified theories \cite{Georgi:1974yf}, and in \cite{Susskind:1978ms} Leonard Susskind introduced Technicolour to solve the hierarchy problem, attributing its formulation to Kenneth G. Wilson. Indeed in a 1970 paper \cite{Wilson:1970ag} he mentioned the absence of scalar particle in Nature relating it with the absence of symmetries to protect them.
 Concerning instead recent literature, for a very clear exposition of the hierarchy problem (together with its supersymmetric solution) see \cite{Lodone:2012kp}, for an interesting analogy in the way of posing it see \cite{Rychkov:2011br}.}
 
 The issue stems from the fact that any dimensionful parameter in the Lagrangian \textit{which is not protected by a symmetry} receives a contribution from radiative corrections, which is proportional to the highest energy scale $\Lambda_{\text{NP}}$ felt by that parameter, $\Lambda_{\text{NP}}$ being either the mass of a particle or the typical energy of an interaction. Thus one would expect the measured value of this parameter to be of the order of this high scale.
 
 In the Standard Model this translates in the statement that the Higgs boson mass $m_h$ is not stable under radiative corrections, \textit{if $h$ couples to some physics at much higher energies}. For example, a generic coupling with a fermion of the form $y\, h \bar{f}_L f_R + \text{h.c.}$ leads $m_h$ to start running at energies $\mu$ higher than the fermion mass $m_f$ like
 \begin{equation}
  \frac{d m_h^2(\mu)}{d \log \mu} = -\frac{3 y^2}{4 \pi^2} m_f^2\,,
 \end{equation}
 so that one would expect the Higgs mass to be roughly of the same order of the fermion mass, \textit{barring a small coupling $y$}, as well as accidental cancellations between this and the other contributions (that are proportional e.g. to $m_h^2$ or to the mass of some other particles $h$ couples to). The example is useful in raising the question: how precise these cancellations must be, in order to have an Higgs mass of 125 GeV? In other words, how much \textit{fine-tuning} is required?
 If we define the fine-tuning 
 $\Delta$ as \cite{Barbieri:1987fn}
 \begin{equation}
 \Delta = \frac{d \log m_h^2 (m_h^2)}{d \log m_h^2 (\Lambda_{\text{NP}}^2)} \approx \frac{\Lambda_{\text{NP}}^2}{m_h^2}\,,
  \label{FineTuning}
 \end{equation}
then for a generic new physics at the Planck scale (after all the Higgs boson feels the gravitational interaction), the answer is that different contributions must cancel to the precision of a part over $\Delta \sim 10^{34}$. A rather impressive conspiracy!

This kind of fine-tuning is not to be confused with the fact of being uneasy with very small numbers in the Lagrangian, independently of the radiative stability of those parameters, in the spirit of Dirac \cite{Dirac:1937ti}. A small parameter which is radiatively stable, like e.g. the light generations yukawas, is fine-tuned in the sense of Dirac, but does not constitute a hierarchy problem. This statement will be made even clearer by the following discussion.

Let us now be more specific about the fact that setting the parameter $m$ to zero in the Higgs potential \ref{eq:Higgs_Pot} does not add symmetry to the theory. Let us take again the Yukawa couplings as an example: for any of them that we set to zero the SM Lagrangian gains a chiral symmetry (a $U(1)_L\times U(1)_R$ for each of the Dirac fields in \eqref{SMmass}).
Since this symmetry is preserved at the perturbative level by quantum corrections \footnote{Non-perturbatively this symmetry is anomalous, but this has no impact on our argument. 
}, then the running of each yukawa must be proportional to the yukawa itself, because if we set it to zero it cannot be generated by quantum corrections. If they are small at a scale, the running keeps them small at other scales.
Parameters with this property are often referred to be ``technically natural''.
On the contrary $m$ will be radiatively generated by any other mass scale in the Lagrangian, so the general expectation for its value is to be of the order of that scale.
In this respect note that also setting the Higgs quartic coupling $\lambda$ to zero does not add any symmetry, and in fact the running of $\lambda$ is proportional to all the other fundamental parameters of the SM (but this does not pose any fine-tuning problem, since $\lambda$ is adimensional). 
A couple of remarks to avoid possible sources of confusion are in order: the fact that, if the SM only is considered, the running of $m$ is proportional to $m$ itself is just a consequence of the fact that it is the only dimensionful parameter of the theory. Moreover, the fact that setting $m$ to zero restores scale invariance is in general not a good argument against the hierarchy problem, since to define the problem at all one needs another higher scale, that would generically badly break scale invariance.

We conclude by mentioning the other hierarchy problem of Nature, the one of the Cosmological constant $\Lambda_{\text{CC}}$. If one adds gravity to the picture, then a constant term in the Higgs potential \eqref{eq:Higgs_Pot} cannot be reabsorbed anymore in a redefinition of $V$, and one would expect it to be of the order of the Planck scale to the fourth power. If this constant term is interpreted as the Dark Energy of the  Standard Model of cosmology (i.e. a fluid with negative pressure, responsible for the observed accelerated expansion of the universe), then the fine-tuning needed to reproduce the observation of an expanding universe would be of the order of roughly one part over $10^{120}$ (or $10^{60}$ if one relates it to SUSY breaking). We will come back to this later in this section.

\end{itemize}


\subsection{Solutions I: New physics close to the Fermi scale}
\label{sec:NP_closeFermiScale}
 The ``natural'' way to solve the hierarchy problem is that of adding new physics at a scale $\Lambda_{\text{NP}}$ very close to the Fermi scale, which solves the issue for energies above $\Lambda_{\text{NP}}$ no matter what NP is, for example thanks to a symmetry or to a change of regime of the theory (e.g. the theory becomes strongly coupled at a scale, above which the ``composite'' Higgs boson degree of freedom is substituted by more fundamental objects). Weak scale Supersymmetry and composite Higgs models are examples of a solution respectively of the first and second kind. We will discuss them in some details in the following Chapters (also the ``classicalization'' of the Higgs field \cite{Dvali:2010jz} can be ascribed to the second kind of solutions).
 Another possibility to solve the hierarchy problem in a natural way consists in achieving an exponentional suppression of the Planck scale, in such a way that the hierarchy is transferred to the exponent and thus significantly lowered. Examples of this kind are provided by Large Extra Dimensions\cite{ArkaniHamed:1998rs,Antoniadis:1998ig}, and Warped Extra Dimensions\cite{Randall:1999ee,Randall:1999vf}, where the last have been shown to be equivalent to a four dimensional strongly coupled conformal field theory.
 
 The requirement of naturalness is not a strict theoretical necessity, but it is more than a very reasonable solution to the hierarchy problem. As pointed out by 't Hooft himself in \cite{'tHooft:1979bh}, this concept is related to the assumption that to describe physics at a certain scale, it is not necessary to know much details of physics at shorter distances. This attitude has driven progress in physics since centuries. For example to formulate predictive theories for the motion of macroscopic objects, it is not necessary to know the behaviour of their elementary constituents: ballistics and Newton's theory of gravitation were formulated well before the advent of quantum mechanics. In other words, the naturalness criterion is strictly tied to reductionism. We cannot abandon the former without rethinking the latter.
 In a more modern language \cite{Giudice:2008bi}, abandoning naturalness would challenge the effective field theory approach used to describe the low energy effects of a high energy unknown theory, that till nowadays has worked astonishingly well. Reductionism is not at all a dogma though. In Science there are several phenomena that are not understood by reducing a system to its elementary constituents, and rather require a treatment of it as a whole. Examples of this kind are self-organized criticality\cite{PhysRevLett.59.381} and dynamical evolution of complex networks\cite{Dorogovtsev2008}. We will come back to the possible relation of the first one with the hierarchy problem later in this section.
 
 With the run at 8 TeV of LHC now concluded, no direct evidence for NP has emerged from any experiment.
 This is pushing the scale $\Lambda_{\text{NP}}$ back into fine tuned territories, generating what is known as the ``little hierarchy problem'': current bounds from direct NP searches are now imposing a fine-tuning of the order of a part over $\Delta \sim 10^{1 \div 2}$. This of course represents a much more acceptable amount with respect to $10^{34}$, and Nature already gives us examples of fine-tunings of similar order.

 Whatever the specific natural solution to the hierarchy problem, the lack of direct evidences for it is not the only (high) price to pay. The fact of having new physics so close to the Fermi scale makes it immediate to raise another question: why are not we seeing this NP in some indirect measurements? 
 In fact, this is another acute challenge to model building of natural theories.
 
 \subsection{Precision constraints to Solutions I}
 To attack this issue, it is useful to rephrase it in a more precise, yet general language.\\
 The basic success of the SM predictions in precision observables, in light of the presence of new physics at some high scale $\Lambda$, can be summarized as follows. If one describes possible deviations from the above predictions by a phenomenological effective Lagrangian of the form
 \begin{equation}
  \Delta \mathcal{L} = \sum_{i} \frac{c_i}{\Lambda_i^2} \mathcal{O}_i\,+ \text{h.c.}\,,
  \label{eq:L_precision}
 \end{equation}
 where $\mathcal{O}_i$ are generic gauge invariant operators of dimension six, obtained by integrating out the new degrees of freedom appearing above the scale $\Lambda_i$, then one can find lower limits on the scales $\Lambda_i$, depending on the specific operator under consideration. Unless they introduce qualitative new effects, operators of dimension higher than six are less important, because they are suppressed by higher powers of the NP scale. Assuming the adimensional coefficients $c_i$ to be of order one, which is a reasonable assumption for a generic new physics, these limits can even reach the level of thousands of TeV\footnote{In this sense, precision physics experiments indirectely probe energy scales that are much higher than those explored by the LHC.}. How can one reconcile this with the expectation of NP at much lower scales? Below we give an overview of the challenges posed by three different kind of indirect measurements: Higgs couplings, electroweak precision tests and flavour.

 \subsubsection{Higgs couplings}
 Any new physics that contributes to the Higgs mass will unavoidably contribute to its couplings. Qualitatively the bigger the contribution to the former, the bigger the one to the latters. In other words the more a theory is natural, the more we expect deviations in the Higgs couplings. A proper quantification of this statement depends of course on the specific theory. Generically, given a certain amount of fine-tuning, the deviations in the Higgs couplings will be smaller in theories where all these modifications appear at the loop level (like some realizations of Supersymmetry\footnote{A first quantitative analysis of this case was carried out in\cite{Arvanitaki:2011ck}.}) than in those where this is not the case (like composite Higgs models or other Supersymmetric cases).
 
 Among indirect measurements, these are the ones more directly related to naturalness, and are a way of testing it that was unavailable just one year ago. However, given the current experimental precision, the challenges they pose to model building are usually not tougher than the ones posed by direct searches. This statement is definitely worth some specifications: we will provide them when discussing composite Higgs models, and carry out a deeper study for the case of Supersymmetry.

 \subsubsection{Electroweak precision tests}
 Electroweak precision tests is a generic name for a number of measurements aimed at testing the validity of the SM picture of electroweak symmetry breaking at the quantum level. 
 A key role is played by several observables measured by LEP at the $Z$ peak, due to their precision. Recently they attracted attention since the two-loop SM prediction for the $Z$ partial width into bottom quarks $R_b$\cite{Freitas:2012sy} resulted in a disagreement with the LEP measurement by $2.4$ standard deviations \cite{Baak:2012kk}. Together with the long standing $2.5 \sigma$ deviation in the forward-backward asymmetry of the bottom quark at the $Z$ pole, $A_{FB}^b$, they consitute the more acute deviations in the SM fit of EW data \cite{Baak:2013ppa}. Concerning new physics, if new fields with different $SU(2)_L$ quantum numbers mix with the SM ones after EWSB, they will leave an imprint in such observables, as we will explore in Chapter \ref{cha:CHM} in the case of composite Higgs models.
 Many other measurements can be classified under the name of EWPT, for example the ratio between the coupling of the $W$ boson to quarks and leptons, the $W$ mass and the weak angle $\theta_w$. The last two are relevant e.g. in light of the SM tree-level relation $\rho = 1$\eqref{eq:rho_SM}.
 
 A convenient and standard way to parametrize NP effects in EW precision observables is via the so-called ``oblique parameters'' introduced in \cite{Peskin:1990zt,Altarelli:1990zd,Altarelli:1991fk,Peskin:1991sw} (see also the more recent \cite{Barbieri:2004qk}). Let us write the vacuum polarization amplitudes of the EW gauge bosons like
\begin{equation}
 \mathcal{L}_{\text{vac-pol}} = -\frac{1}{2} W_3 \Pi_{33}(q^2) W_3 -\frac{1}{2} B \Pi_{00}(q^2) B - W_3 \Pi_{30}(q^2) B - W^+ \Pi_{WW}(q^2) W^- \,, 
 \label{eq:SandT_def}
\end{equation}
where Lorentz indices are contracted via $q_\mu q_\nu/q^2 - g_{\mu\nu}$, and let us expand them for small momenta up to $O(q^2)$. One ends in this way with 8 independent quantities $\Pi(0)$ and $\Pi^\prime(0)$, three of which are fixed in terms of $v$, $g$ and $g_Y$. Then other two relations are necessary to have a massless photon, and one is left with 3 independent parameters, that are predicted in the SM. Two of them are defined as
\begin{equation}
\hat{S} = \frac{g}{g_Y} \Pi_{30}^\prime (0) = \frac{\alpha_{\text{em}}}{4 \sin^2 \theta_w}\,S, \qquad \hat{T} = \frac{\Pi_{33}(0)-\Pi_{WW}(0)}{m_W^2} = \alpha_{\text em} T\,,
 \label{eq:ST_vacuumAmps}
\end{equation}
the third one, $\hat{U} = \Pi^\prime_{33}(0)-\Pi^\prime_{WW}(0)$, is in general expected to be suppressed 
 with respect to $\hat{T}$. From the above definitions the relation $\hat{T} = \rho -1$ follows.
 In terms of an effective operators description, the leading effects are encoded in the effective Lagrangian
 \begin{equation}
 \mathcal{L}_H = \frac{1}{v^2}\left[ c_H \left| \phi^\dagger D_\mu \phi \right|^2\, +\, \frac{c_{WB}}{g\, g_Y}\, \left( \phi^\dagger \sigma^a \phi\right) W_{\mu\nu}^a B_{\mu\nu}\right]
  \label{eq:ST_effOperators}
 \end{equation}
and correspond to $\hat{S} = 2 c_{WB}/\tan\theta_w$ and $\hat{T} = -c_H$, to be compared with the constraints resulting from a global fit to EW precision observables \cite{Baak:2013ppa} 
\begin{equation}
 S = \frac{4 \sin^2 \theta_w}{\alpha_{\text{em}}} \,\hat{S} = 0.03 \pm 0.10\,, \quad \qquad T = \frac{1}{\alpha_{\text{em}}}\, \hat{T}= 0.05 \pm 0.12\,.
 \label{eq:ST_fit_constraints}
\end{equation}
Note that, while $S$ preserves custodial symmetry (if $g_Y = 0$), $T$ breaks it. In fact the largest one-loop contribution to $\rho$ in the SM is proportional to the top Yukawa coupling, that is the main source of $SO(4)$ breaking other than $g_Y$ (more precisely, its difference with the bottom one)
. Analogously, any NP respecting or weakly breaking the custodial symmetry will yield small effects in the $T$ parameter. We will show this explicitly in the following of this thesis, especially for composite Higgs models.
As we will discuss below, reasoning in terms of symmetries can help also for the constraints posed by flavour and CP violating observables.

\subsubsection{The new physics flavour puzzle}
\label{subsec:NP_flavourpuzzle}
The basic success of the Cabibbo Kobayashi Maskawa (CKM) picture of flavour and CP violations is often summarized by observing that lower limits on the scales $\Lambda_i$ in \eqref{eq:L_precision} are in many cases of the order of $10^3 \div 10^4$ TeV, and reach $10^5$ TeV in the case of the contribution of the operator $\mathcal{O}_{\text{LR}} = (\bar{s}_L d_R)(\bar{s}_R d_L)$ to $\epsilon_K$\cite{Isidori:2010kg}. 
This implies that, \textit{if some new physics appears at a scale $\Lambda$ below $\sim 10^4$} TeV, then the flavour and CP structure of the NP theory has to be highly non trivial. The reason why we put emphasis on this is that even if the hierarchy problem is not cured by some NP close to the Fermi scale, any other kind of NP appearing below that range cannot ignore this issue.
 The link with natural theories can actually be made stricter by noting that the dominant corrections to the Higgs mass come from NP coupled to the top quark, since $y_t$ is the largest of the Yukawa couplings. So a low fine tuning strictly requires just this NP to lie not far from the Fermi Scale, while the one related with the first two generations could become relevant at higher energies. This is definitely a non-trivial flavour structure, and an acute challenge for model building of natural theories.
 On the contrary the physics related to the solution of other SM problems (e.g. Dark Matter), could well be independent of the flavour structure (e.g. the Neutralino in SUSY).

A possibility to address the NP flavour puzzle is the requirement for new physics to display some feature, so that the effective Lagrangian
\begin{equation}
 \Delta \mathcal{L} = \sum_i \frac{c_i}{\Lambda^2} \xi_i \,\mathcal{O}_i\,+ \text{h.c.}\,,
\label{FS_LNPsimm}
\end{equation}
where $\xi_i$ are small parameters controlled by that feature, is in agreement with all current data for coefficients $c_i$ of $O(1)$. With $\Lambda$ sufficiently close to the Fermi scale, this might leave room for new observable effects. Such effects would indeed be very welcome in most extensions of the SM in the EWSB sector and, if observed, might help to shed light on a possible theory of flavour.

What can the feature responsible for the smallness of the $\xi_i$'s be?
The first example we give is that of a flavour symmetry.
The most popular attempt in this direction is the so called Minimal Flavour Violation paradigm \cite{Chivukula:1987py,Hall:1990ac,D'Ambrosio:2002ex}: the Yukawa couplings are promoted to \textit{spurions} transforming as $Y_u \sim (3,\bar{3},1)$ and $Y_d \sim (3,1,\bar{3})$ under $U(3)^3 = U(3)_q \times U(3)_u \times U(3)_d$, so that the SM is formally invariant under this symmetry. Then also NP effects are assumed to be formally invariant via the only use of the spurions $Y_{u,d}$. In this way one obtains parameters $\xi_i$ equal to some power of the CKM matrix elements (depending on the specific operator $\mathcal{O}_i$), in such a way that a scale $\Lambda_{\text{NP}}\sim 10$ TeV is in agreement with all experimental data.
 
Other possibilities that do not invoke any flavour symmetry do exist. One that has attracted a lot of attention recently is ``partial compositeness'' in composite Higgs models\cite{Kaplan:1991dc,Contino:2006nn}. This dynamical feature can be just seen as a rescaling of the fields wave functions \cite{Davidson:2007si}, composite Higgs and extra dimensional models being popular explicit examples that realize it. We will see it in action in the Chapter devoted to composite Higgs models. While addressing also the SM flavour puzzle, generically this mechanism does not give enough suppression of some FCNC processes. As a relevant example, in composite Higgs models with partial compositeness, even neglecting the severe $\epsilon_K$ bound one needs fermions heavier than a TeV (implying a fine tuning at the per-mille level)
, unless a flavour symmetry in the strong sector is introduced\cite{Barbieri:2012tu}.

 Till now we did not mention leptons. Flavour and CP violation processes in the charged lepton sector are extremely powerful in constraining NP scenarios, thanks to null searches for $\mu \to e \gamma$, $\mu \to e$ conversion in nuclei etc.. For example models of partial compositeness would need resonances above $\sim 100$ TeV to be in agreement with data \cite{KerenZur:2012fr}, making some other mechanism, like e.g. a flavour symmetry, necessary if one wishes to describe leptons and quarks in a symmetric way.

\subsection{Solutions II: New physics far from the Fermi scale}
 Could one satisfy all the direct and precision bounds on new physics by just abandoning the view that this NP has to appear close to the Fermi scale? What is the price to pay for this?
 First of all, another paradigm to attack the hierarchy problem of the Fermi scale is needed. Within this other paradigm, one can then ask how it is possible, or if it is possible at all, to solve the other problems of the SM. All this keeping in mind the other, likely profound consequences this shift might imply.
 
 Currently, the ``established'' alternative to the naturalness criterion is the introduction of the concept of multiverse. Very recently, a third attitude started to gain more attention. We will summarize both in the following.\footnote{We stress here that they do not require NP to be far from the Fermi scale, but just allow to potentially solve the hierarchy problem without NP close by.}

 \subsubsection{Multiverse}
 It could be that the search to find a deeper reason for the values taken by some of the SM parameters is just a red herring. Kepler used platonic solids to explain the number of planets in the solar system and the relative distances of their orbits. Now we know that there is no deep physical message in these values: rather they are just an accident, arising from the fact that the Universe contains a large number of possible accidents where to choose from.
 To extend this view to a fundamental theory if Nature, one has to admit the existence of many Universes (i.e. a ``Multiverse'', see e.g. \cite{Wilczek:2013lra} for a recent general discussion with references), each one with different values of some of the ``fundamental'' SM parameters, like the Fermi scale. 
 String theory and inflation can give a theoretical motivation for the existence, in some sense, of such a Multiverse.
 To gain appeal, such a proposal needs a mechanism to select a specific Universe among the many possible ones, in particular to explain the value of the Fermi scale that sets the hierarchy problem.
\begin{itemize}
 \item Anthropic selection\\
 This mechanism, somewhere referred to as ``tipicality'', is based on the observation that the values of some of the parameters we observe are just the ones that allow for our existence as observers at all. This line of reasoning has been proposed in 1987 by Weinberg \cite{Weinberg:1987dv} to predict a value for the vacuum energy, i.e. the cosmological constant.
 The argument stems from the observation that with a small change in its value, in any direction, galaxy formation would have been too slow/fast for planets to form at all. The proposal is then that the cosmological constant should have the maximum value compatible with the existence of observers.
 When a decade later the universe was actually measured to have an accelerated expansion\cite{Riess:1998cb,Perlmutter:1998np} this proposal gained much more attention: the value of $\Lambda_{\text{CC}}$ predicted by Weinberg was just a factor of $10\div100$ larger than the measured one.
 Moreover, it is relevant that no natural solution is known for this hierarchy problem. In this sense the discussion of natural solutions we carried out above implicitely assumed the following specification about the cosmological constant: ``Quantum gravity is not understood anyhow so we exclude it from our naturalness requirements'', to put it with the words 't Hooft himself used in his seminal paper \cite{'tHooft:1979bh}.
 
 It is not yet clear whether an anthropic argument can work also for the hierarchy problem of the Fermi scale, the reason (maybe a bit simplistic) being that it is difficult, if not impossible, to unambiguosly identify the conditions for the existence of an observer, in terms of the SM parameters. For example it has been shown in \cite{Agrawal:1997gf} that an increase of $v$ by a factor of five would prevent atom formation. However in this study $v$ only is let to vary, keeping the other parameters fixed, and thus ignoring possible ``flat directions'' for the anthropic criterion in the SM parameter space (note that also the anthropic explanation of the cosmological constant is not strictly free from flat directions).
 Moreover, such a line of reasoning cannot help explaining the value of some other SM parameters, that nowadays appear completely unrelated to the existence of life, like the mixing angles of neutrinos and heavy quarks or the $\theta$ giving rise to the strong CP problem. Note also that the eventual discovery of another light scalar particle, with little role in the EWSB mechanism, would appear much harder to be explained with an anthropic argument. In light of the observations of this last paragraph, it is interesting to explore other selection mechanisms on the multiverse.

 \item Non-anthropic selection\\
 The measured value of the Higgs mass makes the SM a consistent theory up to the Planck scale. None of its couplings become non-perturbative before that scale, nor the Higgs potential develops an instability such that the universe would have already decayed. The Universe would be metastable, but its lifetime would be way longer than any meaningful astrophysical scale. Note also that for each of the experimental problems of the SM one can imagine solutions that either involve physics well above the instability scale (i.e the one at which $\lambda$ becomes negative), or do not significantly modify the shape of the Higgs potential.
 This gives the possibility to extrapolate the SM couplings up to the Planck scale and try to get some physical message, as we did in \cite{Buttazzo:2013uya}.
 
 First of all it is interesting to note how some of them live close to the border of a transition between different phases.
This is displayed in Fig. \ref{fig:criticality} for the Planck scale values of the top Yukawa coupling $y_t$ (left) and Higgs mass parameter $m$ (right) versus the Higgs quartic coupling $\lambda$, where in the left hand figure we show a broad range for the parameters as allowed by perturbativity. The messages are that $y_t$ and $\lambda$ choose to live at the bottom of the metastability funnel, very close to the instability region, and the Higgs potential parameters in the vicinity of the $\lambda = m = 0$ tri-critical point.
 \begin{figure}[t]
  \begin{center}
   \includegraphics[width=0.46\textwidth]{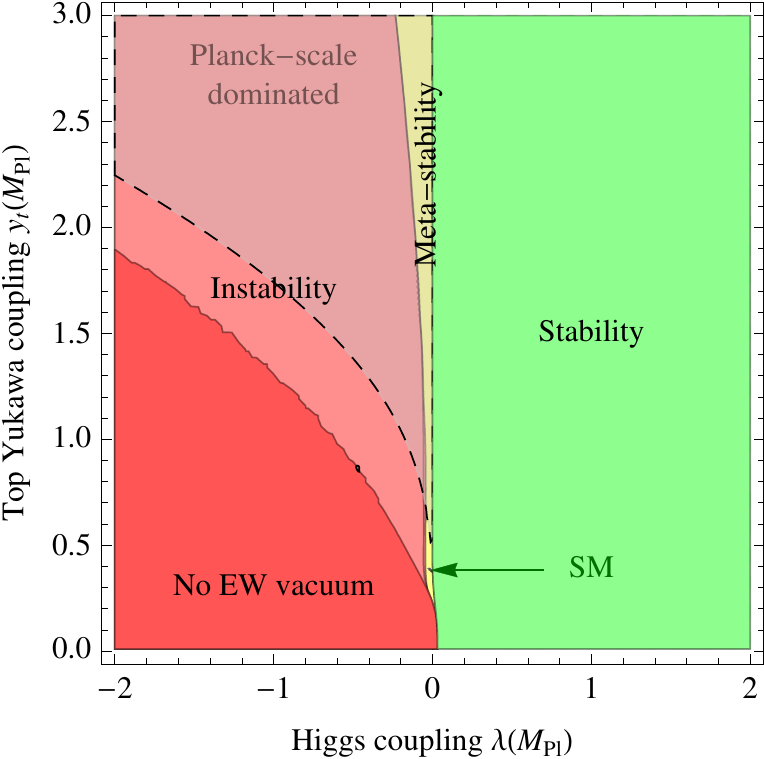}\hfill
   \includegraphics[width=0.50\textwidth]{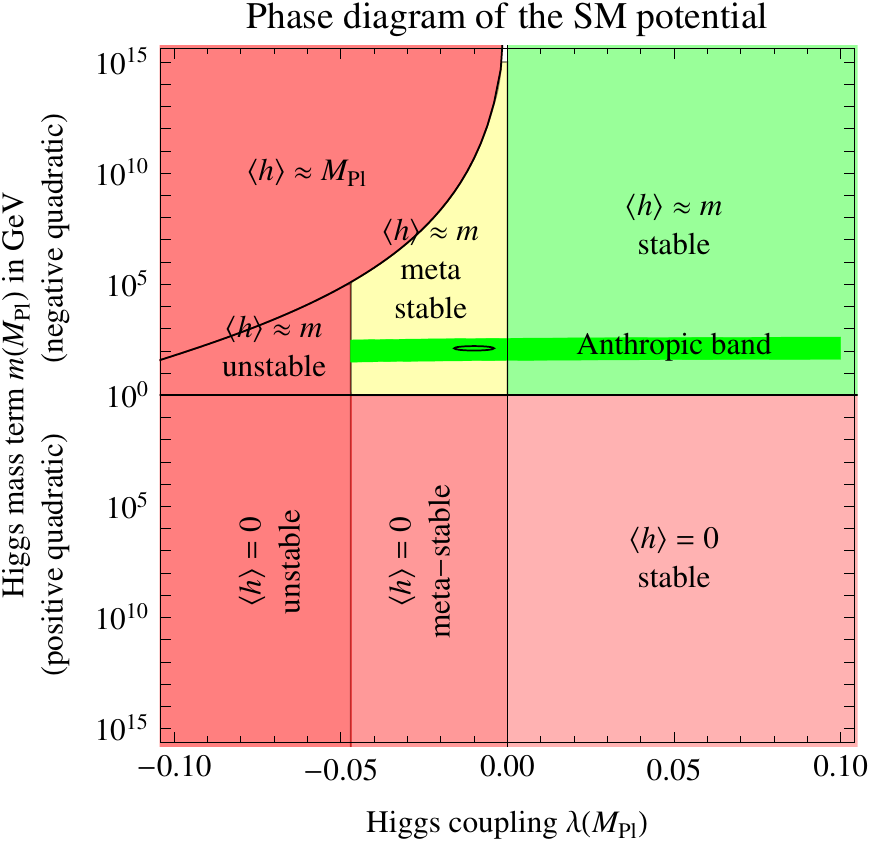}
  \end{center}
 \caption{Phase diagram of the SM in terms of the Planck scale values of $y_t$ and $\lambda$ (left) and of the Higgs potential parameters $m$ and $\lambda$ (right). Left: ``No EW vacuum'' corresponds to a negative $\lambda$ at the weak scale, ``Planck scale dominated'' to the case where the instability scale is larger than the Planck scale. Right: on the vertical axis we plot $|m|$ in the case of negative (above) and positive (below, $v = 0$) Higgs quadratic term; the darker green band shows how the anthropic arguments of \cite{Agrawal:1997gf} bound the Higgs potential parameters.}
 \label{fig:criticality}
 \end{figure}
 As remarked in \cite{Giudice:2006sn} and evident from Figure \ref{fig:criticality} right, also Higgs naturalness can be viewed as a problem of near-criticality between two phases (i.e. why is the Higgs bilinear $m$ carefully selected just to place our universe at the edge between the broken and unbroken EW phases?). This leads to the speculation that, within the multiverse, critical points are attractors. If this vision is correct, the probability density in the multiverse is peaked around the boundaries between different phases, and generic universes are likely to live near critical lines.
 Then, near-criticality would be the result of probability distributions in the multiverse, and would not necessarily follow from anthropic considerations\footnote{Anthropic considerations could re-enter in explaining why some points are attractors. The two concepts are not mutually exclusive.}. In this picture, the Higgs parameters found in our universe are not at all special. On the contrary, they correspond to the most likely occurrence in the multiverse.
 
 We conclude this discussion with the remark that near-criticality is not the only proposed alternative to anthropic selection of parameters that scan on a multiverse: for example one can find ways to statistically prefer particular ranges of the values of $\lambda$, $y_t$ and the gauge couplings at some high scale, see e.g. \cite{Buttazzo:2013uya}.
 
\end{itemize}

\subsubsection{Further steps}
 
 When discussing natural theories, we implicitely assumed something else would cure the hierarchy problem of the cosmological constant. This could well be some selection mechanism on a multiverse. Without it, one could hope that a concrete theory of quantum gravity will do the job.
 In a similar spirit, one could take a (big) step further and assume that the final theory of gravity does not provide radiative corrections to the Higgs mass, even if the Higgs couples to it
 , see e.g. \cite{Shaposhnikov:2012zz}. Of course the SM has to be extended in some way to account for the strongest experimental evidences against it, namely neutrino oscillations, Dark Matter and the baryon asymmetry. This can perhaps be done without reintroducing a hierarchy problem by adding three right handed neutrinos ($\nu$MSM) with masses in the KeV-MeV range\cite{Asaka:2005pn,Shaposhnikov:2007nj}. This attitude towards the hierarchy problem was put forward by Shaposhnikov and collaborators almost ten years ago, mainly to justify their $\nu$MSM models.
 
 Very recently this ``Finite Naturalness'' criterion has instead been considered as a starting point, from which to derive phenomenological consequences \cite{Farina:2013mla}. For example grand unified theories at a high scale would not satisfy such a requirement, because they make the Higgs mass sensitive to such high scale. This criterion also does not offer any additional clue to solve the cosmological constant hierarchy problem. On the contrary, as we already said, Dark Matter, neutrino masses and oscillations and baryogenesis could all be explained by a natural new physics. It is not clear whether other issues, especially related to cosmology (e.g. inflation) can be accomodated within a criterion of this kind.
 On the side of quantum gravity, a first explorative theoretical study appeared this year \cite{Dubovsky:2013ira}, where a toy model of quantum gravity that does not give corrections to the Higgs mass is given. It is also interesting to note that this framework could allow for the presence of extra light scalars. 

Note that here one is not disregarding naturalness nor changing its definition, as in \eqref{FineTuning}. The two (big) differences with respect to Solutions I are:
\begin{itemize}
 \item assuming that gravity poses no problem,
 \item assuming the knowledge of whatever new physics enters at length scales shorter than the ones explored so far.
\end{itemize}

\subsection{Remarks and motivation}


 The very fact of pushing new physics at scales much higher than a TeV makes it more difficult to think about ways to falsify a theory, at least at high energy experiments like the LHC. Some predictions of these theories \textit{could} show up there (like e.g. long lived gluino in split SUSY \cite{ArkaniHamed:2004fb,Giudice:2004tc}), but the non observation of such signals would not falsify the theory. 
 Also, it appears more difficult to test the way such a higher energy theory could solve some of the other SM problems. For example, how to get a clue on the theory of flavour, if it does not predict any deviation accessible at current experiments?
 Other relevant criteria of course exist, like elegance and ``reduction of input parameters'', but relying only on them would make the process of scientific inquiry much less powerful.
 Concerning more specifically the hypothesis of the Multiverse, there is currently not a clue of an experimental way to unambiguously test its presence.

Not finding any clear sign of new physics at LHC14 nor at other precision experiments might have a profound impact on the way we view fundamental physics, not only from the point of view of reductionism that we discussed earlier in Section \ref{sec:NP_closeFermiScale}.
Also, the issue of whether naturalness has been a good guiding principle is now mainly an experimental one.\footnote{Strictly speaking, the absence of new physics at the LHC14 would just require natural models to be a little bit more fine-tuned, and the amount of fine tuning necessary to give up is, to some extent, a matter of subjective taste.}
For these reasons, it is of great importance to understand how experimental results impact on natural theories, and in turn how these theories could show up at current and future facilities. 
This is the general motivation of the work presented in this thesis. More detailed motivations specific to the different subjects are presented in the beginning of Chapters \ref{cha:EFT}, \ref{cha:U2_SUSY}, \ref{cha:SUSY_Higgs} and \ref{cha:CHM}.

\part{Flavour physics}
\label{part:flavour}

\chapter{A closer look at the CKM picture}
\label{cha:SMCKM}

Let us start by an intuitive explanation of why CP violation is related to complex coefficients of the Lagrangian. In the charged current Lagrangian \eqref{SMChargedCurrent}, a CP transformation exchanges the operators
\begin{equation}
 \bar{u}_{Li} \gamma^\mu d_{Lj}\, W^+_\mu \; \longleftrightarrow \; \bar{d}_{Lj} \gamma^\mu u_{Li}\, W^-_\mu\, ,
\end{equation}
but leaves their coefficients $V_{ij}$ and $V_{ij}^*$ unchanged. This means that CP is a symmetry of $\mathcal{L}_{\text{ch.c.}}$ if $V_{ij} = V_{ij}^*$, which is not the case, as we showed with the parameter counting of the previous section. Apart from a possible effect in the strong interactions (see next section), in the SM Lagrangian there are no other complex parameters, this has the relevant implication that all the CP violation is encoded in the SM in the single physical phase contained in $V$.

\section{The CKM matrix}

The form of the CKM matrix $V$ is not unique, but can be reduced to a minimal one by fields redefinition, with three angles and one phase, $\delta_{\text{CKM}}$. 
Note that in the two generations case one could rotate away all the phases, leaving a theory without CP violation\footnote{The idea of extending the Cabibbo matrix \cite{Cabibbo:1963yz} from two to three generations was first put forward by Kobayashi and Maskawa in 1973 \cite{Kobayashi:1973fv}.}.

The elements of $V$ can be written with the self-explanatory notation
\begin{equation}\label{CKMselfexplanatory}
V = \begin{pmatrix}
V_{ud} & V_{us} & V_{ub} \vspace{.3cm}\\
V_{cd} & V_{cs} & V_{cb}\vspace{.3cm}\\
V_{td} & V_{ts} & V_{tb}
\end{pmatrix}.
\end{equation}
The unitarity of $V$ implies $\sum_i V_{ij} V_{ik}^* = \delta_{jk}$ and $\sum_j V_{ij} V_{kj}^* = \delta_{ik}$. Each of the six vanishing relations can be geometrically represented in the complex plane as a triangle. Among them, those that are obtained by taking scalar products of neighboring rows and columns do not have edges of comparable size, so that the most commonly used relation is
\begin{equation}
V_{ud} V_{ub}^* + V_{cd} V_{cb}^* + V_{td} V_{tb}^* = 0\,,
\label{CKMunitaritytriangleRelation}
\end{equation}
to which the term ``unitarity triangle'' usually refers. The analogous one involving a sum over the down quarks is not used, since the measurement of $V_{ts}$ comes from loop mediated processes in the SM (the tree-level couplings of the top quark are poorly known), and is therefore both less precise and, form a theoretical point of view, less reliable than the measurements of $V_{cd}$ and $V_{cb}$, which are extracted from processes induced by tree-level diagrams in the SM \cite{Beringer:1900zz}. All the unitarity triangles have the same area $|J|/2$, where $J$ is the so-called Jarlskog invariant \cite{Jarlskog:1985ht}, defined by $\text{Im} \left(V_{ij}V_{kl}V_{il}^*V_{kj}^* \right) = J \,\epsilon_{ikm} \epsilon_{jln}$ ($i,\dots,n = 1,2,3$). The usefulness of the Jarlskog invariant stems also from the fact that is independent of the parametrization chosen for $V$. A more immediate expression for it can be e.g.
\begin{equation}J = \text{Im}\left(V_{us}^* V_{ud} V_{ts} V_{td}^* \right) = - \text{Im}\left(V_{us}^* V_{ud} V_{cs} V_{cd}^* \right)\,,
\label{CKMJarlskog}
\end{equation}
where in the last equality we have made use of the unitarity relations of $V$. Indeed any quadrilinear product of the CKM elements that is invariant under phase redefinitions, is either real or has an imaginary part equal to $J$.

\begin{figure}[tbp]
\begin{center}
\includegraphics[width=.6\textwidth]{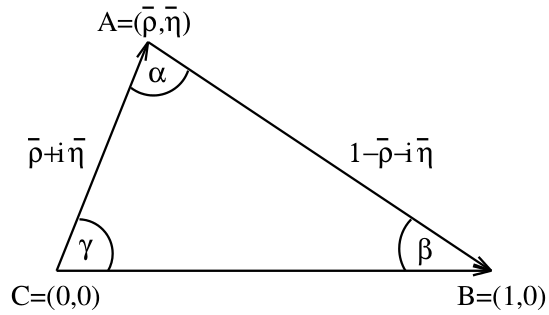}
\end{center}
\caption{Unitarity triangle in the complex $(\bar{\rho}, \bar{\eta})$ plane.}
\label{fig:UnitarityTriangle}
\end{figure}

The three real and one imaginary physical parameters in the CKM matrix can be made explicit by choosing a parametrization. A very useful one is the Wolfenstein parametrization \cite{Wolfenstein:1983yz}, where the four mixing parameters are $(\lambda, A, \rho, \eta)$, defined by
\begin{equation}
 \lambda = \frac{\left| V_{us}\right|}{\sqrt{\left| V_{ud}\right|^2 + \left| V_{us}\right|^2} }\,, \qquad A \lambda^2 = \lambda \left| \frac{V_{cb}}{V_{us}}\right|\,, \qquad A \lambda^3 (\rho +  i \eta) = V_{ub}^*\,,
 \label{CKMWolfensteinParameters}
\end{equation}
with $\lambda \simeq 0.22$ ($= \sin \theta_C$, with $\theta_C$ usually referred to as the ``Cabibbo angle''), playing the role of an expansion parameter and $\eta$ representing the CP violating phase. Up to order $\lambda^5$, one can write \cite{Buras:1994ec}:
 \begin{equation}\label{CKMWolfenstein}
V = \begin{pmatrix}
1 - \frac{1}{2} \lambda^2 - \frac{1}{8} \lambda^4 & \lambda & A \lambda^3 (\rho - i \eta) \vspace{.3cm}\\
-\lambda + \frac{1}{2} A^2 \lambda^5 \left[ 1 - 2 (\rho + i \eta) \right] & 1 - \frac{1}{2} \lambda^2 - \frac{1}{8} \lambda^4 (1 + 4 A^2)& A \lambda^2\vspace{.3cm}\\
A \lambda^3 \left[ 1-(1-\frac{1}{2}\lambda^2) (\rho + i \eta) \right] & - A \lambda^2 + \frac{1}{2} A \lambda^4 \left[ 1 - 2 (\rho + i \eta)\right] & 1 - \frac{1}{2} A^2 \lambda^4
\end{pmatrix}.
\end{equation}
In terms of this parametrization one has $J = \lambda^6 A^2 \eta + O(\lambda^8)$. Moreover it is convenient to introduce the quantities $\bar{\rho} = \rho \left( 1-\lambda^2/2 \right)$ and $\bar{\eta} = \eta \left( 1-\lambda^2/2 \right)$, for which the relation
\begin{equation}
 \bar{\rho} + i \bar{\eta} = - \frac{V_{ud} V_{ub}^*}{V_{cd} V_{cb}^*}
\end{equation}
is independent of the phase convention. This proves very useful in displaying the unitarity triangle \eqref{CKMunitaritytriangleRelation}: by dividing each side by the best-known one, $V_{cd} V_{cb}^*$, one ends up with a triangle whose vertices are exactly $(0,0)$, $(1,0)$ and $(\bar{\rho}, \bar{\eta})$, see Figure \ref{fig:UnitarityTriangle}.
The three angles $\alpha$, $\beta$ and $\gamma$ are defined as
\begin{align}
\alpha = \text{arg}\left(-\frac{V_{td}V_{tb}^*}{V_{ud} V_{ub}^*} \right)\,, \quad  \beta = \text{arg}\left(-\frac{V_{cd}V_{cb}^*}{V_{td} V_{tb}^*} \right)\,, \quad  \gamma = \text{arg}\left(-\frac{V_{ud}V_{ub}^*}{V_{cd} V_{cb}^*} \right)\,.
\end{align}
See \cite{Buras:1994ec} for their expression in terms of $\bar{\rho}$, $\bar{\eta}$. They are physical quantities and can be measured by CP asymmetries in $B$ decays. Finally, it is also useful to define the small angles arising from the relation $ V_{us} V_{ub}^* + V_{cs} V_{cb}^* + V_{ts} V_{tb}^* = 0$,
\begin{equation}
\beta_s = \text{arg}\left(-\frac{V_{ts}V_{tb}^*}{V_{cs} V_{cb}^*} \right) \quad \text{and} \quad \beta_K = \text{arg}\left(-\frac{V_{cs}V_{cd}^*}{V_{us} V_{ud}^*} \right)\,,
\end{equation}
since LHCb is starting to probe the magnitude of the first one.\\

\subsection{Testing the CKM picture}
The last twelve years signed a new era in our understanding of flavour and CP violation. This has been possible thanks to the many measurements of rates, mass splittings and CP asymmetries in $B$ decays in the two $B$ factories, BaBar and Belle, in the two Tevatron detectors, CDF and D0, and recently at the LHCb. The following relevant  question can today be answered:
\begin{enumerate}
 \item Is the CKM mechanism a consistent description of flavour and CP violation?
 \item Is the CKM mechanism the dominant source of the observed flavour and CP violating phenomena?
\end{enumerate}
We will first assume the SM and test the overall consistency of the various measurements, answering the
first question. Then we will go one step further and answer the same question model-independenlty, namely allowing for new physics to contribute to some relevant processes. This will provide an answer to the second question. In doing so, we will make use of the unitarity triangle defined in the previous section, as is usually done in the literature.

Before going on, it is useful to make a distinction between \textit{direct} and \textit{indirect} measurements of the CKM matrix elements. The first are related to SM tree level processes, and are expected to hold almost model-independently. It is in fact very difficult to imagine 
new physics contributions to tree level FV and CPV processes, that have not shown up in any of the other processes where the SM is very well tested, like e.g. in the electroweak sector. Indirect measurements are instead related to SM loop processes, and thus are sensitive to new physics. This is for example the case for processes induced by $B-\bar{B}$ mixing, where the leading SM contribution comes from an EW box diagram, which is therefore $O(g^4)$ suppressed, and is furthermore proportional to the small combination $(V_{td}^* V_{tb})^2$. Consequently, inconsistencies among indirect measurements, or between indirect and direct ones, can give evidence for new physics.
The distinction between direct and indirect measurements can be equivalently formulated in terms of the distinction between \textit{flavour changing charged current} (FCCC) and \textit{flavour changing neutral current} (FCNC) processes, since in the SM tree level flavour violation is mediated only via charged currents.

\subsection{Self consistency of the CKM picture}
 As already anticipated, here we assume the CKM matrix to be the only source of flavour and CP violation. The values of the Wolfenstein parameters $\lambda$ and $A$ defined in \eqref{CKMWolfensteinParameters} are known from tree-level measurements related to $K \to \pi \ell \nu$ and $b \to c \ell \nu$ respectively. The latest results reported by the CKMfitter group \cite{CKMfitter:2013fpcp} are:
 \begin{equation}
 \lambda = 0.2246^{+0.0019}_{-0.0001}\, ,\qquad A = 0.823^{+0.012}_{-0.033}\,.
  \label{FITlambdaA}
 \end{equation}
The fit result for the $\bar{\rho}$ and $\bar{\eta}$ relies on measurement of both tree level ($\gamma$, $\alpha$ and $|V_{ub}|$) and loop level ($\epsilon_K$, $\sin 2 \beta$, $\Delta m_d$ and $\Delta m_s$), induced processes. The latest CKMfitter results are \cite{CKMfitter:2013fpcp}
\begin{equation}
\bar{\rho} = 0.129^{+0.018}_{-0.009}\,, \qquad \bar{\eta} = 0.348^{+0.012}_{-0.012}\,.
 \label{FITrhoeta}
\end{equation}
\begin{figure}[tbp]
\begin{center}
\includegraphics[width=.65\textwidth]{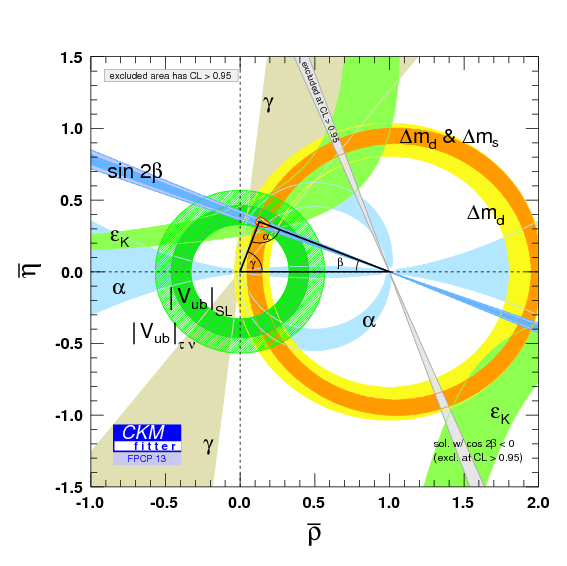}
\end{center}
\caption{Allowed regions in the $(\bar{\rho}, \bar{\eta})$ plane. Superimposed are the individual constraints from the various observables discussed in the text.}
\label{fig:CKMfitterTotal}
\end{figure}
Figure \ref{fig:CKMfitterTotal} shows how the various measurements used for the fit can be visualized as functions of the parameters $\bar{\rho}$ and $\bar{\eta}$, and makes the impressive consistency of the CKM fit to data evident. To further convince ourselves that the KM picture of CP violation is consistent with measurements, we find it convenient to show in Figure \ref{fig:CKMfitterCPVC} the allowed regions in the $(\bar{\rho}, \bar{\eta})$, as determined by only CP conserving and CP violating observables separately.
\begin{figure}[tbp]
\begin{center}
\includegraphics[width=0.49\textwidth]{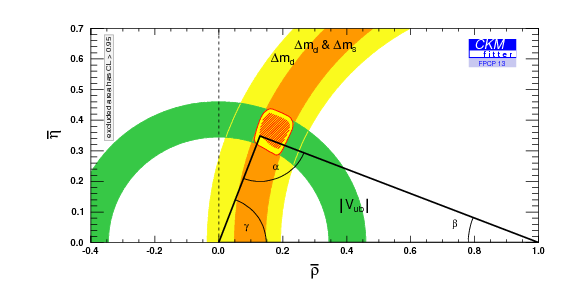}
\includegraphics[width=0.5\textwidth]{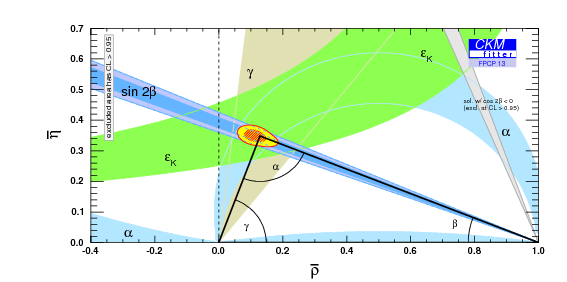}
\end{center}
\caption{Allowed regions in the $(\bar{\rho}, \bar{\eta})$ plane from CP violating (\textit{left panel}) and CP conserving (\textit{right panel}) observables.}
\label{fig:CKMfitterCPVC}
\end{figure} 
We conclude by giving a quick list of the observables used for the above fits\footnote{For a very nice review about CP violation in meson decays see \cite{Nir:2005js}, for a more recent and general one see \cite{Nir:2010jr}. For more complete, ``classic'' references (where e.g. a detailed discussion of the matrix elements and of $\epsilon_K$ can be found) see \cite{Buchalla:1995vs,Buras:1998raa}. For an updated discussion of the extraction of the fit parameters from the various observables see the review about the CKM matrix in \cite{Beringer:1900zz}.} 
:
\begin{itemize}
 \item the rates of various $B \to D K$ decays, which depend on the phase $\gamma = \text{arg}\left(\frac{\rho + i \eta}{\rho^2 + \eta^2}\right)$;
 \item the rates of various $B \to \pi \pi, \rho \pi, \rho \rho$ decays, depending on the phase $\alpha = \pi - \beta - \gamma$;
 \item the rates of charmless semileptonic $B$ decays, depending on $|V_{ub}|^2 \propto \rho^2 + \eta^2$;
 \item the CP asymmetry in $B_d \to \psi K_S$, $S_{\psi K_S} = \sin 2 \beta = \frac{2 \eta (1-\rho)}{(1-\rho)^2 + \eta^2}$;
 \item the ratio between the mass splittings in the neutral $B_d$ and $B_s$ mesons, $\Delta m_d/\Delta m_s$, sensitive to $|V_{td}/V_{ts}|^2 = \lambda^2 (1- \rho^2 + \eta^2)$;
 \item indirect CP violation in $K \to \pi \pi$ decays, $\epsilon_K$, depending in a complicated way on $\rho$ and $\eta$.
\end{itemize}
It can be worth to stress that, in the choice of the observables used in the fits, an important role is played by the knowledge of long-distance QCD effects. Progress in this field will have immediate consequences on our understanding of flavour and CP violation.

\subsection{Dominance of the CKM mechanism}
\begin{figure}[tbp]
\begin{center}
\includegraphics[width=0.46\textwidth]{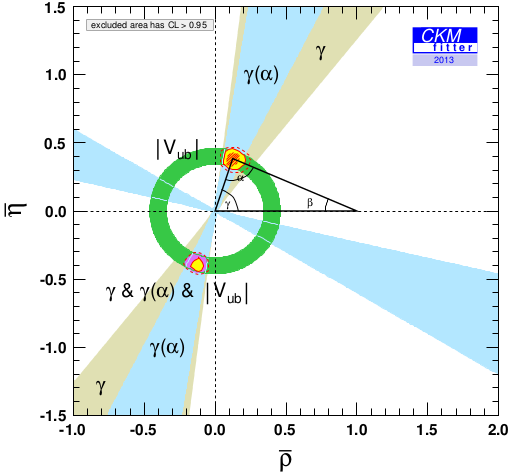}
\includegraphics[width=0.53\textwidth]{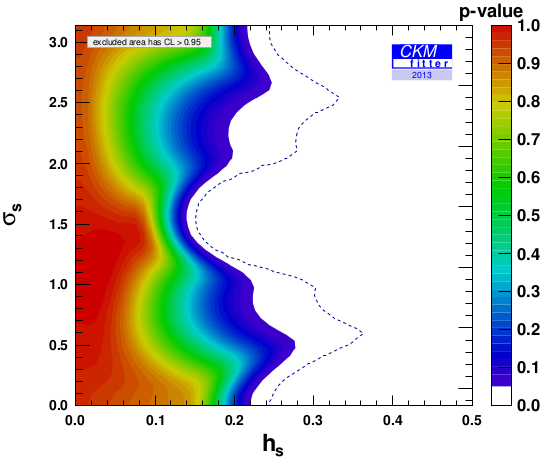}
\end{center}
\caption{Fit results allowing for NP in $B_d-\bar{B}_d$ and $B_s-\bar{B}_s$ mixing, and assuming that tree level processes are dominated by the SM. Left: only observables not sensitive to NP in $\Delta F = 2$ processes. Right: all observables listed in Table I of \cite{Charles:2013aka}, from which these figures are taken. The dotted curve shows the 99.7 \% C.L. contour.
}
\label{fig:CKMfitterNP}
\end{figure}

To answer the second question, we assume the SM to dominate charged-current tree level processes, which as we discussed is a very plausible assumption. In loop level processes we allow for new physics contributions, in addition to the usual SM ones depending on $\rho$ and $\eta$. Our goal is twofold:
\begin{itemize}
 \item first, we want to determine whether $\eta = 0$ is allowed, in which case the KM mechanism would not be effective (but one can already imagine how likely this possibility is, due to the impressive results of the CKM fit);
 \item secondly, we want to see whether a sizeable NP contribution is allowed in certain processes, in which case the CKM mechanism would not be the dominant source of flavour and/or CP violation for the process under examination (a possibility which is even expected if new physics is relevant at energies not far from the Fermi scale).
\end{itemize}
Of course, for the above program to be viable we need to add a limited number of parameters, in such a way that the observables at disposal are still more than the total number of parameters we want to fit, which includes also $\bar{\rho}$ and $\bar{\eta}$. This is possible if we limit to the case of $B_d - \bar{B}_d$ and $B_s - \bar{B}_s$ mixings, where NP effects in the relative amplitude can be parametrized as
\begin{equation}
\left(M_{12}^{d,s}\right)^{\text{NP}} = h_{d,s} \,e^{2 i \sigma_{d,s}} \left(M_{12}^{d,s}\right)^\text{SM} ,
 \label{NPinBBbar}
\end{equation}
The result of this procedure is evident from Figures \ref{fig:CKMfitterNP}.
The one on the left is drawn taking into account only constraints from observables that are not affected from NP in $\Delta F = 2$ processes. It indicates that the KM mechanism is at work, the region with $\bar{\rho} < 0, \bar{\eta} < 0$ is excluded at 68.2\% C.L., but allowed at 95.5\% C.L. (for details see Ref. \cite{Charles:2013aka}). The one on the right implies that the CKM picture is the dominant source of flavour and CP violation, and that the size of the new physics contribution to $B_s - \bar{B}_s$ mixing is quite constrained, $ h_s \lesssim 0.2$. Similar results are obtained for $B_d - \bar{B}_d$ mixing.


\section{Tensions in the CKM unitarity fit}
\label{sec:CKM_tensions}
\begin{figure}[tbp]
\begin{center}
\includegraphics[width=1\textwidth]{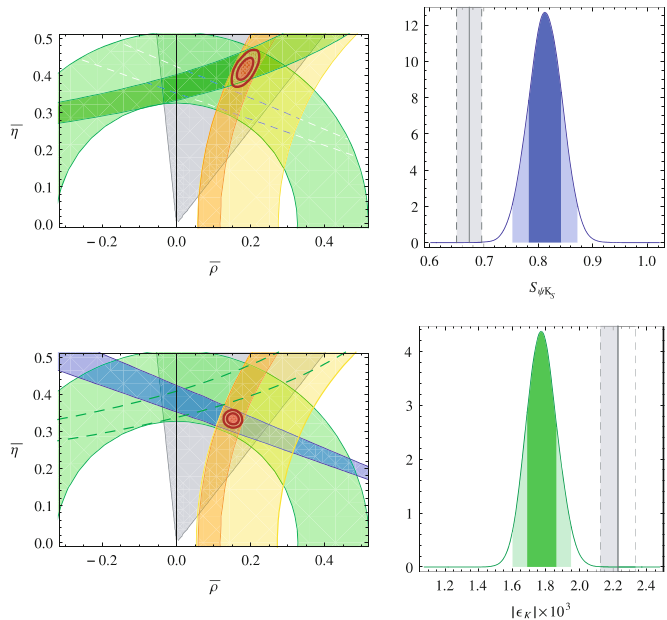}
\end{center}
\caption{Results of two global fits of the CKM matrix using tree-level and $\Delta F = 2$ observables, excluding $S_{\psi K_S} = \sin 2 \beta$ (\textit{top row}) or $\epsilon_K$ (\textit{bottom row}), as of 2011. The bands in the \textit{left} panels correspond to 2 $\sigma$ errors. The dashed bands in the \textit{right} panels correspond to 1 $\sigma$ errors.}
\label{fig:CKMTensions}
\end{figure}

The CKM picture is not free of some small inconsistencies, which we now briefly discuss. We first mention the long-standing issue of the determination of $|V_{ub}|$, for which inclusive observables yields the value $|V_{ub}| = (4.41 \pm 0.15^{+0.15}_{-0.19})\times 10^{-3}$, while exclusive ones give\footnote{The numbers quoted are those reported in the PDG 2012 \cite{Beringer:1900zz}, the PDG 2013 has not updated them yet. In particular they do not include the recent Belle measurement of BR$(B\to\tau \nu)$ \cite{Adachi:2012mm}, which has the effect of decreasing the tension. The most recent values used by the UTfit collaboration for their fit are respectively $|V_{ub}| = (4.40 \pm 0.31)\times 10^{-3}$ and $|V_{ub}| = (3.42 \pm 0.22)\times 10^{-3}$ \cite{Bona:UTfit}.} $|V_{ub}| = (3.23 \pm 0.31) \times 10^{-3}$. Due to the fact this is related to tree level SM processes, a frequent attitude is to either choose one of the two values or an average of the two, instead of trying an explanation of the discrepancy 
via use of some NP model. This is the same attitude we will stick to in our discussion of the $U(2)^3$ phenomenology in the following chapters.\\
Another $\sim 2 \; \sigma$ inconsistency in the unitarity triangle fit is the one between $\epsilon_K$, $S_{\psi K_S}$ and $\Delta m_d/\Delta m_s$ \cite{Lunghi:2008aa, Buras:2008nn, Altmannshofer:2009ne, Lunghi:2010gv, Bevan:2010gi}. To visualize it, a global fit to the CKM matrix is performed in \cite{Barbieri:2011ci}, where one observable is removed from the fit. In this way a prediction for it is obtained, to be compared to its experimental value. The results of this procedure (for whose details we refer to \cite{Barbieri:2011ci}) are displayed in Figure \ref{fig:CKMTensions} for illustrative purposes. Note in fact that recent measurements and improvements in lattice calculations made this tension milder than it was at the time these figures were drawn.
The left panels show the 2 $\sigma$ bounds of the individual constraints in the $(\bar{\rho},\bar{\eta})$ plane, together with the favoured regions for $\bar{\rho}$ and $\bar{\eta}$. The dashed lines show the band of the ``unusued'' constraint, which in both cases clearly deviates from the region preferred by the fit. The same deviations are shown in a more evident way in the right panels, where the predicitions of the fit for $S_{\psi K_S}$ and $\epsilon_K$ are compared with the respective experimental values. This tension could be solved by NP contributions to at least one of the above observables, for example in the $U(2)^3$ framework, as we will see, this possibility is exploitable both for $S_{\psi K_S}$ and $\epsilon_K$, but not for $\Delta m_d/\Delta m_s$\footnote{This is why the analogous procedure is not given for $\Delta m_d/\Delta m_s$.}
.

The experimental situation is continuing to bring new key information to the current picture of flavour and CP. For example the LHCb experiment at CERN is eagerly waited to measure $\beta_s$ and BR$(B_{d,s} \to \mu^+ \mu^-)$ more precisely, and to explore CPV in $D-\bar{D}$ mixing, CP asymmetries in $b \to s \mu \mu$, and direct CPV in $D$ decays, just to mention some of the ``hot'' processes. Also ATLAS and CMS are competitive for some specific observables, like e.g. the aforementioned BR$(B_{d,s} \to \mu^+ \mu^-)$.

\section{Electric dipole moment of the neutron}
We conclude this section about the CKM picture by sketching how the SM contribution to the neutron EDM $d_n$ is by far lower than the current experimental limit. This is relevant since it makes EDMs an excellent probe of new physics, so that in many cases they provide the most severe bounds on generic CPV structures. For simplicity we estimate the SM contribution to the quark EDMs, which is sufficient since we are interested only in an order of magnitude (o.o.m.) analysis.

To produce an electric dipole moments for the light quarks in the SM, one must look at contributions that give a non-vanishing phase in the effective operator $\mathcal{L}_{\text{dip}} = \mu_q e^{i \delta_q} (\bar{q}_L i \sigma_{\mu \nu} q_R) e\,F_{\mu\nu}$, the quark EDM being $d_q \propto \mu_q \sin \delta_q \,e$. One loop $W$-exchange diagrams are excluded, since they are proportional to the absolute value squared of the relevant CKM matrix element. One then has to go to two loop order and exchange two $W$'s, but it turns out that all EDMs as well as ChromoEDMs (CEDMs) vanish exaclty at the two-loop level \cite{Shabalin:1978rs}. The relevant diagrams are then the three-loop ones with a two-$W$ exchange, which have been computed in \cite{Khriplovich:1985jr,Czarnecki:1997bu}. The dominant contribution is the following one to the down quark EDM,
\begin{equation}
d_d \simeq \frac{e \, \alpha_s}{108 \pi^5} m_d m_c^2 G_F^2 \,J\, \ln^2\Big(\frac{m_b^2}{m_c^2}\Big)\,\ln\Big(\frac{m_W^2}{m_b^2}\Big),
 \label{quarkEDM}
\end{equation}
where the presence of $J$ comes from the fact that a two $W$'s exchange imply the presence of a quadrilinear product of the CKM elements.
The above result leads to the prediction $d_d \simeq 10^{-34} \,e$~cm, to be compared with the experimental 90\% C.L. limit $d_n < 2.9 \times 10^{-26} \,e$~cm \cite{Baker:2006ts}. The SM contribution to the neutron EDM is however dominated by long distance effects, the most recent estimation of them \cite{Mannel:2012qk} resulting in $d_n \simeq 10^{-31}  e\, \text{cm}$, well below current and foreseen experimental sensitivities (the most optimistic projections aiming at $\sim 10^{-28} \,e$~cm \cite{Bodek:2008gr,Beck:2011gw}.).
More details about the relations between $d_n$ and the relevant operators can be found in Section \ref{sec:NeutronEDM}, where we discuss the neutron EDM as a constraint on a $U(2)^3$-symmetric new physics.

\chapter[$U(2)^3$ and its phenomenology]{A $U(2)^3$ flavour symmetry and its phenomenology}
\label{cha:EFT}


The discussion of Section \ref{subsec:NP_flavourpuzzle} has pointed to flavour symmetries as a possible solution to the NP flavour puzzle. In particular, we have mentioned that assuming a $U(3)^3$ symmetry (MFV) allows to reconcile the presence of NP at the TeV scale with a natural solution of the hierarchy problem.
Why then the need to go beyond the MFV paradigm? First of all, the $U(3)^3$ symmetry is already badly broken in the SM, so that it appears more natural to take as a starting point a symmetry which is instead preserved to a good level of approximation. In other words, $U(3)^3$ does not address at all the SM flavour puzzle, if not having a fundamental representation which is three-dimensional.
Also, the MFV solution to the NP flavour puzzle leaves little space for observables deviations from the CKM picture of the Standard Model. One can think of other reasons not to be satisfied with $U(3)^3$, like the fact that it does not consent to separate the potential NP energy scale associated with the third generation of quarks from the one associated with the first two. This possibility is very welcome in NP theories that solve the hierarchy problem in a natural way, in light of current collider constraints. 
Also, as we shall see, standard MFV precludes the possibility of achieving a dynamical suppression of the EDMs.

An attempt to cope with these issues already pursued in the literature is the reduction of $U(3)^3$ to a $U(2)$ acting on the first two generations of quarks, irrespective of their chiralities \cite{Pomarol:1995xc,Barbieri:1995uv}. 
However, while providing a rationale for explaining both the quark's hierarchies and the smallness of EDMs, the consideration of a single $U(2)$ does not yield to enough suppression of the right-handed currents contribution to the $\epsilon_K$ parameter \cite{Barbieri:2011ci,Barbieri:1997tu}.

The considerations developed so far motivate us to study the flavour symmetry $U(2)^3 = U(2)_q \times U(2)_u \times U(2)_d$, exhibited by the SM if one neglects the masses of the quarks of the first two generations, as well as their mixing with the third generation ones. This framework was first introduced in Ref. \cite{Barbieri:2011ci} in the context of Supersymmetry. The superiority of $U(2)^3$ with respect to $U(3)^3$ stems at this level from the observed pattern of quark masses and mixings, which makes  $U(2)^3$ a good approximate symmetry of the SM Lagrangian, broken at most by an amount of order a few $\times 10^{-2}$. This is the size of $V_{cb}$,  comparable to or bigger than the mass ratios $m_{c, u}/m_t$ or $m_{s, d}/m_b$. $U(3)^3$ on the contrary is badly broken at least by the top Yukawa coupling.
Of course $U(2)^3$ cannot be an exact symmetry of the Lagrangian, but has to be broken in some specific directions. In the following we will define and discuss our choice for these breaking directions, and then study the phenomenology it implies.

The program we will pursue in this Chapter can be formulated in terms of the following questions.
Taking an Effective Field Theory (EFT) point of view, which are the limits on the possible size of new flavour changing and CP violating interactions consistent with the current observations? 
Which signals can one in turn expect in foreseen experiments? How can this picture be extended to the lepton sector?

\section{$U(2)^3$}
\label{sec:U2intro}
To describe the breaking of $U(2)^3$ we assume that it is encoded in a few small dimensionless parameters. Their origin is unknown and may be different, for example, in different models of EWSB, but we require that they have  definite transformation properties under  $U(2)^3$ itself, so that the overall Lagrangian, fundamental or effective as it may be, remains formally invariant. This is what we mean by saying that $U(2)^3$ is broken in specific directions. Along these lines, the simplest way to give masses to both the up and down quarks of the first two generations is to introduce two (sets of) parameters $\Delta Y_u$, $\Delta Y_d$,  transforming as 
$\Delta Y_u = (2, \bar{2}, 1)$,
$\Delta Y_d = (2, 1, \bar{2})$
under $U(2)_q\times U(2)_u\times U(2)_d$. If these {\it bi-doublets} were the only breaking parameters, the third generation, made of singlets under $U(2)^3$, would not   communicate with the first two generations at all. For this to happen one needs single doublets, at least one,  under any  of the three $U(2)$'s. The only such doublet that can explain the observed small mixing between the third and the first two generations, in terms of a correspondingly small parameter, transforms under $U(2)_q\times U(2)_u\times U(2)_d$ as 
$ \V = (2,1,1)$.
A single doublet under $U(2)_u$ or $U(2)_d$ instead of $U(2)_q$ would have to be of order unity.
This is the minimal amount of spurions needed for a realistic description of quark masses and mixings, we call this Minimal $U(2)^3$.

One can then add two extra spurions $\Vu = (1,2,1)$ and $\Vd = (1,1,2)$, and complete in this way the picture by considering all the possible breaking terms of $U(2)^3$ entering the quark mass terms
\begin{equation}
\lambda_t (\qLbar \V)t_R, \quad \lambda_t \qLbar \Delta Y_u \uR,
\quad \lambda_t \bar{q}_{3L} (\Vu^\dagger \uR),
\label{Yuk_u}
\end{equation}
\begin{equation}
\lambda_b(\qLbar \V)b_R, \quad \lambda_b \qLbar \Delta Y_d \dR,
\quad \lambda_b \bar{q}_{3L} (\Vd^\dagger \dR),
\label{Yuk_d}
\end{equation}
where $\qL, \uR, \dR$  stand for doublets under $U(2)_q, U(2)_u, U(2)_d$ respectively\footnote{In (\ref{Yuk_d}) we have factored out as a common factor the bottom  Yukawa coupling $\lambda_b$, which in principle requires an explanation since $\lambda_b$ is relatively small. One possibility is to consider a symmetry, either continuous or discrete, acting in the same way on all the right-handed $d$-type quarks, broken by the small parameter $\lambda_b$.}. This completion we call Generic $U(2)^3$.
To summarize, we assume that $U(2)^3$ is an approximate symmetry of the flavour sector of the SM only weakly broken in the directions (by the {\it spurions})
\begin{equation}
\Delta Y_u = (2, 2, 1),\quad \Delta Y_d = (2, 1, 2), \quad \V = (2, 1, 1), \quad \Vu = (1, 2, 1), \quad \Vd = (1, 1, 2),
\label{spurions}
\end{equation}
such that every term in  (\ref{Yuk_u}) and (\ref{Yuk_d}) is formally invariant.

\subsection{Physical parameters and CKM matrix}
\label{sec:spurions}
By $U(2)^3$ transformations it is possible and useful to restrict and define the physical parameters appearing in (\ref{spurions}).  In Minimal $U(2)^3$ we choose:
\begin{equation}
\label{eq:MU2_spurions}
\V = \begin{pmatrix}0\\ \epsilon_L\end{pmatrix},\qquad \Delta Y_u = L_{12}^u\,\Delta Y_u^{\rm diag},\qquad
 \Delta Y_d = \Phi_L L_{12}^d\,\Delta Y_d^{\rm diag},
 \end{equation}
 where $\epsilon_L$ is a real parameter, $L_{12}^{u,d}$ are rotation matrices in the space of the first two generations with angles $\theta_L^{u,d}$ and $\Phi_L = {\rm diag}\big(e^{i\phi},1\big)$, i.e. four parameters in total. Incidentally this shows that, if CP violation only resides in $\V, \Delta Y_u, \Delta Y_d$, there is a single physical phase, $\phi$, which gives rise to the CKM phase.
 
 Similarly in Generic $U(2)^3$ we set:
 \begin{equation}\label{genericV}
\V = \begin{pmatrix}0\\ \epsilon_L\end{pmatrix},\qquad \Vu = \begin{pmatrix}0\\ \epsilon^u_R\end{pmatrix},\qquad \Vd = \begin{pmatrix}0\\ \epsilon_R^d\end{pmatrix},
\end{equation}
\begin{align}\label{genericY}
\Delta Y_u &= L_{12}^u\,\Delta Y_u^{\rm diag}\,\Phi_R^u R_{12}^u, & \Delta Y_d &= \Phi_L L_{12}^d\,\Delta Y_d^{\rm diag}\,\Phi_R^d R_{12}^d,\\
\Phi_L &= {\rm diag}\big(e^{i\phi},1\big), & \Phi_R^{u, d} &= {\rm diag}\big(e^{i\phi_1^{u,d}}, e^{i\phi_2^{u,d}}\big),
\end{align}
 which adds to the four parameters of Minimal $U(2)^3$ four real parameters, $\epsilon_R^{u,d}, \theta_R^{u,d}$ and four phases, $\phi_{1,2}^{u,d}$. For later convenience we define $s^{u,d}_L=\sin{\theta_L^{u,d}}$ and $s^{u,d}_R=\sin{\theta_R^{u,d}}$.
 
 The next step consists in writing down the mass terms for the up and down-type quarks, invariant under  $U(2)^3$, and in diagonalizing them\footnote{See Appendix \ref{app:bilinears} for a detailed analysis of their digonalization, together with all the possible quark bilinears appearing in effective operators, relevant for the next chapter.}, which can be done perturbatively by taking into account the smallness of $\epsilon_L, \epsilon_R^{u,d}$ and  $\Delta Y_{u, d}^{\rm diag}$. As a consequence, to a sufficient approximation the unitary transformations that bring these mass matrices to diagonal form are influenced on the left side only by the four parameters of  Minimal $U(2)^3$, $\epsilon_L, \theta_L^{u,d}, \phi$, whereas those on the right side depend on the extra parameters of Generic $U(2)^3$, $\epsilon_R^{u,d}, \theta_R^{u,d}, \phi_{1,2}^{u,d}$. In turn this leads to a unique form of the standard CKM matrix
 \begin{equation}\label{CKM}
V = \begin{pmatrix}
c^u_L c^d_L & \lambda & s^u_L s\,e^{-i\delta}\\
-\lambda & c^u_L c^d_L & c^u_L s\\
-s^d_L s\,e^{i(\delta - \phi)} & -c^d_L s & 1
\end{pmatrix},
\end{equation}
where $s\sim O(\epsilon_L)$, $c^{u,d}_L=\cos{\theta_L^{u,d}}$ and
$s^u_L c^d_L - s^d_L c^u_L e^{i\phi} = \lambda e^{i\delta}$.
Using this parametrization of the CKM matrix, a direct fit of the tree-level flavour observables, presumably not influenced by new physics,  results in
\begin{align}
s^u_L &= 0.086\pm0.003
\,,&
s^d_L &= -0.22\pm0.01
\,,\\
s &= 0.0411 \pm 0.0005
\,,&
\phi &= (-97\pm 9)^\circ
\,.
\end{align}
At this stage, the extra ``right-handed'' parameters present in Generic $U(2)^3$ are unconstrained.

\subsection{Preparing the ground for an EFT analysis}

As outlined in the Introduction, here we are interested in considering from an EFT point of view the leading flavour-violating  operators that are consistent with the $U(2)^3$ symmetry, only broken by the {\it spurions} in (\ref{Yuk_u}) and (\ref{Yuk_d}). Their general form can be summarized in
\begin{equation}
\Delta \mathcal{L} = \Delta \mathcal{L}^{4f}_{L}  + \Delta \mathcal{L}_\text{mag}  + \Delta \mathcal{L}^{4f}_{R}  + \Delta \mathcal{L}^{4f}_{LR}\, ,
\label{eq:genL}
\end{equation}
 where $\Delta \mathcal{L}^{4f}_{L, R, LR} $ are the sets of four-fermion operators with flavour violation respectively in the left-handed sector, in the right-handed sector and in both, whereas $\Delta \mathcal{L}_\text{mag}$ contains the chirality-breaking dipole operators. As we will see, sizeable contributions to $\Delta \mathcal{L}^{4f}_{R}$ and $\Delta \mathcal{L}^{4f}_{LR}$ are absent in Minimal $U(2)^3$, with the notable exception of $\epsilon_K'$ we will discuss in the end of Section \ref{sec:DF1Minimal}.
 
 In what follows, we will write each single term in \eqref{eq:genL} as
\begin{equation}
\Delta \mathcal{L}_{} = \frac{1}{\Lambda^2} \sum_i C_i \mathcal{O}_i ~+\text{h.c.}
,
\end{equation}
where the operators $\mathcal{O}_i$ relevant for the process under examination will be specified case by case. Their derivation, both in Minimal and in Generic $U(2)^3$, is given in Appendix \ref{sec:Diagonalization}, which also contains the full list of interaction bilinears.

\section{Current bounds and possible new effects in Minimal $U(2)^3$}
\label{sec:DF12Minimal}

\subsection{$\Delta F=2$ processes}
\label{sec:DF2Minimal}

The relevant $\Delta F=2$ operators generated in the $U(2)^3$ framework read
\begin{align}
&\Delta \mathcal{L}_{L}^{4f, \Delta S = 2} =
\frac{c_{LL}^K }{\Lambda^2} \xi_{ds}^2 \frac{1}{2}\left({\bar d}_L \gamma_\mu  s_L \right)^2 +\text{h.c.}\,,\label{eq:LeffDS2}\\
& \Delta \mathcal{L}_{L}^{4f, \Delta B = 2} = \sum_{i=d,s}
\frac{c_{LL}^B e^{i\phi_B}}{\Lambda^2} \xi_{ib}^2\frac{1}{2}\left({\bar d}_L^i \gamma_\mu  b_L \right)^2
+\text{h.c.}\,,
\label{eq:LeffDB2}
\end{align}
where $\xi_{ib}= V_{tb}V_{ti}^*$, $\xi_{ds}= V_{ts}V_{td}^*$ and $c_{LL}^{K,B}$ are real, model dependent parameters that can be of $O(1)$, with phases made explicit wherever present. The $U(3)^3$ case at low $\tan\beta$ is recovered for $c_{LL}^K=c_{LL}^B$ and $\phi_B=0$. The observables in $K$, $B_d$ and $B_s$ meson mixing are modified as
\begin{align}
\epsilon_K&=\epsilon_K^\text{SM(tt)}\left(1+h_K\right) +\epsilon_K^\text{SM(tc+cc)} ,
\label{eq:epsKxF}\\
S_{\psi K_S} &=\sin\left(2\beta + \text{arg}\left(1+h_B e^{i\phi_B}\right)\right) ,
\label{eq:Spk} \\
S_{\psi\phi} &=\sin\left(2|\beta_s| - \text{arg}\left(1+h_B e^{i\phi_B}\right)\right) ,
\label{eq:SpsiphixF}\\
\Delta M_d &=\Delta M_d^\text{SM}\,\left|1+h_B e^{i\phi_B}\right| ,
\label{eq:DMdxF}\\
\frac{\Delta M_d}{\Delta M_s} &= \frac{\Delta M_d^\text{SM}}{\Delta M_s^\text{SM}} \,,
\label{eq:MdMs}
\end{align}
where
\begin{align}
h_{K,B} &= c_{LL}^{K,B}
\frac{4s_w^4}{\alpha_{em}^2S_0(x_t)}
\frac{m_W^2}{\Lambda^2}
\approx
1.08\,c_{LL}^{K,B}
\left[ \frac{3\,\text{TeV}}{\Lambda} \right]^2
\,.
\end{align}
In the special case of supersymmetry with dominance of gluino contributions, as we will see in Section \ref{sec:DF2Susy}, one has $h_K=x^2F_0>0$ and $h_B=x F_0$, where $F_0$ is a positive loop function (given in \eqref{F0}) and $x$ an $O(1)$ mixing parameter.

To confront the effective operators (\ref{eq:LeffDS2},\ref{eq:LeffDB2}) with the data, the dependence of the $\Delta F=2$ observables on the CKM matrix elements has to be taken into account. To this end, we performed global fits of the CKM Wolfenstein parameters $A$, $\lambda$, $\bar\rho$ and $\bar\eta$ as well as the coefficients $c_{LL}^{K,B}$ and the phase $\phi_B$ to the set of experimental observables collected in the left-hand column of Tab.~\ref{tab:inputs}, by means of a Markov Chain Monte Carlo, assuming all errors to be Gaussian.\footnote{We report here the fit as of spring 2012, when it was originally performed. Since then, a new more precise determination of some lattice parameters has been obtained in \cite{Carrasco:2013zta}. It resulted in the values $f_{B_s}\sqrt{\hat B_s} = (262\pm10)$ MeV and $\xi = 1.225\pm0.031$, to be compared with the old ones shown in Tab. \ref{tab:inputs}. 
}

\begin{figure}[h!]
\centering
\includegraphics[width=\textwidth]{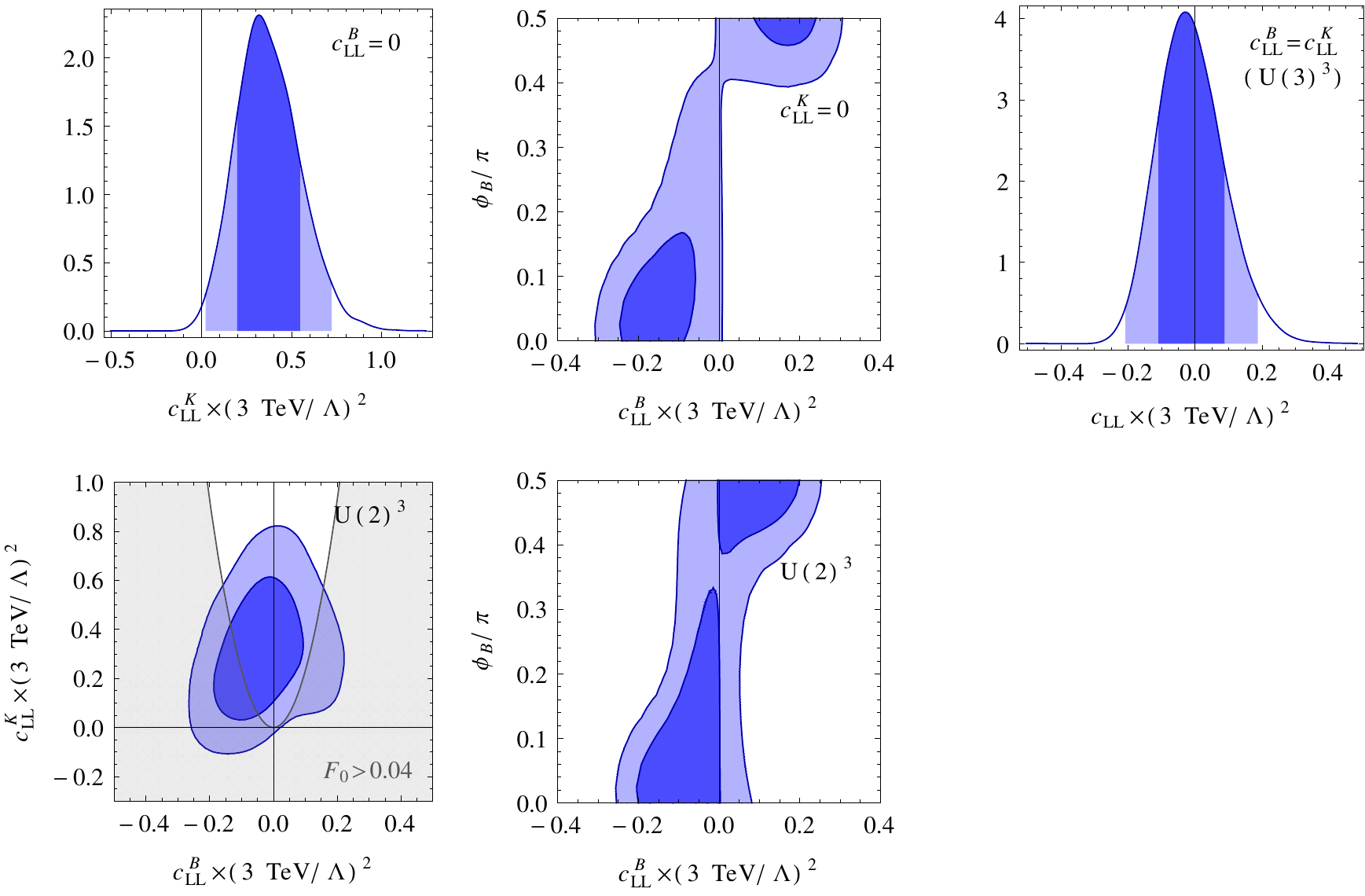}
\caption{Fit predictions (68 and 95\% Bayesian credible regions) in $\Delta F=2$ fits with $c_{LL}^B=0$ (top left), $c_{LL}^K=0$ (top centre), $c_{LL}^B=c_{LL}^K$, $\phi_B=0$ (top right, relevant to $U(3)^3$) and with all 3 parameters independent (bottom). The gray region in the bottom left plot is disfavoured by direct searches only in the SUSY case with dominance of gluino contributions.}
\label{fig:DF2fits}
\end{figure}

The results of four different fits are shown in Fig.~\ref{fig:DF2fits}. The top left panel shows the fit prediction for $c_{LL}^K$ in a fit with $c_{LL}^B=0$. The top centre panel shows the fit prediction in the $c_{LL}^B$--$\phi_B$ plane in a fit with $c_{LL}^K=0$. The preference for non-SM values of the parameters in both cases arises from the tension in the SM CKM fit between $\epsilon_K$ (when using the experimental data for $V_{cb}$ and $\sin2\beta$ as inputs) and $S_{\psi K_S}=\sin2\beta$ \cite{Lunghi:2008aa,Buras:2008nn,Altmannshofer:2009ne,Lunghi:2010gv,Bevan:2010gi,Brod:2011ty}. As shown in Section \ref{cha:SMCKM}, this tension can be solved either by increasing $\epsilon_K$ (as in the first case) or by decreasing $S_{\psi K_S}$ by means of a new physics contribution to the $B_d$ mixing phase (as in the second case). In the second case, also a positive contribution to $S_{\psi\phi}=-\sin2\phi_s$ is generated.
The top right panel shows the fit prediction in a fit where $\phi_B=0$ and $c_{LL}^B=c_{LL}^K\equiv c_{LL}$, i.e. the $U(3)^3$ or MFV limit. In that case, a positive $c_{LL}$ cannot solve the CKM tension, since it would lead to an increase not only in $\epsilon_K$, but also in $\Delta M_{d,s}$.

The two plots in the bottom row of Fig.~\ref{fig:DF2fits} show the projections onto the $c_{LL}^K$--$c_{LL}^B$ and $c_{LL}^B$--$\phi_B$ planes of the fit with all 3 parameters in (\ref{eq:LeffDS2},\ref{eq:LeffDB2}) non-zero. Since both solutions to the CKM tension now compete with each other, the individual parameters are less constrained individually.
In the case of supersymmetry with dominance of gluino contributions, as mentioned above, the modification of the $B_{d,s}$ and $K$ mixing amplitudes is correlated by the common loop function $F_0$, which depends on the gluino and left-handed sbottom masses. Taking into account direct constraints from LHC on the sbottom and gluino masses, one finds that values of $F_0$ above about 0.04 are disfavoured. This constraint is shown in the bottom-left panel of Fig.~\ref{fig:DF2fits} as a gray region. However, we note that this bound (but not the correlation predicted by $U(2)^3$) is invalidated once chargino contributions dominate.

\begin{table}[tb]
\renewcommand{\arraystretch}{1.0}
 \begin{center}
\begin{tabular}{llllll}
\hline
$|V_{ud}|$ & $0.97425(22)$ &\cite{Hardy:2008gy}& $f_K$  & $(155.8\pm1.7)$ MeV & \cite{Laiho:2009eu}\\
$|V_{us}|$ & $0.2254(13)$ &\cite{Antonelli:2010yf}& $\hat B_K$ & $0.737\pm0.020$ &\cite{Laiho:2009eu} \\
$|V_{cb}|$ & $(40.6\pm1.3)\times10^{-3}$ &\cite{Nakamura:2010zzi}& $\kappa_\epsilon$ & $0.94\pm0.02$ & \cite{Buras:2010pza}\\
$|V_{ub}|$ & $(3.97\pm0.45)\times10^{-3}$ &\cite{Kowalewski:2011zz}& $f_{B_s}\sqrt{\hat B_s}$  & $(288\pm15)$ MeV &\cite{Lunghi:2011xy}\\
$\gamma_{\rm CKM}$ & $(74\pm11)^\circ$ &\cite{Bevan:2010gi}& $\xi$ & $1.237\pm0.032$ &\cite{Laiho:2009eu}\\
$|\epsilon_K|$ & $(2.229\pm0.010)\times10^{-3}$ &\cite{Nakamura:2010zzi} &$\eta_{tt}$&$0.5765(65)$&\cite{Buras:1990fn}\\ 
$S_{\psi K_S}$ & $0.673\pm0.023$ &\cite{Asner:2010qj} &$\eta_{ct}$&$0.496(47)$&\cite{Brod:2010mj}\\
$\Delta M_d$ & $(0.507\pm0.004)\,\text{ps}^{-1}$ &\cite{Asner:2010qj} &$\eta_{cc}$&$1.87(76)$&\cite{Brod:2011ty}\\
$\Delta M_s/\Delta M_d$ & $(35.05\pm0.42)$ &\cite{Abulencia:2006ze,Asner:2010qj} &&&\\
$\phi_s$ & $-0.002 \pm 0.087$ & \cite{LHCb-TALK-2012-029} &&&\\
\hline
 \end{tabular}
 \end{center}
\caption{Observables and hadronic parameters used as input to the $\Delta F=2$ fits.}
\label{tab:inputs}
\end{table}

\begin{figure}
\begin{center}
\includegraphics[width=0.6\textwidth]{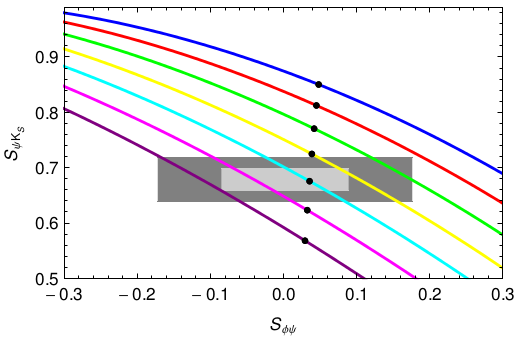}
 \label{fig:BurasVub}
 \caption{$S_{\psi\phi}$ vs. $S_{\psi K_S}$ for different values of $|V_{ub}|$.
 From top to bottom: $|V_{ub}|= 0.0046$ (blue), 0.0043 (red), 0.0040 (green), 0.0037 (yellow),
0.0034 (cyan), 0.0031 (magenta), 0.0028 (purple). Black dots: $\phi_B = 0$. Light and dark gray: experimental 1 and 2 $\sigma$ regions.}
\end{center}
\end{figure}

Before moving to $\Delta F = 1$ processes, we come back to the fact that solving the CKM tension via a new $B_{d,s}$ mixing phase implies a positive contribution to $S_{\psi\phi}$. This conclusion holds if one considers the central value for $|V_{ub}|$, as we did.
In light of the experimental tensions in its determination (see Section \ref{sec:CKM_tensions} for a discussion), one could have instead let this parameter free to vary in a given range, as the authors of \cite{Buras:2012sd} did. Their result is that for small values of $|V_{ub}|$ allowed by the CKM fit, also a negative contribution to $S_{\psi\phi}$ can be generated in the $U(2)^3$ framework. Their Fig. \ref{fig:BurasVub} makes this evident
, where one can also visualize the $U(3)^3$ case for comparison, represented by black dots. In other words, a triple correlation $S_{\psi\phi}$--$S_{\psi K_S}$--$|V_{ub}|$ is obtained: a precise measurements of two of these observables would result in a prediction for the third one, that could allow to test the $U(2)^3$ picture.

Finally, a solution of the tension could be tested by improvements in the experimental determination of $\gamma_{\rm CKM}$, as evident e.g. from Figures \ref{fig:CKMfitterTotal} and \ref{fig:CKMTensions}.

\subsection{$\Delta F=1$ processes}\label{sec:DF1Minimal}

The $U(2)^3$ predictions for $\Delta F=1$ processes are more model-dependent because a larger number of operators is relevant. In addition, the main prediction of universality of $b\to s$ and $b\to d$ amplitudes (but not $s\to d$ amplitudes) is not well tested. Firstly, current data are better for $b\to s$ decays compared to $b\to d$ decays. Secondly, the only clean $s\to d$ processes are $K\to\pi\nu\bar\nu$ decays, but $b\to q\nu\bar\nu$ processes have not been observed yet.
Thus, in the following we will only take into account data from inclusive and exclusive $b\to s$ decays, making use of the results of \cite{Altmannshofer:2011gn}, and present the constraints on the effective operators ($v = 246$ GeV)
\begin{align}
&\Delta \mathcal{L}_\text{mag}^{\Delta B = 1} =\sum_{i=d,s} \xi_{ib} m_{b}
\bigg[
\frac{c_{7\gamma}e^{i\phi_{7\gamma}}}{\Lambda^2} \left( {\bar d^i}_L \sigma_{\mu\nu}  b_R\right) e F^{\mu\nu}
+
\frac{c_{8g}e^{i\phi_{8g}}}{\Lambda^2} \left( {\bar d^i}_L \sigma_{\mu\nu} T^a b_R \right) g_s G^{\mu\nu\,a}\bigg] +\text{h.c.}\,,
\label{eq:LeffDB1mag}\\
& \Delta \mathcal{L}_L^{4f,\Delta B = 1} = \sum_{i=d,s} \xi_{ib}
\bigg[\frac{c_Le^{i\phi_{L}}}{\Lambda^2} \left( {\bar d}_L^i \gamma_{\mu} b_L \right) \left( {\bar l}_L \gamma_{\mu} l_L \right)+
\frac{c_Re^{i\phi_{R}}}{\Lambda^2} \left( {\bar d}_L^i \gamma_{\mu} b_L \right) \left( {\bar e}_R \gamma_{\mu} e_R \right) \nonumber\\
& \qquad \qquad \qquad \qquad+ \frac{c_He^{i\phi_{H}}}{\Lambda^2} \frac{v^2}{2} \left( {\bar d}_L^i  \gamma_\mu b_L \right)\frac{g}{c_w}Z^{\mu}
\bigg]
+\text{h.c.}\,,
\label{eq:LeffDB1L}
\end{align}
where the coefficients $c_{7\gamma,8g}, c_{L,R,H}$ are real and in general all the coefficients in~\eqref{eq:LeffDB1mag},\eqref{eq:LeffDB1L} can be relevant and of $O(1)$, while in the supersymmetric case only $c_{7\gamma}$ and $c_{8g}$ are relevant (see Section \ref{sec:BdecaySusy}).
Since the chromomagnetic penguin operator enters the observables considered in the following only through operator mixing with the electromagnetic one, we will ignore $c_{8g}$ in the following.

\begin{figure}[tb]
\centering
\includegraphics[width=0.68\textwidth]{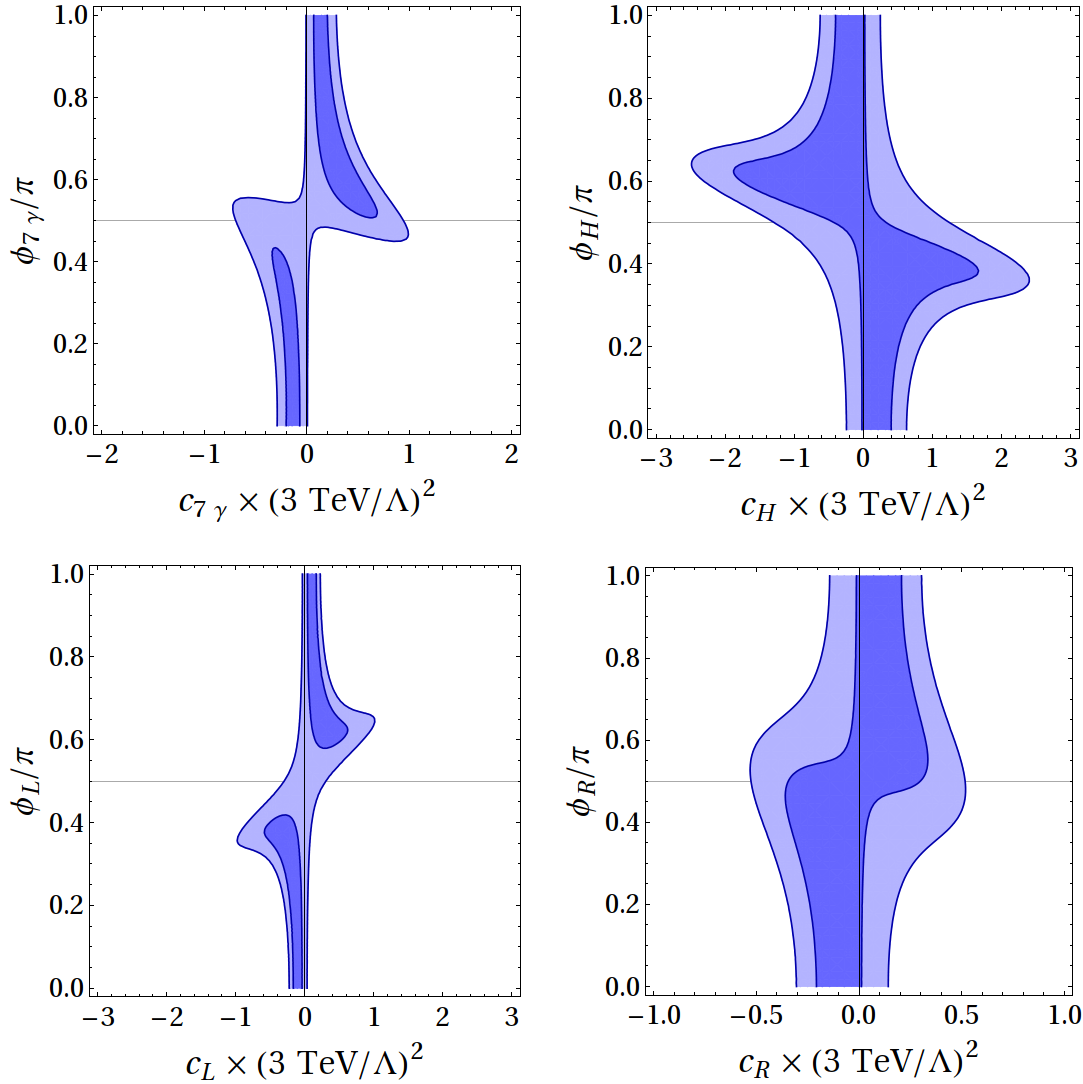}
\caption{68 and 95\% C.L. allowed regions for the $\Delta F=1$ coefficients in $U(2)^3$, using the results of a global analysis of inclusive and exclusive $b\to s$ decays, as well as $B_s \to \mu \mu$ \cite{Altmannshofer:2013foa}. Courtesy of David M. Straub.}
\label{fig:DF1fits}
\end{figure}

Fig.~\ref{fig:DF1fits} shows the constraints on the coefficients of the four operators in~\eqref{eq:LeffDB1mag},\eqref{eq:LeffDB1L} and their phases. They are updated with the analysis of Ref. \cite{Altmannshofer:2013foa}, which includes the most recent results for $B_s \to \mu\mu$ and the various $b\to s$ decays. The constraints are particularly strong for the magnetic penguin operator and the semi-leptonic left-left vector operator. In the first case, this is due to the $B\to X_s\gamma$ branching ratio, in the second case due to the angular observables in $B\to K^*\mu^+ \mu^-$.
Interestingly, in both cases, the constraint is weaker for maximal phases of the new physics contribution, since the interference with the (real) SM contribution is minimized in that case. For the two operators in the right-hand column of Fig.~\ref{fig:DF1fits}, this effect is slightly less pronounced. The reason is that both coefficients are accidentally small in the SM: in the first case due to the small $Z$ coupling to charged leptons proportional to $(1-4s_w^2)$ and in the second case due to $C_9(\mu_b)\approx-C_{10}(\mu_b)$ (in the convention of \cite{Altmannshofer:2011gn}).

To ease the interpretation of Fig.~\ref{fig:DF2fits},~\ref{fig:DF1fits}, all the coefficients are normalized, as explicitly indicated, to a scale $\Lambda = 3$ TeV, which might represent both the scale of a new strong interaction responsible for EWSB, $\Lambda_s \approx 4\pi v$, or the effective scale from loops involving the exchange of some new weakly interacting particle(s) of mass of $O(v)$. Interestingly the fits of the current flavour data are generally consistent with coefficients of order unity, at least if there exist sizable non vanishing phases, when they are allowed. Note in this respect that $U(3)^3$ at low $\tan{\beta}$ does not allow phases in $c_{L, R, H}$ in Fig.~\ref{fig:DF1fits}, with correspondingly more stringent constraints especially on $c_L$.
A possible interpretation of these Fig.~\ref{fig:DF2fits},~\ref{fig:DF1fits} is that the current flavour data are at the level of probing the $U(2)^3$ hypothesis in a region of parameter space relevant to several new theories of EWSB. Needless to say, the presence of phases in flavour-diagonal chirality breaking operators has to be consistent with the limits coming from the neutron Electric Dipole Moment. This observation will find quantitative support in Section \ref{sec:NeutronEDM}. It should also be noted that the coefficients of the flavour diagonal operators, analogue of the ones in \eqref{eq:LeffDB1L}, are limited by the ElectroWeak Precision Tests at a similar level to the ones shown in Fig. \ref{fig:DF1fits}.

\subsubsection{Analysis of $\epsilon'/\epsilon$} 
An observable that has not been included in the previous analysis and is actually relevant in other contexts as well, like in Minimal Flavour Violation (MFV) \cite{D'Ambrosio:2002ex}, is direct CP violation in $K$ decays, as summarized in the parameter $\epsilon^\prime$.
 Either in $U(2)^3$ or in MFV, a contribution to $\epsilon^\prime$ arises from the operators
 \begin{equation}\label{operatorsepsilonprime}
 \Delta \mathcal{L}^{4f, \Delta S = 1}_{LR} = \frac{1}{\Lambda^2} \xi_{ds} (c^d_5 \mathcal{O}^d_5 + c^u_5 \mathcal{O}^u_5 + c^d_6 \mathcal{O}^d_6 + c^u_6 \mathcal{O}^u_6) + \text{h.c.}, ~\xi_{ds} = V_{td}V_{ts}^*\, ,
 \end{equation}
 where
 \begin{align}
 \mathcal{O}_5^q = ( \bar{d}_L \gamma_\mu s_L ) (\bar{q}_R \gamma_\mu q_R), \qquad
 \mathcal{O}_6^q = ( \bar{d}_L^\alpha \gamma_\mu s_L^\beta) ( \bar{q}_R^\beta \gamma_\mu q_R^\alpha),\qquad q=u,d.
\end{align}

 The dominant contribution to the $\epsilon'$ parameter reads
\begin{equation}
\left|\frac{\epsilon'}{\epsilon}\right| \simeq \frac{\left| \text{Im}A_2 \right|}{\sqrt{2} \, |\epsilon| \,\text{Re}A_0} \,,
\end{equation}
where $A_i=A(K \to (\pi \pi)_{I=i})$. 
Using $\langle (\pi \pi)_{I=2} | \mathcal{O}_i^u + \mathcal{O}_i^d | K \rangle \simeq 0$ from isospin conservation and neglecting contributions from other operators, which are subleading,  we obtain
\begin{align}
 \text{Im}A_2 = \frac{1}{\Lambda^2}\left[ \left( C_5^d-C_5^u\right) \langle (\pi \pi)_{I=2} | \mathcal{O}_5^d| K \rangle + \left( C_6^d-C_6^u\right) \langle (\pi \pi)_{I=2} | \mathcal{O}_6^d| K \rangle \right].
\end{align}
From \cite{Bosch:1999wr} we have at the scale $\mu = m_c$
\begin{align}
\langle (\pi \pi)_{I=2} | \mathcal{O}_5^d | K \rangle & \simeq -\frac{1}{6 \sqrt{3}}\big( m_K^2 \rho^2 - m_K^2 + m_\pi^2\big) f_\pi  B_7^{(3/2)}(m_c), \nonumber \\
\langle (\pi \pi)_{I=2} | \mathcal{O}_6^d | K \rangle & \simeq -\frac{1}{2 \sqrt{3}} \big( m_K^2 \rho^2 - \frac{1}{6} ( m_K^2 - m_\pi^2) \big)  f_\pi B_8^{(3/2)}(m_c) ,
\end{align}
where $\rho = m_K/m_s$. In the following we set $B_7^{(3/2)}(m_c) =  B_8^{(3/2)}(m_c) =1$.

The coefficients $C^{(3/2)}_i = C_i^d-C_i^u$ at the low scale $\mu$ read in terms of those at the high scale $\Lambda$\cite{Buchalla:1995vs}
\begin{align}
C^{(3/2)}_5 (m_c) &= \eta_5 C^{(3/2)}_5 (\Lambda), \nonumber \\
C^{(3/2)}_6 (m_c) &= \eta_6 C^{(3/2)}_6(\Lambda) + \frac{1}{3}(\eta_6 - \eta_5) C^{(3/2)}_5(\Lambda),
\end{align}
where
\begin{align}
\eta_5 &= \left(\frac{\alpha_s(\Lambda)}{\alpha_s(m_t)}\right)^{\frac{3}{21}} \left(\frac{\alpha_s(m_t)}{\alpha_s(m_b)}\right)^{\frac{3}{23}}\left(\frac{\alpha_s(m_b)}{\alpha_s(m_c)}\right)^{\frac{3}{25}}  \simeq 0.82\,,\notag \\
\eta_6 &= \left(\frac{\alpha_s(\Lambda)}{\alpha_s(m_t)}\right)^{-\frac{24}{21}} \left(\frac{\alpha_s(m_t)}{\alpha_s(m_b)}\right)^{-\frac{24}{23}} \left(\frac{\alpha_s(m_b)}{\alpha_s(m_c)}\right)^{-\frac{24}{25}} \simeq 4.83 \,.
\end{align}

Requiring the extra contribution from $\Delta \mathcal{L}^{4f, \Delta S = 1}_L$ to respect $|\epsilon^\prime/\epsilon | < |\epsilon^\prime/\epsilon |_\text{exp} \simeq 1.7\times 10^{-3}$, we obtain 
\begin{align}\label{boundepsilonprime}
c_5^{u,d} &\lesssim 0.4 \left(\frac{\Lambda}{3~\text{TeV}}\right)^2, & c_6^{u,d} &\lesssim 0.13 \left(\frac{\Lambda}{3~\text{TeV}}\right)^2.
\end{align}
Taking into account the uncertainties in the estimate of the SM contribution to $\epsilon^\prime/\epsilon$, which could cancel against a new physics contribution, as well as the uncertainties in the $B_i$ parameters, this bound might perhaps be relaxed by a factor of a few\footnote{Note that in Supersymmetry the heaviness of the first generation squark circulating in the box loop suppresses the coefficients $c_5^{u,d}$ and $c_6^{u,d}$.}.

\subsection{Up quark sector in Minimal $U(2)^3$}\label{sec:up}
An analogous analyisis can be performed for operators involving the up quarks. However if these operators are weighted by the same scale $\Lambda$ as for the down quarks, they are phenomenologically irrelevant unless some of the relative dimensionless coefficients are at least one order of magnitude bigger than the ones in Eqs.~(\ref{eq:LeffDS2}-\ref{eq:LeffDB2}) and (\ref{eq:LeffDB1mag}-\ref{eq:LeffDB1L}). This is in particular the case for operators contributing to $D-\bar{D}$ mixing, to direct CP violation in $D$-decays or to top decays, $t\rightarrow c \gamma$ or $t\rightarrow c Z$, and to the top chromo-electric dipole moment (CEDM), to which now we turn our attention.

Within our setup, the relevant effective operators for the above processes are:
\begin{align}
\mathcal H_{\rm LL}^{\rm D}  & = \frac{c_{LL}^D }{\Lambda^2}\, \xi_{uc}^2 \, \dfrac{1}{2}\left({\bar u}_L
\gamma_\mu  c_L \right)^2, \label{eq:Dmixing} \\
\mathcal H_{\rm cb}^{\rm D}  & = \frac{c_{g}^{D} e^{i \phi_{g}^{D}}}{\Lambda^2} \,m_c \,\xi_{uc}
\left( {\bar u}_L \sigma_{\mu \nu} c_R \right) g_s G_{\mu \nu},
\label{eq:Ddecay}\\
\mathcal H_{\rm cb}^{\rm t,\, \alpha} & = \frac{c_{\alpha}^{t} e^{i \phi_{\alpha}^{t}}}{\Lambda^2}
\,m_t \,\xi_{ct} \left( {\bar c}_L \sigma_{\mu \nu} t_R \right) O^\alpha_{\mu \nu},
\quad O^\alpha_{\mu \nu} = e F_{\mu \nu}, ~\frac{g}{c_w} Z_{\mu \nu},
\label{eq:topcb} \\
\mathcal H_{\rm cc}^{\rm t} & = \frac{c_\text{cc}^{t} e^{i \phi_\text{cc}^{t}}}{\Lambda^2}\frac{v^2}{2} \,\xi_{ct}
\left({\bar c}_L \gamma_\mu t_L \right) \frac{g}{c_w} Z_\mu,
\label{eq:topcc}\\
\mathcal H_{\rm dm}^t & = \frac{c_\text{dm} e^{i \phi_\text{dm}}}{\Lambda^2}\,m_t
\left( {\bar t}_L \sigma_{\mu \nu} t_R \right) g_s G_{\mu \nu},
\end{align}
where
$\xi_{uc} = V_{ub}V_{cb}^*$, $\xi_{ct} = V_{cb}V_{tb}^*$ and $c_{LL}^{D}$, $c_g^D$, $c_{\alpha}^{t}$, $c_\text{cc}^{t}$,
$c_\text{dm}$ are real parameters, with the phases made explicit wherever present.
All these coefficients are model dependent and, in principle, can be of $O(1)$.
Since $\Lambda \gg v$, the requirement of $SU(2)_L$ invariance correlates $c_{LL}^{D}$, $c_\text{cc}^{t}$ and
$\phi_\text{cc}^{t}$ with the analogous parameters in the down sector. One can easily see they have to be equal
within a few percent, and so they have to respect bounds similar to those for $c_{LL}^K$, $c_H$ and
$\phi_H$ (see Figures \ref{fig:DF2fits} and \ref{fig:DF1fits}).

\subsubsection{$D$ mixing and decay}

In the neutral $D$ meson system the SM short distance contribution to the mixing is orders of magnitudes
below the long distance one, thus complicating the theoretical calculation of the mass and width splittings
$x$ and $y$ (see \cite{Gedalia:2009kh}, also for a discussion of the relevant parameters).
Despite the above uncertainties, many studies (see \cite{Falk:2001hx, Falk:2004wg} and references therein)
indicate that the standard model could naturally account for
the values $x \sim y \sim 1 \%$, thus explaining the measured $95\%$ CL intervals
$x \in [0.19,0.97]\%$, $y \in [0.54,1.05]\%$ \cite{Asner:2010qj}.

Here, like in \cite{Gedalia:2009kh, Isidori:2011qw}, we take the conservative approach
of using the above data as upper bounds to constrain new physics contributions.
Referring to the analysis carried out in \cite{Gedalia:2009kh, Isidori:2011qw}, within our framework it turns
out that the most effective bound is the one on the coefficient in the operator $\mathcal H_{\rm LL}^{\rm D}$.
In our notation it reads
\beq
{c_{LL}^D}^2 \left(\dfrac{3 \,\text{TeV}}{\Lambda}\right)^2 < 90,
\label{eq:Dmix}
\eeq
so to saturate it we would need values of $c_{LL}^D$ that are excluded, since they would imply a too large contribution to
$\Delta F = 2$ observables in the down sector.

Let us now turn our attention to direct CP violation in $D$ decays \cite{Aaij:2011in}. The quantity of interest is the difference between the time-intagrated CP asymmetries in the decays
$D^0 \to K^+ K^-$ and $D^0 \to \pi^+ \pi^-$, for which the HFAG reports the world average $\Delta$A$_\text{CP} = A_{KK} - A_{\pi \pi} = -(0.33 \pm 0.12)$ \cite{Amhis:2012bh}.
In reference \cite{Isidori:2011qw} all the possible effective operators contributing to the asymmetry are considered, while respecting at the same time the bounds coming from $D$-$\bar{D}$ mixing and from $\epsilon_K'/\epsilon_K$.
Following that analysis, the only operator that can give a relevant contribution in Minimal $U(2)^3$ is
$\mathcal H_{\rm cb}^{\rm D}$, $\Delta a_\text{CP}$ being proportional to the imaginary part of the relative coefficient.
Referring to the estimations carried out in \cite{Isidori:2011qw} and \cite{Giudice:2012qq} for the hadronic
matrix elements\footnote{A more explicit discussion of how to extract bounds from this observable, on $\mathcal H_{\rm cb}^{\rm D}$ and other operators, is presented in \ref{sec:U2gen_DeltaAcp}.}, the measured value of $\Delta$A$_\text{CP}$ imposes an upper bound
\beq
c_g^D \sin ({\rm arg} \xi_{uc} + \phi_g^D) \left( \frac{3\, \text{TeV}}{\Lambda}\right)^2 \lesssim 20,
\label{eq:Ddevay}
\eeq
a value out of reach if we want to keep the parameter $c_g^D$ to be of order one.

\subsubsection{Top FCNC and dipole moments}

The LHC sensitivity at 14 TeV with $100 \, \text{fb}^{-1}$ of data is expected to be (at $95 \%$ CL)
\cite{Carvalho:2007yi}:
$\text{BR}(t \to c\, Z, \,u\, Z) \simeq 5.5 \times 10^{-5}$ and $\text{BR}(t \to c\, \gamma,\, u\,\gamma)
\simeq 1.2 \times 10^{-5}$. Here we concentrate on the charm channels, since both in the SM and in our framework
the up ones are CKM suppressed. In the SM, $\text{BR}(t \to c\, Z,\, c\, \gamma)$ can be estimated to be
of order $ (m_b^2/m_W^2)^2 \, |V_{cb}|^2 \, \alpha^2/s_w^2 \sim 10^{-12}$, so that an experimental observation
will be a clear signal of new physics. To estimate the $U(2)^3$ effects for these processes, we
follow the analysis carried out in \cite{Fox:2007in}. The dominant contributions are those given by the operators
$\mathcal H_{\rm cb}^{\rm t,\, \gamma}$ for $t \to c \gamma$, and $\mathcal H_{\rm cb}^{\rm t,\, Z}$ and
$\mathcal H_{\rm cc}^{\rm t}$ for $t \to c Z$. We obtain
\begin{align}
\text{BR}(t \to c\, \gamma) \simeq & \; 1.7 \times 10^{-8} \left(\dfrac{3 \, \text{TeV}}{\Lambda}\right)^4
{c_{\gamma}^t}^2,\\
\text{BR}(t \to c\, Z) \simeq & \; 8.5 \times 10^{-8} \left(\dfrac{3 \, \text{TeV}}{\Lambda}\right)^4 
\left(0.61 \, {c_Z^t}^2 + 0.39 \, {c_\text{cc}^t}^2 + 0.83 \, c_Z^t \, c_\text{cc}^t \cos (\phi_\text{cc}^t - \phi_Z^t)\right),
\label{eq:topBR}
\end{align}
leading us to conclude that any non-zero evidence for these decays at the LHC could not be explained in our setup,
unless we allow the dimensionless coefficients to take values more than one order of magnitude bigger
than the corresponding ones in the down sector (actually this could be possible only for $c_Z^t$ and $c_{\gamma}^t$
but not for $c_\text{cc}^t$, because of its correlation with $c_H$ and of the bounds of Fig.~\ref{fig:DF1fits}).

The recent analysis carried out in \cite{Kamenik:2011dk} has improved previous bounds \cite{CorderoCid:2007uc}
on the top CEDM $\tilde{d}_t$
by two orders of magnitude, via previously unnoticed contributions of $\tilde{d}_t$ to the neutron
electric dipole moment (EDM).
In deriving this bound, the authors of \cite{Kamenik:2011dk} have assumed
the up and down quark EDMs $d_{u,d}$ and CEDMs $\tilde{d}_{u,d}$ to be negligible.
This is relevant in our context if
\begin{itemize}
 \item we allow for generic phases outside the spurions $\V$,
 $\Delta Y_u$ and $\Delta Y_d$,
 \item we assume that some other mechanism is responsible for making $d_{u,d}$ and $\tilde{d}_{u,d}$
 negligible. Notice that this is actually the case in SUSY with heavier first two generations, where
 on the contrary there is no further suppression of $\tilde{d}_t$ with respect to the EFT natural estimate.
\end{itemize}
Then,
the bound given in \cite{Kamenik:2011dk} imposes
\beq
c_\text{dm} |\sin \phi_\text{dm}|\, \left( \dfrac{3\, \text{TeV}}{\Lambda} \right)^2 < 0.6 \,,
\label{topCEDM}
\eeq
so that future experimental improvements in the determination of the neutron EDM
will start to challenge the $U(2)^3$ scenario with CP violating phases outside the spurions, if the
hypothesis of negligible $d_{u,d}$ and $\tilde{d}_{u,d}$ is realized.

\subsection{Electric dipole moment of the neutron}
\label{sec:NeutronEDM}
The presence of phases in flavour-diagonal chirality breaking operators has to be consistent with the limits coming from the neutron electric dipole moment (EDM). To make the statement more precise, we derive a quantitative bound for the Minimal breaking case, also for comparison with the General breaking case, analyzed in the next section.

The relevant contraints come from the CP violating contributions to the operators
\begin{align}\label{EDM}
\Delta\mathcal{L}_\text{mag}^{\Delta F=0} &= \frac{1}{\Lambda^2} \left[
\tilde c_u^g e^{i\tilde \phi_u^g} m_u(\bar u_L\sigma_{\mu\nu}T^a u_R)
+ \tilde c_d^g e^{i\tilde\phi_d^g} m_d(\bar d_L\sigma_{\mu\nu}T^a d_R)
\right] g_s G^{\mu\nu}_a \nonumber \\
 &+
 \frac{1}{\Lambda^2} \left[
\tilde c_u^\gamma e^{i\tilde\phi_u^\gamma} m_u(\bar u_L\sigma_{\mu\nu} u_R)
+ \tilde c_d^\gamma e^{i\tilde\phi_d^\gamma} m_d(\bar d_L\sigma_{\mu\nu} d_R)
\right] eF^{\mu\nu} +\text{h.c.}\, ,
\end{align}
where we have made all the phases explicit. 
In terms of the coefficients of \eqref{EDM}, the up and down quark electric dipole moments (EDMs) and chromoelectric dipole moments (CEDMs), defined as in \cite{Pospelov:2000bw}, are
\begin{align}
d_{q} = 2 e \, \frac{m_q}{\Lambda^2}\,\tilde c_{q}^{\gamma}\sin(\tilde\phi_i^{\gamma}),\qquad \tilde d_q = 2 \, \frac{m_q}{\Lambda^2}\, \tilde c_q^g\sin(\tilde \phi_q^g),\qquad q=u,d.
\end{align}
The contribution to the neutron EDM reads \cite{Pospelov:2000bw}
\begin{equation} \label{neutron}
d_n = (1 \pm 0.5) \left( 1.4( d_d - \tfrac{1}{4} d_u) + 1.1 e ( \tilde{d}_d + \tfrac{1}{2} \tilde{d}_u)  \right) \, ,
\end{equation}
where all the coefficients are defined at a hadronic scale of 1~GeV.

Taking into account the RG evolution between 3~TeV and the hadronic scale, the 90\% C.L. experimental bound
$|d_n| < 2.9 \times 10^{-26} ~e\,\text{cm}$ \cite{Baker:2006ts}
implies for the parameters at the high scale
\begin{align}
\tilde c_{u}^{\gamma}\sin(\tilde\phi_u^{\gamma}) &\lesssim 1.9 \times 10^{-2}\left(\frac{\Lambda}{3~\text{TeV}}\right)^2, & \tilde c_{d}^{\gamma}\sin(\tilde\phi_d^{\gamma}) \lesssim 2.4 \times 10^{-3}\left(\frac{\Lambda}{3~\text{TeV}}\right)^2,\\
\tilde{c}_{u}^g\sin(\tilde\phi_u^{g}) &\lesssim 7.1 \times 10^{-3}\left(\frac{\Lambda}{3~\text{TeV}}\right)^2, & \tilde{c}_{d}^g\sin(\tilde\phi_d^{g}) \lesssim 1.8 \times 10^{-3}\left(\frac{\Lambda}{3~\text{TeV}}\right)^2.
\end{align}
Note that the bounds are automatically satisfied if one does not allow for phases outside the spurions. Otherwise, even with generic phases $\tilde \phi_{u,d}^{\gamma,g}$, the smallness of the coefficients can be explained taking the new physics scale related to the first generation decoupled from the scale $\Lambda$ of EWSB, as allowed by the $U(2)^3$ symmetry.
This can be realized in concrete models such as supersymmetry, where the operators in \eqref{EDM} come from Feynman diagrams involving the exchange of heavy first generation partners. Note that this possibility is absent in MFV, making $U(2)^3$ a somehow more attractive framework also from the point of view of flavour blind CP violation.

\section{Extra-features of Generic $U(2)^3$}

Generic $U(2)^3$, introducing physical rotations in the right handed sector as well, gives rise to extra flavour and CP violating contributions in \eqref{eq:genL}. The most significant of them are contained in  $\Delta \mathcal{L}_\text{mag}$ and in $\Delta \mathcal{L}^{4f}_{LR}$ of \eqref{eq:genL}. In the following, we first discuss the relevant new effects with respect to Minimal~$U(2)^3$, which show up in $\Delta C = 1$, $\Delta S = 1$ and $\Delta S = 2$ observables, as well as in flavour conserving electric dipole moments. We then see how in $B$ and $t$ decays and in $D$-$\bar{D}$ mixing the new effects are at most analogous in magnitude to those of the Minimal breaking case.

\subsection{$\Delta C=1$: $D$ decays}
\label{sec:U2gen_DeltaAcp}

CP asymmetries in $D$ decays receive contributions from chromo-magnetic dipole operators with both chiralities,
\begin{equation}
\Delta \mathcal{L}^{\Delta C=1}_\text{mag} = \frac{1}{\Lambda^2} c_D^g e^{i \phi_D^g}
\zeta_{uc} \left[
e^{-i\phi_2^u} \frac{\epsilon^u_R}{\epsilon_L}\, \mathcal{O}_8
+
e^{i\phi_1^u} \frac{s^u_R}{s^u_L} \frac{\epsilon_R^u}{\epsilon_L}\, \mathcal{O}_8'
\right] + \text{h.c.}\\
\end{equation}
where
\begin{equation}
\mathcal{O}_8 = m_t(\bar u_L\sigma_{\mu\nu}T^a c_R)g_s G^{\mu\nu}_a,\qquad \mathcal{O}_8' = m_t(\bar u_R\sigma_{\mu\nu}T^a c_L)g_s G^{\mu\nu}_a,
\end{equation}
and with $\phi_D^{g}$ we account for the possibility of CP violating phases outside the spurions (see Appendix \ref{app:bilinears} for details). 
Most notably, the recently observed CP asymmetry difference between $D\to KK$ and $D\to \pi\pi$ decays provide the stronger constraint to new physics contributions to the chromo-magnetic operators. Following \cite{Isidori:2011qw,Giudice:2012qq} we write at the scale $\mu = m_c$
\begin{equation}
\Delta A_\text{CP} \simeq - \frac{2}{\lambda} \Big[\text{Im}(V_{cb}^* V_{ub}) \text{Im}\left(\Delta R^{\text{SM}} \right) + \frac{1}{\Lambda^2}
\Big(
\text{Im}\big( C_8\big) \text{Im}\big( \Delta R^{\text{NP}} \big) +
\text{Im} ( C_8' ) \text{Im}\big( {\Delta R'}^{\text{NP}} \big) 
\Big) \Big]
,
\label{DeltaAcp}
\end{equation}
where the Cabibbo angle $\lambda$ is defined in \eqref{CKM}, $\Delta R^{(\prime)\text{SM},\text{NP}} = R_K^{(\prime)\text{SM},\text{NP}} + R_{\pi}^{(\prime)\text{SM}, \text{NP}}$, $R_{K,\pi}^{\text{SM}}$ are the ratios between the subleading and the dominant SM hadronic matrix elements, and
\begin{align}
\label{Dhadr}
{R_{K}^{(\prime)\text{NP}}} &\simeq V_{cs}^* V_{us} \frac{\langle K^+ K^- | \mathcal{O}_8^{(\prime)}| D\rangle}{\langle K^+ K^- | \mathcal{L}^{\text{SM}}_{eff}| D \rangle} \sim 0.1 \times \frac{4 \pi^2 m_t}{m_c} \frac{\sqrt{2}}{G_F},\\
R_{\pi}^{(\prime)\text{NP}} &\simeq V_{cd}^* V_{ud} \frac{\langle \pi^+ \pi^- | \mathcal{O}_8^{(\prime)}| D\rangle}{\langle \pi^+ \pi^- | \mathcal{L}^{\text{SM}}_{eff}| D \rangle} \simeq R_K^{(\prime)\text{NP}}.
\end{align}
In our estimates we will assume maximal strong phases, which imply $|\text{Im}\, \Delta R^{(\prime)\text{NP}}| \simeq 2 R_{K}^{\text(\prime){NP}}$. The SM contribution can be naively estimated to be $\Delta R^{\text{SM}} \sim \alpha_s(m_c)/\pi \sim 0.1$, but larger values from long distance contributions could arise.\footnote{At the time these results were published, the experimental central value of $\Delta A_\text{CP}$ was a factor of two larger than the current one and more than $3 \sigma$ away from zero \cite{Aaij:2011in,CDF-Note-10784}, making a possible explanation within the SM an open issue (see e.g \cite{Cheng:2012wr,Brod:2012ud,Isidori:2012yx}).}.
\\Requiring the new physics contribution to $\Delta A_\text{CP}$ to be less than updated world average \cite{Amhis:2012bh}, $\Delta A_\text{CP}^\text{exp}=(-0.33\pm0.12)\%$, 
implies
\begin{align}
c_D^{g}\, \frac{\epsilon^u_R}{\epsilon_L}\, \frac{\sin\left(\delta -\phi_2^u + \phi_D^{g}\right)}{\sin \delta}
&\lesssim 0.18
\left( \frac{\Lambda}{3\, {\rm TeV}} \right)^2
,
&
c_D^{g}\, \frac{s^u_R}{s^u_L} \frac{\epsilon^u_R}{\epsilon_L}\, \frac{\sin(\delta + \phi_1^u - \phi_D^{g})}{\sin \delta}
&\lesssim 0.18
\left( \frac{\Lambda}{3\, {\rm TeV}} \right)^2
.
\label{eq:DC1bound}
\end{align}
The bound can be saturated without violating indirect constraints on these operators arising from $\epsilon'$ or $D$-$\bar D$ mixing due to weak operator mixing \cite{Isidori:2011qw}. We stress that the bounds in \eqref{eq:DC1bound} carry an order one uncertainty coming from the normalized matrix elements $R_{\pi,K}$.

\subsection{$\Delta F=0$: neutron EDM}

In the flavour conserving case, important constraints arise from the up and down quark electric dipole moments (EDMs) and chromo-electric dipole moments (CEDMs). In addition to \eqref{EDM} there are new contributions coming from the CP violating part of the operators
\begin{align}\label{EDMright}
 \Delta \mathcal{L}^{\Delta F=0}_\text{mag} &= \frac{m_t}{\Lambda^2} \xi_{uu}\,e^{-i \phi_1^u} \frac{s^u_R}{s^u_L} \frac{\epsilon^u_R}{\epsilon_L}\left[
c_u^g e^{i\phi_u^g} (\bar u_L\sigma_{\mu\nu}T^a u_R)g_sG^{\mu\nu}_a
+ c_u^\gamma e^{i\phi_u^\gamma} (\bar u_L\sigma_{\mu\nu}u_R)eF^{\mu\nu}
\right] \nonumber \\
 &+
 \frac{m_b}{\Lambda^2} \xi_{dd} \,e^{-i \phi_1^d} \frac{s^d_R}{s^d_L} \frac{\epsilon^d_R}{\epsilon_L}\left[
c_d^g e^{i\phi_d^g} (\bar d_L\sigma_{\mu\nu}T^a d_R)g_sG^{\mu\nu}_a
+ c_d^\gamma e^{i\phi_d^\gamma} (\bar d_L\sigma_{\mu\nu}d_R)eF^{\mu\nu}
\right] + \text{h.c.}\, ,
\end{align}
where we remind that $\phi_1^{u,d}$ are non zero even if there are no CP phases outside the spurions. The new contributions to the quark (C)EDMs are
\begin{equation}
d_u = 2e\frac{m_t}{\Lambda^2}\xi_{uu}\frac{s^u_R}{s^u_L}\frac{\epsilon^u_R}{\epsilon_L}c_u^{\gamma}\sin(\phi_u^{\gamma} - \phi_1^u),\qquad \tilde d_u = 2\frac{m_t}{\Lambda^2}\xi_{uu} \frac{s^u_R}{s^u_L}\frac{\epsilon^u_R}{\epsilon_L}c_u^{g}\sin(\phi_u^{g} - \phi_1^u), \qquad (u\leftrightarrow d).
\end{equation}
From \eqref{neutron}, considering again the running of the Wilson coefficients from 3 TeV down to the hadronic scale of 1 GeV, the experimental bound on the neutron EDM implies for the parameters at the high scale
\begin{align}
c_u^{\gamma}\,|\sin (\phi_u^{\gamma}-\phi_1^u )| \frac{s^u_R}{s^u_L}\,\frac{\epsilon^u_R}{\epsilon_L}\lesssim 1.2 \times 10^{-2}
\left( \frac{\Lambda}{3\, {\rm TeV}} \right)^2
,
\nonumber\\
c_d^{\gamma}\,|\sin (\phi_d^{\gamma}-\phi_1^d )|  \frac{s^d_R}{s^d_L}\,\frac{\epsilon^d_R}{\epsilon_L}\lesssim 3.2 \times 10^{-2}
\left( \frac{\Lambda}{3\, {\rm TeV}} \right)^2
,
\nonumber\\
c_u^{g}\,|\sin (\phi_u^{g}-\phi_1^u )|  \frac{s^u_R}{s^u_L}\,\frac{\epsilon^u_R}{\epsilon_L}\lesssim 4.4 \times 10^{-3}
\left( \frac{\Lambda}{3\, {\rm TeV}} \right)^2
,
\nonumber\\
c_d^{g}\,|\sin (\phi_d^{g}-\phi_1^d )|  \frac{s^d_R}{s^d_L}\,\frac{\epsilon^d_R}{\epsilon_L}\lesssim 2.5 \times 10^{-2}
\left( \frac{\Lambda}{3\, {\rm TeV}} \right)^2
.
\label{eq:DF0bound}
\end{align}

Notice that since the operators of \eqref{EDMright} are generated through the right-handed mixings with the third generation, the coefficients $c_{u,d}^{\gamma, g}$ can no longer be
suppressed by the large scale associated with the first generations quarks (such as the mass scale of scalar or fermionic first generation quark partners in SUSY or composite Higgs models) as in the Minimal case, and the bounds above will constrain $s^u_R\epsilon^u_R$ and $s^d_R\epsilon^d_R$.

\subsection{$\Delta S=1$: $\epsilon'/\epsilon$}

The $s\to d$ chromomagnetic dipole
\begin{align}
 \Delta\mathcal{L}^{\Delta S = 1}_\text{mag} &= 
 \frac{m_t}{\Lambda^2} {c}_K^g e^{i(\phi_K^g-\phi_2^d)}\lambda_b \xi_{ds} \frac{\epsilon^d_R}{\epsilon_L} \left( \bar{d}_L \sigma_{\mu\nu} T^a s_R \right) g_s G_{\mu\nu}^a
\end{align}
contributes to $\epsilon'$. Following the analysis in \cite{Mertens:2011ts}, one obtains the bound
\begin{equation}
{c}_K^g \frac{\sin (\beta +\phi_K^g - \phi_2^d)}{\sin \beta}\,\frac{\epsilon^d_R}{\epsilon_L}
\lesssim 0.7 \left(\frac{\Lambda}{3 \,{\rm TeV}} \right)^2 \,.
\label{eq:DS1bound}
\end{equation}
Furthermore, in addition to the LR four fermion operators in \eqref{operatorsepsilonprime}, there is a contribution to $\epsilon'$ also from the operators with exchanged chiralities
\begin{equation}
\Delta \mathcal{L}^{4f, \Delta S = 1}_{LR} = \frac{1}{\Lambda^2} \xi_{ds} \frac{s^d_R}{s^d_L} \left(\frac{\epsilon^d_R}{\epsilon_L}\right)^2 e^{i(\phi_1^d-\phi_2^d)}(c^{\prime d}_5 \mathcal{O}^{\prime d}_5 + c^{\prime u}_5 \mathcal{O}^{\prime u}_5 + c^{\prime d}_6 \mathcal{O}^{\prime d}_6 + c^{\prime u}_6 \mathcal{O}^{\prime u}_6) + \text{h.c.}\, ,
\end{equation}
where
\begin{align}
 \mathcal{O}_5^{\prime q} = ( \bar{d}_R \gamma_\mu s_R ) (\bar{q}_L \gamma_\mu q_L), \qquad
 \mathcal{O}_6^{\prime q} = ( \bar{d}_R^\alpha \gamma_\mu s_R^\beta) ( \bar{q}_L^\beta \gamma_\mu q_L^\alpha),\qquad q=u,d.
\end{align}
From \eqref{boundepsilonprime} one gets
\begin{align}
c_5^{\prime u,d} \frac{\sin (\beta +\phi_1^d - \phi_2^d)}{\sin \beta}\, \frac{s^d_R}{s^d_L} \left(\frac{\epsilon^d_R}{\epsilon_L}\right)^2
&\lesssim 0.4\left(\frac{\Lambda}{3~\text{TeV}}\right)^2,\\
c_6^{\prime u,d} \frac{\sin (\beta +\phi_1^d - \phi_2^d)}{\sin \beta}\, \frac{s^d_R}{s^d_L} \left(\frac{\epsilon^d_R}{\epsilon_L}\right)^2
&\lesssim 0.13\left(\frac{\Lambda}{3~\text{TeV}}\right)^2,
\end{align}
which is not particularly relevant since a stronger bound on the combination $(s^d_R/s^d_L)(\epsilon^d_R/\epsilon_L)^2$ comes from $\epsilon_K$.

\subsection{$\Delta S=2$: $\epsilon_K$}
   
Finally, the only relevant new effect contained in $\Delta \mathcal{L}^{4f}_{LR}$ arises from $\Delta S=2$ operators contributing to $\epsilon_K$, which are enhanced by a chiral factor and by renormalization group effects. The relevant operators are
\begin{equation}
\Delta\mathcal{L}^{\Delta S=2}_{LR} = \frac{1}{\Lambda^2}\frac{s^d_R}{s^d_L} \left(\frac{\epsilon^d_R}{\epsilon_L}\right)^2\xi_{ds}^2 e^{i(\phi_1^d-\phi_2^d)}\left[ c_K^{SLR}\lambda_b^2  \left(\bar{d}_L s_R\right) \left(\bar{d}_R s_L \right) + c_K^{VLR} \left(\bar{d}_L \gamma_\mu s_L \right)\left(\bar{d}_R \gamma_\mu s_R \right)\right]
.
\end{equation}
Using bounds from \cite{Isidori:2010kg}, one gets
\begin{equation}
 c_K^{VLR} \frac{\sin (2 \beta +\phi_1^d - \phi_2^d)}{\sin 2 \beta}\, \frac{s^d_R}{s^d_L} \left(\frac{\epsilon^d_R}{\epsilon_L}\right)^2
\lesssim 6 \times 10^{-3} \left(\frac{\Lambda}{3 \,{\rm TeV}} \right)^2\,.
\label{eq:DS2bound}
\end{equation}

\subsection{$D$ mixing, $B$ and top FCNCs}

In the $D$ and $B$ systems there are no enhancements of the matrix elements of the operators in $\Delta \mathcal{L}_{LR}^{4f}$ and $\Delta \mathcal{L}_{R}^{4f}$, unlike what happens for $K$ mesons. Moreover the new contributions to these operators are all suppressed by some powers of $\epsilon^{u,d}_R/\epsilon_L$ (see Appendix \ref{app:bilinears}). Therefore they are all subleading with respect to those of Minimal $U(2)^3$, once we take into account the bounds from the other observables that we have discussed. An analogous suppression holds also for the operators that contain chirality breaking bilinears involving one third generation quark, relevant for $B$ and top FCNCs. 
A four fermion operator of the form $(\bar{u}_
L c_R)(\bar{u}_R c_L)$
might in principle be relevant for $D$-$\bar{D}$ mixing. However, taking into account the bounds on $\epsilon^u_R/\epsilon_L$,  this new contribution gives effects of the same size of those already present in Minimal $U(2)^3$ and far from the current sensitivity.
Consequently the phenomenology of $B$ decays is the same for Minimal and Generic $U(2)^3$. The only difference  in the latter is that CP violating effects are generated also if we set to zero the phases outside the spurions, though suppressed by at least one power of $\epsilon^d_R/\epsilon_L$.

Concerning the up quark sector, 
given the future expected sensitivities for top FCNCs \cite{Carvalho:2007yi} and CPV in $D$-$\bar{D}$ mixing \cite{Aushev:2010bq, Merk:2011zz}, within the $U(2)^3$ framework we continue to expect no significant effects in these processes (see Appendix \ref{sec:up} for the size of the largest contributions). We stress that, while an observation of a flavour changing top decay at LHC would generically put the $U(2)^3$ framework into trouble, a hypothetical observation of CP violation in $D$-$\bar{D}$ mixing would call for a careful discussion of the long distance contribution.

\subsection{Comparison of bounds}

\begin{figure}[tb]
\centering
\includegraphics[width=0.8\textwidth]{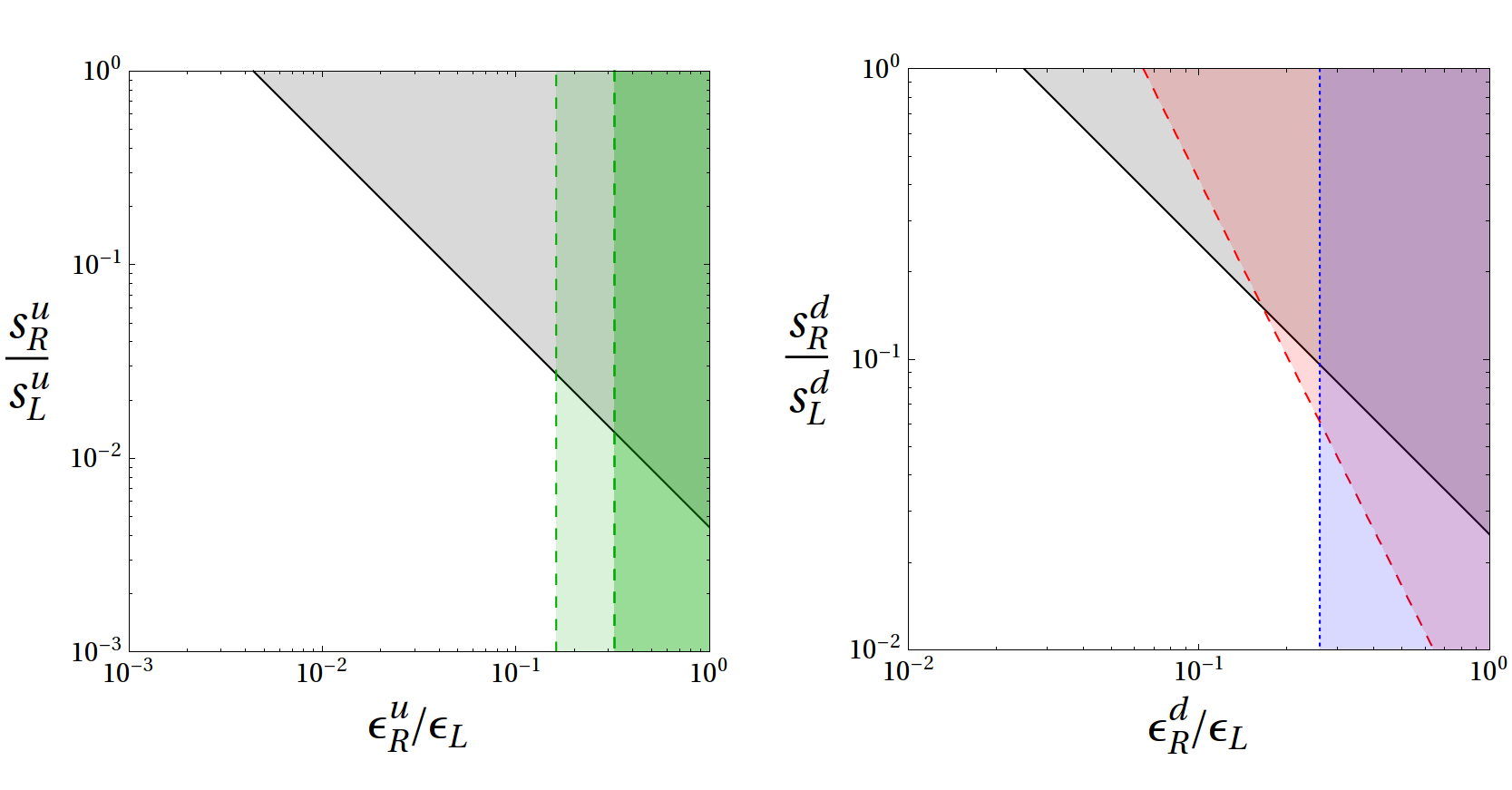}
\caption{Bounds on the free parameters of Generic $U(2)^3$ breaking, normalized to the parameters present in Minimal $U(2)^3$ breaking (determined by the CKM), with maximal phases.
The black solid line in both plots shows the bound from the neutron EDM (the shaded region is disfavoured at 90\% C.L.). In the left-hand plot, the green dashed line on the left corresponds to the current bound from $\Delta A_\text{CP}$, the one on the right to the old experimental central value. The darker shaded region is disfavoured, while in the lighter region, new physics could account for the old large experimental value. In the right-hand plot, the red dashed line shows the bound from $\epsilon_K$ and the blue dotted line the one from $\epsilon'$.
}
\label{fig:bounds}
\end{figure}

The bounds in Eqs. (\ref{eq:DC1bound}), (\ref{eq:DF0bound}), (\ref{eq:DS1bound}) and (\ref{eq:DS2bound}) constrain the $U(2)^3$ breaking parameters $\epsilon^{u,d}_R$ and $s^{u,d}_R$ for given values of the model-dependent parameters $c_i^{\alpha}$ and phases. Assuming all the real parameters to be unity and all the phases to be such as to maximize the corresponding bounds on the $U(2)^3$ breaking parameters, to be conservative, Fig.~\ref{fig:bounds} compares the strength of the bounds from the different observables. One can make the following observations:
\begin{itemize}
\item $\Delta A_\text{CP}$ could be due to new physics compatible with $U(2)^3$ if $\epsilon^u_R\sim0.1\epsilon_L$. However, with phases that maximize all the constraints, the bound on the up-quark CEDM then requires the angle $s^u_R$ to be more than one order of magnitude smaller than the corresponding ``left-handed'' angle $s^u_L$, whose size is determined by the CKM matrix.
\item If $\epsilon^d_R\lesssim0.1\epsilon_L$, bounds from the kaon system and the down quark (C)EDM are satisfied even without a considerable alignment of the $\Delta Y_d$ spurion.
\end{itemize}
Needless to say, in concrete models the relative strength of these bounds could vary by factors of a few.

\section{Comparison with $U(3)^3$ at small and large $\tan{\beta}$}

It is useful to compare the expectations of the $U(2)^3$ symmetry suitably broken as described in Section~\ref{sec:U2intro} with the case of $U(3)_q\times U(3)_u\times U(3)_d$ broken in the directions $Y_u= (3, \bar{3},1)$ and $Y_d = (3, 1, \bar{3})$, as we now briefly recall from the literature \cite{D'Ambrosio:2002ex,Feldmann:2008ja,Kagan:2009bn,Isidori:2012ts}.

First, by sole $U(3)^3$ transformations, one can set without loss of generality
\begin{equation}
Y_u = V_0^\dagger  Y_u^\text{diag}, ~~Y_d = Y_d^\text{diag},
\label{yud_3}
\end{equation}
where $ Y_u^\text{diag}, Y_d^\text{diag}$ are real diagonal matrices and $V_0$ is a unitary matrix\footnote{We use the notation $V_0$ to distinguish it from the CKM matrix $V$ which differs in general by order one corrections from $V_0$.} dependent on one single phase. Therefore the CKM phase is the only source of CP violation if no new phase is born outside of $Y_u$ or $Y_d$, no matter what the value of $\tan{\beta}$ is. 

As in the $U(2)^3$ case, to determine the relevant flavour violating operators one has to reduce the kinetic terms to canonical form and the mass matrices to real diagonal form. In turn this depends on the value of $\tan{\beta}$ which determines the need to include or not powers of $Y_d Y_d^\dagger $  in effective operators. For moderate $\tan{\beta}$ the relevant flavour violating quark bilinears in the physical basis have the approximate form ($ i,j = d,s,b$):
\begin{equation}
A \,\bar{d}_{Li}\sigma_{\mu \nu}(V^\dagger  \mathcal{I}_3 V Y_d^\text{diag})_{ij}d_{Rj},~~
B \,\bar{d}_{Li}\gamma_{\mu}  (V^\dagger  \mathcal{I}_3 V)_{ij} d_{Lj},
\end{equation}
where $A$ is a complex parameter and $B$ a real one. This leads to the following set of relevant operators ($\xi_{ij} = V_{tj} V_{ti}^*$):

i) $\Delta F =2$
\begin{equation}
C_{LL}  \xi_{ij}^2 \frac{1}{2}(\bar{d}_{Li}\gamma_\mu d_{Lj})^2,
\end{equation}

ii) $\Delta F =1$, chirality breaking ($\alpha = \gamma, G$):
\begin{equation}
C^\alpha e^{i \chi^\alpha} \xi_{ij}  (\bar{d}_{Li}\sigma_{\mu \nu} m_j d_{Rj})  O^\alpha_{\mu \nu}, ~~
O^\alpha_{\mu \nu}= e F_{\mu\nu}, ~ g_s G_{\mu \nu},
\end{equation}

iii) $\Delta F =1$, chirality conserving ($\beta = L, R, H$):
\begin{equation}
C^\beta  \xi_{ij}(\bar{d}_{Li}\gamma_\mu d_{Lj})O^\beta_\mu, ~~
O^\beta_\mu = (\bar{l}_L\gamma_\mu l_L),~(\bar{e}_R\gamma_\mu e_R),~(H^\dagger D_\mu H),
\end{equation}
with $C_{LL}, C^\alpha, C^\beta$ real.

At large $\tan{\beta}$  both powers of $Y_u Y_u^\dagger $ and $Y_d Y_d^\dagger $ are relevant in effective operators \cite{Feldmann:2008ja,Kagan:2009bn,Colangelo:2008qp}. The fact that  $\lambda_t$ and $\lambda_b$ are both of $O(1)$ leads to the the breaking of $U(3)^3$ down to $U(2)^3$. Note  that this is not enough to conclude that $U(3)^3$ at large  $\lambda_b$ leads to the same pattern of flavour violation described in the previous section. For this to happen one needs that the breaking directions of $U(2)^3$ be as in (\ref{spurions}). However, due to the fact that $U(3)^3$ is only broken in the $Y_{u, d}$ directions, this is  the case. To see this, after suitable $U(3)^3$ transformations, one can write $Y_{u, d}$ as \cite{Feldmann:2008ja,Kagan:2009bn}
\begin{equation}
Y_{u,d} =e^{\pm i \hat{\chi}/2}
\,\lambda_{t,b}
\begin{pmatrix}
\Delta Y_{u,d} & 0 \\
0 & 1
\end{pmatrix},
\end{equation}
where  $\Delta Y_{u,d}$ are $2\times 2$ matrices as in (\ref{spurions}) and $ \hat{\chi}$ is a  hermitian $3\times 3$ matrix with $\chi$ a 2-component vector,
\begin{equation}
 \hat{\chi} =
\begin{pmatrix}
0 & \boldsymbol{\chi} \\
\boldsymbol{\chi}^\dagger & 0
\end{pmatrix},
\end{equation}
which determines the misalignment of $Y_{u,d}$  in the $1 3$ and $2 3$ directions, known to be small, of order $|V_{cb}|$, from the CKM matrix. Expanding in $ \hat{\chi}$, 
\begin{equation}
Y_{u,d} \approx
\lambda_{t,b}
\begin{pmatrix}
\Delta Y_{u,d} & \pm i\boldsymbol{\chi}/2 \\
\pm i\boldsymbol{\chi}^\dagger\Delta Y_{u,d}/2 & 1
\end{pmatrix},
\label{yudmatrix}
\end{equation}
which shows that $\boldsymbol{\chi}$ plays the same role as $\V$. For example the term in the bottom left of (\ref{yudmatrix}), of subleading order, is given in the language of (\ref{Yuk_u}, \ref{Yuk_d}) by $\bar{q}_3 (\V^\dagger  \Delta Y_d \ELdR)$ or $\bar{q}_3 (\V^\dagger  \Delta Y_u \ELuR)$\footnote{As in the case of Eq.~(\ref{yud_3}), note that $Y_{u,d}$ in (\ref{yudmatrix}) are not in general proportional to the physical $u, d$ mass matrices, due to the presence of $Y_uY_u^\dagger $ and $Y_dY_d^\dagger $ corrections both in the mass terms themselves and in the kinetic terms.}.
Therefore $U(3)^3$ at large $\tan{\beta}$ leads in general to the same effective operators as in (\ref{eq:LeffDS2},\ref{eq:LeffDB2}) and (\ref{eq:LeffDB1mag},\ref{eq:LeffDB1L}), apart from the characteristic $\tan{\beta}$-dependence of the various coefficients, which should also show up both in flavour violating and in suitable flavour conserving amplitudes.

\section{Extension to the charged lepton sector}
\label{sec:U2leptons}

If one tries to extend the considerations developed so far to the lepton sector, one faces two problems. First, while the hierarchy of charged lepton masses is comparable to the mass hierarchies in the quark sector, the leptonic charged-current mixing matrix does not exhibit the hierarchical pattern of the CKM matrix. Secondly, in the likely possibility that the observed neutrinos are of Majorana type and are light because of a small mixing to heavy right-handed neutrinos, the relevant parameters in the Yukawa couplings of the  lepton sector,  $H \bar{l}_L Y_e e_R$ and $\tilde{H} \bar{l}_L Y_\nu \nu_R$,  are augmented, relative to the ones in the quark sector, by the presence of the right-handed neutrino mass matrix, $\nu_R^T M \nu_R$, which is unknown.
To overcome these problems we make the following two hypotheses:
\begin{itemize}
\item we suppose that the charged leptons, thorough  $Y_e$, behave in a similar way to the quarks, whereas $Y_\nu$ and $M$ are responsible for the anomalously large neutrino mixing angles;
\item we assume that $Y_\nu$ has no significant influence on flavour physics near the Fermi scale in spite of the presence at this scale of new degrees of freedom carrying flavour indices, like sneutrinos or heavy composite leptons. One can imagine many reasons for this to be the case, like e.g. in the discussion of leptons in composite Higgs models of Chapter \ref{cha:CHM}\footnote{For an implementation of $U(2)^3$ in Supersymmetry that includes also neutrinos see \cite{Blankenburg:2012nx}.}.
\end{itemize}

Extending $U(3)^3$ and $U(2)^3$ to the leptonic sector, we consider respectively a $U(3)_l\times U(3)_e$ and a $U(2)_l\times U(2)_e$ symmetry. Here comes another significant difference between the two cases: in the $U(3)_l\times U(3)_e$ case, with $Y_e$ transforming as a $(3, \bar{3})$, there is no new flavour changing phenomenon at the Fermi scale other than the  leptonic charged-current mixing matrix, since $Y_e$  can be diagonalized by a $U(3)_l\times U(3)_e$  transformation.
On the contrary, let us assume that the  $U(2)_l\times U(2)_e$ symmetry be  broken {\em weakly} by the spurions
\begin{equation}
\Delta Y_e = (2,\bar 2) \,,
\qquad
\Ve = (2,1) \,.
\label{lept_spur}
\end{equation}
By proceeding in the same way as for the quarks in Section~\ref{sec:U2intro}, one can set
\begin{equation}
\Ve^T = (0, \eta), ~~ \Delta Y_e = R_{12}^e \Delta Y_e^{diag}
\end{equation}
and see the occurrence of flavour changing bilinears with two important differences relative to the quark case. 
Firstly, one cannot relate the size of $\eta$ or of the angle $\theta_e$ in $R_{12}^e$ to the mixing matrix in the leptonic charged current. Secondly, due to the importance of $\mu \rightarrow e \gamma$, one has to include in the expansion of the chirality breaking bilinears (see Appendix \ref{app:bilinears} for the analogous procedure in the quark sector), subleading terms like $(\ELlLbar \Ve) (\Ve^\dagger \Delta Y_e \ELeR)$, and perform the rotation to the physical basis to the corresponding order in the spurions.

The flavour changing dimension six effective operators in the lepton sector can then be written as
\begin{enumerate}[i)]
\item $\tau\to\mu,e$, chirality breaking:
\begin{equation}
c_\tau \zeta_{i\tau} m_\tau \left(\bar{e}_{Li}\sigma_{\mu \nu} \tau_R\right) e F_{\mu\nu},
\label{taumue}
\end{equation}
\item $\mu\to e$, chirality breaking:
\begin{equation}
c_\mu \zeta_{e\mu} m_\mu \left(\bar{e}_{L}\sigma_{\mu \nu} \mu_R\right) e F_{\mu\nu},
\label{mue}
\end{equation}
\item $\tau\to\mu,e$, chirality conserving:
\begin{equation}
c^\beta_\tau \zeta_{i\tau} \left(\bar{e}_{Li}\gamma_\mu \tau_L\right) O^\beta_\mu, ~~
\end{equation}
\begin{equation}
O^\beta_\mu = \left(\bar{l}_L\gamma_\mu l_L\right),~\left(\bar{e}_R\gamma_\mu e_R\right),
~\left(\bar{q}_L\gamma_\mu q_L\right),~\left(\bar{u}_R\gamma_\mu u_R\right),~\left(\bar{d}_R\gamma_\mu d_R\right),
~\left(H^\dagger D_\mu H\right),
\end{equation}
\item $\mu\to e$, chirality conserving:
\begin{equation}
c^\beta_\mu  \zeta_{e\mu}\left(\bar{e}_{L}\gamma_\mu \mu_L\right)O^\beta_\mu, ~~
\end{equation}
\end{enumerate}
where $\zeta_{ij}= U_{eL}^{3i*}U_{eL}^{3j}$ and $|U_{eL}|\simeq R_{12}^eR_{23}^e$ is the left-handed charged lepton Yukawa diagonalization matrix.
All these coefficients are  model dependent and, in principle, of similar order.

\subsection{Lepton Flavour Violation in $U(2)^2$}\label{sec:applfv}

Here we list the contributions to LFV observables induced by the operators
\begin{align}
\mathcal H_\text{eff} =
&\sum_{j>i}\frac{\zeta_{ij}}{\Lambda^2}\bigg[
c_j m_{e_j} (\bar{e}_{Li}\sigma_{\mu \nu} e_{Rj}) e F_{\mu\nu}+
c_j^{l}(\bar{e}_{Li} \gamma_\mu e_{Lj})(\bar e_{Li} \gamma^\mu e_{Li})+
c_j^{e}(\bar{e}_{Li} \gamma_\mu e_{Lj})(\bar e_{Ri} \gamma^\mu e_{Ri})
\\
&+
c_j^{u}(\bar{e}_{Li} \gamma_\mu e_{Lj})(\bar u \gamma^\mu u)+
c_j^{d}(\bar{e}_{Li} \gamma_\mu e_{Lj})(\bar d \gamma^\mu d)
\bigg]
+\text{h.c.}
\end{align}
The $l^j\to l^i\gamma$ branching ratio is given by
\begin{equation}
 \text{BR}( l^j \rightarrow  l^i \gamma)
= \frac{192\pi^3 \alpha}{G_F^2} \frac{|\zeta_{ij} c_j|^2}{\Lambda^4}
\;b^{ij}\,,
\end{equation}
where $b^{ij}=\text{BR}( l^j\rightarrow l^i\nu\bar\nu)$.
The $ l^j \rightarrow  l^i\bar l^i l^i$ branching ratio reads \cite{Hisano:1995cp,Arganda:2005ji}
\begin{multline}
\text{BR}( l^j \rightarrow 3 l^i)=\frac{1}{2G_F^2}b^{ij}\frac{|\zeta_{ij}|^2}{\Lambda^4}
\bigg[
e^4|c_j|^2\left(16\ln\frac{m_{ l_j}}{m_{ l_i}}-22\right)
+\frac{1}{2}|c^l_j|^2
+\frac{1}{4}|c^e_j|^2
\\
+e^2\left(2c_j c^{l*}_j+c_j c^{e*}_j+\text{h.c.}\right)
\bigg] .
\end{multline}
For $\mu$-$e$ conversion, one obtains \cite{Barbieri:1995tw,Hisano:1995cp}
\begin{equation}
\Gamma(\mu\to e)=
\frac{\alpha^3}{4\pi^2}
\frac{Z_\text{eff}^4}{Z} |F(q)|^2 m_\mu^5
\frac{|\zeta_{ij}|^2}{\Lambda^4}
\left|
(2Z+N)c_j^u+(2N+Z)c_j^d
-
2Ze^2 c_j
\right|^2 \,.
\end{equation}
In the case of ${}^{48}_{22}$Ti, one has $Z_\text{eff}=17.6$ and $|F(q^2)|=0.54$ and the conversion rate is defined as 
\begin{equation}
\text{CR}(\mu\text{ Ti}\to e\text{ Ti}) =
\frac{\Gamma(\mu\text{ Ti}\to e\text{ Ti})}{\Gamma(\mu\text{ Ti}\to \text{capture})},
\end{equation}
where the capture rate is $\Gamma(\mu\text{ Ti}\to \text{capture})=(2.590 \pm 0.012)\times10^6 ~\text{s}^{-1}$.

\subsection{Bounds on the $U(2)^3$ coefficients}

\begin{table}[tbp]
\begin{center}
\renewcommand{\arraystretch}{1.5}
\renewcommand\tabcolsep{5pt}
\begin{tabular}{ccccccccc}
\hline
$\mu\to e\gamma$ & $2.4\times10^{-12}$ &\cite{Adam:2011ch}& $\mu\to 3e$ & $1.0\times10^{-12}$ &\cite{Bellgardt:1987du}& $\mu\to e \text{ (Ti)}$ & $6.1\times10^{-13}$ &\cite{Wintz:1998rp} \\
$\tau\to e\gamma$ & $3.3\times10^{-8}$ &\cite{:2009tk}& $\tau\to 3e$ &$2.7\times10^{-8}$ &\cite{Hayasaka:2010np}&&\\
$\tau\to \mu\gamma$ & $4.3\times10^{-8}$ &\cite{:2009tk}& $\tau\to 3\mu$ & $2.1\times10^{-8}$ &\cite{Hayasaka:2010np}&&\\
\hline
\end{tabular}
\end{center}
\caption{90\% C.L. experimental upper bounds on the branching ratios of 6 LFV decays and on the $\mu\to e$ conversion rate in Titanium.}
\label{tab:lfvexp}
\end{table}

\begin{table}[tbp]
\begin{center}
\renewcommand{\arraystretch}{1.5}
\begin{tabular}{lcccl}
\hline
operator & $|\widetilde c_{e\mu}|$ & $|\widetilde c_ {e\tau}|$ & $|\widetilde c_ {\mu\tau}|$ & constrained from\\
\hline
$m_{e_j} (\bar{e}_{Li}\sigma_{\mu \nu} e_{Rj}) e F_{\mu\nu}$    & 0.07 & 0.79 & 0.2 & $l_j\to l_i\gamma$\\
$(\bar{e}_{Li} \gamma_\mu e_{Lj})(\bar e_{Li} \gamma^\mu e_{Li})$ & 0.6 & 9.4 & 1.8 & $l_j\to 3l_i$ \\
$(\bar{e}_{Li} \gamma_\mu e_{Lj})(\bar e_{Ri} \gamma^\mu e_{Ri})$ & 0.9 & 13 & 2.6 & $l_j\to 3l_i$ \\
$(\bar{e}_{Li} \gamma_\mu e_{Lj})(\bar u \gamma^\mu u)$  & 0.03 &--&-- & $\mu\to e \text{ (Ti)}$\\
$(\bar{e}_{Li} \gamma_\mu e_{Lj})(\bar d \gamma^\mu d)$  & 0.03 &--&-- & $\mu\to e \text{ (Ti)}$\\
\hline
\end{tabular}
\renewcommand{\arraystretch}{1.0}
\end{center}
\caption{90\% C.L. upper bounds on the reduced coefficients defined in (\ref{eq:ctilde}). The last column lists the processes giving the strongest constraint on the respective operators.}
\label{tab:lfvbounds}
\end{table}

The current bounds on LFV processes are collected in Tab.~\ref{tab:lfvexp}. Using them, bounds can be set on the above coefficients, making assumptions about the mass scale of new physics and the size of the mixing angles $\zeta_{ij}$. 
In Tab.~\ref{tab:lfvbounds}, we show the bounds on the dimensionless reduced coefficients
\begin{equation}
\widetilde c_{ij} = c_{j} \times \left[\frac{3\,\text{TeV}}{\Lambda}\right]^2\left[\frac{\zeta_{ij}}{V_{ti}V_{tj}^*}\right],
\label{eq:ctilde}
\end{equation}
where the indices $i,j = e, \mu, \tau$ refer to the specific flavour transitions.
For $\Lambda = 3$ TeV some of these bounds are significant, especially from $\mu$-decay processes. Note however that the normalization of  the $\zeta_{ij}$ to the corresponding products of the  CKM matrix elements should only be taken as indicative. Note furthermore that the operators contributing to $\mu\to e$ conversion are suppressed in specific models by explicit gauge coupling factors (supersymmetry) or by small mass mixing terms (composite Higgs models, see below).


%
%
%
%

\section{Summary and partial conclusions}
\begin{table}[tb]
\renewcommand{\arraystretch}{1.5}
 \begin{center}
\begin{tabular}{lcccccc}
&\multicolumn{2}{c}{Chirality conserving} & \multicolumn{2}{c}{Chirality breaking}\\
\hline
& $\Delta B = 1,2$ & $\Delta S = 1,2$ & $\Delta B = 1$ & $\Delta C = 1$ 
\\\hline
$U(3)^3$ moderate $t_\beta$ & \multicolumn{2}{l}{\hspace{.61cm}\ovalbox{$\mathbbm R\qquad\qquad\;\; \mathbbm R$}} & $\mathbbm C$ & 0 
\\
Minimal $U(2)^3$, $U(3)^3$ large $t_\beta$  & $\mathbbm C$ & $\mathbbm R$ & $\mathbbm C$ & 0 
\\
Generic $U(2)^3$ & $\mathbbm C$ & $\mathbbm C$ & $\mathbbm C$ & $\mathbbm C$ 
\\\hline
 \end{tabular}
 \end{center}
\caption{Expected new physics effects in $U(3)^3$ and both Minimal and Generic $U(2)^3$, for chirality conserving and chirality breaking $\Delta F=1,2$ FCNC operators in the $B$, $K$, $D$ systems. $\mathbbm R$ denotes possible effects, but aligned in phase with the SM, $\mathbbm C$ denotes possible effects with a new phase, and 0 means no or negligible effects. In $U(3)^3$ with moderate $\tan\beta$ an additional feature is that the effects in $b\to q$ ($q = d, s$) and $s\to d$ transitions are perfectly correlated.
}
\label{tab:u2u3}
\end{table} 
 
A suitably broken $U(2)^3$ flavour symmetry may allow for deviations from the CKM picture of flavour and CP violations related to new physics at the ElectroWeak scale and waiting to be discovered. 
We have defined a Minimal and a Generic $U(2)^3$ case, depending on the fact that one takes a minimal set of breaking {\it spurions} or one allows all the possible terms contributing to the quark mass terms. 
Using an EFT approach to Minimal $U(2)^3$ it is possible to write down an effective Lagrangian
 \begin{equation}
\Delta \mathcal{L} = \sum_i \frac{c_i \xi_i}{(4\pi v)^2} \mathcal{O}_i ~+\text{h.c.}
\label{ideal2}
\end{equation}
where the $\xi_i$ are suitable combinations of the standard CKM matrix elements and $|c_i| = 0.2$ to $1$ consistently with current experimental constraints. This remains true even after the inclusion of the constraint coming from direct CP violation in K decays (the $\epsilon^\prime$ parameter) which had escaped  attention so far, to the best of our knowledge, in the EFT context. If these considerations are of any guidance,  the main observables that deserve attention, in view of conceivable experimental progress, are CP violation in the mixing of the $B_s$ system, rates and/or asymmetries in $B$ decays, like $b\rightarrow s(d)\gamma$, $b\rightarrow s(d) \ell\bar{\ell}$, $b\to s(d)\nu\bar{\nu}$  and in $K\rightarrow \pi \nu\bar{\nu}$ decays, either charged or neutral. A precise determination of $|V_{ub}|$ and $\gamma_{\rm CKM}$ would also play a key role.
 
Generic $U(2)^3$ introduces new parameters which do not have a correspondence with the ones of standard CKM. 
As such, even insisting on an effective scale at 3 TeV, one cannot predict the size of the extra effects introduced in  Generic $U(2)^3$.
We have seen, however, where the main constraints on the new parameters come from: in the up sector from  CP asymmetries in $D$ decays and from the neutron EDM and, in the down sector, also from the neutron EDM and from CP violation in the Kaon system. 
Always with an effective scale at 3 TeV and barring cancellations among  phases, the size of the new breaking terms included in Generic $U(2)^3$ have to be smaller than the corresponding ones in Minimal $U(2)^3$. 
Both Minimal and Generic $U(2)^3$ are unlikely to give rise to any sizeable effect neither in top FCNCs nor in CP violation in $D$-$\bar{D}$ mixing at forseen experiments.

Lepton Flavour Violation is in many respects the next due subject although with a major difficulty: the peculiar properties of the neutrino mixing matrix, quite different from the quark one, and, perhaps not unrelated, the weaker information available in the lepton sector relative to the quark sector due to the possible role of the right-handed neutrino mass matrix.
Altogether we find it conceivable that the observed neutrino masses and mixings have a very high energy origin with little impact on Fermi scale physics. Taking this view, we have on the contrary assumed that the charged leptons may behave with respect to flavour in a similar way to the quarks, with a natural extension of $U(2)^3$ to include as well a $U(2)_l\times U(2)_e$ extra approximate symmetry. 
While this can only be a qualitative picture, since we lack any direct information on the relevant mixing matrix, it gives nevertheless a possible coherent description of LFV signal in the TeV range. One interesting feature characteristic of $U(2)_l\times U(2)_e$, suitably broken as in (\ref{lept_spur}), is in the comparison between the $\mu\rightarrow e$ and the $\tau\rightarrow \mu$ transitions, with the chirality breaking operators respectively proportional to $\zeta_{\mu \tau} m_\tau$ and $\zeta_{e\mu} m_\mu$.

Suppose that a significant deviation from the SM emerged in the experiments to come, which could be accounted for in the effective framework described above. How could one tell that $U(2)^3$ is the relevant approximate symmetry, without uncovering by direct production the underlying dynamics (supersymmetry, a new strong interaction or whatever)? The best way would be to study in $B$ decays the correlation between the $s$ and the $d$ quarks in the final state, which would have to be the same as in the SM.
 Such correlation is in fact also expected in MFV. However, the only way to have effects in MFV similar to the ones discussed above in Minimal $U(2)^3$ requires the presence of two Higgs doublets, one coupled to the up quarks and one to the down quarks, with large values of the usual $\tan{\beta}$ parameter \cite{D'Ambrosio:2002ex,Kagan:2009bn}. This, in turn, would have other characteristic effects  not necessarily expected in Minimal $U(2)^3$. 
 In the case of small $\tan\beta$ or with one Higgs doublet only, distinguishing MFV from $U(2)^3$ would be straightforward by means of the additional effects discussed in \cite{Barbieri:2012uh}, like CP violation in $B_s$ mixing or non-universal contributions to $B\to K\nu\bar\nu$ vs.\ $K\to\pi\nu\bar\nu$ decays. A synthetic description of new physics effects in Minimal $U(2)^3$ and in Generic $U(2)^3$ is given in Table \ref{tab:u2u3}
 and compared with MFV (i.e. $U(3)^3$ at moderate or large $\tan{\beta}$). On top of the qualitative differences shown in Table \ref{tab:u2u3}, the size of the possible effects is significantly more constrained in $U(3)^3$ at moderate  $\tan{\beta}$ than in all other cases.

\part{Supersymmetry}
\label{part:SUSY}

\chapter{Supersymmetry}
\label{cha:SUSYintro}

In 1967 Coleman and Mandula \cite{Coleman:1967ad} proved that the most general symmetry structure that any quantum field theory can respect is the direct product of Poincar\'e covariance and some internal symmetry, otherwise the $S$ matrix would be trivial, i.e. not enough degrees of freedom would be available to describe the physical processes that we observe.
It is possible to evade this no-go theorem by relaxing some of its assumptions. One of them is that the symmetry algebra involves only commutators, i.e. its generators are bosonic: contrary to causality, locality etc. this assumption seems not to have any particular physical reason. Indeed one can extend the definition of a Lie algebra (like the Poincar\'e algebra) to the one of a graded Lie algebra, which involves also anticommutators. Supersymmetry in four spacetime dimensions \cite{Wess:1974tw,Wess:1973kz} is a graded Lie Algebra of grade 1, i.e. a spacetime symmetry where spin-1/2 generators $Q$ are introduced, where
\begin{equation}
 \label{eq:SUSY_generators}
 Q |fermion\rangle = |boson\rangle\,
\end{equation}
and vice versa. In 1975 Haag, Lopuszanski and Sohnius \cite{Haag:1974qh} prooved that higher grades would not be consistent, given the other physical assumptions of the Coleman-Mandula theorem.

After this theoretical motivation, we limit ourselves to the case of $N=1$ Supersymmetry, where $N$ is the number of distinct copies of the generators of supersymmetric transformations, since this appears to be the only phenomenologically viable possibility in 4 dimensions. 

\section{Supersymmetric theories of Nature}
In this section we give a general overview of Supersymmetry (SUSY), and we refer to the reviews \cite{Martin:1997ns,Terning:2006bq,Derendinger:1990tj} for more complete presentations.
\begin{itemize}
 \item {\bf Particle content}\\
 By definition, each particle falls in an irreducible representation of the Supersymmetric group, called ``supermultiplet''. Using the fact that the $Q$'s commute with the generator of spacetime translations $P_\mu$ and the spin-statistics theorem, one can show that each supermultiplet contains an equal number of fermionic and bosonic degrees of freedom, with the relative states usually called ``superpartners'' of each other. Since Supersymmetry is in a direct product with the gauge symmetries of the theory, superpartners have exactly the same gauge-transformation properties. It is useful and sufficient to introduce two kinds of supermultiplets, each one containing two bosonic and two fermionic degrees of freedom: i) chiral supermultiplets, consisting of a Weyl fermion $\psi$ and a complex scalar field $\phi$, and ii) gauge (or vector) supermultiplets, consisting of a massless spin-1 boson $A_\mu^a$ and a spin-1/2 Weyl fermion $\lambda^a$, usually referred to as ``gaugino'' (since it transforms as 
the adjoint representation of the gauge group, as its gauge boson superpartner).
 
 The immediate way to let the Standard Model particles fit into supermultiplets is to associate to each SM field a superpartner (which is usually denoted with the same letter, with the addition of a tilde ``~$\widetilde{  }$~'' on top of it).
 Each SM fermion \eqref{SMfermions} is then associated with the corresponding ``sfermion'' into a chiral supermultiplet, each gauge boson with a ``gaugino'' into a vector supermultiplet, and the Higgs boson with a ``Higgsino'' into a chiral supermultiplet.
 One can see that only one Higgs boson is not a viable possibility (at this level it is sufficient to recognize that it would ruin anomaly cancellation), rendering necessary at least another Higgs doublet with opposite hypercharge\footnote{There are actually two other reasons for another Higgs field to be necessary: i) since the superpotential (see the following of this section) of the theory has to be analytic in the superfields, it is needed to generate Yukawa couplings in the down-quark and in the lepton sectors; ii) if one wants to 
realize EWSB in a Supersymmetric way three spinors are needed (one for $\widetilde{Z}$ and two for $\widetilde{W}^\pm$), and with only one Higgsino only two of them are available.}.
 The SM Higgs boson is then a linear combination of the two supersymmetric Higgses $H_u$ and $H_d$, their vacuum expectation values are related by $v^2 = v_u^2 + v_d^2$, and the parameter $\tan \beta = v_u/v_d$ is usually introduced\footnote{Usually one considers values of $\tan \beta \geq 1$, since smaller ones would generically induce the top Yukawa coupling to blow up before the GUT scale.}.
 Is this the most economical field content obtainable when building a Supersymmetric Standard Model? If two SM fermionic and bosonic fields had the same quantum numbers (remember that SUSY and gauge symmetries commute), then they could belong to the same supermutiplet. The only available possibility would then be to identify the SM-Higgs with a left-handed sneutrino, but this would induce the same anomaly non-cancellation, plus some phenomenological problems whose solution would spoil the minimality which in this spirit motivates the choice, see e.g. the recent \cite{Riva:2012hz} and references therein.
 The choice of doubling the fields (and adding a Higgs doublet) is then the one with the minimal field content, defining what is known as the Minimal Supersymmetric Standard Model (MSSM). Other non-minimal possibilities have been studied in the literature, for example the Next-to-Minimal Supersymmetric Standard Model (NMSSM), where an extra gauge singlet is added to the particle content of the theory. We will motivate and discuss this particular case in more detail in the following.
 \newpage
 \item{\bf Supersymmetric Lagrangian}\\
 Here we give the recipe for building a supersymmetric Lagrangian starting from a given particle content. The kinetic and gauge terms for all the fields are the usual ones, built with standard covariant derivatives. Supersymmetry invariance forces some of the possible additional terms to have coefficients proportional to the gauge couplings, namely it gives rise to
 \begin{equation}
 \mathcal{L}_{\text{g}} = \sum_a \left[\sqrt{2} g_a (\phi^* T^a \psi \lambda_a + \text{h.c.}) + \frac{1}{2} g_a^2 (\phi^* T^a \phi)^2\right]\,.
  \label{SUSYgauge}
 \end{equation}
The other interactions not controlled by the gauge couplings are most conveniently expressed in terms of a function $W$ of the scalar fields $\phi$ (or, equivalently, of the supermultiplets), called the \textit{superpotential},
 \begin{equation}
 W = L^i \phi_i + \frac{1}{2} M^{ij} \phi_i \phi_j + \frac{1}{6}y^{ijk} \phi_i \phi_j \phi_k\,.
  \label{superpotential}
 \end{equation}
This is the most general form of the superpotential that allow to build a renormalizable interaction Lagrangian which preserves supersymmetry (the latter requirement forcing $W$ to be an analytic function of the complex variable $\phi$), which has the form
\begin{equation}
\mathcal{L}_{\text{int}} = - W^i W_i^* - \frac{1}{2}(W^{ij} \psi_i \psi_j + \text{h.c.})\, ,
 \label{SUSYint}
\end{equation}
where $W^i$ and $W^{ij}$ are the first and second derivatives of $W$, with respect to the fields $\phi_i$ and $\phi_i$, $\phi_j$. One can easily show that fermions and bosons have exactly the same squared mass matrix $M^*_{ik} M^{kj}$, and that the $y_{ijk}$'s give rise to Yukawa-like interactions between fermions and bosons, plus some terms in the scalar potential of the theory.

A fundamental property of Supersymmetry are the \textit{non-renormalization theorems} which, for our purposes, coincide with the statement that the renormalization of the $W$ parameters is limited to wavefunction renormalization, i.e. $W$ does not renormalize. This implies that the radiative corrections to each parameter of the superpotential be proportional to the parameter itself: if at a given input scale a parameter is small, its running will keep it small also at the other energy scales. In particular, parameters with the dimension of a mass do not receive corrections proportional to the square of the highest energy scale they couple to, if all the way up to this scale Supersymmetry is preserved
.

In order to know the whole SUSY-preserving Lagrangian of the MSSM, thanks to the recipe defined above it is sufficient to give the superpotential
\begin{equation}
W_{\text{MSSM}} = \widetilde{q}_L^\dagger H_u Y_u \widetilde{u}_R + \widetilde{q}_L^\dagger H_d Y_d \widetilde{d}_R +  \widetilde{\ell}_L^\dagger H_d Y_e \widetilde{e}_R + \mu H_u H_d\, ,
 \label{WMSSM}
\end{equation}
where we have not included all the terms that violate either lepton or baryon number, since e.g. they would generically induce too large contributions to proton decay. To justify the absence of such terms a symmetry is usually invoked, the so-called ``$R$-parity'', under which all the SM particles and the Higgs bosons are even, while their superpartners are odd. The introduction of this parity has (at least) two other very important phenomenological consequences: i) the lightest supersymmetric particle (LSP) is stable, thus providing a possible candidate for particle Dark Matter, and ii) superparticles are always produced in pairs at colliders. 

At this level each SM particle would have exactly the same mass of its corresponding superpartner, a possibility evidentely ruled out by the fact that no supersymmetric particle has been observed yet.

 \item{\bf Supersymmetry breaking Lagrangian}\\
In order to take into account this lack of direct evidence, Supersymmetry must be broken. Here we do not say anything about the particular breaking mechanism, for which many examples exist in the literature, and report the recipe to build a SUSY breaking, or ``soft'' Lagrangian. Our ignorance can be paremetrized by adding the following extra terms:
\begin{equation}
\mathcal{L}_{\text{soft}} = - \left( \frac{1}{2} M_a \lambda^a \lambda^a + \frac{1}{6} A^{ijk} \phi_i \phi_j \phi_k + \frac{1}{2} b^{ij} \phi_i \phi_j + \text{h.c.}\right) - (m^2)^{ij} \phi_j^* \phi_i\,,
 \label{SUSYsoft}
\end{equation}
where for simplicity we omitted terms that are absent in the MSSM. 
All the parameters in the ``soft'' Lagrangian have at least the dimension of a mass: this aspect is extremely important since (thanks to the Symanzik theorem) it does not invalidate  the non-renormalization theorems, as one would generically expect because of the breaking of Supersymmetry. A consequence is for example that the dimensionful parameter $\mu$ appearing in the superpotential will not become very large by radiative corrections involving the masses of heavy unknown particles, i.e. it is technically natural.
On the contrary, radiative corrections to the soft parameters do not benefit of a similar protection, and their running will be proportional to combinations of all the soft Lagrangian parameters. Since the Fermi scale emerges as a combination of these parameters, taking too heavy sparticles would introduce a little hierarchy problem. In particular, avoiding too much fine-tuning requires the stops to be light, since their mass is the one giving the bigger contributions to the quadratic term of the Higgs potential. We will come back to this in the next section. Note finally that avoiding fine-tuning requires also the SUSY-preserving parameter $\mu$ to be of the same order of the soft parameters, despite its a priori different origin. This issue is known as the $\mu$-problem. 

Concerning more specifically the MSSM, the introduction of $\mathcal{L}_{\text{soft}}$ gives masses $M_{1,2,3}$ to the gauginos, and adds terms of the type $(m^2)^{ij}$ and $b$ to the Higgs potential, which is the only one allowing a $b$-term. The $A$-terms and soft masses $m^2$ appear in slepton and squark interactions, each one thus being a $3 \times 3$ complex matrix in flavour space.
\end{itemize}

\subsection{Some phenomenological features}
From the phenomenological point of view, the MSSM constitutes a particular attractive model for many reasons. We already mentioned that the more immediate way to forbid proton decay, $R$-parity, naturally provides a stable Dark Matter candidate. Another appealing feature is that gauge couplings unification happens at energies of order $10^{15\div16}$ GeV with a much better precision than in the SM. Note that the above two properties still hold if a gauge singlet $S$ is added to the Higgs sector as in the NMSSM.

\begin{figure}[t]
 \begin{center}
  \includegraphics[width = 0.6\textwidth]{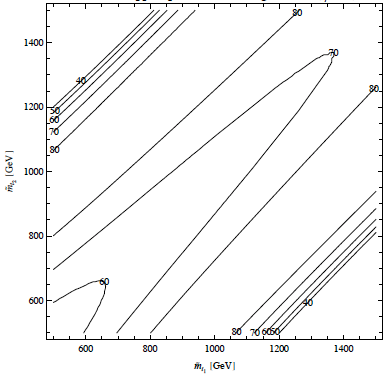}
 \end{center}
\label{fig:Delta_t}
\caption{Isolines of the radiative contribution $\Delta_t$ to the Higgs mass, as a function of the physical stop masses, for the case of maximal stop mixing and $\tan \beta = 4$.}
\end{figure}

More specific to the MSSM is the expectation of a light Higgs boson, in fact the tree-level relation $m_h < m_Z |\cos 2 \beta|$ holds. This upper bound can be lifted by radiative corrections: their effect is to a very good approximation given by the top-stop loop contributions to the quartic $H_u$ coupling, that modifies the Higgs mass upper bound to
 \begin{equation}
 m_h^2 < m_Z^2 \cos^2 2 \beta + \Delta_t^2\,.
  \label{eq:SUSY_Higgs_mass}
 \end{equation}
The dominant contribution to $\Delta_t$ is given by
\begin{equation}
 \Delta_t^2 = \frac{3}{(4 \pi)^2}\frac{m_t^4}{v^2} \left[\log{\frac{m_{\tilde t}^2}{m_t^2}} + \frac{X_t^2}{m_{\tilde t}^2}\left(1-\frac{X_t^2}{12 m_{\tilde t}^2} \right) \right]\,,
 \label{eq:topstop}
\end{equation}
where $m_{\tilde t}^2 = m_{Q_3} m_{u_3}$, with $m_{Q_3,u_3}^2$ soft masses of the left and right stops respectively, and we have chosen $v = 174$ GeV. The stop mixing parameter $X_t = A_t - \mu \cot \beta$, with $A_t$ is the stop $A$-term, maximizes the value of $\Delta_t$ for $X_t =\sqrt{6} \, m_{\tilde t}$.
To obtain a Higgs mass of $\sim 126$ GeV, one needs values of $\Delta_t$ higher than $\sim 85$ GeV. The value of $\Delta_t$ as a function of the physical stop masses is displayed in Figure \ref{fig:Delta_t} for the case of maximal mixing. There also subleading two-loop contributions not made explicit in \ref{eq:topstop} have been taken into account \cite{Carena:1998gk}, that have the effect of lowering the value \ref{eq:topstop} by a few GeV.
Figure \ref{fig:Delta_t} clearly indicates that, in order to reproduce the Higgs mass, stops heavier than a TeV are needed, and this lower bound increases in the multi TeV range if one deviates from the value $X_t =\sqrt{6} \, m_{\tilde t}$.

It is interesting to note that the measured value of the Higgs mass also sets an upper bound on the scale where to expect sfermions, i.e. the SUSY breaking scale, of course depending on the assumptions made on the rest of the spectrum. We refer the reader to \cite{Giudice:2011cg} for a thorough discussion. Here we just mention that for low values of $\tan \beta$ this scale can even be above $10^3$ TeV, while already for $\tan \beta \gtrsim 4$ it cannot be larger than~$\sim 10$~TeV.

\section{Supersymmetry and the hierarchy problem}
\label{sec:NatSUSY}

\subsection{Natural Supersymmetry}
The supersymmetric solution to the hierarchy problem constitutes the main reason to expect sparticles to show up at the LHC.
Let us take the $Z$ boson mass as a definition of the weak scale, which in the MSSM at largish $\tan \beta$ is given by
\begin{equation}
m_Z^2 = -2 (m_{H_u}^2 + |\mu|^2)\,.
 \label{eq:SUSY_Zmass}
\end{equation}
This clearly points our attention to the parameters $\mu$ and $m_{H_u}$ as prominent ones in defining the weak scale
. The former as we saw is technically natural thanks to the non renormalization theorems, the latter instead has no protection and in fact receives corrections proportional to the other soft masses, in particular to the stop one
\begin{equation}
\delta m_{H_u}^2 \simeq -\frac{3 y_t^2}{8 \pi^2} \left(m_{Q_3}^2 + m_{u_3}^2 + |A_t|^2 \right) \log{\frac{\Lambda}{m_{\tilde t}^2}}\,,
 \label{eq:mHu}
\end{equation}
where $m_{Q_3}^2$ and $m_{u_3}^2$ are the soft masses of the left and right stops respectively, and $\Lambda$ is the scale at which the soft masses fade away. 
The next subdominant one-loop correction to $m_{H_u}^2$ has an opposite sign, and is proportional to the Wino mass parameter $M_2$. More important, at least for not too large $M_2$, is the two-loop gluino contribution, that enters via its correction to the stop masses
\begin{equation}
\delta m_{Q_3, u_3}^2 \simeq \frac{8 \alpha_s}{3 \pi^2} |M_3|^2 \log{\frac{\Lambda}{M_3}}\,,
 \label{eq:mstop_gluino}
\end{equation}
where $M_3$ is the gluino mass.
To achieve a natural solution to the hierarchy problem in SUSY, i.e. not to have a big fine-tuning in \ref{eq:SUSY_Zmass}, one then needs $\mu$, $m_{Q_3, u_3}$ and $M_3$ to be under control\footnote{
For a recent and complete analysis, extending also to the NMSSM case, see \cite{Hall:2011aa}.}. This corresponds to a spectrum with light Higgsinos, stops, left sbottom and, to a lesser extent, a light gluino, while the other sparticles can be heavier. For example squarks of the first two generations could reach values of tens of TeV before reintroducing fine-tuning problems.
This is the kind of spectrum that defines the so-called ``Natural Supersymmetry'', first proposed in connection with the hierarchy problem already two decades ago\cite{Dimopoulos:1995mi,Cohen:1996vb}, and recently reappraised (see e.g. \cite{Barbieri:2009ev,Papucci:2011wy}) in more explicit connection with superpartners searches at the LHC.

\begin{figure}[th]
 \begin{center}
  \includegraphics[width = 0.51\textwidth]{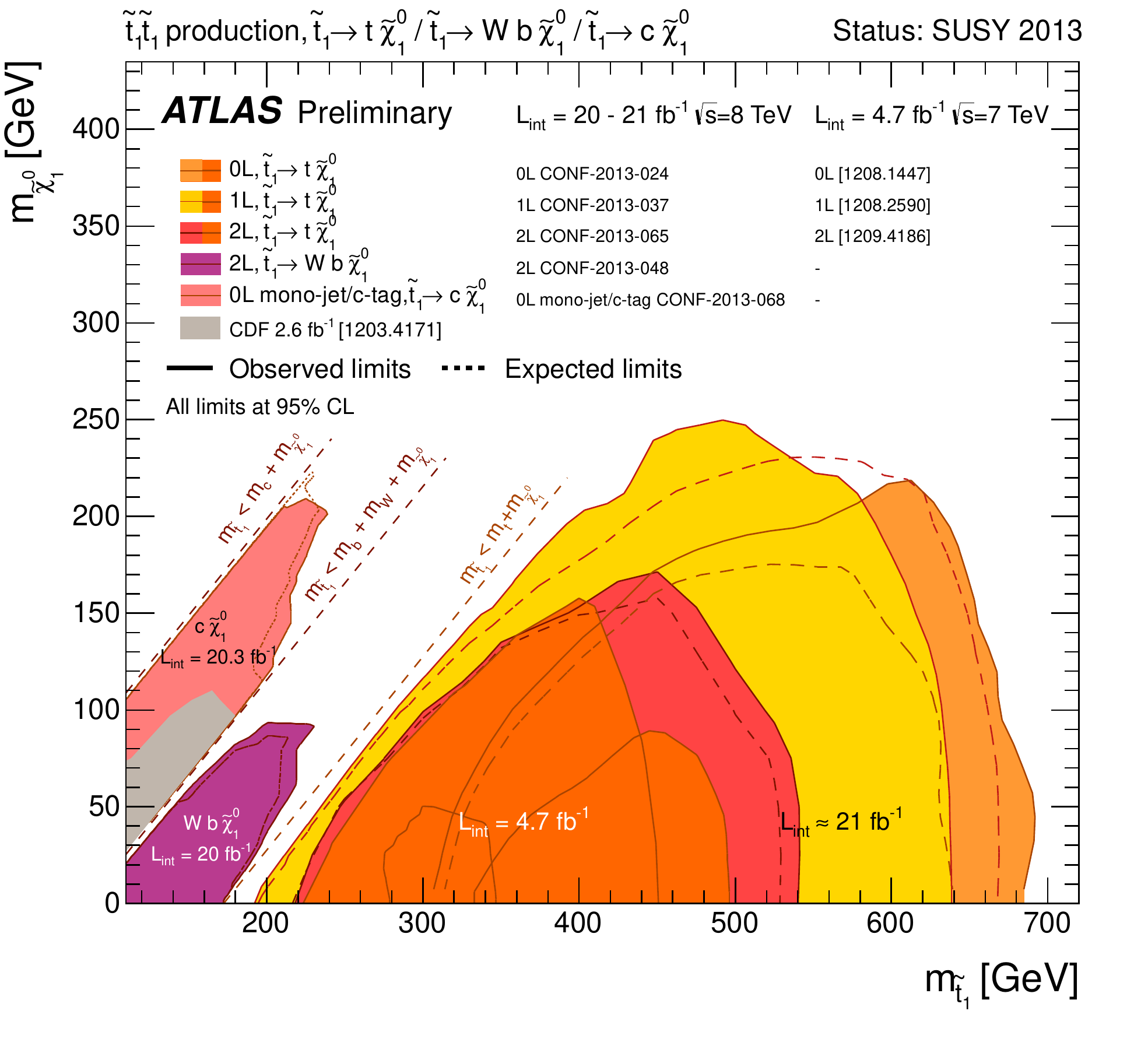}\hfill \includegraphics[width = 0.49\textwidth]{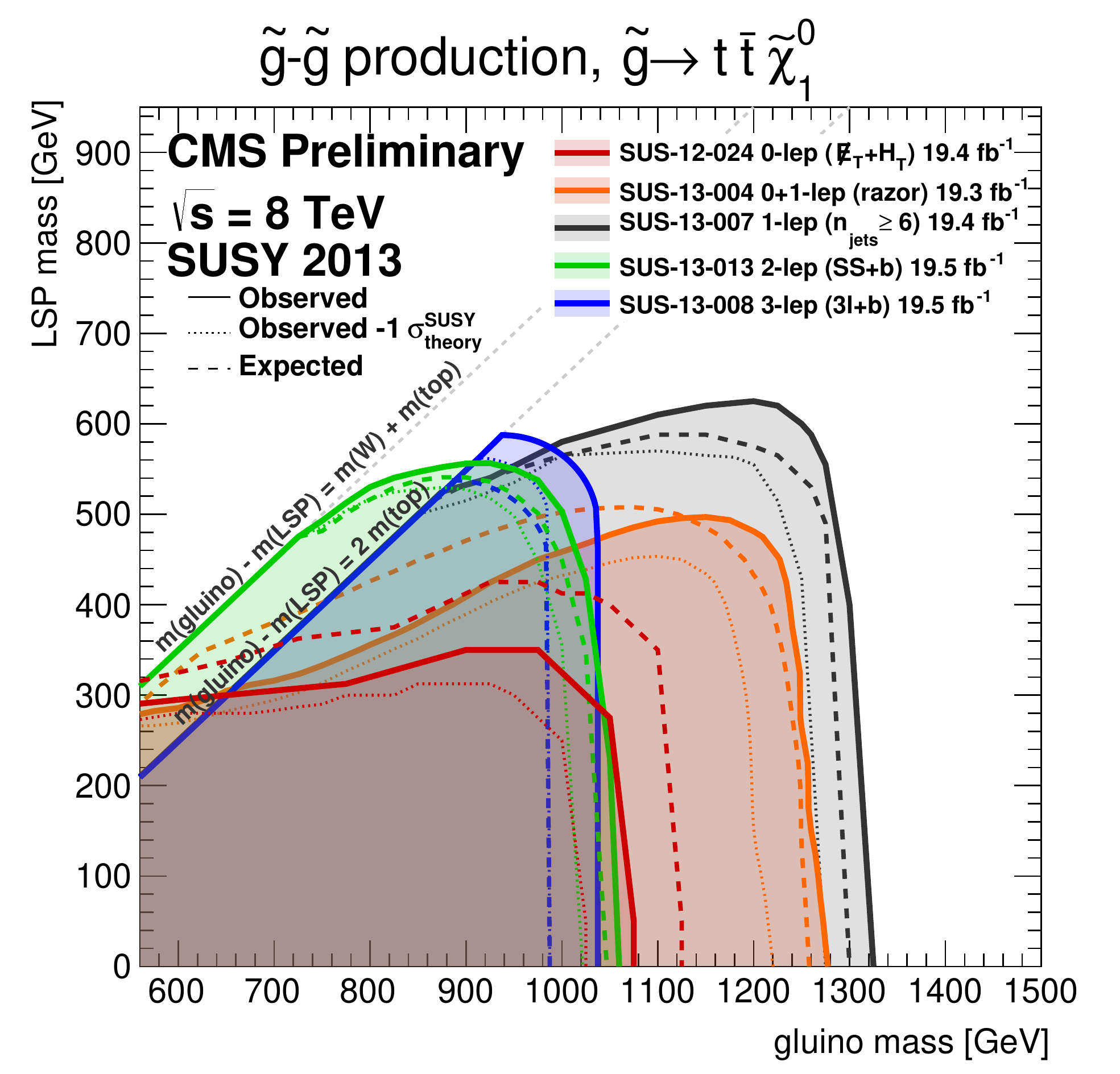}
 \end{center}
\caption{Observed exclusions from various searches in the LSP-stop (ATLAS \cite{Boyd:1596524}, left) and LSP-gluino (CMS\cite{Richman:boh}, right) masses planes.}
\label{fig:stopgluino_searches}
\end{figure}

How is this scenario constrained by the LHC?
Lower limits on the masses of stops and gluinos are reaching values of $\sim 600\div700$ GeV and $\sim 1.3$ TeV respectively, as can be seen from Fig. \ref{fig:stopgluino_searches}. These searches rely on some assumptions. In particular they would be evaded if e.g. the neutralino were close in mass to the stop, see e.g. \cite{Delgado:2012eu}.
On the contrary in CMS and ATLAS the mass reach in the searches of Higgsinos does not go beyond $\sim 300$ GeV, again with possible caveats.

In any case the fine-tuning induced by the null direct searches is less severe than the one required in the MSSM in order to accomodate a 126 GeV Higgs boson. As shown in Fig. \ref{fig:Delta_t} this implied stops heavier than a TeV, resulting in a cancellation of a part over $10^{2\div3}$ in the different contributions to \eqref{eq:SUSY_Zmass}.


\section{The NMSSM as a most natural scenario}
\label{sec:NMSSM_natural}

It is not difficult to think about ways to increase the Higgs mass at tree level and allow in such a way for lighter stops, e.g. via adding an extra gauge group or a singlet. In the following we will be interested in a solution of the second kind, the NMSSM, where as already said a gauge singlet $S$ is added to the MSSM particle spectrum. The most general superpotential is then
\begin{equation}
 W_{\text{NMSSM}} = W_{\text{MSSM}} + \lambda S H_u H_d + f(S)\,,
 \label{eq:W_NMSSM}
\end{equation}
where $f$ is a polynomial containing up to cubic terms in $S$. This results in
\begin{equation}
 m_h < \, m_Z^2 \cos^2 2\beta + \Delta_t^2 + \lambda^2 v^2 \sin^2 2\beta,
\end{equation}
to be compared with \eqref{eq:SUSY_Higgs_mass}, and where we chose $v = 174$ GeV. The additional contribution to this relation is crucial in lowering the Higgs mass sensitivity to the stop masses and, as a consequence, its fine-tuning. The maximal contribution to the Higgs mass is reached for $\tan \beta \sim 1$, furthermore values of $\lambda$ close or greater than one lower the fine-tuning of the Fermi scale \cite{Barbieri:2006bg}.
This last statement can be understood parametrically by looking at the dependence of the weak scale $v$ on $m_{H_u}$, the parameter where the biggest loop corrections enter. Considering as an example the scale invariant NMSSM ($f(S) = \kappa S^3$ and $\mu = 0$\footnote{Note that this specific NMSSM provides a solution to the $\mu$-problem.}), one has $\text{d}v/\text{d}m_{H_u}^2 \simeq \kappa/(\lambda^3 \tan 2 \beta)$, while in the MSSM this was fixed to $\text{d}v/\text{d}m_{H_u}^2 \simeq -2 v^2/m_Z^2$.
As pointed out in \cite{Hall:2011aa}, a minimally tuned case appears to be the NMSSM with a $\lambda >1$, where the too big contribution to the Higgs mass is compensated by a non-negligible mixing with the Singlet. 
However this very compensation can itself reintroduce a tuned cancellation in the determination of $m_h$. 
A quantification depends on the precise definition chosen, as an example in \cite{Gherghetta:2012gb} they considered a scale invariant NMSSM, defined the fine-tuning as the product of the one in $v$ and the one in $m_h$\footnote{This definition maximizes fine-tuning, in the sense that it considers $m_h$ and $v$ as completely independent variables.}, and obtained results at the level of 5\% or better for $\lambda \approx 1$ and stop masses up to 1.2 TeV, again well above current direct bounds.

In summary, the case of moderate $\tan \beta$ and $\lambda \gtrsim 1$ emerges as a most natural scenario. 
On the other hand a well known objection
to the NMSSM at $\lambda \gtrsim 1$ is its compatibility with gauge coupling unification. Requiring $\lambda$ to
stay semi-perturbative up to the GUT scale bounds $\lambda$ at the weak scale to be below about 0.7 \cite{Espinosa:1991gr}.
There are however several ways \cite{Harnik:2003rs,Chang:2004db,Birkedal:2004zx,Delgado:2005fq,Gherghetta:2011wc,Craig:2011ev,Csaki:2011xn,Hardy:2012ef} in which $\lambda$ could go to $1\div 1.5$  without spoiling unification nor affecting  the  consequences at the weak scale of the NMSSM Lagrangian, like e.g. adding a strongly interacting sector at the scale where $\lambda$ becomes non perturbative. One other simple possibility, based on \cite{Barbieri:2013hxa}, is illustrated in Section \ref{sec:Lambda_model}.

Here we underline the importance of investigating to which extent such values of $\lambda$ are excluded by experiments, mainly Higgs signal strength measurements, and where such a most natural scenario is more likely to show up. These issues will be addressed in Chapter~\ref{cha:SUSY_Higgs}.

\section{The SUSY flavour and CP problems}
\label{sec:SUSY_flavCP}

While the SUSY-preserving Lagrangian adds to the SM only one parameter, $\mu$ (and the quartic Higgs terms in the potential are fixed by gauge invariance), the SUSY-breaking one $\mathcal{L}_{\text{soft}}$ adds 103 new physical parameters, among which 42 new CP violating phases. The majority of these new parameters appears in the soft masses $m^2$ and in the $A$-terms, so that in general one expects the CKM picture of flavour and CP violation to be completely ruined. This is an issue all concrete supersymmetric models have to deal with, and that, as we will see, for SUSY near the Fermi scale is solved within the $U(2)^3$ framework. For this reason, we dedicate the rest of this section to state the problem in a more precise, yet synthetic way. Note that the discussion of flavour and CP in the whole section is valid for both the MSSM and the NMSSM. \\
\begin{itemize}
 \item {\bf SUSY CP problem}\\
 The SUSY CP problem is usually defined as the one determined by the severe experimental bounds on the nuclear and leptonic electric dipole moments. Since these CP violating observables are flavour preserving, for simplicity we describe the problem in a supersymmetric model with universal soft terms (i.e. where each trilinear coupling $A_i$ is proportional to the respective Yukawa matrix $Y_i$ via the same constant $a$, the gaugino masses $M$ are all equal, and soft sfermion masses are proportional to the identity).\\ In this simplified framework four independent new phases appear, coming from $\mu$ in the SUSY-preserving Lagrangian and from $b$, $a$ and $M$ in the SUSY-breaking one. One can show that two of them can be removed by suitable field rotations, leaving the two physical ones
 \begin{equation}
 \phi_A = \text{arg}\left( a^* M\right) \qquad \text{and} \qquad \phi_B = \text{arg}\left(M \mu b^* \right)\,.
  \label{SUSYphases}
 \end{equation}
Here we are interested in their impact on electric and chromoelectric dipole moments. If one considers for example those of the first generation quarks, generated via one-loop gluino diagrams, one obtains a contribution to the neutron EDM which reads (we set $a = m_{\tilde{q}}$ for simplicity)\cite{Fischler:1992ha}
\begin{equation}
d_n \simeq 3 \left(\frac{1 \text{TeV}}{m_{\tilde{q}}} \right)^2 \sin \phi_{A,B} 10^{-25} \;e \,cm\,,
 \label{dNSUSY}
\end{equation}
and whose typical size is one order of magnitude bigger than the experimental $90 \%$ C.L. bound $|d_n| < 2.9 \times 10^{-26}\;e \,cm$. Expression \eqref{dNSUSY} already suggests two possible ways out for this problem:
\begin{itemize}
\item[i)] Heavy squarks (actually only the heaviness of the first generation squarks is needed);
\item[ii)] Small flavour blind phases.
\end{itemize}
Of course by relaxing the assumptions on the soft terms one obtains more physical phases which, barring cancellation between different contributions, generically make the bounds stronger.
 \item {\bf SUSY flavour problem}\\
 All the new sources of flavour violation, with respect to the Standard Model, are introduced via the soft masses $m^2$ and the trilinear couplings $A$ in the SUSY-breaking Lagrangian \eqref{SUSYsoft}. If the flavour structure of these terms is generic, then loop diagrams involving gauginos and sfermions induce FCNC processes at levels that are orders of magnitude above the experimentally allowed ranges.
 
 Let us focus on the quark sector. Here four fermion FV operators generated from squark-gluino loops generically show the dependence $\sim \left( \delta^q_{MN} \delta^q_{PQ}\right)^2/m^2_{\tilde{q}}$, where $m^2_{\tilde{q}}$ is a typical squark mass scale and $\left(\delta^q_{MN}\right)_{ij}$ are adimensional non-diagonal entries of the $6 \times 6$ $q$-squarks mass matrix ($M,\dots,Q = L,R$). They can be written as
 \begin{equation}
 \left(\delta_{MM}^q\right)_{ij} = \sum_{k=1}^3 \frac{\left(\Delta m^2_{\widetilde{q}_M}\right)_k}{\bar{m}^2_{\widetilde{q}_M}} \left(W_M^q\right)_{ik} \left(W_M^q\right)^*_{jk}\, ,
 \label{SUSYdeltas}
 \end{equation}
 where $W_{L,R}$ are the mixing matrices appearing in the gluino-squark-quark vertices $\left( \bar{q}_{L,R} \, W_{L,R} \, \widetilde{q}_{L,R} \right) \widetilde{g}$, $\bar{m}_{\widetilde{q}_M}$ is the average soft mass of the $\widetilde{q}_M$ squarks, and $\left(\Delta m^2_{\tilde{q}_M}\right)_k = m^2_{\tilde{q}_k} - \bar{m}^2_{\widetilde{q}_M}$. A similar expression holds for $\left(\delta_{MN}^q\right)_{ij}$, with $M \neq N$.
 
 To give an idea of the size of the stronger bounds, let us mention that taking into account processes induced by $K-\bar{K}$ and $D-\bar{D}$ mixing, for $M_{\tilde{g}} = m_{\tilde{q}} = 1$ TeV one obtains \cite{Isidori:2010kg} $\left(\delta^d_{LR}\right)_{12} \lesssim 10^{-4}$ and $ \left(\delta^u_{LR}\right)_{12} \lesssim 10^{-2}$. In the down quark sector, analogous bounds are weaker by $\sim$ one o.o.m. for $\left(\delta^d_{LL}\right)_{12}$, and by $\sim$ two o.o.m. for $\left(\delta^d_{MN}\right)_{i3}$.
 
The above discussion is useful in indicating what the possible solutions of the SUSY flavour problem are:
\begin{itemize}
\item[i)] Heavy squarks;
\item[ii)] Degenerate squark masses, so that a GIM suppression mechanism is active;
\item[iii)] Alignment in flavour space between the Yukawa matrices and the ``soft'' matrices $m_{\tilde{q}}^2$ and $A$, which implies $|W| \ll 1$;
\item[iv)] Any combination of the above three. For example in a framework with little comunication between the third and the first two generations (i.e. small mixing angles in the relative entries of the $W$ matrices), only the heaviness of the first two generation squark could be needed.
\end{itemize}
Approximate CP conservation would not be a solution, since the stronger bounds from CP conserving processes are one o.o.m. smaller than those coming from CP violating ones, but still very relevant.\\
\end{itemize}


\chapter{Supersymmetry and $U(2)^3$}
\label{cha:U2_SUSY}

The discussion of the previous Chapter shows that supersymmetry as a solution to the hierarchy problem needs some flavour structure for at least two phenomenological reasons\footnote{Note that in order for the Yukawa couplings not to receive too big radiative corrections from the soft sector, one usually assumes each $A$-term to be proportional to the respective yukawa. This assumption can just follow from imposing a flavour symmetry at some higher scale.}. The first is the hierarchical spectrum of Section \ref{sec:NatSUSY}, that e.g. could not be accomodated within a $U(3)^3$ symmetry. Heavier first two generation squarks would also be very welcome to provide a dynamical explanation for the non observation of fundamental EDMs. The second is the severe SUSY flavour problem. Then it appears natural to implement the $U(2)^3$ framework, as discussed in Chapter \ref{cha:EFT}, in the specific case of Supersymmetry. In this Chapter we explore the phenomenological consequences of this embedding.

\section{$U(2)^3$ in Supersymmetry}

We assume the flavour symmetry breaking in the soft squark masses and $A$ terms to be controlled by the same spurions of \eqref{spurions}. 
The analysis of $U(2)^3$ in Supersymmetry will be limited to the Minimal $U(2)^3$ case.
After rotating to the mass basis both the quarks and their superpartners, the supersymmetric mixing matrices appearing in the vertices
$\left( \bar{d}_{L,R} \, W^{L,R} \, \tilde{d}_{L,R} \right) \tilde{g}$ take the correlated forms (in the same basis where the CKM matrix read as in \eqref{CKM})
\begin{equation}
W^L = \left(\begin{array}{ccc}
 c_L^d &  \kappa^* & - \kappa^* s_L e^{i \gamma_L}  \\
- \kappa  &  c_L^d & -c_L^d s_L e^{i \gamma_L}   \\
  0  & s_L e^{-i \gamma_L} & 1
\end{array}\right), \qquad
W^R = \mathbbm{1},
\end{equation}
where $\kappa = s_L^d e^{i (\delta + \varphi)}$, the new parameter $s_L > 0$ is of order $\lambda^2 \sim \epsilon$,
like $s$, and $\gamma_L$ is an independent CP violating phase, which is zero if we do not allow for phases outside the spurions.
$W_R$ is equal to the identity to an accuracy which is sufficient to our purposes and, in particular, to avoid a too large contribution to $\epsilon_K$
.

\section{$\Delta F = 2$ observables and CKM fit tension}
\label{sec:DF2Susy}
\begin{figure}[tbp]
\begin{center}
\includegraphics[width=0.5\textwidth]{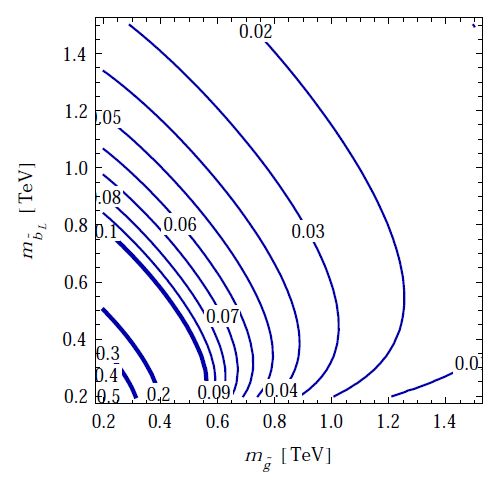}
\includegraphics[width=0.49\textwidth]{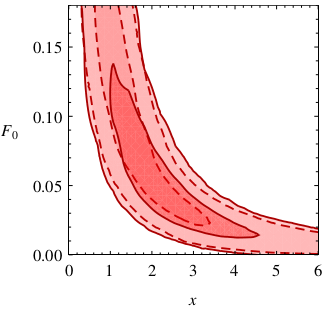}
\end{center}
\caption{\textit{Left}: Value of the loop function defined in \eqref{F0} as a funtion of the gluino and the left-handed sbottom masses. \textit{Right}: Correlation among the preferred values of $x = |\xi_L|^2$ and $F_0$, as obtained from the global fit. The dashed contours correspond to the $68\%$ and $90\%$ C.L. regions in the fit without $S_{\psi \phi}$, the solid contours to the fit inlcuding the LHCb result \cite{LHCb:2011aa} for $S_{\psi \phi}$. Figure taken from \cite{Straub:2011fs}.}
\label{fig:F0}
\end{figure}

\begin{figure}[tbp]
\begin{center}
\includegraphics[width=0.8\textwidth]{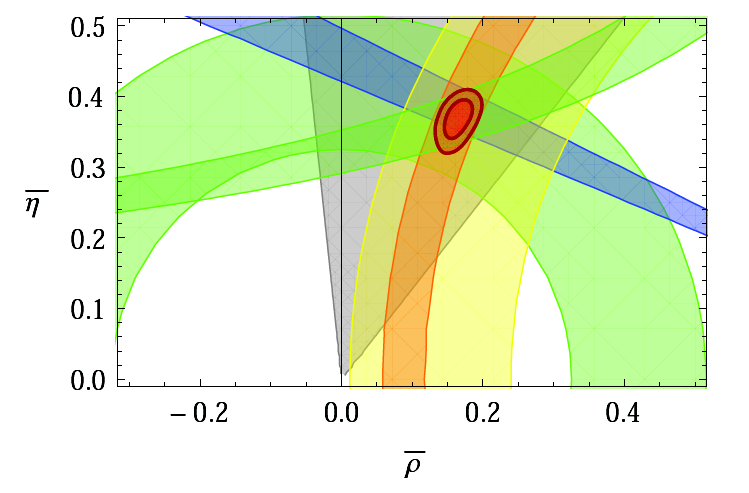}
\end{center}
\caption{Result of the global fit with the inclusion of the corrections as in Eqs. \eqref{SusyEpsilon},\eqref{SusySpsiK},\eqref{SusyDeltaMb}.}
\label{fig:CKMsusy}
\end{figure}

The mixing amplitudes in the kaon and $B_{d,s}$ systems,
including SM and gluino-mediated contributions,
read:
\begin{align}
&M_{12}^K  = (M_{12}^K)^\text{SM, tt} \left(1 + |\xi_L|^4 F_0\right) + (M_{12}^K)^\text{SM, tc + cc},\label{MKsusy}\\
&M_{12}^{B_{d,s}} = (M_{12}^{B_{d,s}})^\text{SM} \left(1+\xi_L^2 F_0\right), \label{MBsusy}
\end{align}
where
\begin{equation}
\xi_L = (c_L^d s_L/|V_{ts}|)\, e^{i \gamma_L}
\label{eq:xi_L}
\end{equation}
and $F_0(m_{\tilde{b}}, m_{\tilde{g}})$ is a positive function of the sbottom and gluino masses,
\begin{align}
&F_0  = \frac{2}{3} \left( \frac{g_s}{g} \right)^4 \frac{m_W^2}{m^2_{\tilde{b}}} \frac{1}{S_0(x_t)} 
\left[ f_0(x_g) + \mathcal{O}\left( \frac{m^2_{\tilde{b}}}{m^2_{\tilde{h}}}\right) \right]\,, & \,x_g  = \frac{m^2_{\tilde{g}}}{m^2_{\tilde{b}}}\,, \label{F0}\\
&f_0(x)  = \frac{11 + 8 x - 19 x^2 + 26 x \log{x} + 4 x^2 \log{x}}{3 (1-x)^3} \,,
& f_0(1) = 1\, ,
\end{align}
where $S_0(x_t = m_t^2/m_W^2) \simeq 2.4$ is the SM one-loop electroweak coefficient function. Note that, since the SM and gluino-mediated contributions generate the same $\Delta F = 2$ effective operator, all non-perturbative effects and long-distance QCD corrections have been factorized. The typical size of $F_0$, as a funtion of the gluino and sbottom mass, is shown in the left panel of Fig. \ref{fig:F0}.
\\Using \eqref{MKsusy}, \eqref{MBsusy} one obtains correlated
expressions for the indirect CP violating parameter $\epsilon_K$ and the mixing-induced CP asymmetries in
$B^0 \to \psi K_S$ and $B^0 \to \psi \phi$ decays:
\begin{align}
\epsilon_K &= \epsilon^\text{SM, tt}_K \left(1 + |\xi_L|^4 F_0\right) +
\epsilon^\text{SM, tc + cc}_K, \vspace{.2 cm}\label{SusyEpsilon}\\
S_{\psi K_S} &=\sin\left(2\beta + \phi_\Delta\right), \vspace{.2 cm}\label{SusySpsiK}\\
\Delta M_{d,s} &= \Delta M_{d,s}^{\text{SM}} \times \left| 1 + \xi_L^2 F_0\right| \label{SusyDeltaMb}\\
S_{\psi\phi} &=\sin\left(2|\beta_s| - \phi_\Delta\right) \label{SusySpsiphi},
\end{align}
where $\beta$ and $\beta_s$ are the SM mixing phases and
\begin{equation}
\phi_\Delta=\text{arg}\!\left(1 + \xi_L^2 F_0\right).
\label{eq:phi_Delta}
\end{equation}
The above effects are interesting in light of the tension in the CKM description of flavour and CP  violation, namely among $\epsilon_K$, $S_{\psi K_S}$ and $\Delta M_d/\Delta M_s$ \cite{Lunghi:2008aa, Buras:2008nn, Altmannshofer:2009ne, Lunghi:2010gv, Bevan:2010gi}. This tension can be solved in our framework by the new contributions we obtain to the first two observables. The improvement in the CKM unitarity fit is evident by comparing the result in the $(\bar{\rho}, \bar{\eta})$ plane of Fig. \ref{fig:CKMsusy} with the one of Fig. \ref{fig:CKMTensions} in Chapter 2. More precisely $\chi^2/\text{N}_\text{dof} = 0.7/2$ is found, compared to $\chi^2/\text{N}_\text{dof} = 9.8/5$ for the full SM fit.\footnote{This fit, as well as all the phenomenology discussed in this Chapter, does not take into account this year's new lattice and experimental results, see Section \ref{sec:DF12Minimal}.}
Note that a specific prediction of the model (with the assumption of gluino dominance) is a definite sign of the correction to $\epsilon_K$, which is the one needed in order to solve the CKM tension.\\
A global fit is then performed, varying the SUSY parameters $\xi_L$, $F_0$ and $\gamma_L$ together with
the four parameters of the CKM matrix, and using all the observables listed in Table \ref{tab:inputs} as constraints.
This leads, at the $90 \%$ C.L., to the preferred intervals
\begin{equation}
|\xi_L| \in [0.8, \, 2.1], \quad
\phi_\Delta \in [-9^\circ,\, -1^\circ], \quad
\gamma_L \in [-86^\circ, \,-25^\circ] \; \text{ or } \;\gamma_L \in [94^\circ, \,155^\circ], 
\label{DeltaF2}
\end{equation}
and to the predictions $S_{\psi \phi} \in [0.05, \,0.20]$ and  $m_{\tilde{g}}, \; m_{\tilde{b}} \lesssim 1.5$ TeV
(the latter due to $F_0 \neq 0$, see the right panel of Fig. \ref{fig:F0}, the former only if $\phi_s$ is not included in the fit).
The favoured value for $S_{\psi \phi}$ is within the $1 \sigma$ interval of the LHCb result \cite{LHCb:2011aa}, which was absent at the time of publication of these results.\\
As already stressed in Chapter 2, some of the predictions we make are independent of the specific model under consideration
(Supersymmetry with hierarchical squark masses),
namely (i) the absence of a new phase in $M_{12}^K$,
(ii) the presence of a new phase in $B_{s,d}$ mixing
and (iii) the universality $M_{12}^{B_d} = M_{12}^{B_s}$.

\section{$\Delta F = 1$ observables: $B$-decay CP asymmetries}
\label{sec:BdecaySusy}
The purpose of this section is to extend the analysis of the previous one to $\Delta F=1$ 
processes, i.e.~rare decays, studying in particular possible signatures of CP violation 
correlated with the predicted CP violation in meson mixing ($\Delta F=2$).
Contrary to the  $\Delta F=2$ sector, where the pattern of deviations from the SM previously 
identified is unambiguously dictated by the $U(2)^3$ symmetry, the predictions of $\Delta F=1$ observables are more model dependent. 
In this analysis we concentrate on a framework with moderate values of $\tan\beta$
and, in order to establish a link between 
$\Delta F=2$ and $\Delta F=1$ CP-violating observables, 
we take small flavour-blind phases and assume
the dominance of gluino-mediated flavour-changing amplitudes.

\subsection{$\Delta B=1$ effective Hamiltonian}\label{sec:Heff}

The part of the $b\to s$ effective Hamiltonian sensitive to NP in our setup reads
\begin{align}
 \mathcal{H}_\text{eff} &= -\frac{4 G_F}{\sqrt2} V_{ts}^*V_{tb} \sum_{i=3}^{10} C_i O_i + \text{ h.c.}\,,
\end{align}
\begin{align*} 
  O_3 & =(\bar{s}P_Lb)\tsum_q(\bar{q}P_Lq) \qquad &
  O_4 & =(\bar{s}_{\alpha}P_Lb_{\beta})\tsum_q(\bar{q_{\beta}}P_Lq_{\alpha}) \\
  O_5 & =(\bar{s}P_Lb)\tsum_q(\bar{q}P_Rq) \qquad &
  O_6 & =(\bar{s}_{\alpha}P_Lb_{\beta})\tsum_q(\bar{q_{\beta}}P_Rq_{\alpha}) \\
  O_{7} & =\frac{e}{16\pi^2}m_b(\bar{s}\sigma_{\mu\nu}P_R b)F_{\mu\nu} \qquad  &
  O_{8} & =\frac{g_s}{16\pi^2}m_b(\bar{s}\sigma_{\mu\nu}P_R b)G_{\mu\nu} \\
  O_{9} & =\frac{e^2}{16\pi^2}(\bar{s}\gamma_{\mu}P_L b)(\bar{l}\gamma_{\mu}l) \qquad &
  O_{10} & =\frac{e^2}{16\pi^2}(\bar{s}\gamma_{\mu}P_L b)(\bar{l}\gamma_{\mu}\gamma_5 l) 
\end{align*}

The QCD penguin operators $O_{3\ldots6}$ are relevant for the CP asymmetries in $b\to s\bar ss$ penguin decays to be discussed below. In the case of hierarchical sfermions, the box contributions are mass-suppressed and we only have to consider gluon penguins, which contribute in a universal way as
\begin{equation}
C_3 = C_5 = -\tfrac{1}{3} C_4 = -\tfrac{1}{3} C_6 \equiv  C_G \,.
\end{equation}
The analogous photon penguin can be neglected in non-leptonic decays since it is suppressed by $\alpha_{em}/\alpha_s$, but it can contribute to $C_9$. There is no effect in $C_{10}$, which remains SM-like.

In the MSSM, the $\Delta F=1$ effective Hamiltonian receives contributions from loops involving charginos, neutralinos, charged Higgs bosons or gluinos. In the following, we will concentrate for simplicity on gluino contributions, which are always proportional to the complex $\xi_L$ in $U(2)^3$. Among the remaining contributions, some are proportional to $\xi_L$ while some are real in the absence of flavour blind phases. Their omission does not qualitatively change our predictions for CP asymmetries,  but we stress that the real contributions can have an impact in particular on the branching ratios to be considered below.

The gluino contributions to $C_G$ and $C_9$ are
\begin{equation}
C_G  = - \xi_L \; \dfrac{\alpha_s}{\alpha_2} \; \dfrac{\alpha_s}{4 \pi} \; \dfrac{m_W^2}{m_{\tilde{b}}^2} \;
\dfrac{13}{108} f_G(x_{\tilde{g}})
\,, \qquad
C_9 = \xi_L \; \dfrac{\alpha_s}{\alpha_2} \; \dfrac{m_W^2}{m_{\tilde{b}}^2} \;
\dfrac{2}{27} f_{\gamma}(x_{\tilde{g}})
\,,
\end{equation}
where here and throughout, $x_{\tilde{g}} = m_{\tilde{g}}^2/m_{\tilde{b}}^2$ and all the loop functions are defined such that $f_i(1)=1$ with the exact form given in the end of this section. 

The main difference concerning the magnetic and chromomagnetic Wilson coefficients $C_{7}$ and $C_{8}$ is that here $\tilde{b}_L$-$\tilde{b}_R$
mass insertion contributions are important, while for the other Wilson coefficients they were
chirality-suppressed. The gluino contributions read
\begin{align}
 C_{7} & =  - \xi_L \; \dfrac{\alpha_s}{\alpha_2} \; \dfrac{m_W^2}{m_{\tilde{b}}^2}\;
 \dfrac{1}{27} \left[ f_{7}(x_{\tilde{g}}) + 2 \dfrac{\mu \tan \beta - A_b}{m_{\tilde{g}}}
 g_{7}(x_{\tilde{g}}) \right]
 \,,\\
 C_{8} & =  - \xi_L \; \dfrac{\alpha_s}{\alpha_2} \; \dfrac{m_W^2}{m_{\tilde{b}}^2} \;
 \dfrac{5}{36} \left[ f_{8}(x_{\tilde{g}}) + 2 \dfrac{\mu \tan \beta - A_b}{m_{\tilde{g}}}
 g_{8}(x_{\tilde{g}}) \right]
 \,.
\end{align}

For $m_{\tilde g}=m_{\tilde b}\equiv\tilde m$, we thus find\footnote{For  
$0.5 < m_{\tilde g}/ m_{\tilde b}  < 2$  the relative variation of the numerical coefficients is within $\pm 50\%$.} 
at the scale $m_W$
\begin{align}
 C_G & =  -1.1\times10^{-4}
 \left( \dfrac{500 \GeV}{\tilde m}\right)^2 \xi_L
\,,\\
 C_9 & =  9\times10^{-3} \left( \dfrac{500 \GeV}{\tilde m}\right)^2 \xi_L
\,,\\
 C_{7} & =  -3.4\times10^{-3} \left( \dfrac{500 \GeV}{\tilde m}\right)^2 \xi_L
 \left[1 + 2 \dfrac{\mu \tan \beta - A_b}{\tilde m} \right]
,\\
 C_{8} & =  -1.3\times10^{-2} \left( \dfrac{500 \GeV}{\tilde m}\right)^2 \xi_L
 \left[ 1 + 2 \dfrac{\mu \tan \beta - A_b}{\tilde m}  \right]
.
\end{align}

A model-independent consequence of the $U(2)^3$ symmetry
is that the modification of $b\to s$ and $b\to d$ $\Delta F=1$ amplitudes is {\em universal}, i.e.~only distinguished by the same CKM factors as in the SM (exactly as in the  $U(3)^3$, 
or MFV case~\cite{D'Ambrosio:2002ex}). Therefore, all the expressions for the $b\to s$ Wilson coefficients 
derived in this section are also valid for $b\to d$ processes.

\subsubsection{Loop functions}
\label{sec:LF}

\begin{center}
\begin{tabular}{ll}
$f_G(x) = \dfrac{2(73 - 134 x + 37 x^2)}{39 (x-1)^3} -
\dfrac{2(18 - 27 x + x^3)}{13 (x-1)^4} \ln x$ & $f_G(1) = 1$\\
\\
$f_{\gamma}(x) = -\dfrac{2(2 - 7 x + 11 x^2)}{3(x-1)^3} +
\dfrac{4 x^3}{(x-1)^4} \ln x$ & $f_{\gamma}(1) = 1$\\
\\
$f_{7}(x) = \dfrac{2(-1 + 5 x + 2 x^2)}{(x-1)^3} -
\dfrac{12 x^2}{(x-1)^4} \ln x$ & $f_{7}(1) = 1$\\
\\
$g_{7}(x) = -\dfrac{6 x (1 + 5 x)}{(x-1)^3} +
\dfrac{12 x^2 (2 + x)}{(x-1)^4} \ln x$ & $g_{7}(1) = 1$\\
\\
$f_{8}(x) = \dfrac{-19 - 40 x + 11 x^2}{5 (x-1)^3} -
\dfrac{6 x (-9 + x)}{5 (x-1)^4} \ln x$ & $f_{8}(1) = 1$\\
\\
$g_{8}(x) = \dfrac{12 x (11 + x)}{5 (x-1)^3} +
\dfrac{6 x (-9 - 16 x + x^2)}{5 (x-1)^4} \ln x$ & $g_{8}(1) = 1$
\end{tabular}
\end{center}

\subsection{$B$ physics observables}

\begin{table}[tp]
\renewcommand{\arraystretch}{1.3}
 \begin{center}
\begin{tabular}{llll}
\hline
Observable & SM prediction & Experiment & Future sensitivity\\
\hline
$\text{BR}(B\to X_s\gamma)$ & $(3.15\pm0.23)\times10^{-4}$ \cite{Misiak:2006zs} & $(3.52\pm0.25)\times10^{-4}$ & $\pm0.15\times10^{-4}$  \\
$A_\text{CP}(b\to s\gamma)$ &  $(-0.6\div2.8)\%$ \cite{Benzke:2010tq} & $(-1.2 \pm 2.8) \%$ & $\pm0.5\%$ 
\\
$\text{BR}(B\to X_d\gamma)$ & $(1.54^{+0.26}_{-0.31})\times10^{-5}$ \cite{Crivellin:2011ba} & $(1.41\pm0.49)\times10^{-5}$ &
\\
$S_{\phi K_S}$  & $0.68\pm0.04$ \cite{Buchalla:2005us,Beneke:2005pu} & $0.56^{+0.16}_{-0.18}$ & $\pm0.02$ \\
$S_{\eta' K_S}$ & $0.66\pm0.03$ \cite{Buchalla:2005us,Beneke:2005pu} & $0.59\pm0.07$  & $\pm0.01$ \\
$\langle A_7 \rangle$ & $(3.4\pm0.5)\times10^{-3}$ \cite{Altmannshofer:2008dz} & -- &  \\
$\langle A_8 \rangle$ & $(-2.6\pm0.4)\times10^{-3}$ \cite{Altmannshofer:2008dz} & -- &  \\
\hline
 \end{tabular}
 \end{center}
\renewcommand{\arraystretch}{1}
 \caption{SM predictions, current experimental world averages \cite{Asner:2010qj} and experimental sensitivity at planned experiments \cite{Aushev:2010bq,O'Leary:2010af} for the $B$ physics observables. $<A_{7,8}>$ are suitable angular CPV asymmetries in $B\rightarrow K^* \mu^+ \mu^-$.}
 \label{tab:exp}
\end{table}

\subsubsection{BR($B\to X_q\gamma$)}\label{sec:bsg}

The branching ratio of $B\to X_s\gamma$ is one of the most important flavour constraints in the MSSM in view of the good agreement between theory and experiment. Experimentally, the quantities
\begin{equation}
R_{bq\gamma} = \frac{\text{BR}({B} \rightarrow X_q \gamma)}{\text{BR}({B} \rightarrow X_q \gamma)_\text{SM}}
\end{equation}
are constrained to be
\begin{equation}
R_{bs\gamma}=1.13\pm0.11
\,,\qquad
R_{bd\gamma}=0.92\pm0.40
\,,
\label{eq:Rbqg}
\end{equation}
using the numbers in Table~\ref{tab:exp}. In $U(2)^3$, one has $R_{bs\gamma}\approx R_{bd\gamma}$ just as in MFV, so the $b\to s\gamma$ constraint is more important and we will concentrate on it in the following.

Beyond the SM (but in the absence of right-handed currents), the branching ratio can be written as 
\begin{equation}
\text{BR}({B} \rightarrow X_s \gamma) = R \left[ \left|C_{7}^\text{SM,eff} + C_{7}^\text{NP,eff} \right|^2
+ N(E_{\gamma}) \right],
\end{equation}
where $R = 2.47 \times 10^{-3}$ and $N(E_{\gamma}) = (3.6 \pm 0.6) \times 10^{-3}$ for a photon energy cut-off $E_{\gamma} = 1.6$ \GeV\ \cite{Buras:2011zb}.

Considering only gluino contributions and setting $m_{\tilde g}=m_{\tilde b}\equiv\tilde m$, we find
\begin{align}
R_{b s\gamma}  &= 1 \;+ \; 2.2 \times 10^{-2} 
\left( \dfrac{500 \GeV}{\tilde m}\right)^2
|\xi_L| \cos \gamma_L
\left( 1
+ 2 \dfrac{\mu \tan \beta - A_b}{\tilde m}  \right)
.
\end{align}

As stressed in Section~\ref{sec:Heff}, there are additional real contributions to the Wilson coefficient $C_{7,8}$ that can modify the branching ratio. In particular, there is a $\tan\beta$ enhanced chargino contribution proportional to the stop trilinear coupling, which can interfere constructively or destructively with the SM. Thus, with a certain degree of fine-tuning, the constraints in (\ref{eq:Rbqg}) can always be fulfilled. In our numerical analysis, we will require the branching ratio including {\em only SM and gluino contributions} to be within $3\sigma$ of the experimental measurement.

\subsubsection{$A_\text{CP}(B\to X_s\gamma)$}\label{sec:Absg}

The direct CP asymmetry in $B\to X_s\gamma$
\begin{equation}
A_\text{CP}(B\to X_s\gamma)= \dfrac{\Gamma(\bar{B} \rightarrow X_s \gamma) - \Gamma(B\rightarrow X_{\bar{s}} \gamma)}
{\Gamma(\bar{B} \rightarrow X_s \gamma) + \Gamma(B \rightarrow X_{\bar{s}} \gamma)}
\,,
\end{equation}
already constrained by the $B$ factories as shown in Table~\ref{tab:exp}, will be measured by next generation experiments to a precision of 0.5\%. On the theory side, the recent inclusion of ``resolved photon'' contributions reduced the attainable sensitivity to NP in view of the large non-perturbative effects leading to a SM estimate \cite{Benzke:2010tq}
\begin{equation}
-0.6\% <A_\text{CP}(B\to X_s\gamma)_\text{SM} < 2.8\%
\end{equation}
compared to an earlier estimate \cite{Hurth:2003dk}
\begin{equation}
 A_\text{CP}(B\to X_s\gamma)_\text{SM}^\text{SD} = (0.44^{+0.24}_{-0.14})\% \,.
\end{equation}

In view of these uncertainties and to get an estimate of the size of the NP effects, we will consider the NP contributions to the CP asymmetry ignoring the resolved photon contributions.
It can then be written as
\begin{equation}
 A_\text{CP}(B\to X_s\gamma)_\text{NP}^\text{SD} \times R_{b s\gamma} = -0.29 \text{Im}(C_{7}^{\text{NP}}) + 0.30 \text{Im}(C_{8}^\text{NP})
-0.99 \text{Im}(C_{7}^{\text{NP}*}C_{8}^\text{NP}),
\end{equation}
valid for $E_{\gamma}=1.85 \GeV$.

Setting $m_{\tilde g}=m_{\tilde b}\equiv\tilde m$ and $R_{bs\gamma}=1$, the gluino contributions to the CP asymmetry are
\begin{align}
 A_\text{CP}(B\to X_s\gamma)_\text{NP}^\text{SD} = 
- 1.74 \times 10^{-3} \left( \dfrac{500 \GeV}{\tilde m}\right)^2
 |\xi_L| \sin \gamma_L
\left(1
+ 2 \dfrac{\mu \tan \beta - A_b}{\tilde m} \right)
\end{align}
we note that the CP asymmetry can have either sign due to the two solutions for $\gamma_L$ allowed by $\Delta F=2$, see \eqref{DeltaF2}.

\subsubsection{$B \rightarrow K^* \mu^+ \mu^-$}

Angular CP asymmetries in $B \rightarrow K^* \mu^+ \mu^-$ are sensitive probes of non-standard CP violation and will be measured soon at the LHCb experiment\footnote{For an updated, model independent analysis of the recent data not included here see \cite{Descotes-Genon:2013wba,Altmannshofer:2013foa}.}. In our framework, where right-handed currents are absent, the relevant observables are the T-odd CP asymmetries $A_7$ and $A_8$ as defined in \cite{Altmannshofer:2008dz}.

For these observables, integrated in the low dilepton invariant mass region, we obtain the simple dependence on the Wilson coefficients, as usual to be evaluated at the scale $m_b$,
\begin{align}
\langle A_7 \rangle \times R_\text{BR} & \approx  - 0.71 \; \text{Im} (C_{7}^\text{NP})
\,,\\
\langle A_8 \rangle \times R_\text{BR}  & \approx    0.40 \; \text{Im} (C_{7}^\text{NP}) + 0.03 \; \text{Im} (C_{9}^\text{NP})
\,,
\end{align}
where $R_\text{BR}$ is the ratio between the full result for the CP-averaged branching ratio and the SM one \cite{Barbieri:2011vn}, $R_\text{BR} \approx 1$ in our framework.
Although $\text{Im} (C_{7}^\text{NP})$ and $\text{Im} (C_{9}^\text{NP})$ can be of the same order, the contribution from $C_{9}^\text{NP}$ is numerically suppressed and one will thus still have approximately
\begin{equation}
\langle A_8 \rangle \simeq - 0.56 \langle A_7 \rangle
\,.
\end{equation}
Setting $m_{\tilde g}=m_{\tilde b}\equiv\tilde m$ and $R_\text{BR}=1$, the gluino contributions to $\langle A_7 \rangle$ read
\begin{align}
\langle A_7 \rangle
&
= 2.5 \times10^{-3} \left(
\dfrac{500 \GeV}{\tilde m} \right)^2 
|\xi_L| \sin \gamma_L
\left(1+
2 \dfrac{\mu \tan \beta - A_b}{\tilde m}
\right) \,.
\end{align}

\subsubsection{$S_{\phi K_S}$ and $S_{\eta' K_S}$}

The expression for the mixing-induced CP asymmetries in $B_d$ decays to final CP eigenstates $f$ is
\begin{equation}
S_f = \sin \left( 2 \beta + \phi_\Delta + \delta_f \right),
\end{equation}
where $\phi_\Delta$ is the new phase in $B_{d,s}$ mixing defined in \eqref{eq:phi_Delta}. For the tree-level decay $f=\psi K_S$, $\delta_f=0$, while for the penguin-induced modes $B\to\phi (\eta')K_S$, the contribution from the decay amplitude can be written as \cite{Buchalla:2005us}
\begin{equation}
\delta_f =
2 \text{arg}\bigg(1 + a^u_f e^{i \delta} + \sum_{i \geq 3} b_{i,f}^c C_i^\text{NP}\bigg)
\end{equation}
where $\delta=\gamma_\text{CKM}=\phi_3$ is the usual CKM angle and the $a^u_f$ and $b^c_{i,f}$ can be found in \cite{Buchalla:2005us}.

For the gluino contributions, setting $m_{\tilde g}=m_{\tilde b}\equiv\tilde m$, we obtain
\begin{align}
\sum_{i=3}^6 b_{i,\phi K_S}^c C_i^\text{NP}& = -1.11\times 10^{-2}  \left( \dfrac{500 \GeV}{\tilde m}\right)^2
|\xi_L| e^{i\gamma_L}
\,,\\
\sum_{i=3}^6 b_{i,\eta' K_S}^c C_i^\text{NP}& = -1.10\times 10^{-2}  \left( \dfrac{500 \GeV}{\tilde m}\right)^2
|\xi_L| e^{i\gamma_L}
\,,
\end{align}
\begin{align}
b_{8,\phi K_S}^c C_8^\text{NP}& = -1.82\times 10^{-2}  \left( \dfrac{500 \GeV}{\tilde m}\right)^2
|\xi_L| e^{i\gamma_L} \left(1+ 2\dfrac{\mu \tan \beta - A_b}{\tilde m}\right)
\label{eq:C8phi}
,\\
b_{8,\eta' K_S}^c C_8^\text{NP}& = -1.10\times 10^{-2}  \left( \dfrac{500 \GeV}{\tilde m}\right)^2
|\xi_L| e^{i\gamma_L} \left(1+ 2\dfrac{\mu \tan \beta - A_b}{\tilde m}\right)
\label{eq:C8eta}
.
\end{align}
The effects of the QCD and chromomagnetic penguins in the above expressions are comparable, with the exception of the left-right mixing piece only present for the chromomagnetic ones.

\begin{figure}[tbp]
\begin{center}
\includegraphics[width=0.48\textwidth]{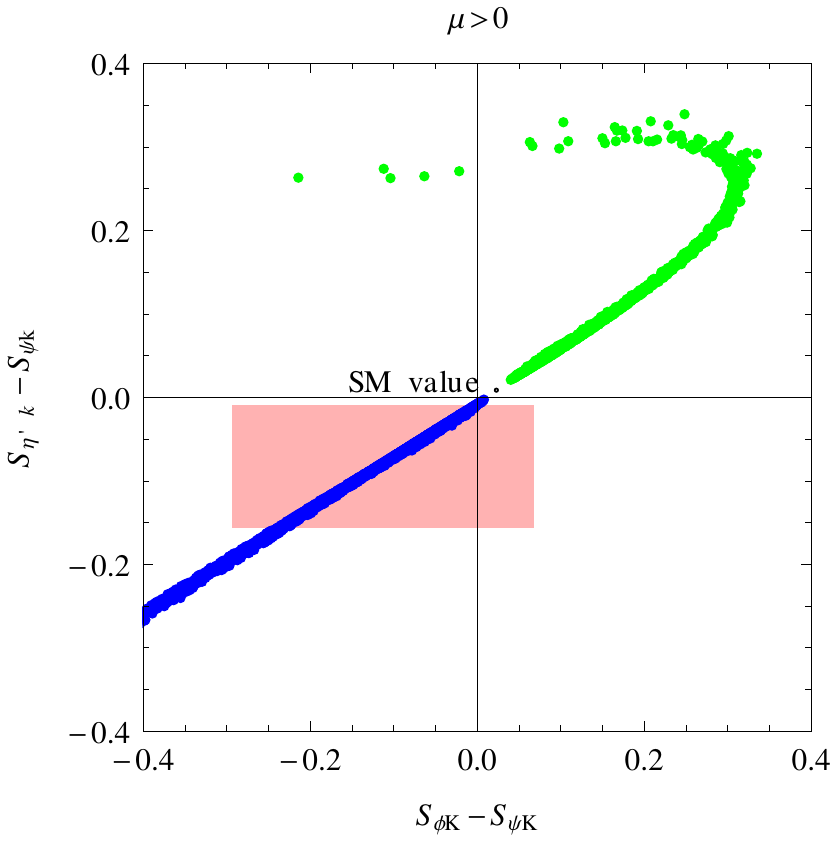}
\includegraphics[width=0.48\textwidth]{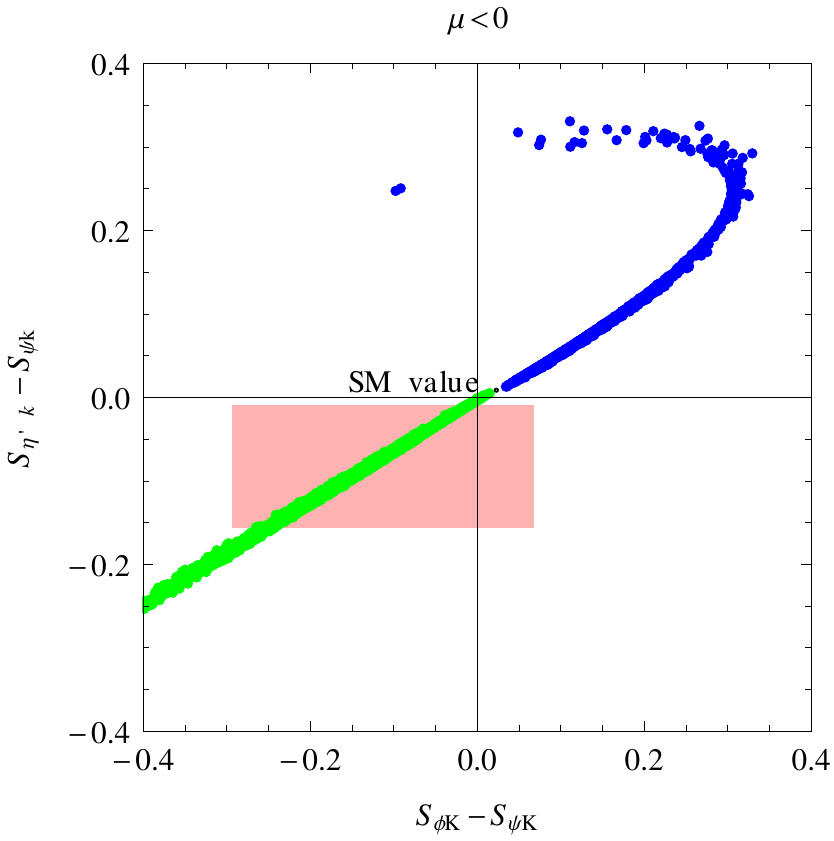}
\end{center}
\caption{Correlation between $S_{\phi K_S}-S_{\psi K_S}$ and $S_{\eta' K_S}-S_{\psi K_S}$ for positive $\mu$ (left) and negative $\mu$ (right), showing points with $\gamma_L>0$ (blue) and $\gamma_L<0$ (green). The shaded region shows the $1\sigma$ experimental ranges.}
\label{fig:S}
\end{figure} 
\begin{figure}[tbp]
\begin{center}
\includegraphics[width=0.48\textwidth]{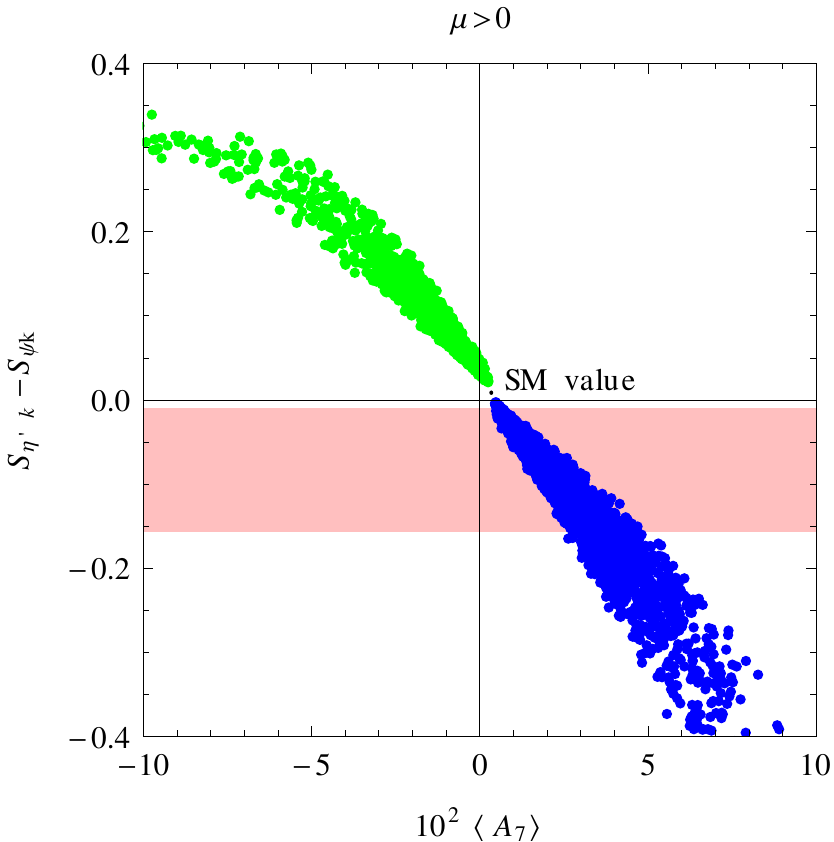}
\includegraphics[width=0.48\textwidth]{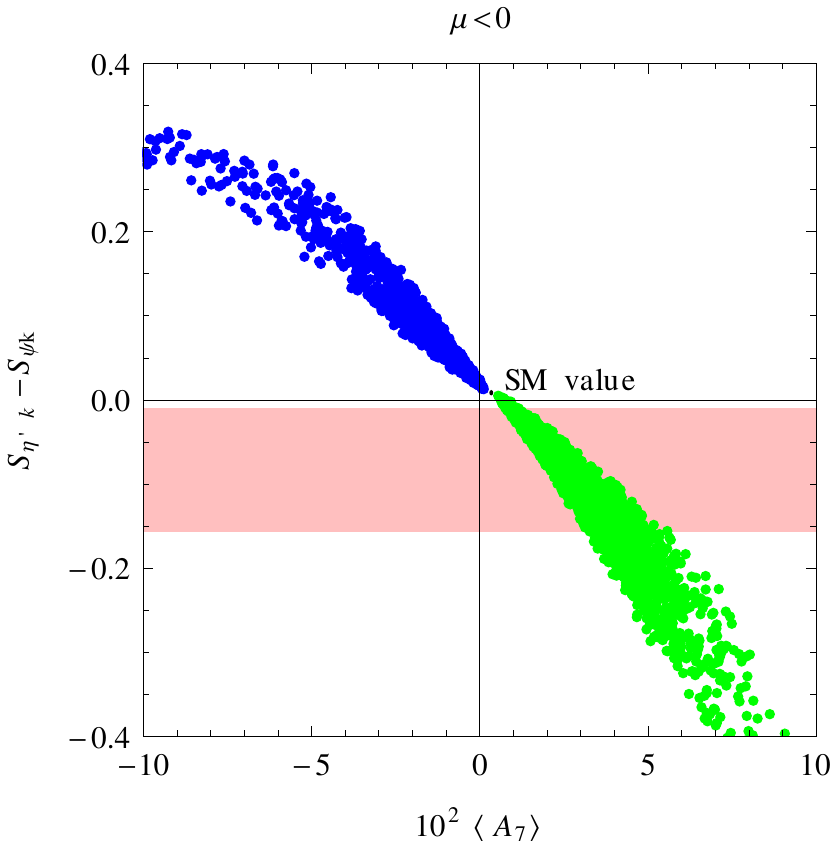}
\end{center}
\caption{Correlation between $\langle A_7 (B\to K^*\mu^+\mu^-)\rangle$ and the difference $S_{\eta' K_S}-S_{\psi K_S}$ for positive $\mu$ (left) and negative $\mu$ (right), showing points with $\gamma_L>0$ (blue) and $\gamma_L<0$ (green). The shaded region is the $1\sigma$ experimental range.
}
\label{fig:A7}
\end{figure} 
\begin{figure}[tbp]
\begin{center}
\includegraphics[width=0.48\textwidth]{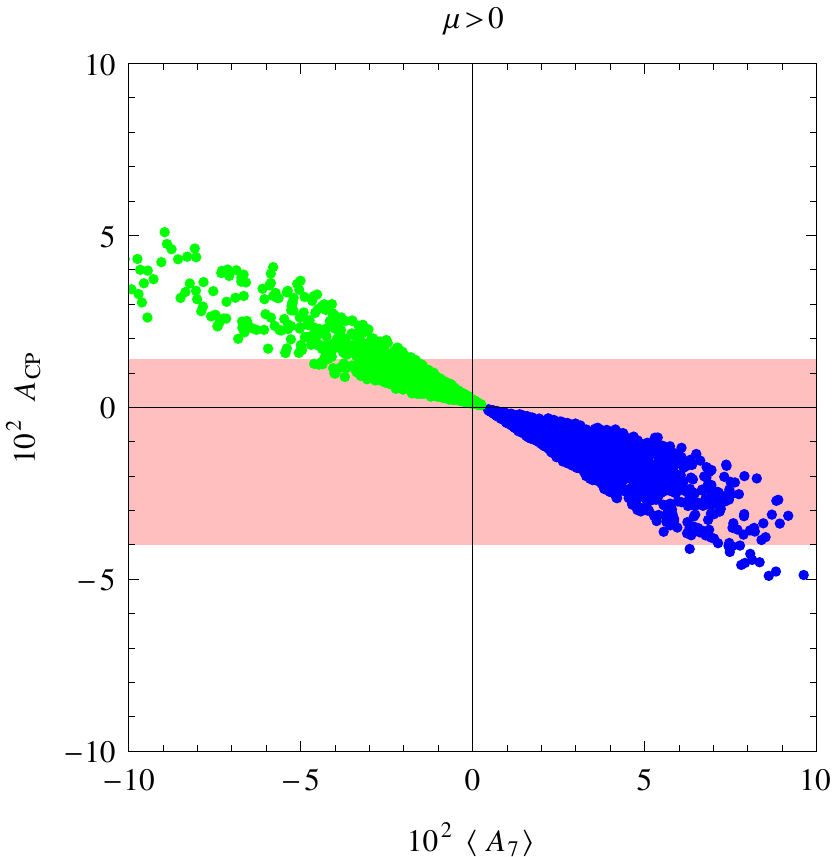}
\includegraphics[width=0.48\textwidth]{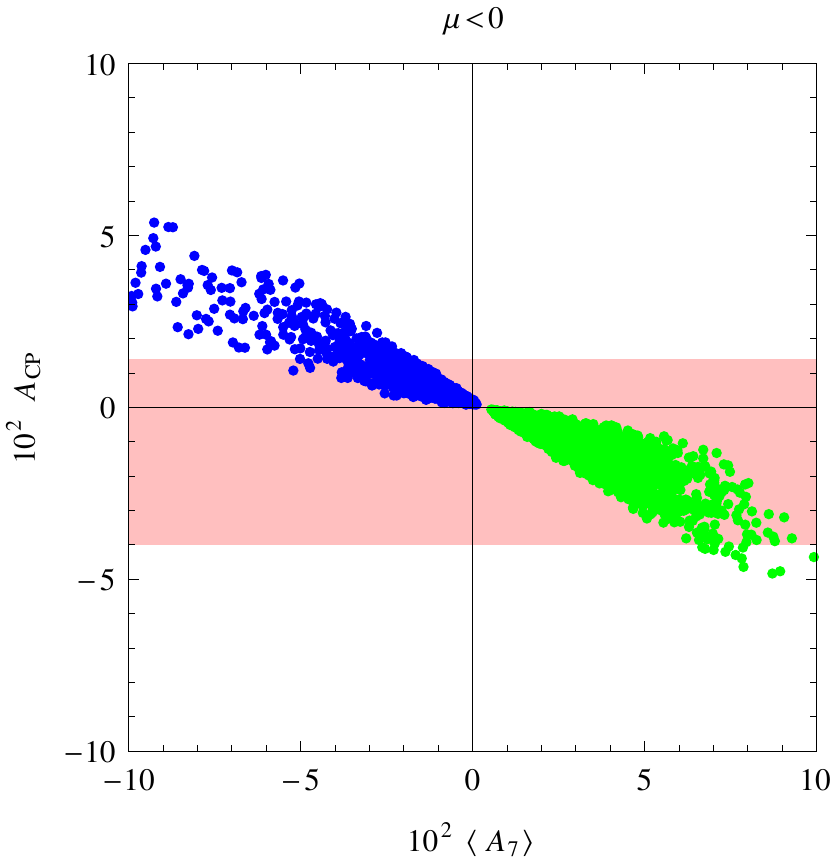}
\end{center}
\caption{Correlation between $\langle A_7 (B\to K^*\mu^+\mu^-)\rangle$ and the NP contributions to the CP asymmetry in $B\to X_s\gamma$ (neglecting long-distance effects) for positive $\mu$ (left) and negative $\mu$ (right), showing points with $\gamma_L>0$ (blue) and $\gamma_L<0$ (green). The shaded region is the $1\sigma$ experimental range for $A_\text{CP}(B\to X_s\gamma)$, which would apply in the absence of long-distance effects (see Section~\ref{sec:Absg}).
}
\label{fig:A7Absg}
\end{figure} 

\subsubsection{Numerical analysis}

In Figures \ref{fig:S} and \ref{fig:A7} we show the correlations between the CP asymmetries, scanning the gluino mass between 0.5 and 1~TeV, the sbottom mass,  the $\mu$ term and $A_b$ between 0.2 and 0.5~TeV and $\tan\beta$ between 2 and 10. We require the $\Delta F=2$ observables to be in the region where the CKM tensions are reduced (cf. \eqref{DeltaF2}).
The maximum size of the effects is mostly limited by the $\text{BR}(B\to  X_s\gamma)$ constraint, which we require to be fulfilled at the $3\sigma$ level including SM and gluino contributions only (cf. the discussion at the end of Sec.~\ref{sec:bsg}).

Figure~\ref{fig:S} shows the correlation between the mixing-induced CP asymmetries $S_{\phi K_S}$ and $S_{\eta' K_S}$ in relation to $S_{\psi K_S}$, effectively subtracting the contribution due to the modified $B_d$ mixing phase.
The experimental $1\sigma$ ranges corresponding to the average in Table~\ref{tab:exp} are shown as shaded regions.
Due to the $\tan\beta$ enhanced terms in (\ref{eq:C8phi},~\ref{eq:C8eta}), large effects are easily possible.
A negative value for these differences, as currently indicated by the central values of the measurements, can be obtained for $\gamma_L>0$ ($\gamma_L<0$) if $\mu>0$ ($\mu<0$). For a given sign of the $\mu$ term, the sign of the $\Delta B=1$ CP asymmetries can thus serve to distinguish between the two solutions for the phase $\gamma_L$ in \eqref{DeltaF2} allowed by the $\Delta F=2$ analysis.

Figure~\ref{fig:A7} shows the correlation between $S_{\eta' K_S}$ and the CP asymmetry $\langle A_7 \rangle$ in $B\to K^*\mu^+\mu^-$.
Values up to $\pm10\%$ would be attainable for $\langle A_7 \rangle$, while the current measurement of $S_{\eta' K_S}$ implies, at the $1\sigma$ level, $0 < \langle A_7 \rangle < 5\%$.

Finally, Figure~\ref{fig:A7Absg} shows the correlation between the CP asymmetry in $B\to X_s\gamma$ and  the new physics contribution to $\langle A_7 \rangle$.
$A_\text{CP}(B\to X_s\gamma)_\text{NP}^\text{SD}$ attains values up to $\pm5\%$. Imposing the $1\sigma$ experimental range allowed for $S_{\eta' K_S}$, this decreases to $-2\% < A_\text{CP}(B\to X_s\gamma)_\text{NP}^\text{SD} < 0 \%$.
In the plots, we also show the $1\sigma$ experimental range for $A_\text{CP}(B\to X_s\gamma)$ (cf. Table~\ref{tab:exp}),
keeping in mind that in the presence of sizable long-distance effects (see Section~\ref{sec:Absg}), the prediction for this observable is modified and the bound cannot be directly compared to our predictions.

\section{Summary and partial conclusions}

In this Chapter we have found particularly useful to consider that an approximate $U(2)^3$ symmetry be operative in determining the full flavour structure of the supersymmetric extension of the SM. Among the appealing features of $U(2)^3$ and an advantage over the standard MFV proposal is that it allows the first two generations of sfermions to be substantially heavier than the third one, which helps to address specifically also the supersymmetric CP problem, without spoiling the supersymmetric solution to the hierarchy problem.\\ With our specific choice for the breaking pattern, characteristic correlations exist between the various $\Delta F = 2$ amplitudes, and one can exploit them to improve the consistency of the fit of the flavour and CP-violating observables. Using the data available at the time this study was performed, one obtains the preferred regions $m_{\tilde{b}}$, $m_{\tilde{g}} \lesssim 1.5$ TeV for the sbottom and gluino masses.

We then have moved to study CP asymmetries in $B$ decays. Even in the absence of flavour-blind phases, we find potentially sizable CP violating contributions to $\Delta B=1$ decay amplitudes. We identify the dominant contributions to arise in the magnetic and chromomagnetic dipole operators due to their sensitivity to chirality violation, with subleading contributions in semi-leptonic and QCD penguin operators. The most promising observables are the mixing-induced CP asymmetries in non-leptonic penguin decays like $B\to\phi K_S$ or $B\to\eta' K_S$, angular CP asymmetries in $B\to K^*\mu^+\mu^-$, and the direct CP asymmetry in $B\to X_s\gamma$ 
(barring potential uncertainties in controlling long-distance effects in the radiative~\cite{Benzke:2010tq} 
and non-leptonic modes~\cite{Buchalla:2005us,Beneke:2005pu}).

Due to the different dependence on the sparticle masses, we cannot predict a clear-cut correlation between CP violating $\Delta F=1$ and $\Delta F=2$ observables. However, we have demonstrated that observable effects in $\Delta F=1$ CP asymmetries are certainly compatible with the pattern of deviations from the SM suggested by $\Delta F=2$ observables, 
if interpreted in terms of this supersymmetric framework.
Interestingly, while we considered a setup without flavour-blind phases, the correlations between $\Delta F=1$ observables turn out to be very similar to those in MFV \cite{Altmannshofer:2008hc,Altmannshofer:2009ne} or in effective MFV \cite{Barbieri:2011vn} with flavour-blind phases. The main difference between the two cases are the CP violating effects in $K$ and $B$ mixing, which occur in $U(2)^3$, but not in (effective) MFV. We view this as an interesting example of the usefulness of correlated studies of $\Delta F=1$ and $\Delta F=2$ observables as a handle to distinguish between models.
Such studies  would become extremely interesting in presence of direct evidences of 
a hierarchical sparticle spectrum from the LHC.


\chapter[Exploring the SUSY Higgs sector]{Exploring the supersymmetric Higgs sector}\label{cha:SUSY_Higgs}

In general terms, to see whether the newly found resonance at 126 GeV is part of an extended Higgs system is a primary task of the current and future experimental studies. This appears to be especially true for the extra Higgs states of the NMSSM, which might be the lightest new particles of a suitable supersymmetric model, except perhaps for the LSP. A particularly important question is how the measurements of the couplings of $h_{\rm LHC}$, current and foreseen, bear on this issue, especially in comparison with the potential of the direct searches of new Higgs states.

Not the least difficulty that one encounters in attacking these problems is the number of parameters that enter the Higgs system of the NMSSM, especially if one does not want to stick to a particular version of it but rather wishes to consider the general case. Here we aim at an analytic understanding of the properties of the Higgs system of the general NMSSM, trying to keep under control as much as possible the complications due to the proliferation of model parameters and avoid the use of benchmark points.\\
The framework we will outline makes possible to describe the impact of the various direct searches in a systematic way, together with the indirect ones in the $h_{\text{LHC}}$ couplings. It will result in setting a possible overall strategy to search for signs of the CP-even extra-states of the NMSSM Higgs sector.

The content of this Chapter is the following. In Section \ref{sec2} we establish some relations between the physical parameters of the CP-even Higgs system valid in the general NMSSM. In Section \ref{sec:S_dec} and \ref{sec:Hdec} we consider two limiting cases in which one of the CP-even scalars is decoupled, determining in each situation the sensitivity of the measurements of the couplings of $h_{\rm LHC}$, current and foreseen, as well as the production cross sections and the branching ratios for the new intermediate scalar.
In Section \ref{sec:HnonDec} we give two examples of analysis of a situation in which none of the Higgses is decoupled. In Section \ref{sec:MSSM} we compare one of these NMSSM cases with the much studied MSSM, using as much as possible the same language. In Section \ref{sec:Lambda_model} we illustrate a possible simple and generic extension of the NMSSM that can make it compatible with standard gauge unification even for a coupling $\lambda \gtrsim 1$.
Section \ref{sec:NMSSM_concl} contains a detailed summary and conclusions.

\section{Physical parameters of the CP-even Higgs\\system in the general NMSSM}
\label{sec2}

Assuming a negligibly small violation of CP in the Higgs sector, we take as a starting point the form of the squared mass matrix of the neutral CP-even Higgs system in the general NMSSM: 
\begin{equation}\label{scalar_mass_matrix}
{\cal M}^2=\left(
\begin{array}{ccc}
m_Z^2 \cos^2\beta+m_A^2 \sin^2\beta & \left(2 v^2 \lambda ^2-m_A^2-m_Z^2\right) \cos\beta \sin\beta &  v M_1  \\
 \left(2 v^2 \lambda ^2-m_A^2-m_Z^2\right) \cos\beta \sin\beta & m_A^2 \cos^2\beta+m_Z^2 \sin^2\beta +\delta_t^2 &  v M_2  \\
  v  M_1 &  v M_2 & M_3^2
\end{array}
\right)
\end{equation}
in the basis $\mathcal{H} = (H_d^0, H_u^0, S)^T$. In this equation 

\begin{equation}
\label{mHcharged}
m_A^2 = m_{H^{\pm}}^2 - m_W^2 +\lambda^2 v^2,
\end{equation}
where $m_{H^{\pm}}$ is the physical mass of the single charged Higgs boson, $v \simeq 174$ GeV, and
\begin{equation}
\delta_t^2 = \Delta_t^2/ \sin^2\beta
\label{delta-t}
\end{equation}
is the well known effect of the top-stop loop corrections to the quartic coupling of $H_u$.
We neglect the analogous correction to \eqref{mHcharged} \cite{Djouadi:2005gj}, which lowers $m_{H^{\pm}}$ by less than 3 GeV for stop masses below 1 TeV. More importantly we have also not included in Eq. \eqref{scalar_mass_matrix} the one loop corrections to the $12$ and $11$ entries, respectively proportional to the first and second power of $(\mu A_t)/\langle m_{\tilde{t}}^2\rangle$, to which we shall return.
We leave unspecified the other parameters in (\ref{scalar_mass_matrix}), $M_1, M_2, M_3$, which are not directly related to physical masses and  depend on the particular NMSSM under consideration.

The vector of the three physical mass eigenstates  $\mathcal{H}_{\rm ph}$ is related to the original scalar fields by
\begin{equation}
\mathcal{H} = R^{12}_{\alpha} R^{23}_{\gamma} R^{13}_{\sigma} \mathcal{H}_{\rm ph} \equiv R \mathcal{H}_{\rm ph},
\label{rotation_matrix}
\end{equation}
where $R^{ij}_\theta$ is the rotation matrix in the $i j$ sector by the angle $\theta = \alpha,\gamma,\sigma$.\\
Defining $ \mathcal{H}_{\rm ph} = (h_3, h_1, h_2)^T$, we have
\begin{equation}
R^T {\cal M}^2 R = \mathrm{diag}(m_{h_3}^2, m_{h_1}^2, m_{h_2}^2).
\label{diag_matrix}
\end{equation}
We identify $h_1$ with the state found at LHC, so that $m_{h_1} = 125.7$ GeV.
From (\ref{rotation_matrix}) $h_1$ is related to the original fields by
\begin{equation}
h_1 = c_{\gamma} (-s_{\alpha} H_d + c_\alpha H_u) + s_{\gamma} S,
\end{equation}
where $s_\theta = \sin{\theta}, c_\theta = \cos{\theta}$. Similar relations, also involving the angle $\sigma$, hold for $h_2$ and $h_3$.

\subsubsection{Couplings of $h_{\text{LHC}}$ and fit of its signal strengths}

These angles determine  the couplings of $h_1 = h_{\text{LHC}}$  to the fermions or to vector boson pairs, $VV = WW, ZZ$, normalized to the corresponding couplings of the SM Higgs boson. Defining  $\delta = \alpha - \beta +\pi/2$, they are given by (see also \cite{Cheung:2013bn,Choi:2012he})
\begin{equation}
\frac{g_{h_1tt}}{g^{\text{SM}}_{htt}}= c_\gamma(c_\delta +\frac{s_\delta}{\tan\beta}),~~\frac{g_{h_1bb}}{g^{\text{SM}}_{hbb}}= c_\gamma(c_\delta -s_\delta \tan\beta ),~~\frac{g_{h_1VV}}{g^{\text{SM}}_{hVV}}=  c_\gamma c_\delta.
\label{h1couplings}
\end{equation}
\begin{figure}
\begin{center}
\includegraphics[width=.48\textwidth]{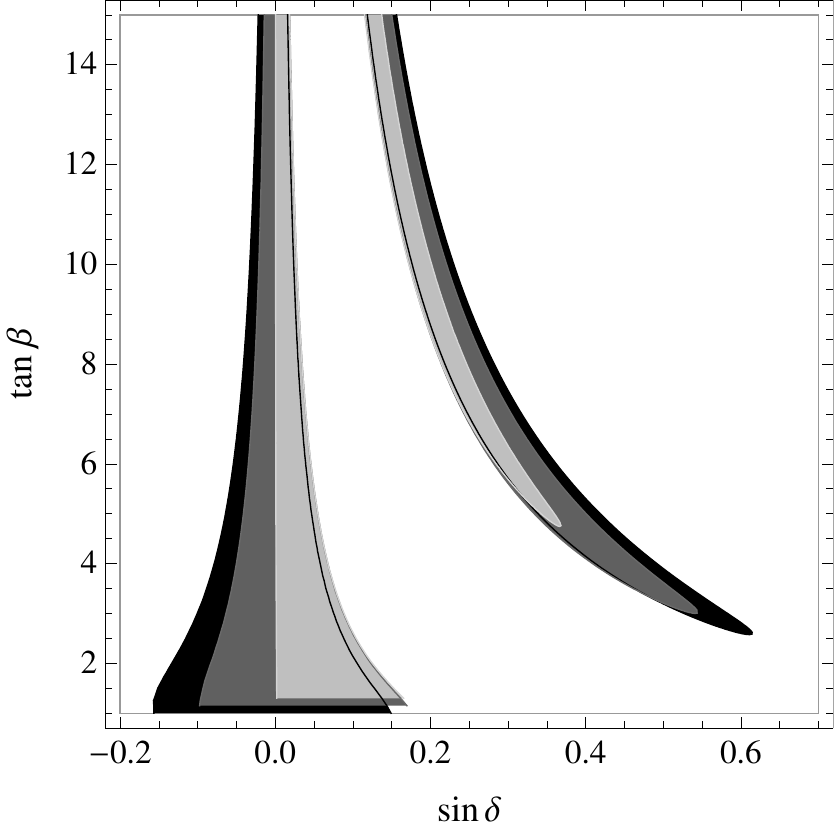}\hfill
\includegraphics[width=.48\textwidth]{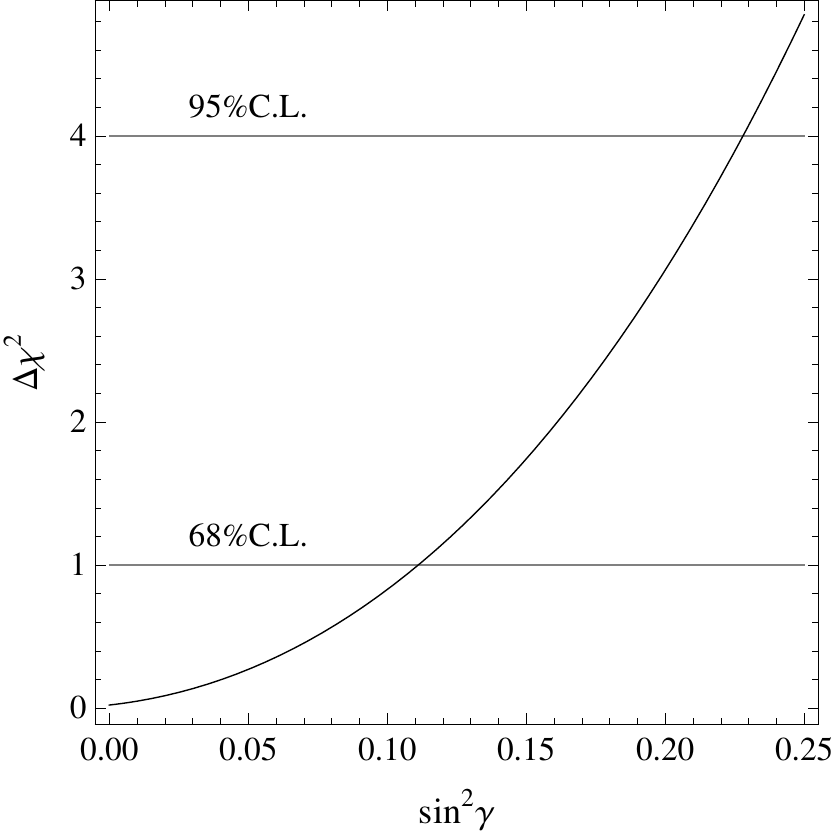}
\caption{\label{fig:FIT}\small Fit of the measured signal strengths of $h_1 = h_{\text{LHC}}$. Left: 3-parameter fit of $\tan \beta$, $s_{\delta}$ and $s_{\gamma}^{2}$. The allowed regions at 95\%C.L. are given for $s_{\gamma}^{2} = 0$ (black), $0.15$ (dark grey), $0.3$ (light grey). The regions overlap in part, but their borders are also shown. Right: fit of $s_{\gamma}^{2}$ in the case of $\delta = 0$.
}
\end{center}
\end{figure}
The fit of all ATLAS \cite{ATLAS}, CMS \cite{CMS} and TeVatron \cite{tevatron:2013} data collected so far on the various signal strengths of $h_{\text{LHC}}$ gives the bounds on $\delta$ for different fixed values of $\gamma$ shown in Figure~\ref{fig:FIT} left, and the bound on $\gamma$ for $\delta= 0$ shown in Figure~\ref{fig:FIT} right. To make this fit, we adapt the code provided by the authors of \cite{Giardino:2013bma}. As stated below, we do not include in this fit any supersymmetric loop effects. Note that in the region of $s_\delta$ close to zero, a larger $s_\gamma^2$ forces $\delta$ to take a larger central value.\\
It is also interesting to explore the consequences of an improvement of such measurements, as foreseen in the coming state of the LHC. To quantify this we consider the impact, on the fit, of the measurements of the signal strengths of $h_{\text{LHC}}$ with the projected errors at LHC14 with $300~\mathrm{fb}^{-1}$ by ATLAS\cite{ATLAS-collaboration:2012iza} and CMS\cite{CMS-14}, shown in Table \ref{tab1}. 
Assuming SM central values for the signal strengths, the projected bound on $\gamma$ is $s_\gamma^2 < 0.15$, while the one on $\delta$ becomes much stricter with respect to the current one.
\begin{table}[h!]
\begin{center}
\begin{tabular}{ccc}
& ATLAS & CMS \\
\hline
$h \to \gamma \gamma$ & 0.16 & 0.15 \\
$h \to Z Z$ & 0.15 & 0.11 \\
$h \to W W$ & 0.30 & 0.14 \\
$V h \to V b \bar{b}$ & -- & 0.17 \\
$h \to \tau \tau$ & 0.24 & 0.11 \\
$h \to \mu \mu$ & 0.52 & -- \\
\end{tabular}
\caption{\label{tab1}Projected uncertainties of the measurements of the signal strengths of $h_{\text{LHC}}$, normalized to the SM, at the 14 TeV LHC with $300~\mathrm{fb}^{-1}$.}
\end{center}
\end{table}

\subsubsection{Analytical expressions for the rotation angles}

The matrix equation (\ref{diag_matrix}) restricted to the $1 2$ sector gives three relations between the mixing angles $\delta\equiv\alpha-\beta +\pi/2, \gamma, \sigma$ and the physical masses $m_{h_1,h_2,h_3}, m_{H^{\pm}}$ for any given value of $\lambda, \tan{\beta}$ and $\Delta_t$.
In terms of the $2\times 2$ submatrix $M^2$ in the $1 2$ sector of ${\cal M}^2$, Eq. (\ref{scalar_mass_matrix}), these relations can be made explicit as\footnote{Notice that Eq.~\eqref{eq:sin:2alpha:general} is completely equivalent to the expression for $\sin 2\alpha$ in Eq. (2.10) of Ref.~\cite{Barbieri:2013hxa}.}

\begin{align}
 s_\gamma^{2} &=  \frac{ \det M^{2} + m_{h_1}^{2} (m_{h_1}^{2} - \tr M^{2})}{(m_{h_1}^{2} - m_{h_2}^{2})  (m_{h_1}^{2} - m_{h_3}^{2}) }, 
  \label{eq:sin:gamma:general}
  \\
  s_\sigma^{2} &= \frac{m_{h_2}^{2} - m_{h_1}^{2}}{m_{h_2}^{2} - m_{h_3}^{2}} \; \frac{ \det M^{2} + m_{h_3}^{2} (m_{h_3}^{2} - \tr M^{2}) }
  { \det M^{2} - m_{h_2}^{2} m_{h_3}^{2} + m_{h_1}^{2} (m_{h_2}^{2} + m_{h_3}^{2} - \tr M^{2}) },
  \label{eq:sin:sigma:general}
  \\
    s_{2\delta}&=
 \Big[ 
    2 s_\sigma c_\sigma s_\gamma \left(m_{h_3}^2-m_{h_2}^2\right) \left(2 \tilde M^2_{11}-m_{h_1}^2c_\gamma^2 -m_{h_2}^2(s_\gamma^2+s_\sigma^2c_\gamma^2) - m_{h_3}^2(c_\sigma^2+s_\gamma^2 s_\sigma^2)\right) \notag
    \\ 
    & +2 \tilde M^2_{12} \left(m_{h_3}^2 \left(c_\sigma^2-s_\gamma^2 s_\sigma^2\right)+m_{h_2}^2 \left(s_\sigma^2-s_\gamma^2  c_\sigma^2\right)-m_{h_1}^2 c_\gamma^2 \right)
  \Big]
  \notag \\
  & \times \Big[ \left(m_{h_3}^2-m_{h_2}^2 s_\gamma^2- m_{h_1}^2 c_\gamma^2\right)^2
  +\left(m_{h_2}^2-m_{h_3}^2\right)^2 c_\gamma^4 s_\sigma^4 \notag\\
   &+2 \left(m_{h_2}^2-m_{h_3}^2\right) \left(m_{h_3}^2+m_{h_2}^2 s_\gamma^2-m_{h_1}^2 \left(1+s_\gamma^2\right)\right) c_\gamma^2 s_\sigma^2 \Big]^{-1}, \label{eq:sin:2alpha:general}
\end{align}
where $s_\theta = \sin{\theta}, c_\theta = \cos{\theta}$ and, in Eq.~\eqref{eq:sin:2alpha:general}, $\tilde M^2= R_{\beta-\pi/2}M^2 R_{\beta-\pi/2}^t$.
These expressions for the mixing angles do not involve the unknown parameters $M_1, M_2, M_3$, which depend on the specific NMSSM. Their values in particular cases may limit the range of the physical parameters $m_{h_1,h_2,h_3}, m_{H^{\pm}}$ and $\alpha, \gamma, \sigma$ but cannot affect Eqs. (\ref{eq:sin:gamma:general}, \ref{eq:sin:sigma:general}, \ref{eq:sin:2alpha:general}). To our knowledge, analytical expressions for the mixing angles in the general NMSSM have first been presented in our \cite{Barbieri:2013hxa}.

\subsubsection{Simplifying assumptions}
To simplify the analysis we consider two limiting cases:
\begin{itemize}
\item $H$ decoupled: In (\ref{scalar_mass_matrix}) $m_A^2 \gg v M_1, v M_2$ or $m_{h_3} \gg m_{h_1,h_2}$ and $\sigma, \delta = \alpha - \beta +\pi/2 \rightarrow 0$,

\item Singlet decoupled: In (\ref{scalar_mass_matrix}) $M_3^2 \gg v M_1, v M_2$ or $m_{h_2} \gg m_{h_1,h_3}$ and $\sigma, \gamma \rightarrow 0$,
\end{itemize}
but we  use (\ref{eq:sin:gamma:general}, \ref{eq:sin:sigma:general}, \ref{eq:sin:2alpha:general}) to control the size of the deviations from the limiting cases when the heavier mass is lowered. We will also dedicate a specific section to the study of the non-decoupled case, and further comment it in the conclusions.


When considering the couplings of the CP-even scalars to SM particles, relevant to their production and decays, we shall not include any supersymmetric loop effect other than the one that gives rise to (\ref{delta-t}). This is motivated by the kind of spectrum outlined in Section \ref{sec:NatSUSY}, with all s-particles at their ``naturalness limit'', and provides in any event a useful well defined reference case.

We also do not include any invisible decay of the  CP-even scalars, such as dark matter,  or into any undetected final state, because of background, like for example a pair of light pseudo-scalars.
To correct for this is straightforward with all branching ratios and signal strengths of $h_{\text{LHC}}$, that will have to be multiplied by a factor $\Gamma_{\rm vis}/(\Gamma_{\rm vis} + \Gamma_{\rm inv})$.
Would this inclusion alter the excluded regions from the measurements of the above signal strengths? The answer we find is different between the $H$- and the Singlet- decoupled cases. In the first one, the inclusion in the fit of the LHC  data of an invisible branching ratio of $h_{\text{LHC}}$, $\mathrm{BR}_{\rm inv}$, leaves essentially unchanged the allowed range for $\delta$ at different $\tan{\beta}$ values, provided $\mathrm{BR}_{\rm inv} \lesssim 0.2$. On the contrary in the Singlet-decoupled case this inclusion would strengthen the bound on the mixing angle to $s^2_\gamma \lesssim (0.22 - 0.78 \mathrm{BR}_{\rm inv})$.

Finally we do not consider in this chapter the two neutral CP-odd scalars, since in the general NMSSM both their masses and their composition in terms of the original fields depend upon extra parameters not related to the masses and the mixings of the CP-even states nor to the mass of the charged Higgs.

\section{Singlet decoupled}
\label{sec:S_dec}

Let us first consider the limit in (\ref{scalar_mass_matrix}) $M_3^2 \gg v M_1, v M_2$, which corresponds to $m_2 \gg m_{1,3}$ and $\sigma, \gamma \rightarrow 0$. 
In this case the three relations that have led to (\ref{eq:sin:gamma:general}, \ref{eq:sin:sigma:general}, \ref{eq:sin:2alpha:general}) become
\begin{align}
\label{sin2alpha}
\sin 2\alpha &= \sin 2\beta \; \frac{2\lambda^2 v^2-m_Z^2-m_A^2|_{m_{h_1}}}{m_A^2|_{m_{h_1}} +m_Z^2 +\delta_t^2 -2m_{h_1}^2},\\
m_{h_3}^2&= m_A^2|_{m_{h_1}}+m_Z^2 +\delta_t^2 -m_{h_1}^2,
\end{align}
where
\begin{equation}\label{mA_mh}
m_A^2\big|_{m_{h_1}}=\frac{\lambda^2v^2(\lambda^2v^2-m_Z^2)\sin^2 2\beta-m_{h_1}^2(m_{h_1}^2-m_Z^2-\delta_t^2)-m_Z^2\delta_t^2 \cos^2\beta}{m_{hh}^2-m_{h_1}^2}.
\end{equation}
Identifying 
$h_1$ with the resonance found at the LHC, this determines $m_{h_3}, m_{H^+} $ and the angle $\delta = \alpha - \beta +\pi/2$ as functions of  $(\tan{\beta}, \lambda, \Delta_t)$.
The couplings of  $h_3$ become
\begin{equation}
\frac{g_{h_3tt}}{g^{\text{SM}}_{htt}}=\sin\delta-\frac{\cos\delta}{\tan\beta},~~\frac{g_{h_3bb}}{g^{\text{SM}}_{hbb}}=\sin\delta+\tan\beta\cos\delta,~~\frac{g_{h_3VV}}{g^{\text{SM}}_{hVV}}= \sin\delta.
\label{h3couplings}
\end{equation}

 From our point of view the main motivation for considering the NMSSM is in the possibility to account for the mass of $h_{\text{LHC}}$ with not too big values of the stop masses. For this reason  we take $ \Delta_t = 75$ GeV, which can be obtained, e.g., for an average stop mass of about 700 GeV \cite{Carena:1995bx}. In turn, as it will be seen momentarily, the consistency of Eqs.  \eqref{sin2alpha}-\eqref{mA_mh} requires not too small values of the coupling $\lambda$.
It turns out in fact that for any value of $ \Delta_t \lesssim 85$ GeV,  the dependence on $\Delta_t$ itself can be neglected, so that $m_{h_3}, m_{H^\pm} $ and  $\delta$ are determined by $\tan{\beta}$ and $\lambda$ only.  For the same reason it is legitimate to neglect the one loop corrections to the $11$ and $12$ entries of the mass matrix, Eq. \eqref{scalar_mass_matrix}, as long as $(\mu A_t)/\langle m_{\tilde{t}}^2\rangle \lesssim 1$, which is again motivated by naturalness.
\begin{figure}[t]
\begin{center}
\includegraphics[width=.48\textwidth]{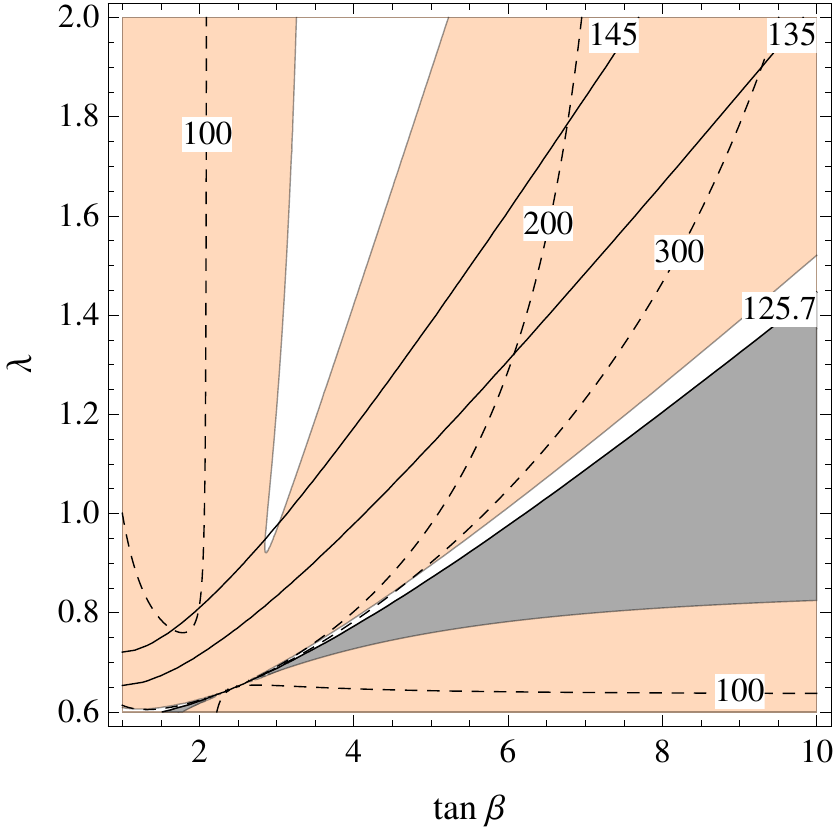}
\caption{\label{fig:mSdecoupled-1}\small Singlet decoupled. Isolines of $m_{hh}$ \eqref{mhh} (solid) and $m_{H^{\pm}}$ (dashed), the grey region is unphysical due to $m_{H^{\pm}}^{\,2} < 0$. The coloured regions are excluded at 95$\%$C.L.}
\end{center}
\end{figure}

\begin{figure}[t!]
\begin{center}
\includegraphics[width=0.48\textwidth]{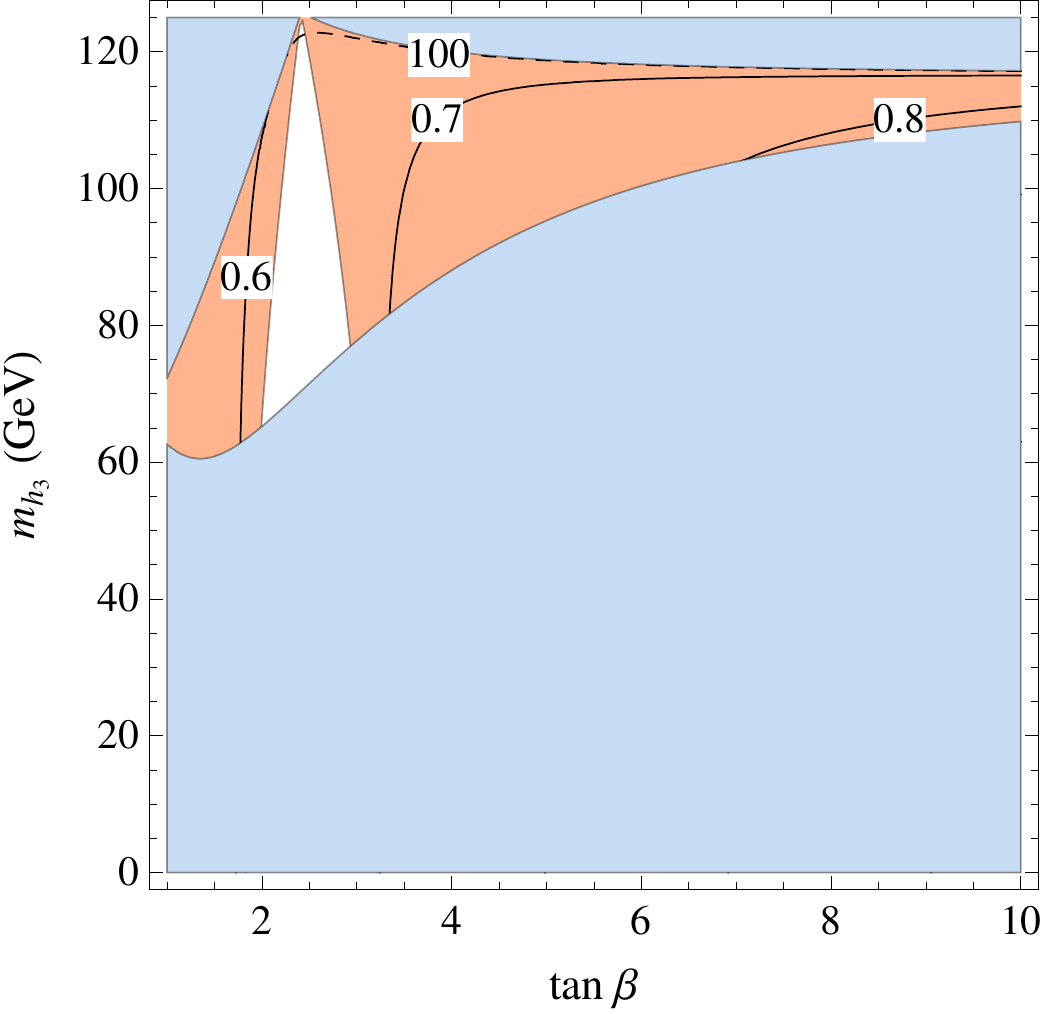}\hfill
\includegraphics[width=0.48\textwidth]{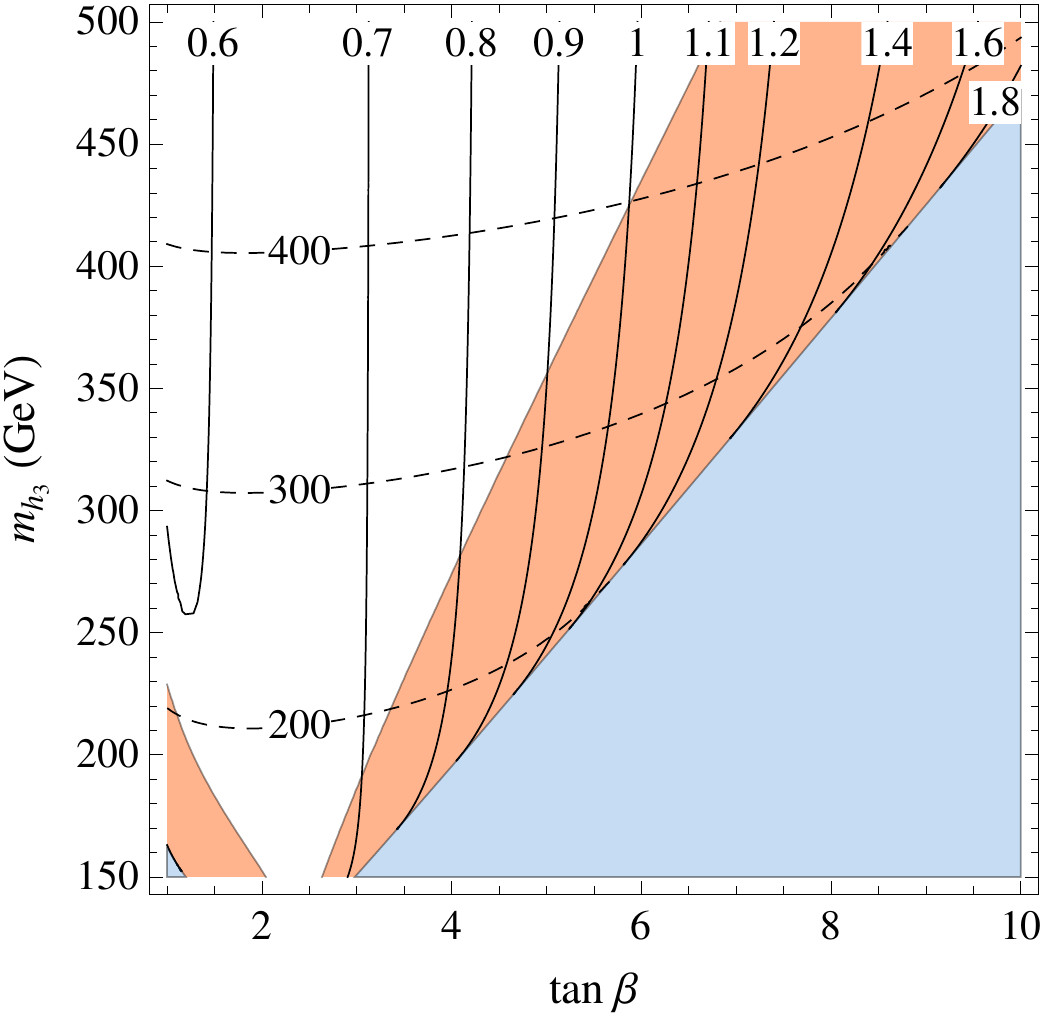}
\caption{\label{fig1} Singlet decoupled. Isolines of $\lambda$ (solid) and $m_{H^\pm}$ (dashed). Left: $h_{\rm LHC}>h_3$. Right: $h_{\rm LHC}<h_3$. The orange region is excluded at 95\%C.L. by the experimental data for the signal strengths of $h_1 = h_{\rm LHC}$. The blue region is unphysical.}
\label{fig:NMSSM_mh3_tbeta_lambda}
\end{center}
\end{figure}

From all this we can represent in Figure~\ref{fig:mSdecoupled-1} the allowed regions in the plane $(\tan{\beta}, \lambda)$, that are determined by a 2-parameter fit of $\tan \beta$, $\sin \delta$. This fit results in an allowed region which is virtually the same as the one with $\gamma = 0$ in Figure~\ref{fig:FIT} left. When inverting $\lambda$ as a function of $\tan\beta, m_{h_3}$, there are two solutions. In Figures~\ref{fig1} we show only the one which corresponds to the narrow allowed region with $m_{hh}$ close to 126 GeV, both for $h_3 < h_{\text{LHC}} (< h_3(=S))$ and for $h_{\text{LHC}} < h_3 (< h_3(=S))$, together with the isolines of $\lambda$ and $m_{H^\pm}$.
The other allowed region in Figure \ref{fig:mSdecoupled-1}, when translated to the ($\tan\beta, m_{h_3}$) plane, corresponds to the other solution for $\lambda$, and is not displayed in Figure~\ref{fig1} left. There $h_{\rm LHC}$ is the lightest CP-even state, and the charged Higgs mass $m_{H^{\pm}}$ is always below 150 GeV, which is disfavored by indirect constraints \cite{Misiak:2006zs}. Moreover note that this region, corresponding to the allowed region with large $\delta$ in Figure~\ref{fig:FIT}, is mainly allowed because of the large error in the measurement of the $b\bar b$ coupling of $h_{\text{LHC}}$. Reducing this error down to about 30\% around $g_{h_1bb}/g_{hbb}^{\rm SM}\simeq 1$ would exclude the region.

 Coming back to Fig. \ref{fig1}, the knowledge of $\delta$ in every point of the $(\tan{\beta}, m_{h_3})$ plane fixes the couplings of $h_3$ and  $ h_{\text{LHC}}$, which allows to draw the currently excluded regions from the measurements of the signal strengths of $h_{\text{LHC}}$. 
Negative searches at LHC of $h_3 \rightarrow \bar{\tau} \tau$ may also exclude a further portion of the 
parameter space for $h_3 > h_{\text{LHC}}$. Note, as anticipated, that in every case $\lambda$ is bound to be above about $0.6$. To go to lower values of $\lambda$ would require considering $\Delta_t \gtrsim  85$ GeV, i.e. heavier stops. On the other hand in this singlet-decoupled case lowering $\lambda$ and raising $\Delta_t $ makes the NMSSM close to the minimal supersymmetric Standard Model (MSSM), to which we shall return.
%
\begin{figure}[h!]
\begin{center}
\includegraphics[width=0.48\textwidth]{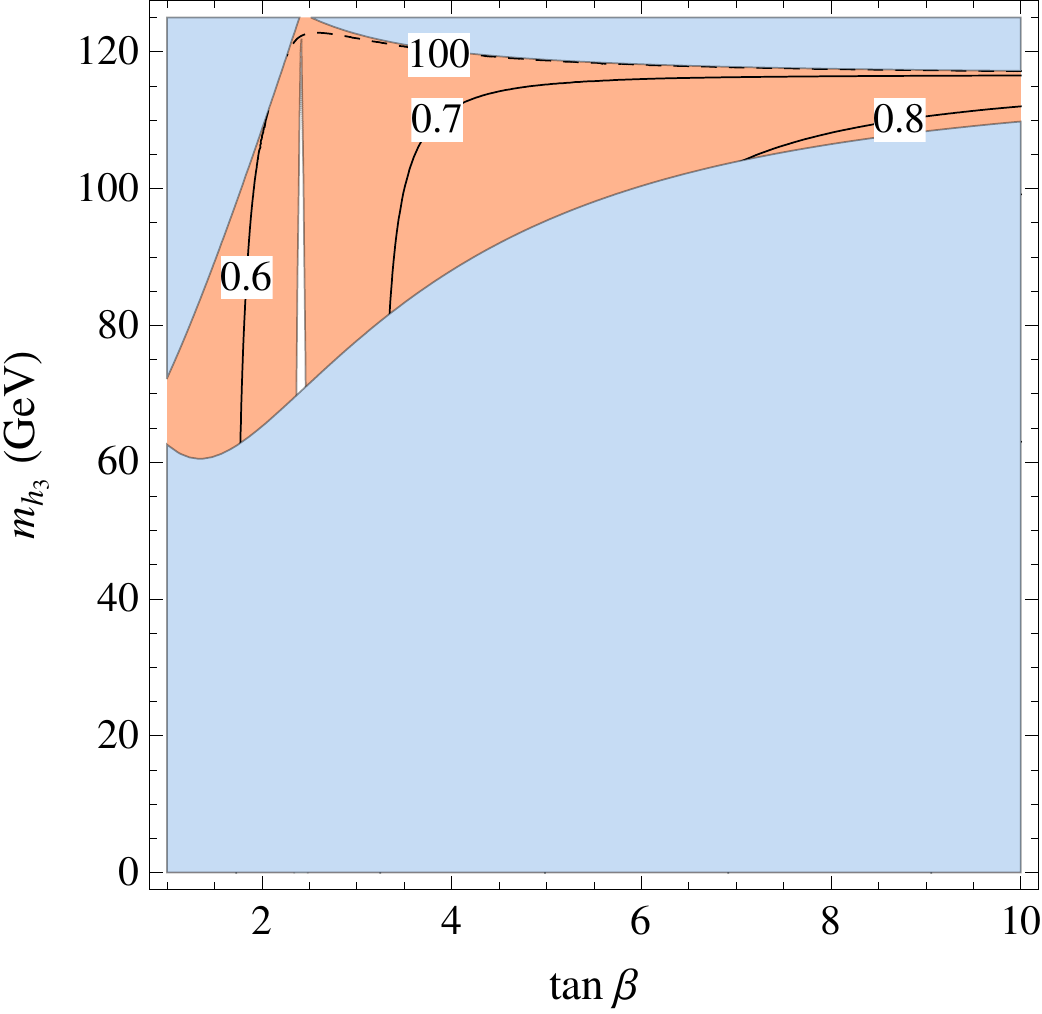}\hfill
\includegraphics[width=0.48\textwidth]{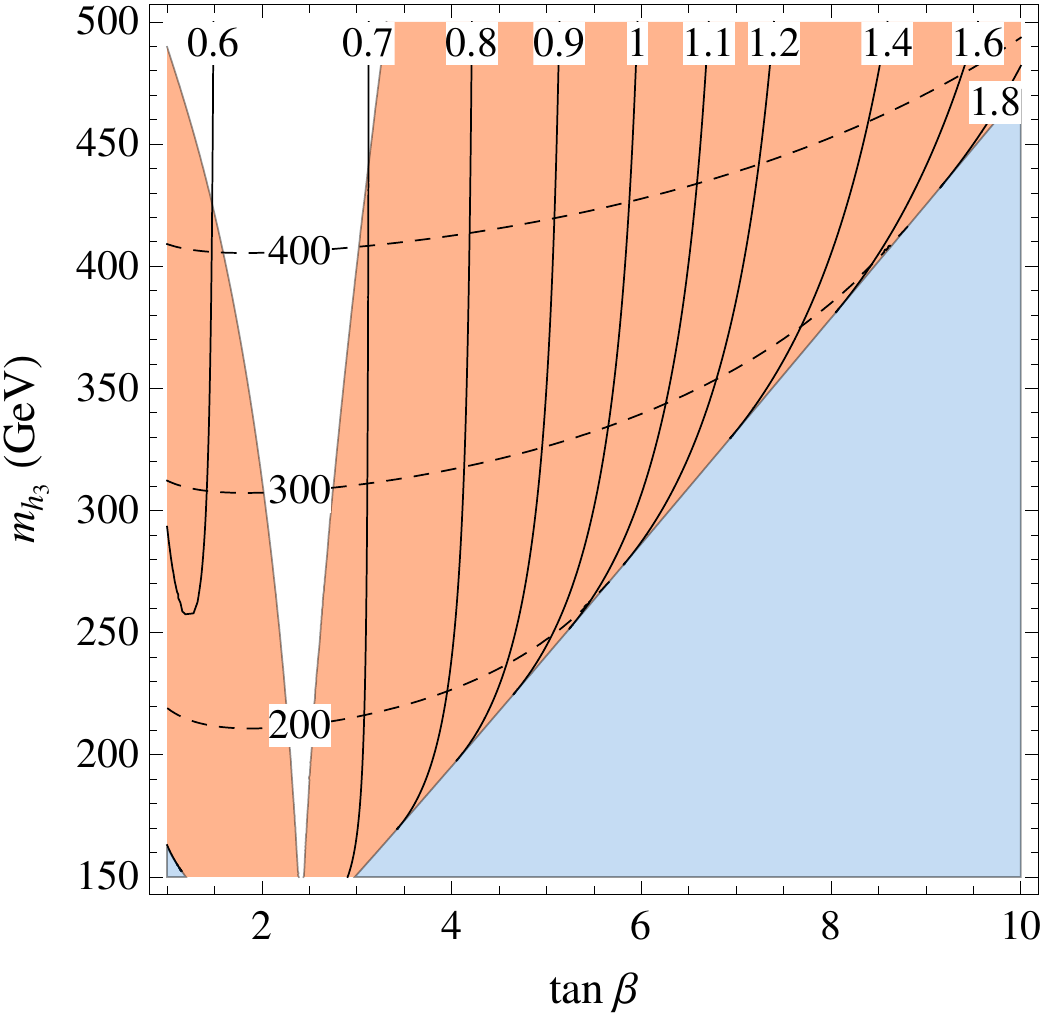}
\caption{\label{fig2} Singlet decoupled. Isolines of $\lambda$ (solid) and $m_{H^\pm}$ (dashed). Left: $h_{\rm LHC}>h_3$. Right: $h_{\rm LHC}<h_3$. The orange region would be excluded at 95\%C.L. by the experimental data for the signal strengths of $h_1 = h_{\rm LHC}$ with SM central values and projected errors at the LHC14 as discussed in the text. The blue region is unphysical.}
\end{center}
\end{figure}

The significant constraint set on Fig. \ref{fig1} by the  current measurements of the signal strengths of $h_{\text{LHC}}$ suggests that an improvement of such measurements, as foreseen in the coming stage of LHC, could lead to an effective exploration of most of the relevant parameter space. 
The result is shown in Fig. \ref{fig2}, again both for $h_3 < h_{\text{LHC}} (< h_2(= S))$ and for $h_{\text{LHC}} < h_3 (< h_2(= S))$, assuming SM central values for the signal strengths.

Needless to say, the direct search of the extra $CP$-even states will be essential either in presence of a possible indirect evidence from the signal strengths or to fully cover the parameter space for $h_3 > h_{\text{LHC}}$. To this end, under the stated assumptions, all production cross sections and branching ratios for the $h_3$ state are determined in every point of the $(\tan{\beta}, m_{h_3})$ plane. As an example, we end the section by discussing this explicitely for the case $h_3 > h_{\text{LHC}}$.

The couplings (\ref{h3couplings}) allow to compute the gluon-fusion production cross section of $h_3$ by means of \cite{Cheung:2013bn}
\begin{figure}[h]
\begin{center}
\includegraphics[width=.48\textwidth]{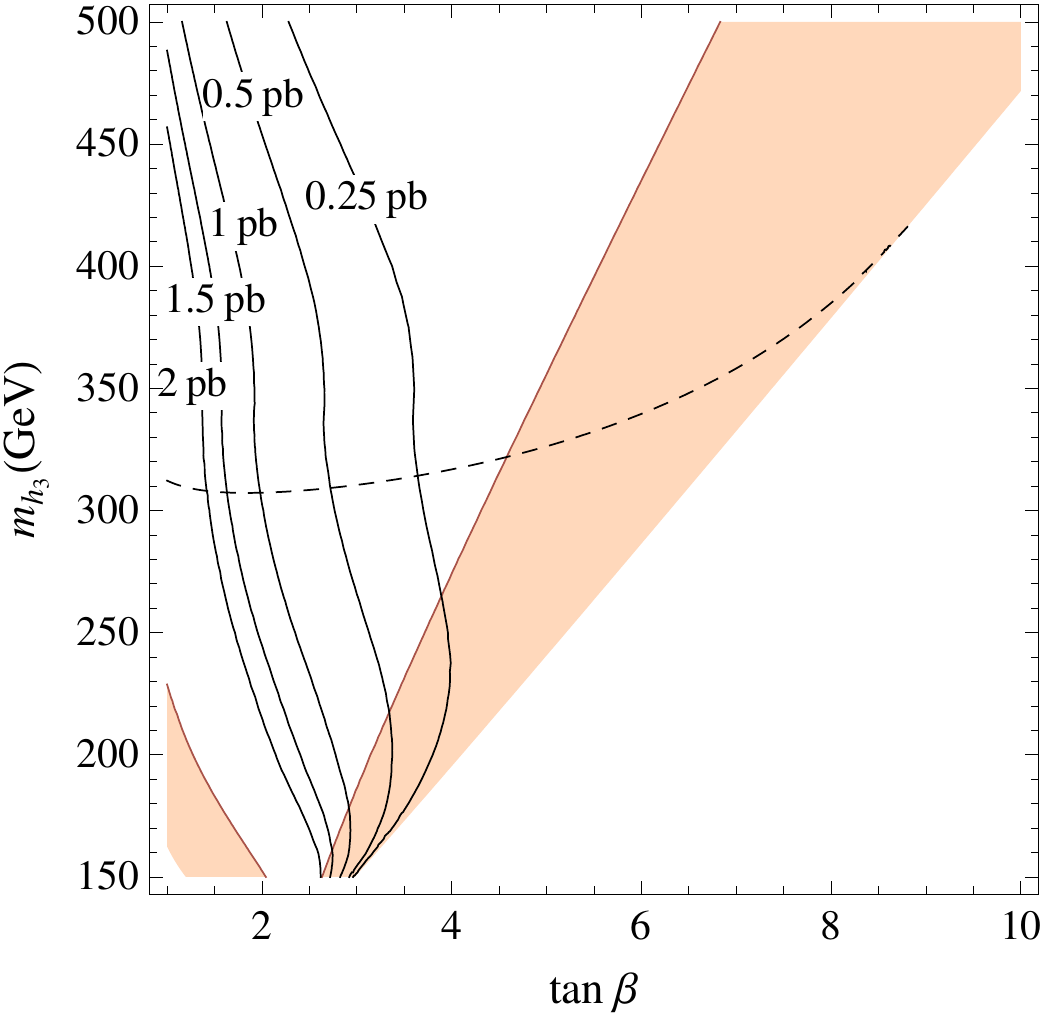}\hfill
\includegraphics[width=.48\textwidth]{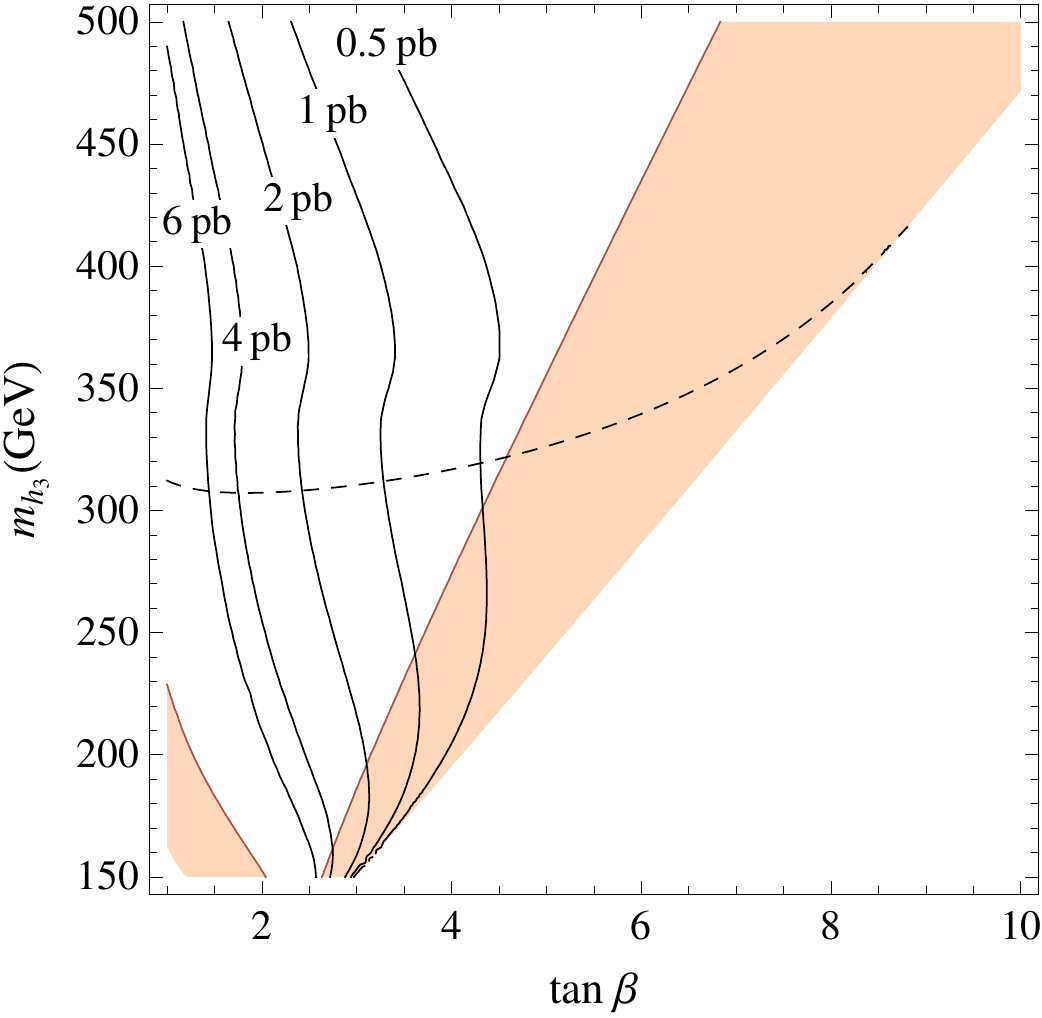}
\caption{\label{fig:mSdecoupled-Xsec}\small Singlet decoupled. Isolines of gluon fusion production cross section $\sigma(gg\to h_3)$. The coloured regions are excluded at 95$\%$C.L., and the dashed line shows $m_{H^\pm}=300$ GeV. Left: LHC8. Right: LHC14.}
\end{center}
\end{figure}
\begin{equation}
\sigma(gg\rightarrow h_3) = \sigma^{\text{SM}}(gg\rightarrow H(m_{h_3}))
\Big|\mathcal{A}_t \frac{g_{h_3tt}}{g^{\text{SM}}_{htt}} + \mathcal{A}_b \frac{g_{h_3bb}}{g^{\text{SM}}_{hbb}}\Big|^2,
\end{equation}
where
\begin{equation}
\mathcal{A}_{t,b} =  \frac{F_{\frac{1}{2}}(\tau_{t,b})}{F_{\frac{1}{2}}(\tau_t) + F_{\frac{1}{2}}(\tau_b)},   ~~~~~~\tau_i = 4 \frac{m_i^2}{m_{h_3}^2},
\end{equation}
and $F_{\frac{1}{2}}(\tau)$ is a one-loop function that can be found e.g. in \cite{Azatov:2012qz,Carmi:2012in}. This cross section is shown in Figure~\ref{fig:mSdecoupled-Xsec}, where we used the values of $\sigma^{\text{SM}}$ at NNLL precision provided in \cite{Dittmaier:2011ti}, and the running masses $m_{t,b}$ at NLO precision. We checked the validity of this choice by performing the same computation both with the use of masses at LO precision and K-factors \cite{Anastasiou:2009kn}, and with the program \texttt{HIGLU} \cite{Spira:1995rr,Spira:1995mt}, finding in both cases an excellent agreement.
\begin{figure}[t!]
\begin{center}
\includegraphics[width=.48\textwidth]{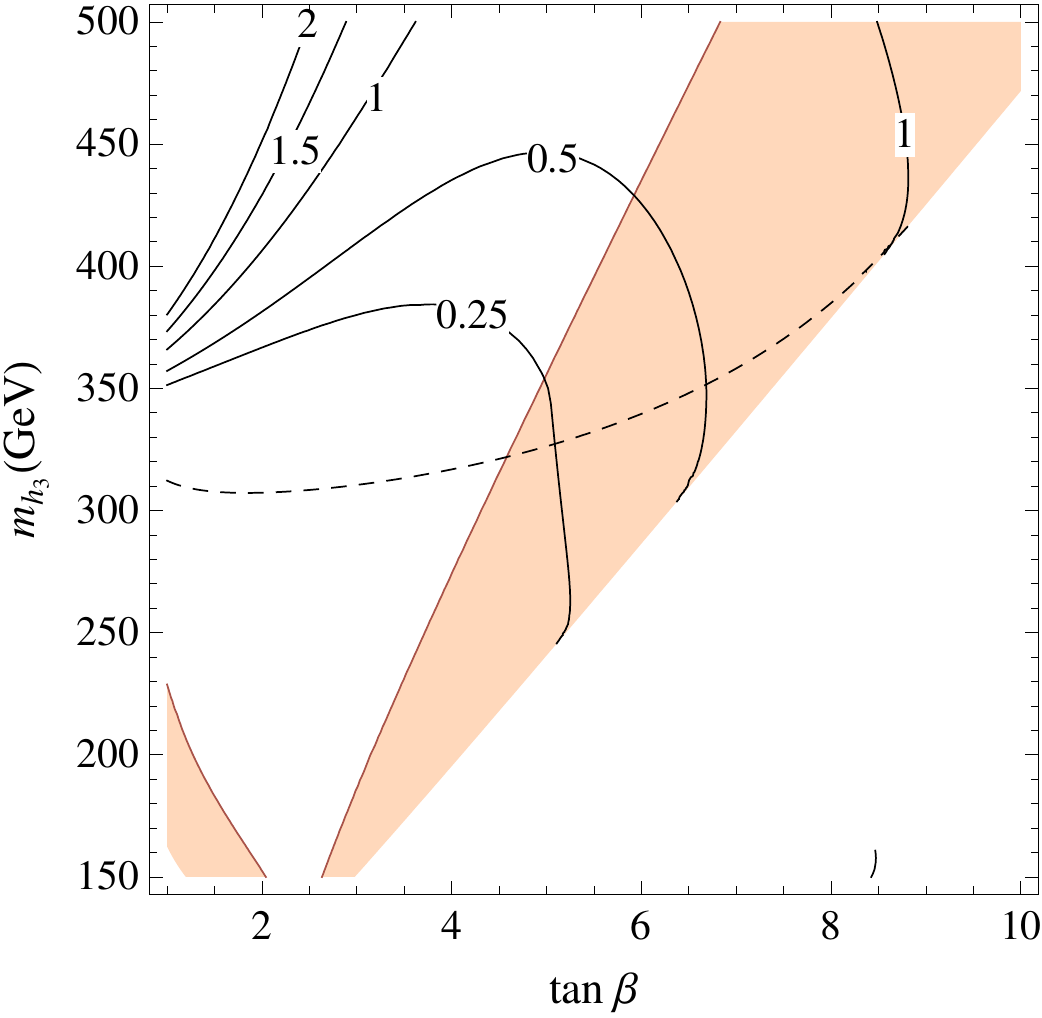}\hfill
\includegraphics[width=.48\textwidth]{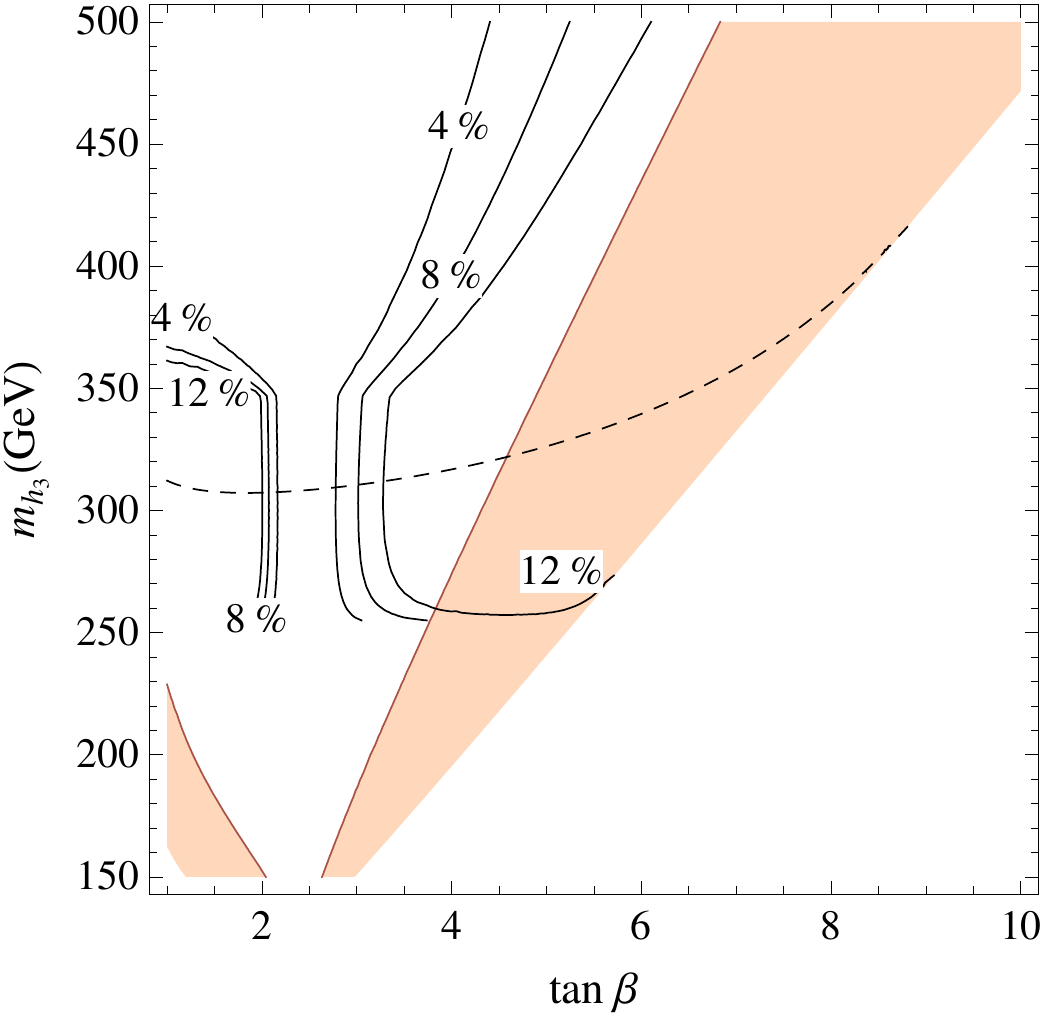}
\caption{\label{fig:mSdecoupled-BRs}\small Singlet decoupled. Left: isolines of the total width $\Gamma_{h_3}$(GeV). Right: isolines of BR$(h_3\!\to \!h h)$. The coloured regions are excluded at 95$\%$C.L., and the dashed line shows $m_{H^\pm}=300$ GeV.}
\vspace{1.2 cm}
\includegraphics[width=.48\textwidth]{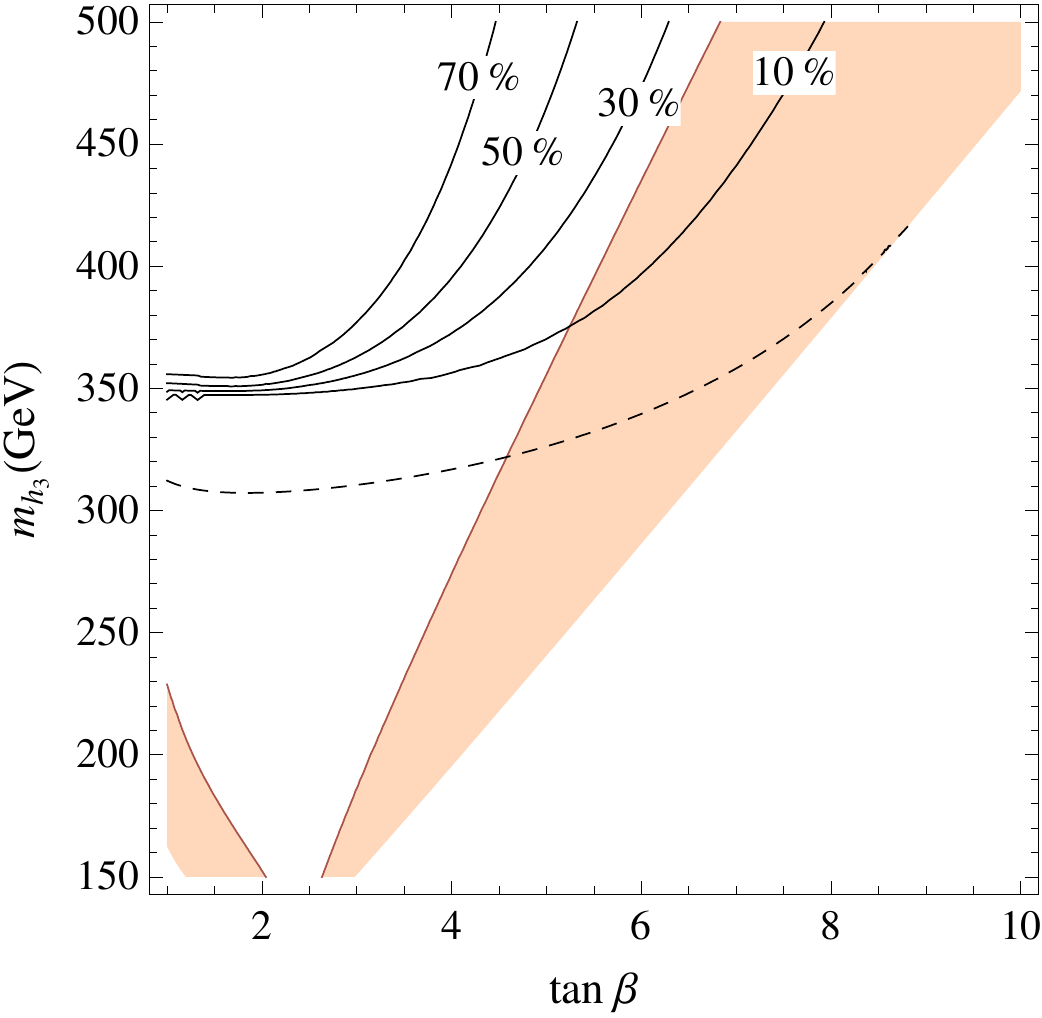}\hfill
\includegraphics[width=.48\textwidth]{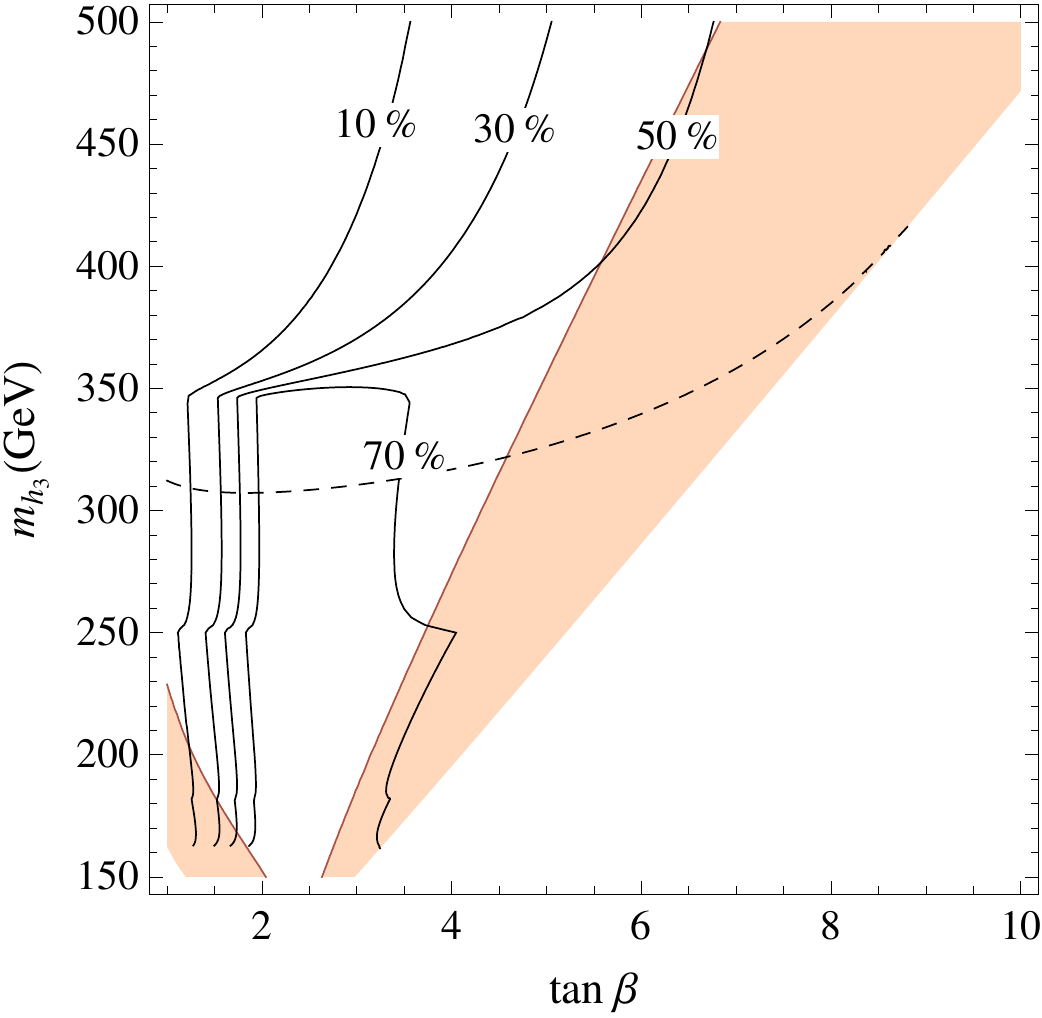}
\caption{\label{fig:mSdecoupled-BRf}\small Singlet decoupled. Left: isolines of BR$(h_3\to t\bar t)$. Right: isolines of BR$(h_3\to b \bar{b})$. The coloured regions are excluded at 95$\%$C.L., and the dashed line shows $m_{H^\pm}=300$ GeV.}
\end{center}
\end{figure}
The coupling of $h_3$ to the lighter state $\displaystyle\frac{g_{h_3 h_1^2}}{2} h_3 h_1^2$ and the triple Higgs coupling $\displaystyle\frac{g_{h_1^3}}{6}h_1^3$ are given by
\begin{align}
g_{h_3h_1^2}&=\frac{1}{2 \sqrt{2} v} \left[(m_Z^2 + v^2 \lambda^2) \sin \delta + 3 (m_Z^2 - \lambda^2 v^2) \sin(4 \beta + 3 \delta)\right] \nonumber \\
 &- \frac{3 \Delta_t^2}{\sqrt{2} v} \frac{\cos(\beta + \delta) \sin^2(\beta+\delta)}{\sin^3\beta},\\
\frac{g_{h_1^3}}{g^{\text{SM}}_{h_1^3}} &= \frac{(m_Z^2 + v^2 \lambda^2) \cos \delta + (m_Z^2 - v^2 \lambda^2) \cos(4 \beta + 3 \delta)}{2 m_{h_1}^2} + \frac{\Delta_t^2}{m_{h_1}^2} \frac{\sin^3(\beta+\delta)}{\sin^3\beta}.
\end{align}
Figures~\ref{fig:mSdecoupled-BRs} and \ref{fig:mSdecoupled-BRf} show the most relevant widths of $h_3$. Note that had we taken a smaller value for $\Delta_t$, which is not inconsistent with direct stop searches (see discussion in Section \ref{sec:NatSUSY}), then the relative importance of the $h_1 h_1$ decay channel would have increased, as found for example in the recent \cite{Lu:2013cta}.

\section{$H$ decoupled}
\label{sec:Hdec}

\begin{figure}
\begin{center}
\includegraphics[width=0.48\textwidth]{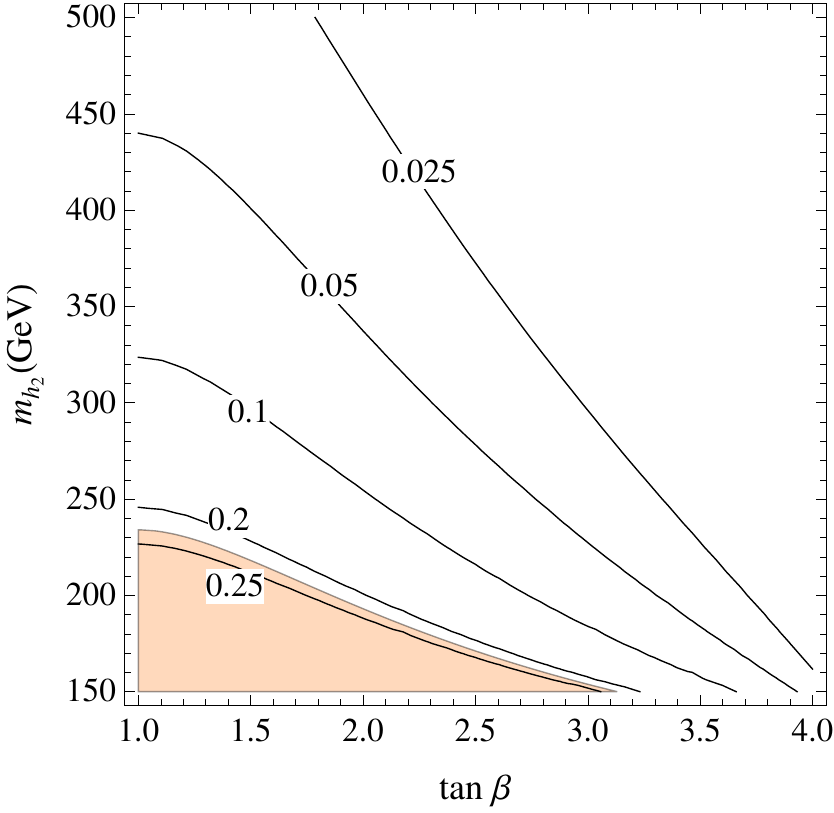}\hfill
\includegraphics[width=0.48\textwidth]{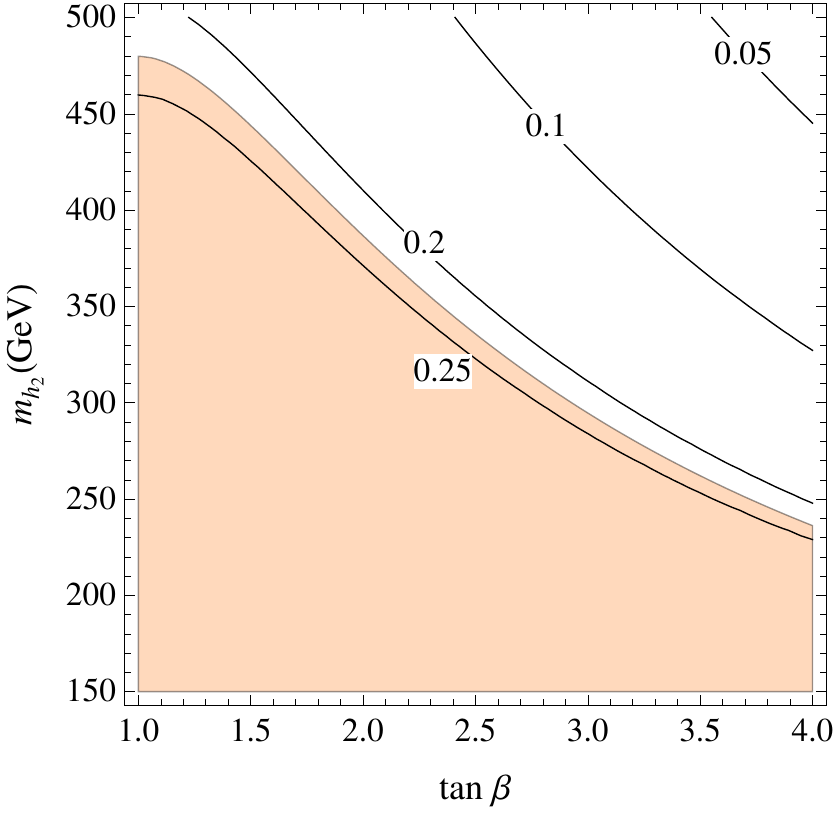}
\caption{\label{fig:mAdecoupled-1}\small $H$ decoupled. Isolines of $\sin^2\gamma$. Left: $\lambda=0.8$. Right: $\lambda=1.4$. The coloured region is excluded at 95\%C.L. by the experimental data for the signal strengths of $h_1 = h_{\rm LHC}$.}
\end{center}\end{figure}

To study this limiting case, it is best to go in the basis $(H, h, s)$ with $H=s_\beta H_d - c_\beta H_u$ and $h=c_\beta H_d + s_\beta H_u$, and let $H\approx h_3$ decouple, so that $\sigma, \delta = \alpha - \beta +\pi/2 \rightarrow 0$. For the remaining non-vanishing angle $\gamma$ one has 
\begin{equation}
\sin^2{\gamma}= \frac{m_{hh}^2-m_{h_1}^2}{m_{h_2}^2-m_{h_1}^2},
\label{sin2gamma}
\end{equation}
where 
\begin{equation}\label{mhh}
m_{hh}^2 = m_Z^2 c_{2\beta}^2 + \lambda^2 v^2 s_{2\beta}^2 + \Delta_t^2
\end{equation}
is the first diagonal entry in the  square mass matrix of  the reduced basis $(h, s)$.

Under the conditions specified in the previous section it is straightforward to see that  the couplings of $h_1 = h_{\text{LHC}}$ and $h_2$ to fermions or to vector boson pairs, $VV = WW, ZZ$, normalized to the same couplings of the SM Higgs boson, are given by
\begin{equation} \label{Hdec_couplings}
\frac{g_{h_1ff}}{g^{\text{SM}}_{hff}} = \frac{g_{h_1VV}}{g^{\text{SM}}_{hVV}}= c_\gamma,  ~~~~~~
\frac{g_{h_2ff}}{g^{\text{SM}}_{hff}}= \frac{g_{h_2VV}}{g^{\text{SM}}_{hVV}}= - s_\gamma.
\end{equation}
As a consequence none of the branching ratios of $h_1$ gets modified with respect to the SM ones, whereas its production cross sections, or the various signal strengths, are reduced by a common factor $c_\gamma^2$ with respect to the SM ones with $m_{h_{\text{SM}}} = m_{h_1}$. 
We recall that the fit of all experimental data collected so far gives the bound on $s_\gamma^2$ shown in Figure~\ref{fig:FIT} right, $s_\gamma^2 < 0.22$, and that the exclusion reach of the next LHC run, with the projected precisions of Table \ref{tab1}, extends to $s_\gamma^2 < 0.15$, in the case SM central values are measured for the $h_{\rm LHC}$ signal strengths.

\subsection{$h_{\rm LHC}$ lightest state}

Upon use of (\ref{sin2gamma}) the impact of this bound on the parameter space is shown in Figure~\ref{fig:mAdecoupled-1} for $\lambda = 0.8$ and 1.4, together with the isolines of different values of $s_\gamma^2$ that might be probed by future improvements in the measurements of the $h_1$ signal strengths. 
Unlike in the singlet-decoupled case, the improvement in the measurements of the signal strengths of $h_{\text{LHC}}$ is not going to play a major role in further probing the allowed regions.\\
Larger values of $\lambda$ already exclude a significant portion of the parameter space at least for moderate $\tan{\beta}$, as preferred by naturalness. Again in this section we are taking a fixed value of $\Delta_t = 75$~GeV in (\ref{delta-t}), and as long as one stays at $\Delta_t \lesssim 85$ GeV and $\lambda \gtrsim 0.8$, in a range of moderate fine tuning, our results do not depend significantly on $\Delta_t$.
\begin{figure}[t!]
\begin{center}
\subfigure[\label{fig:mAdecoupled-Xsec8a}8 TeV, $\lambda=0.8$]{\includegraphics[width=.48\textwidth]{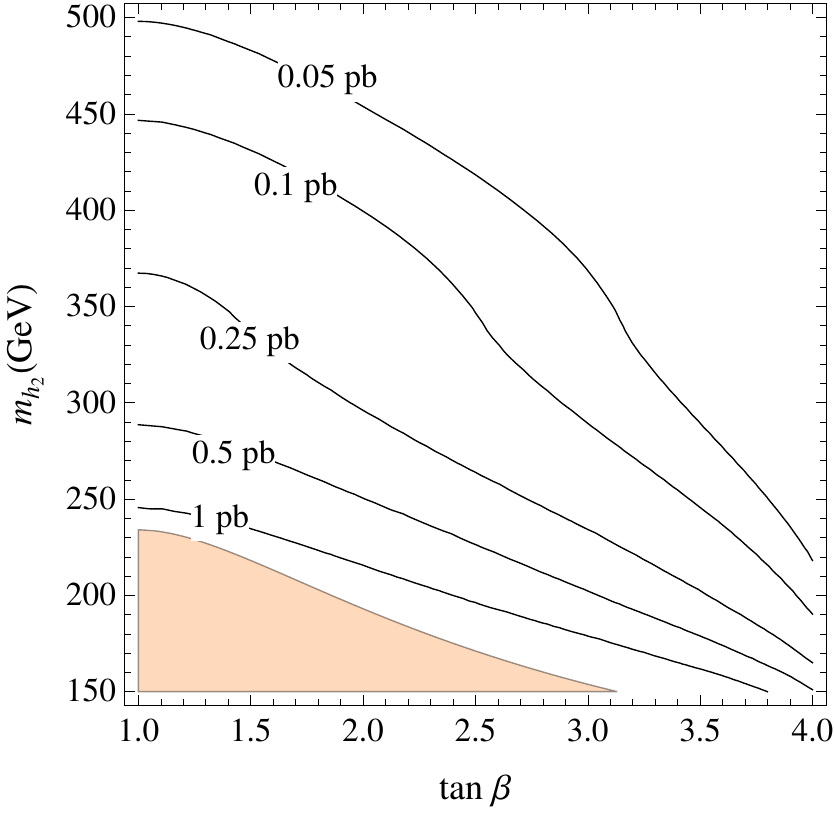}}\hfill
\subfigure[\label{fig:mAdecoupled-Xsec8b}8 TeV, $\lambda=1.4$]{\includegraphics[width=.48\textwidth]{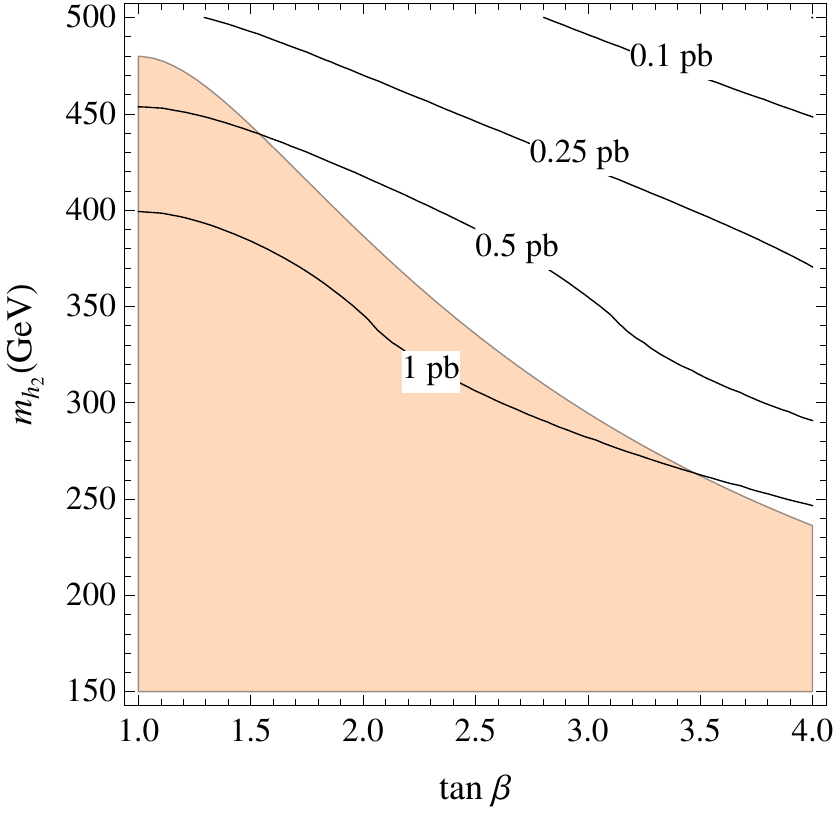}}\vspace{0.8cm}\\
\subfigure[\label{fig:mAdecoupled-Xsec14c}14 TeV, $\lambda=0.8$]{\includegraphics[width=.48\textwidth]{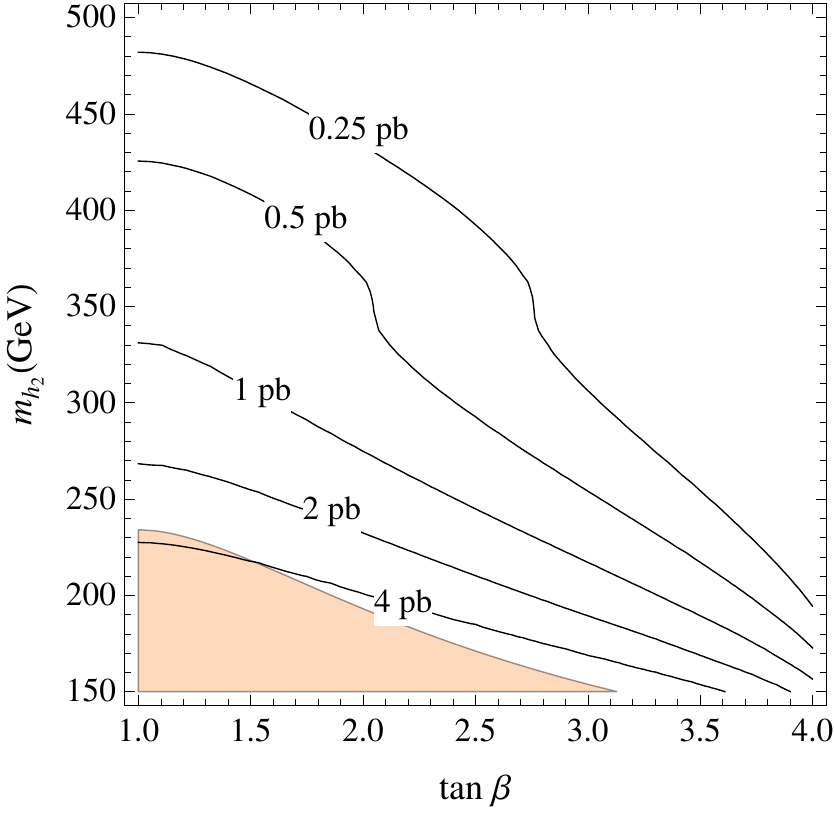}}\hfill
\subfigure[\label{fig:mAdecoupled-Xsec14d}14 TeV, $\lambda=1.4$]{\includegraphics[width=.48\textwidth]{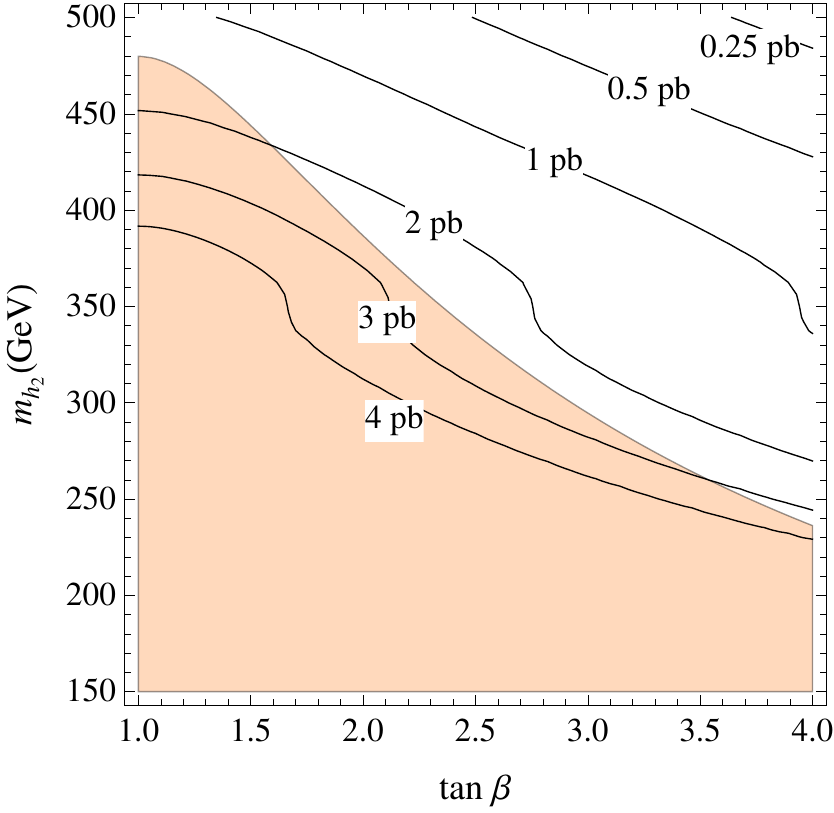}}
\caption{\label{fig:mAdecoupled-Xsec}\small $H$ decoupled. Isolines of gluon fusion cross section $\sigma(gg\to h_2)$ at LHC8 and LHC14, for the values $\lambda=0.8$ and $\lambda=1.4$. The coloured region is excluded at 95$\%$C.L.}
\end{center}
\end{figure}
In the same $(\tan{\beta}, m_{h_2})$ plane of Figure~\ref{fig:mAdecoupled-1} and for the same values of $\lambda$, Figure~\ref{fig:mAdecoupled-Xsec} shows the gluon-fusion production cross sections of $h_2$ at LHC for 8 or 14 TeV c.o.m. energies, where we rescaled by $c_\gamma^2$ the NNLL ones provided in \cite{Dittmaier:2011ti}. All other $h_2$ production cross sections, relative to the gluon-fusion one, scale as in the SM with $m_{h_{\text{SM}}} = m_{h_2}$.

\begin{figure}[t!]
\begin{center}
\includegraphics[width=.48\textwidth]{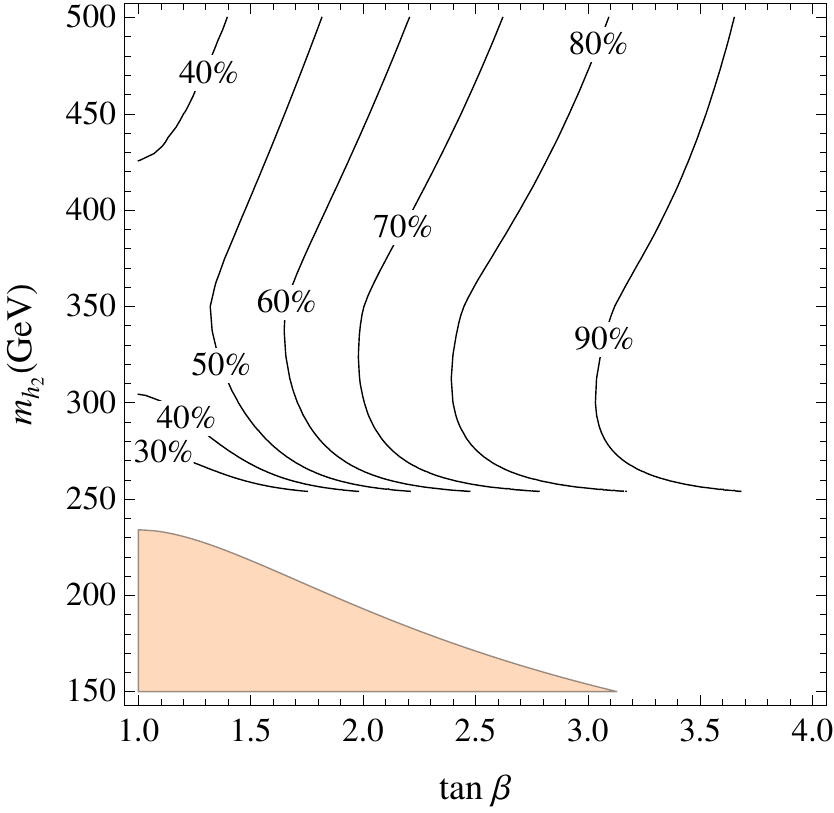}\hfill
\includegraphics[width=.48\textwidth]{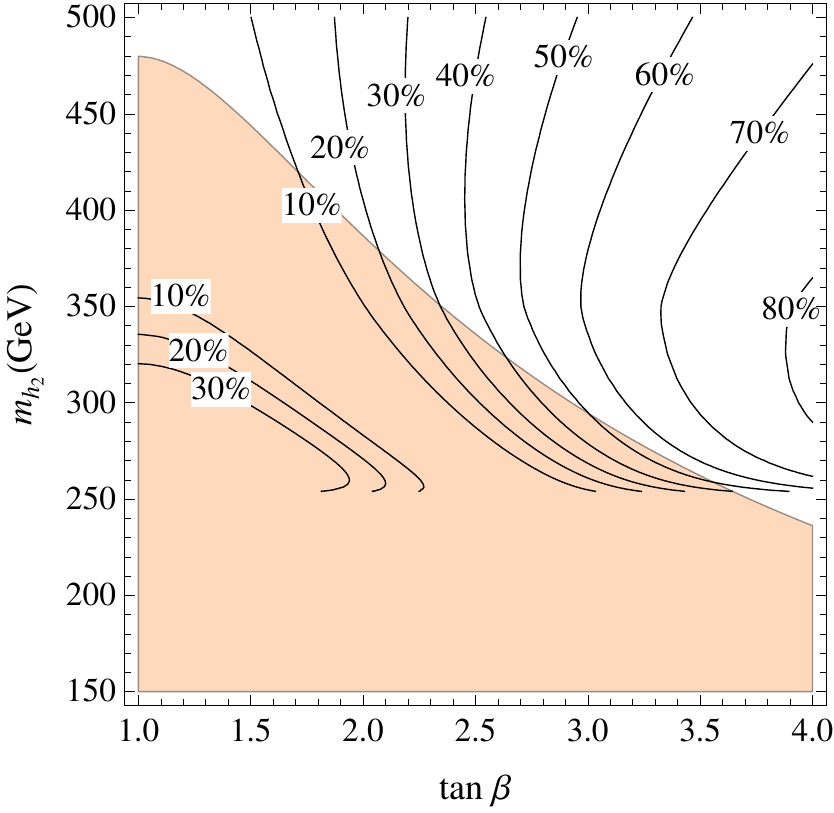}
\caption{\label{fig:mAdecoupled-hh}\small $H$ decoupled. Isolines of BR$(h_2\to h h )$. Left: $\lambda=0.8$ and $v_S=2v$. Right: $\lambda=1.4$ and $v_S=v$. The coloured region is excluded at 95$\%$C.L.}
\vspace{2cm}
\includegraphics[width=.48\textwidth]{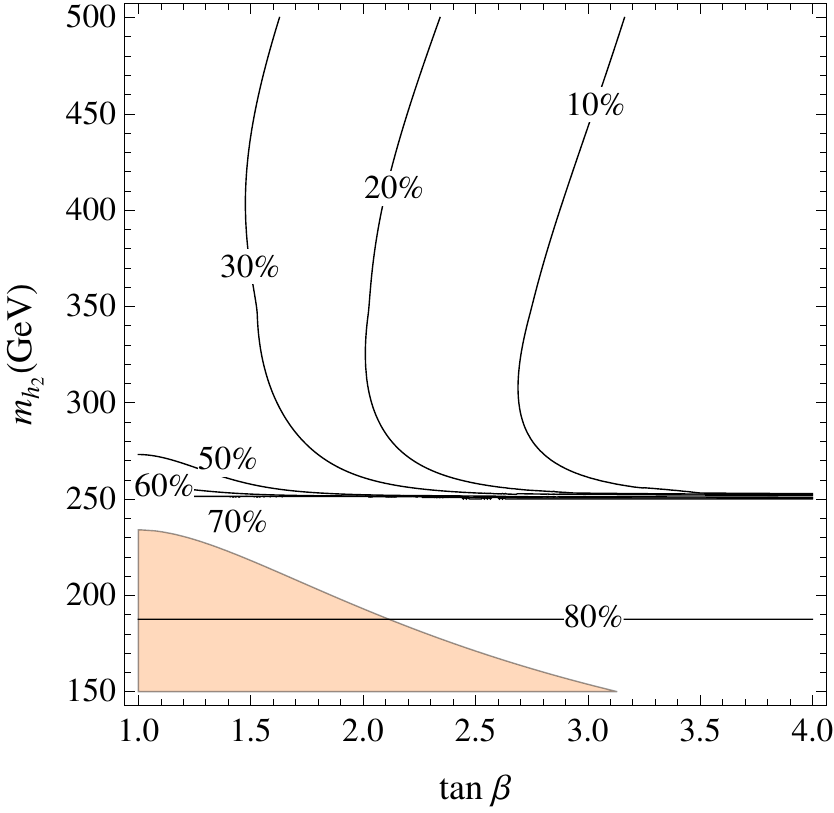}\hfill
\includegraphics[width=.48\textwidth]{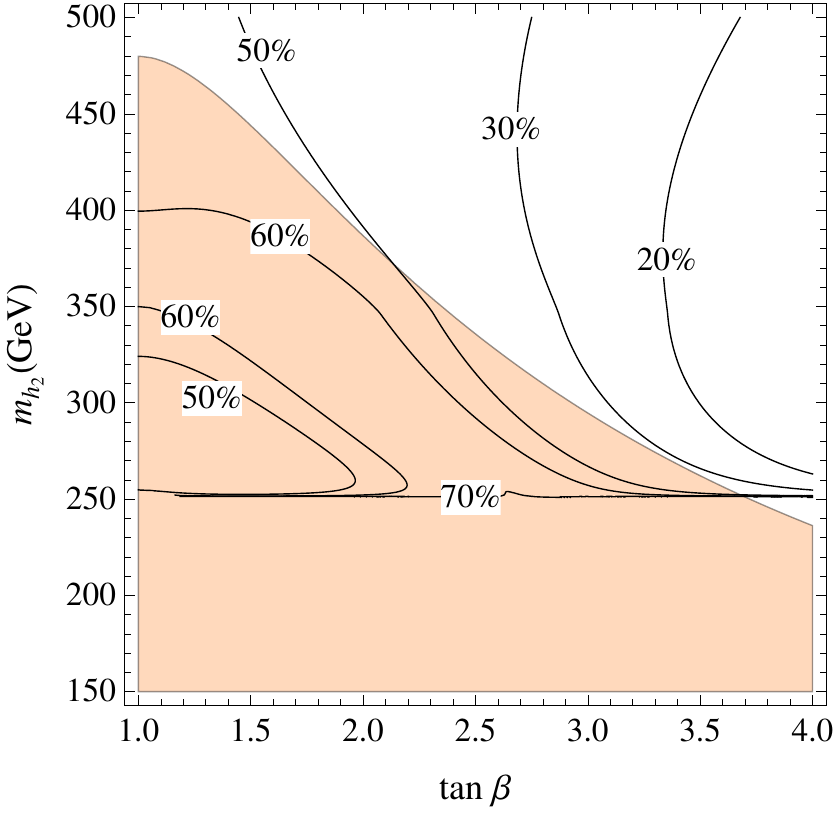}
\caption{\label{fig:mAdecoupled-WW}\small $H$ decoupled. Isolines of BR$(h_2\to W^+W^- )$. Left: $\lambda=0.8$ and $v_S=2v$. Right: $\lambda=1.4$ and $v_S=v$. The coloured region is excluded at 95$\%$C.L.}
\end{center}
\end{figure}\eject
\begin{figure}[t!]
\begin{center}
\includegraphics[width=.48\textwidth]{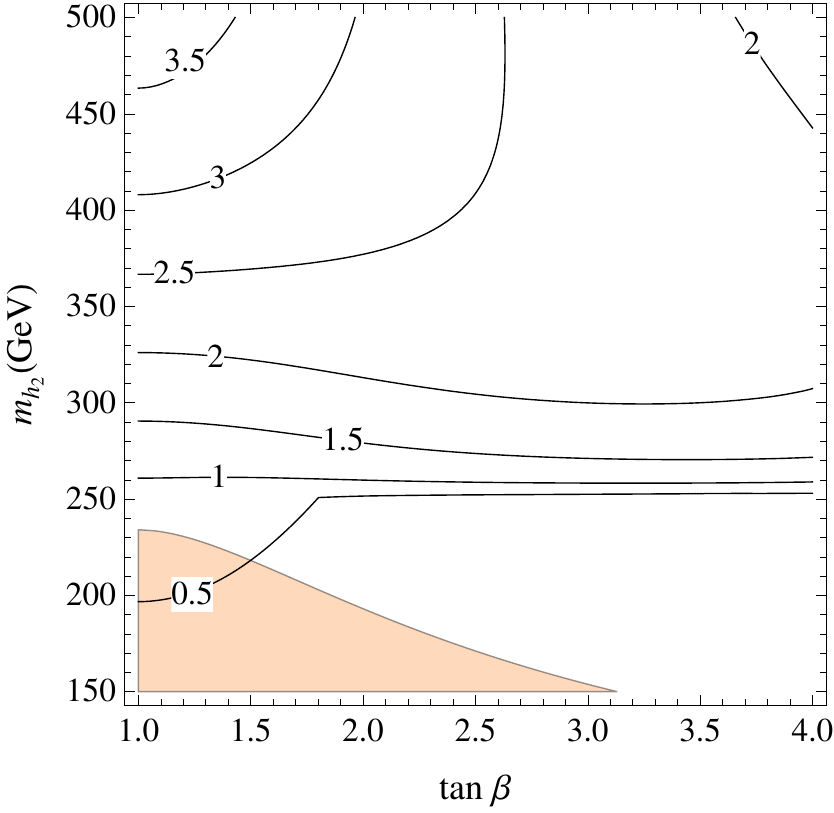}\hfill
\includegraphics[width=.48\textwidth]{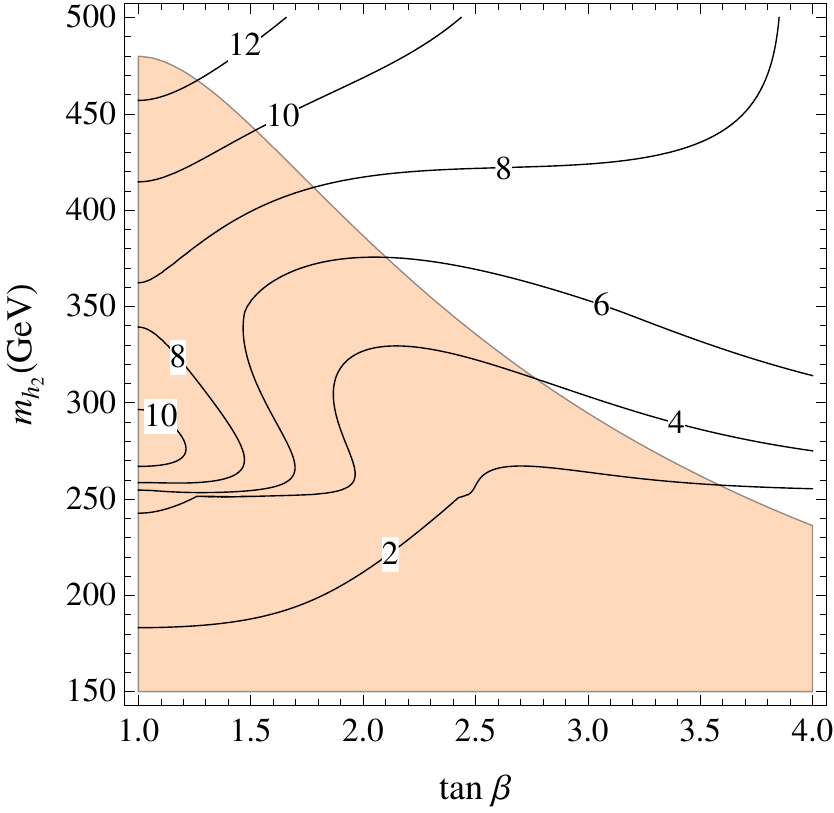}
\caption{\label{fig:mAdecoupled-width}\small $H$ decoupled. Isolines of the total width $\Gamma_{h_2}(\text{GeV})$. Left: $\lambda=0.8$ and $v_S=2v$. Right: $\lambda=1.4$ and $v_S=v$. The coloured region is excluded at 95$\%$C.L.}
\end{center}
\end{figure}
\begin{figure}[h!]
\begin{center}
\includegraphics[width=.48\textwidth]{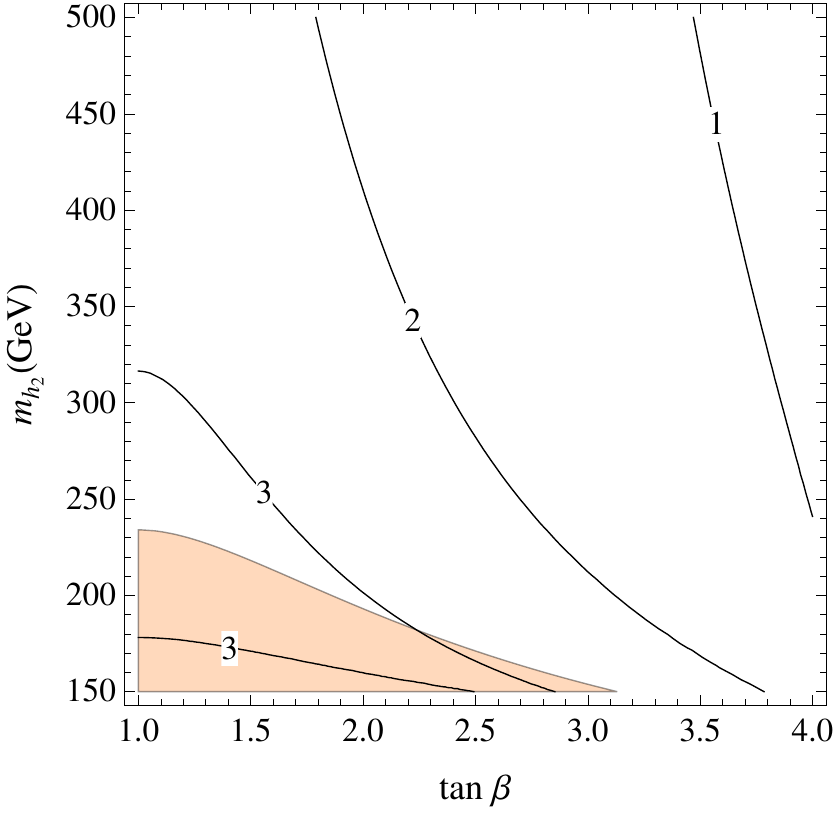}\hfill
\includegraphics[width=.48\textwidth]{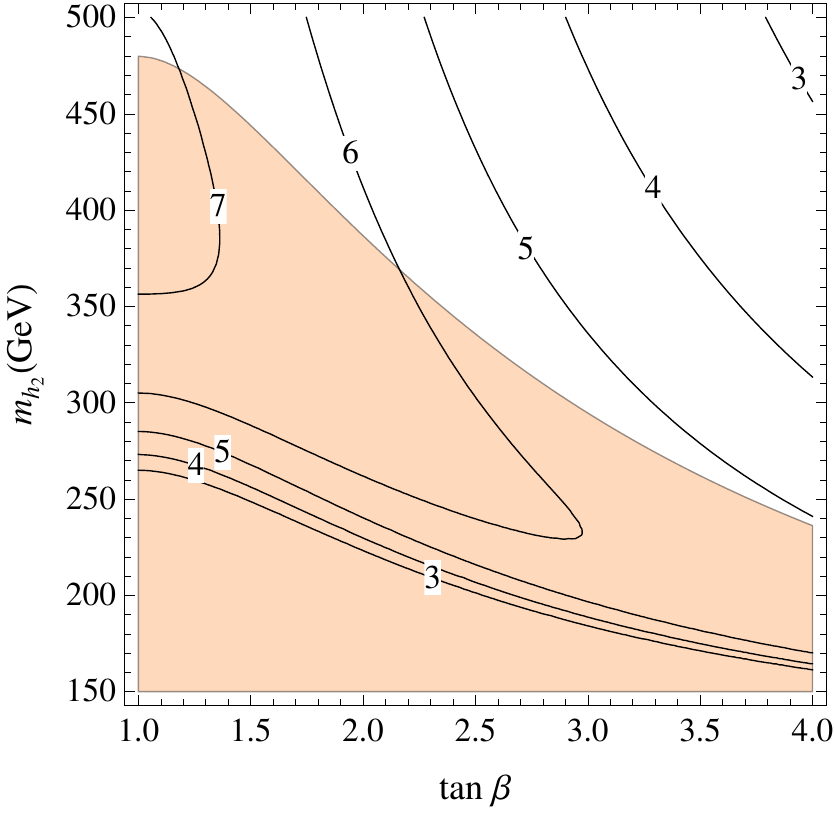}
\caption{\label{fig:mAdecoupled-cubic}\small $H$ decoupled. Isolines of $g_{hhh}/g_{hhh}^{\text{SM}}$. Left: $\lambda=0.8$ and $v_S=2v$. Right: $\lambda=1.4$ and $v_S=v$. The coloured region is excluded at 95$\%$C.L.}
\end{center}
\end{figure}
To determine the decay properties of $h_2$ it is crucial to know its coupling $(g_{h_2h_1^2}/2) h_2h_1^2$ to the lighter state. In the general NMSSM and in the large $m_H$ limit considered in this section, the leading $\lambda^2$-term contribution to this coupling, as well as the one to the cubic $h_1$-coupling $(g_{h_1^3}/6) h_1^3$, are given by
\begin{align}
g_{h_2h_1^2}&=\frac{\lambda ^2 v}{8 \sqrt{2}} \left(4 \frac{v_S}{v} \cos\gamma +12 \frac{v_S}{v} \cos3 \gamma-7  \sin\gamma +12  \cos4 \beta \cos^2\gamma \sin\gamma +9  \sin3 \gamma \right)\notag\\
&- \frac{3}{\sqrt{2}v}\Delta_t^2 \cos^2\gamma\sin\gamma,\\
\frac{g_{h_1^3}}{g^{\text{SM}}_{h_1^3}}&=\frac{\lambda ^2 v^2}{8 m_{h_1}^2} \cos\gamma \left(10 - 4  \cos4 \beta \cos^2\gamma - 6 \cos2 \gamma + 8 \frac{v_S}{v} \sin2 \gamma\right) + \dfrac{\Delta_t^2}{m_{h_1}^2} \cos^3\gamma,
\end{align}
where $v_S$ is the vev of the singlet
. Figures~\ref{fig:mAdecoupled-width} and \ref{fig:mAdecoupled-hh} show the total width of $h_2$ and its branching ratio into a pair of light states for some choices of $v_S$. 
The other most significant decay mode of $h_2$ is into a $W$-pair, with a branching ratio given in Figure~\ref{fig:mAdecoupled-WW}.  Figure~\ref{fig:mAdecoupled-cubic} shows the triple $h_1$-coupling normalized to the SM one.

These results depend on the value taken by $v_S$, in particular we note that the Higgs fit still allows the triple Higgs coupling to get a relative enhancement of a factor of a few (with a negative or positive sign) with respect to the Standard Model one, thus yielding potentially large effects in Higgs pair production cross sections\cite{Baglio:2012np}. Finally, contrary to the S-decoupled case, the effect in BR($h_2 \to h_1 h_1$) of taking a much lighter $\Delta_t$ would be irrelevant.


\subsection{$h_2$ lightest state}
\label{sec4}

\begin{figure}[t]
\begin{center}
\includegraphics[width=0.48\textwidth]{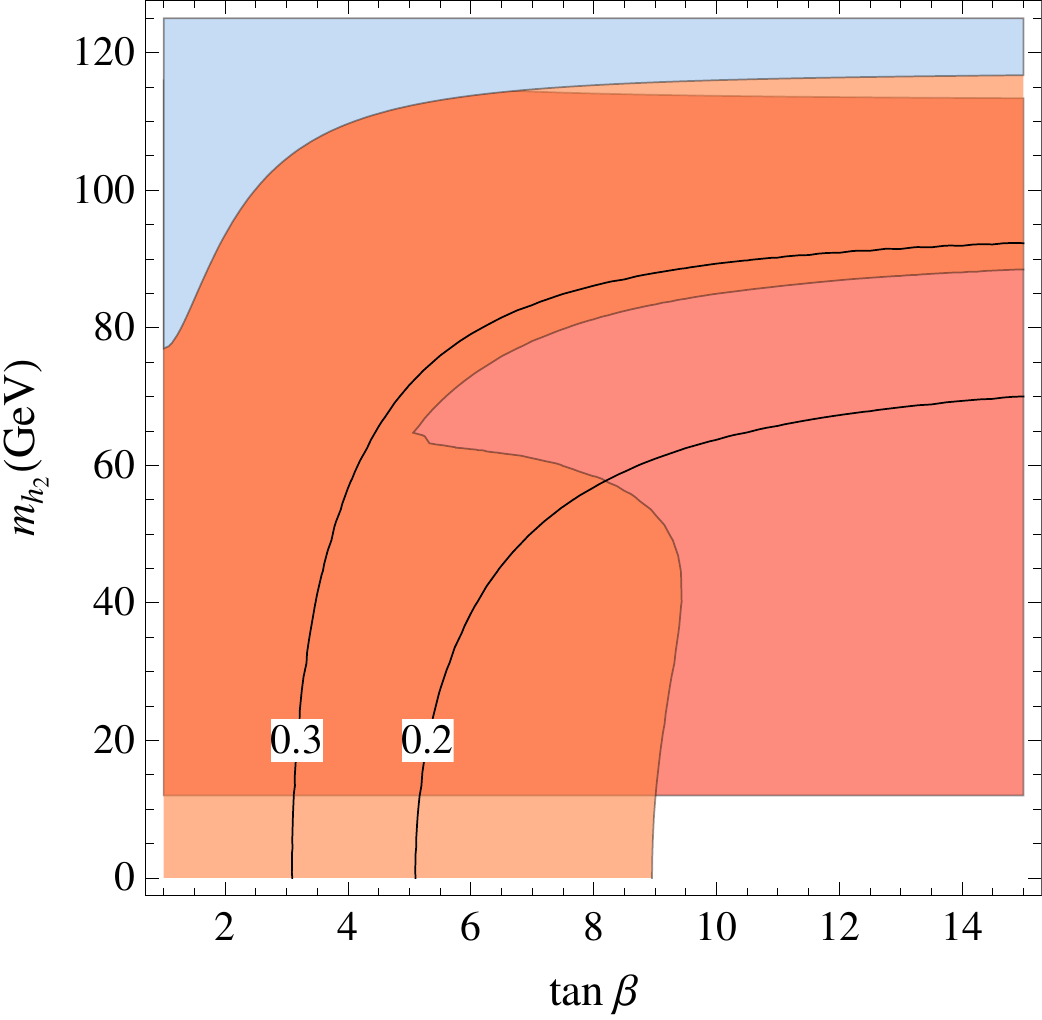}\hfill
\includegraphics[width=0.48\textwidth]{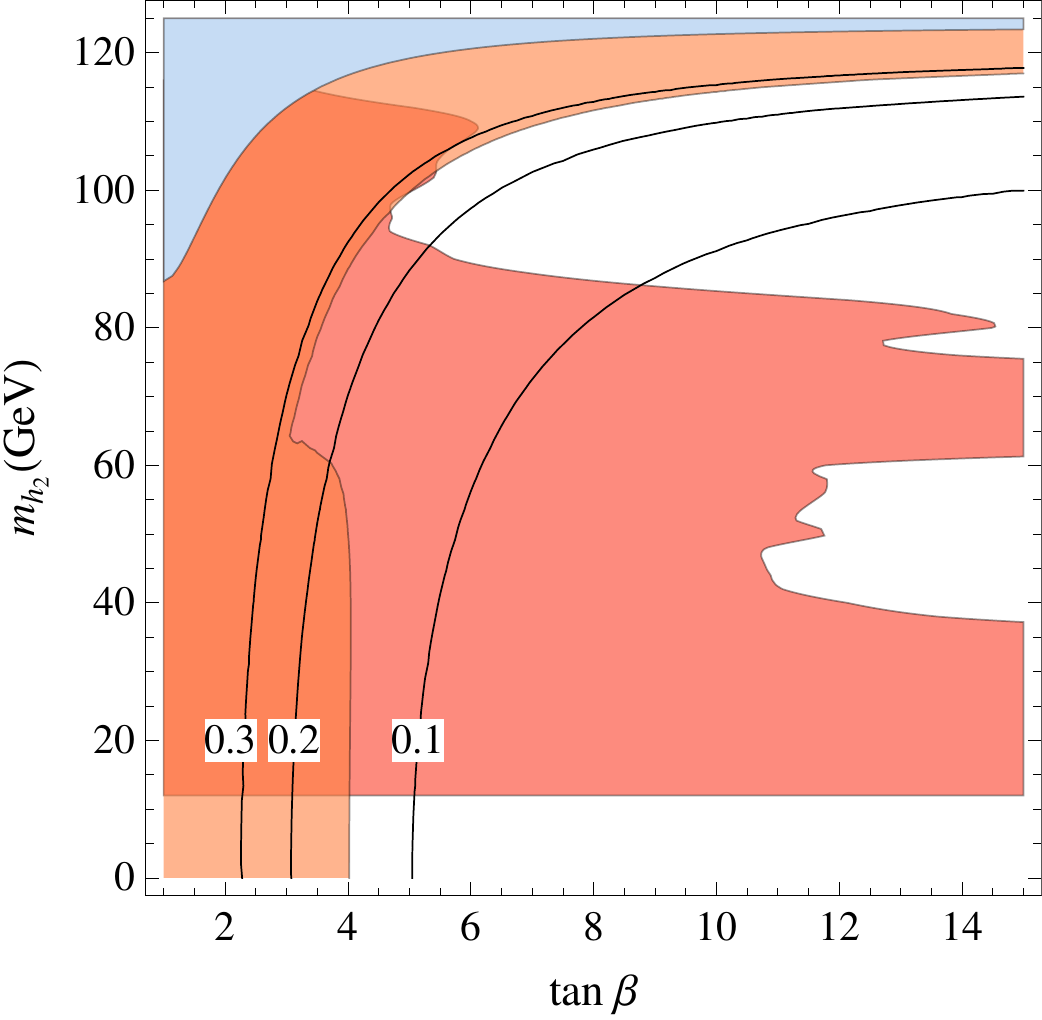}
\caption{\label{fig3} $H$-decoupled. Isolines of $s^2_\gamma$. $\lambda=0.1$ and $v_S=v$. Left: $\Delta_t=75$ GeV. Right: $\Delta_t=85$ GeV. Orange and blue regions as in Fig. \ref{fig:NMSSM_mh3_tbeta_lambda}. The red region is excluded by LEP direct searches for $h_2\to b\bar b$.}
\end{center}
\end{figure}

\begin{figure}[h!]
\begin{center}
\includegraphics[width=0.48\textwidth]{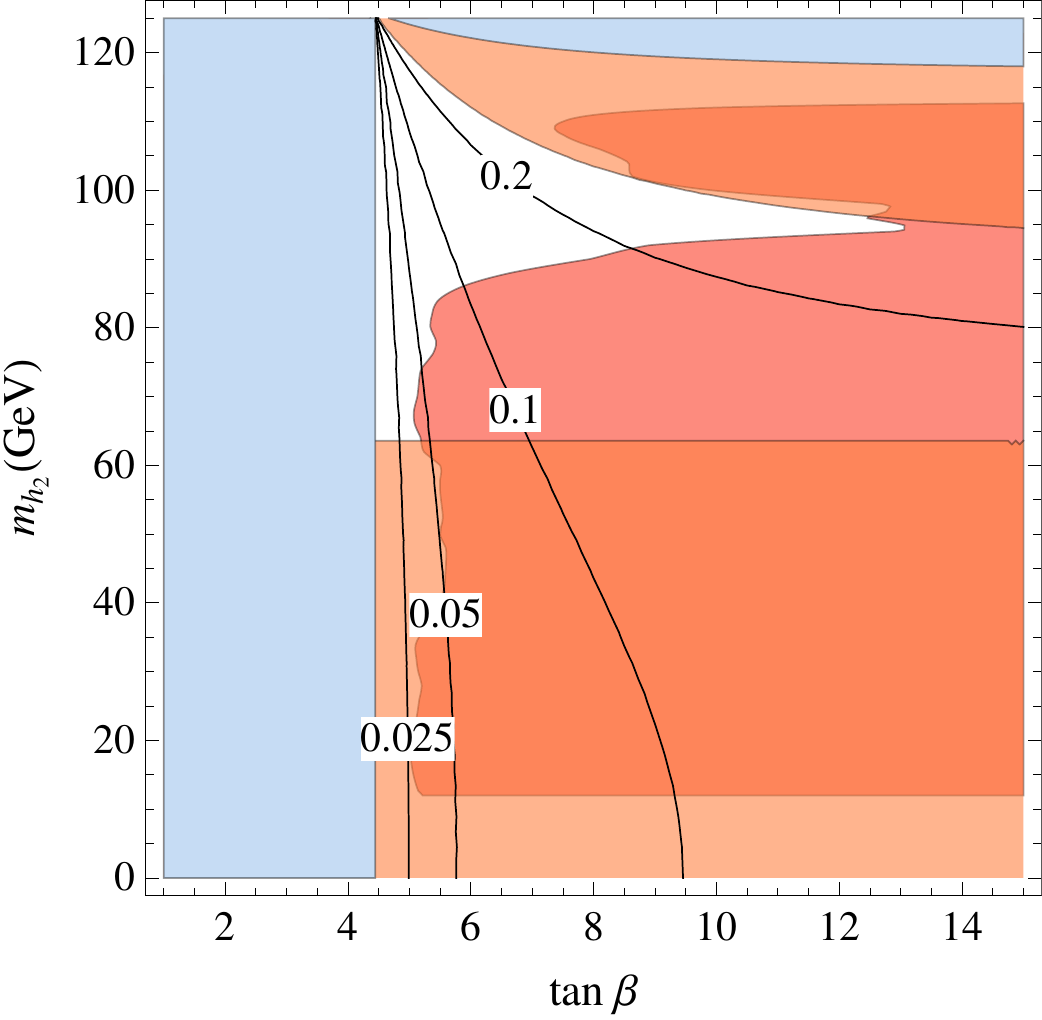}\hfill
\includegraphics[width=0.48\textwidth]{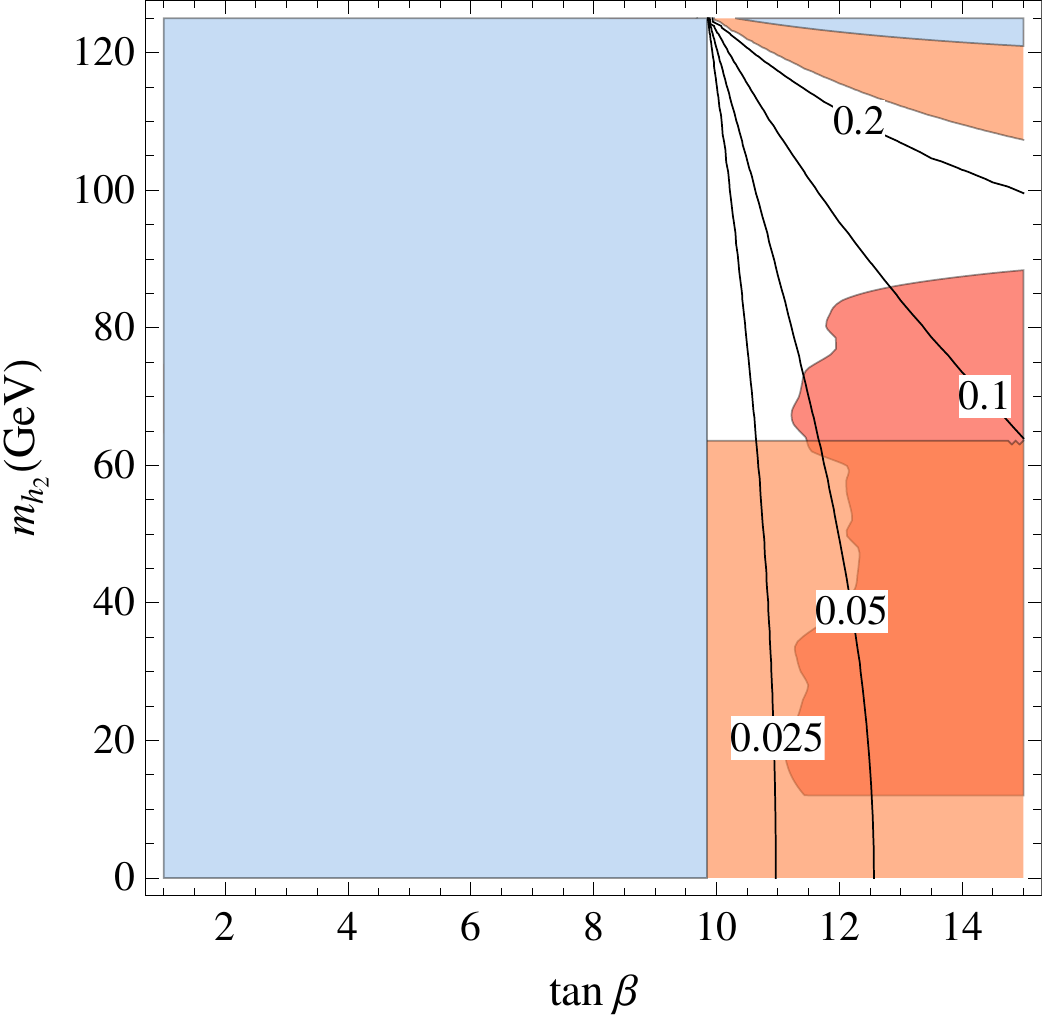}
\caption{\label{fig4} $H$-decoupled. Isolines of $s^2_\gamma$. $\Delta_t=75$ GeV and $v_S=v$. Left: $\lambda=0.8$. Right: $\lambda=1.4$. Orange and blue regions as in Fig. \ref{fig:NMSSM_mh3_tbeta_lambda}. The red region is excluded by LEP direct searches for $h_2\to b\bar b$.}
\end{center}
\end{figure}
Here we concentrate on the case of $h_2 < h_{\text{LHC}} (< h_3(= H))$ and we consider both the low and the large $\lambda$ case.

The low $\lambda$ case $(\lambda = 0.1)$ is shown in Fig. \ref{fig3} for two values of $\Delta_t$ together with the isolines of $s^2_{\gamma}$.
As a consequence of the couplings in \eqref{Hdec_couplings}, for $m_{h_2} > m_{h_{\text{LHC}}}/2$
none of the branching ratios of  $h_1 = h_{\text{LHC}}$ and $h_2$ get modified with respect to the  ones of the SM Higgs boson with the corresponding mass, whereas their production cross sections are reduced by a common factor $c_\gamma^2$ or $s^2_\gamma$ respectively  for  $h_1 = h_{\text{LHC}}$ and $h_2$. As before, the lighter regions in Fig. \ref{fig3} are excluded by the fit of the $h_1$ signal strengths. The red regions are due to the negative searches of $h_2\rightarrow \bar{b} b, \tau^+\tau^-$ at LEP \cite{Schael:2006cr}, which were performed down to $m_{h_2} = 12$ GeV.  We remind that, as in the previous case, we do not include any invisible decay mode except for $h_{\text{LHC}}\rightarrow h_2 h_2$, when kinematically allowed.\footnote{To include $h_{\text{LHC}}\rightarrow h_2 h_2$ we rely on the triple Higgs couplings as computed by retaining only the $\lambda^2$-contributions.
As stated before, this is a defendable approximation for $\lambda$ close to unity, where $h_{\text{LHC}}\rightarrow h_2 h_2$ is important. In the low $\lambda$ case the $\lambda^2$-approximation can only be taken as indicative, but there $h_{\text{LHC}}\rightarrow h_2 h_2$ is less important.}

For $\lambda$  close to unity we take as in the singlet-decoupled case  $\Delta_t = 75$ GeV, but any choice  lower than this would not change the conclusions. The currently allowed region is shown in Fig.~\ref{fig4} for two values of $\lambda$. Note that, for large $\lambda$, no solution is possible at low enough $\tan{\beta}$, since, before mixing, $m^2_{hh}$ in Eq.~\eqref{mhh} has to be below the mass squared of $h_{\text{LHC}}$.

\begin{figure}[t]
\begin{center}
\includegraphics[width=0.48\textwidth]{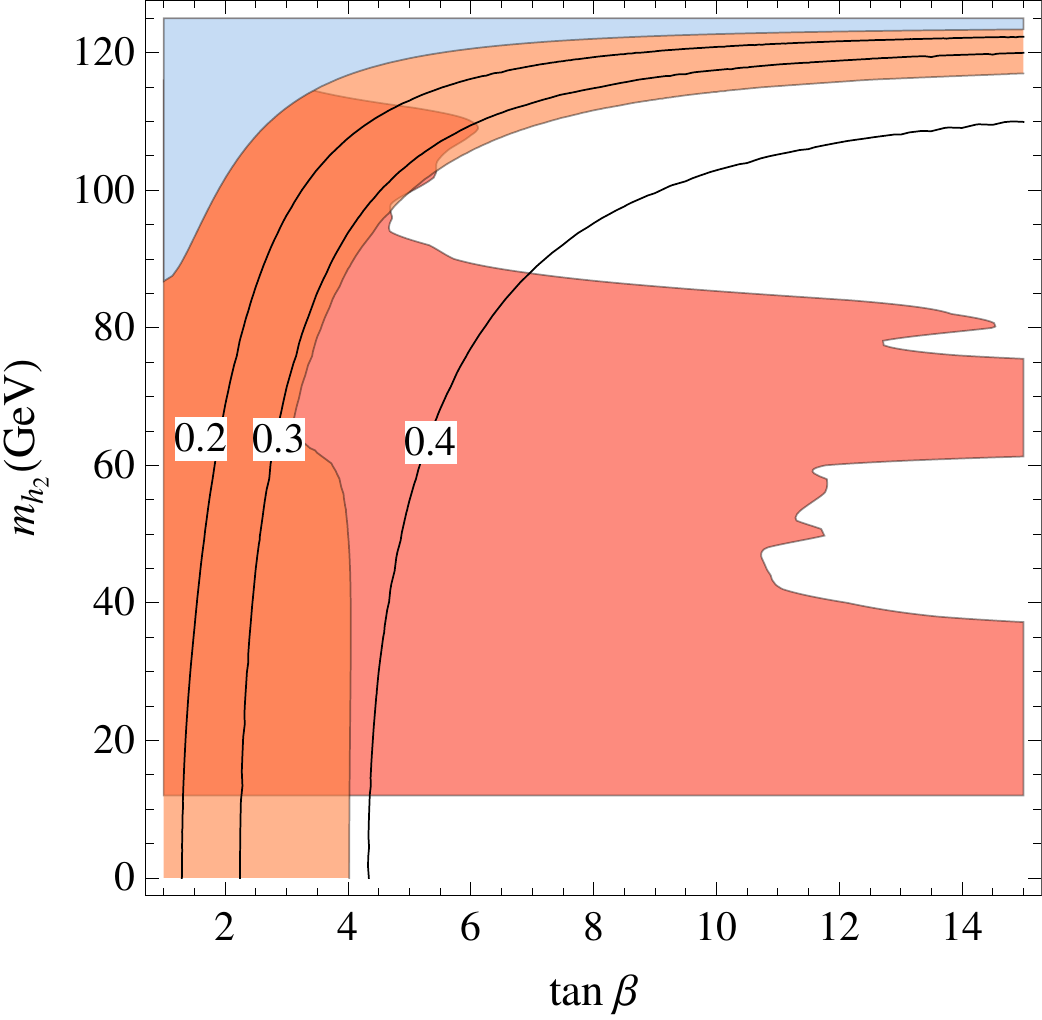}\hfill
\includegraphics[width=0.48\textwidth]{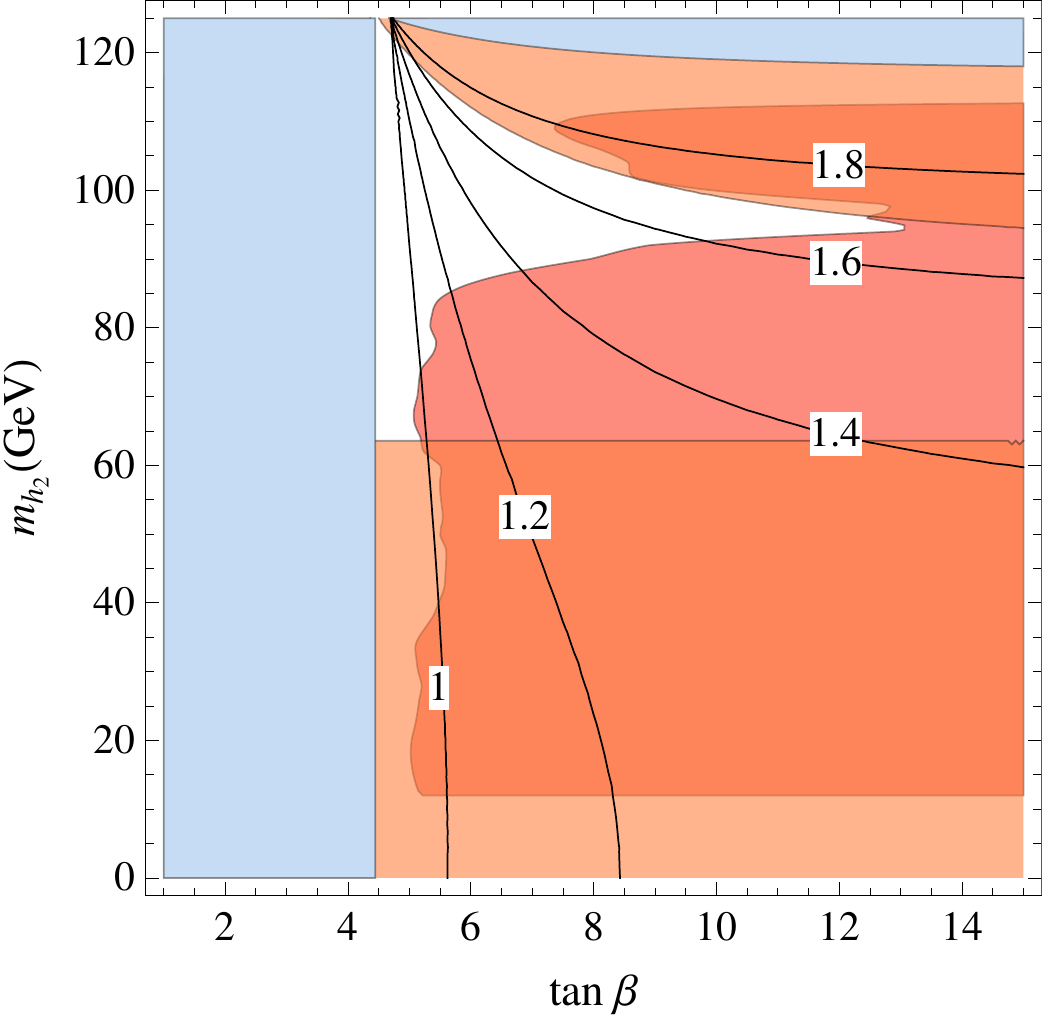}
\caption{\label{fig5} $H$-decoupled. Isolines of $g_{h^3}/g_{h^3}\big|_{\rm SM}$. Left: $\lambda=0.1$, $\Delta_t=85$ GeV and $v_S=v$. Right: $\lambda=0.8$, $\Delta_t=75$ GeV and $v_S=v$. Orange and blue regions as in Fig. \ref{fig:NMSSM_mh3_tbeta_lambda}. The red region is excluded by LEP direct searches for $h_2\to b\bar b$.}
\end{center}
\end{figure}

How will it be possible to explore the regions of parameter space currently still allowed in this $h_2 < h_{\text{LHC}} (< h_3(= H))$ case in view of the reduced couplings of the lighter state? While an improvement in the measurement of the $h_{\rm LHC}$ signal strenghts is not going to play a major role, a significant deviation from the case of the SM can occur in the cubic $h_{\text{LHC}}$-coupling, $g_{h^3_1}$, as shown in Fig.~\ref{fig5}.
As for the $h_{\rm LHC} < h_2$ case, the LHC14 in the high-luminosity regime is expected to get enough sensitivity to be able to see such deviations \cite{ATLAS-collaboration:2012iza,Dolan:2012rv,Goertz:2013kp}. 
At that point, on the other hand,  the searches for directly produced s-partners should have already given some clear indications on the relevance of the entire picture.


\section{Fully mixed case and the $\gamma \gamma$ signal}
\label{sec:HnonDec}

The phenomenological exploration of the situation considered in the previous section could be significantly influenced if the third state, i.e. the doublet $H$, were not fully decoupled. In this case the three angles $\delta$, $\gamma$ and $\sigma$ can all be different from zero, and the three masses $m_{h_2}$, $m_{h_3}$ and $m_{H^\pm}$ are all virtually independent. In the following, also for comparison with the results of Section \ref{sec:Hdec}, we study the $m_{h_2} - \tan\beta$ plane for different reference values of $m_{h_3}$.
Then, using Eqs.~\eqref{eq:sin:gamma:general}-\eqref{eq:sin:2alpha:general}, for fixed values of $\sigma$, $\lambda$ and $\Delta_t$, the two remaining angles $\alpha$ (or $\delta = \alpha - \beta + \pi/2$) and $\gamma$ are determined in any point of the $(\tan{\beta}, m_{h_2})$ plane and so are all the branching ratios of $h_{2,3}$ and of $h_{\text{LHC}}$. More precisely $\delta$ is fixed up to the sign of $s_\sigma c_\sigma s_\gamma$ (see first line of Eq.~\eqref{eq:sin:2alpha:general}), which is the only physical sign that enters the observables we are considering.

\begin{figure}
\begin{center}
\includegraphics[width=.48\textwidth]{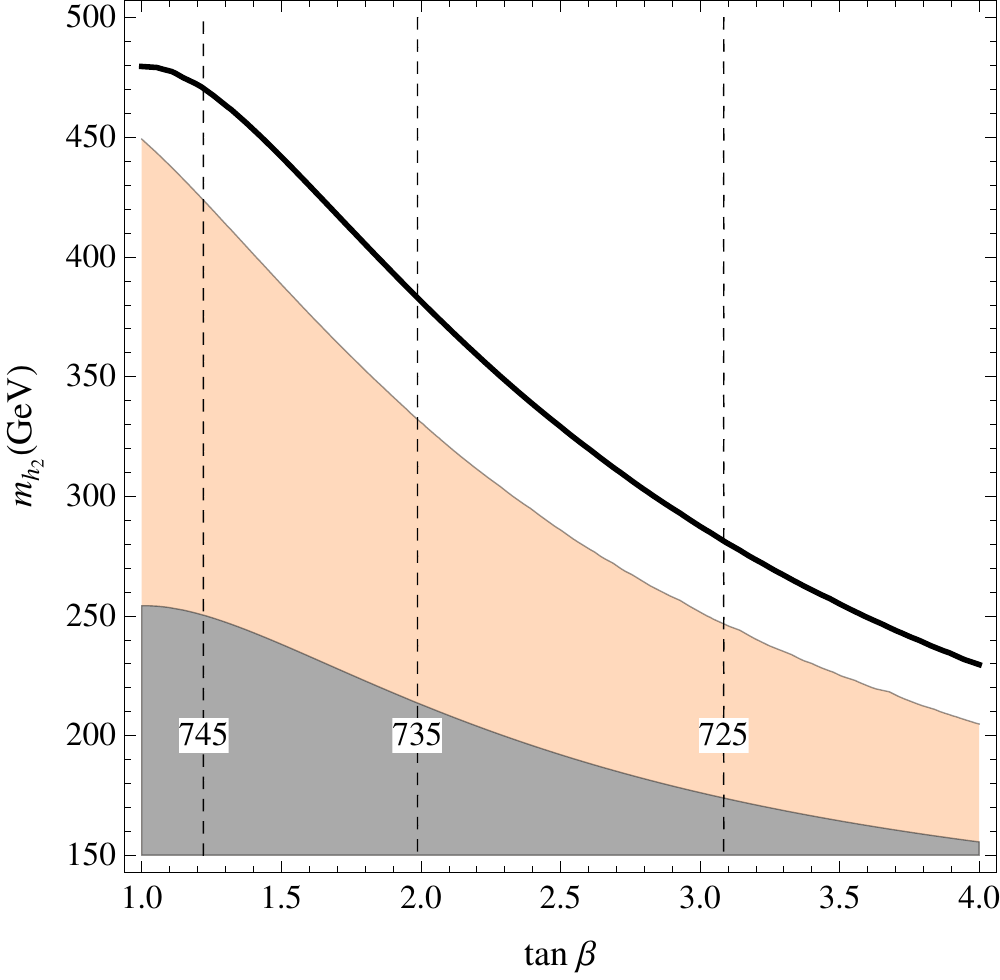}\hfill
\includegraphics[width=.48\textwidth]{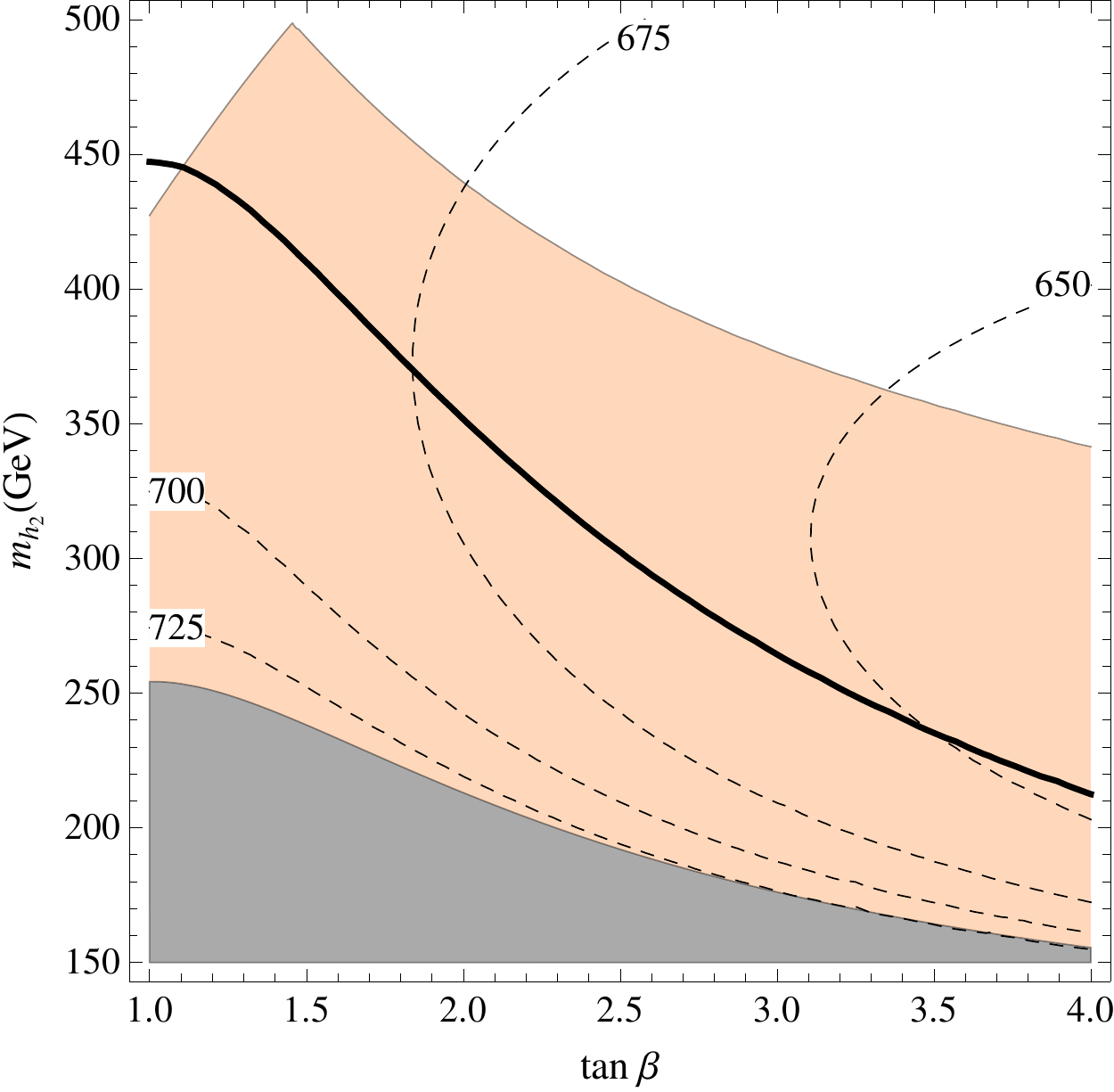}
\caption{\label{fig:mA:quasidecoupled}\small $H$ ``almost decoupled'' with $\lambda = 1.4$ and $m_{h_3} = 750$ GeV. The dashed isolines are for $m_{H^{\pm}}$. Left: $\sin^2\sigma=0$. Right: $\sin^2\sigma=0.25$. The coloured region is excluded at $95\%$C.L. In the grey area there is no solution for $\delta$. The thick line shows the na\"ive exclusion limit from $s_{\gamma}^{2}$ only.}
\end{center}
\end{figure}
As a first example, to see how the impact of the Higgs fit changes, in Figure~\ref{fig:mA:quasidecoupled} we show the excluded regions in the plane $(\tan\beta,m_{h_2})$ for $m_{h_3} = 750$ GeV and $\lambda = 1.4$, setting $s^2_{\sigma}$ to two different values in order to fix $m_{H^\pm}$. When $s^2_{\sigma} = 0$ one recovers the previous $H$ decoupled case in the limit $m_{h_3}\to\infty$. With respect to this case, both $\gamma$ and $\delta$ are free parameters in the fit to the couplings of $h_{\rm LHC}$, and as a consequence the bounds are milder than what is expected from using only $\gamma$. If $s^2_{\sigma} \neq 0$, $h_2$ and $h_3$ are not decoupled, and their masses can not be split too much consistently with all the other constraints. This is reflected in a broader excluded region for low $m_{h_2}$ in Figure~\ref{fig:mA:quasidecoupled} right, where we take $s^2_{\sigma} = 0.25$.

\begin{figure}[t]
\begin{center}
\includegraphics[width=0.48\textwidth]{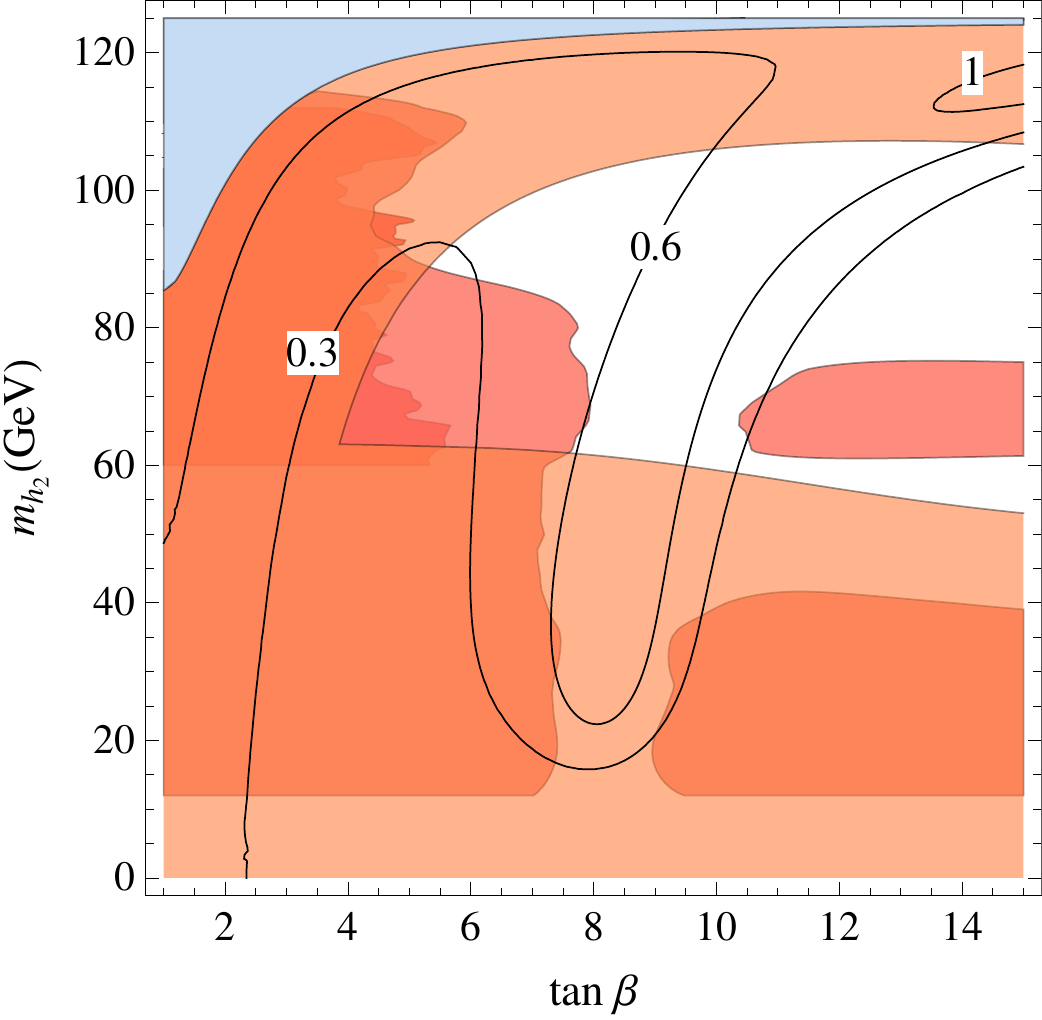}\hfill
\includegraphics[width=0.48\textwidth]{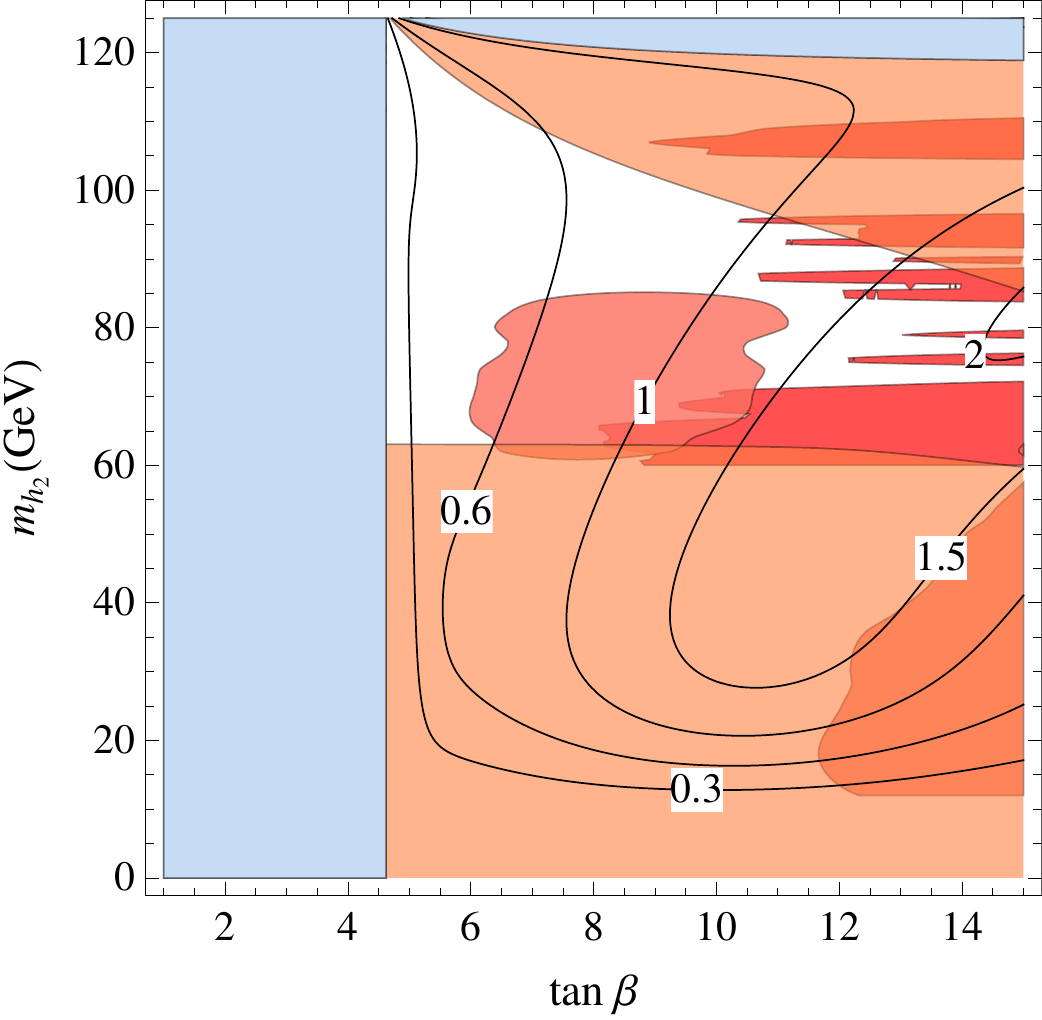}
\caption{\label{fig6} Fully mixed situation. Isolines of the signal strength of $h_2 \to \gamma \gamma$ normalized to the SM. We take  $m_{h_3}=500$ GeV, $s^2_\sigma=0.001$ and $v_s = v$. Left: $\lambda=0.1$, $\Delta_t=85$ GeV. Right: $\lambda=0.8$, $\Delta_t=75$ GeV. Orange and blue regions as in Fig. \ref{fig:NMSSM_mh3_tbeta_lambda}. The red and dark red regions are excluded by LEP direct searches for $h_2\to b\bar b$ and $h_2 \to$ hadrons respectively.}
\end{center}
\end{figure}

As a second example we consider the case of a state $h_2$ lighter than $h_{\text{LHC}}$, lowering $m_{h_3}$ to 500 GeV, to see if it could have an enhanced signal strength into $\gamma \gamma$.
The corresponding situation is represented in Fig.~\ref{fig6}, for two choices of $\lambda$ and $\Delta_t$ (the choice $\lambda = 0.1$ was recently discussed in \cite{Badziak:2013bda}). The sign of $s_\sigma c_\sigma s_\gamma$ has been taken negative in order to suppress BR$(h_2\to b \bar b)$. This constrains $s_\sigma^2$ to be very small in order to leave a region still not excluded by the signal strengths of $h_{\text{LHC}}$, with $\delta$ small and negative. To get a signal strength for $h_2\rightarrow \gamma \gamma$ close to the SM one for the corresponding mass is possible for a small enough value of $s^2_\gamma$, while the dependence on $m_{h_3}$ is weak for values of $m_{h_3}$ greater than 500 GeV. Note that the suppression of the coupling of $h_2$ to $b$-quarks makes it necessary to consider the negative LEP searches for $h_2 \to$ hadrons \cite{Searches:2001aa}, which have been performed down to $m_{h_2} = 60$~GeV.

Looking at the similar problem when $h_2 > h_{\text{LHC}}$, we find it harder to get a $\gamma \gamma$ signal strength close to the SM one, although this might be possible for a rather special choice of the parameters.\footnote{An increasing significance of the excess found by the CMS \cite{CMShint} at 136 GeV would motivate such special choice.} Our purpose here is more to show that in the fully mixed situation the role of the measured signal strengths of $h_{\text{LHC}}$, either current or foreseen, plays a crucial role.

\section{Electroweak Precision Tests}
\label{sec:SUSY_EWPT}

One may ask if the electroweak precision tests (EWPT) set some further constraint on the parameter space explored so far. We have directly checked that this is not the case in any of the different situations illustrated in the various figures. The reason is different in the singlet-decoupled and in the $H$-decoupled cases.

In the $H$-decoupled case the reduced couplings of $h_{\text{LHC}}$ to the weak bosons lead to well known asymptotic formulae for the corrections to the $\hat{S}$ and $\hat{T}$ parameters \cite{Barbieri:2007bh}
\begin{equation}
\Delta \hat S =  + \frac{\alpha}{48\pi s_w^2}s^2_\gamma \log \frac{m_{h_2}^2}{m_{h_{\text{LHC}}}^2}    ,~~\Delta \hat T =      - \frac{3\alpha}{16\pi c_w^2}s^2_\gamma \log \frac{m_{h_2}^2}{m_{h_{\text{LHC}}}^2}
\end{equation}
valid for $m_{h_2}$ sufficiently heavier that $h_{\text{LHC}}$. The correlation of $s^2_\gamma$ with $m_{h_2}$ given in Eq. \eqref{sin2gamma} leads therefore to a rapid decoupling of these effects. 
The one loop effect on $\hat{S}$ and $\hat{T}$ becomes also vanishingly small as 
 $m_{h_2}$ and  $h_{\text{LHC}}$ get close to each other, since in the degenerate limit any mixing can be redefined away and only the standard doublet contributes as in the SM.

In the singlet-decoupled case the mixing between the two doublets can in principle lead to more important effects, which are however limited by the constraint on the mixing angle $\alpha$ or the closeness to zero of $\delta = \alpha - \beta +\pi/2$ already demanded by the measurements of the signal strengths of $h_{\text{LHC}}$.\footnote{Notice that in the fully mixed situation there may be relevant regions of the parameter space still allowed by the fit with a largish $\delta$ (see e.g. Fig. \ref{fig:FIT}).  This could further constrain the small allowed regions, but the precise contributions to the EWPT depend on the value of the masses of the $CP$-odd scalars, which in the generic NMSSM are controlled by further parameters.}
Since in the $\delta = 0$ limit every extra effect on $\hat{S}$ and $\hat{T}$  vanishes, this explains why the EWPT do not impose further constraints on the parameter space that we have considered.

\section{The MSSM for comparison}
\label{sec:MSSM}

\begin{figure}[t]
\begin{center}
\includegraphics[width=.48\textwidth]{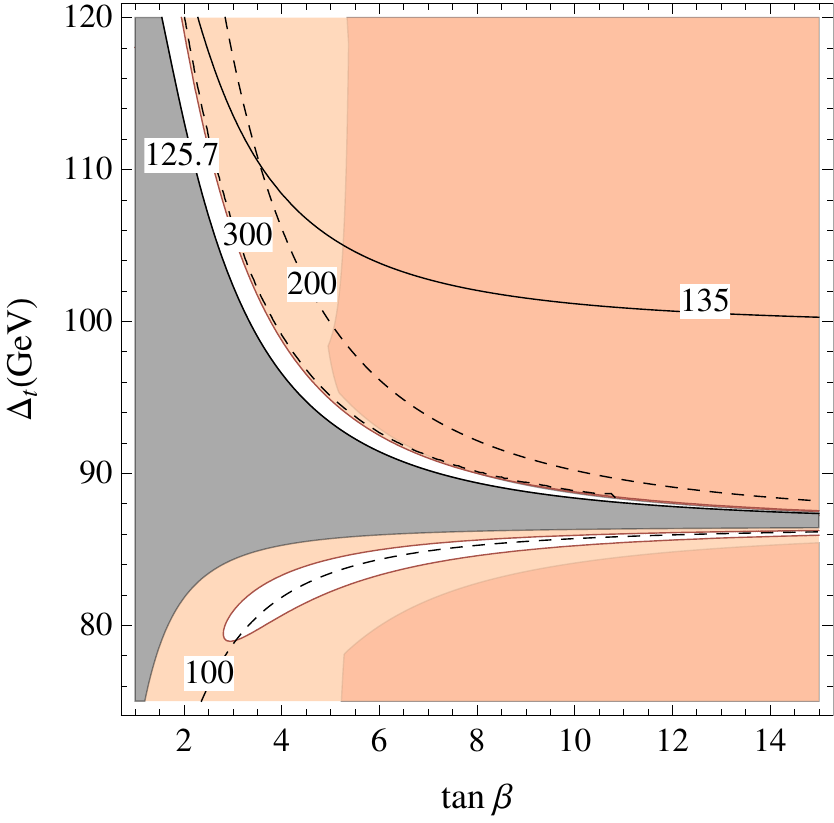}
\caption{\label{fig:MSSM-1}\small MSSM. Isolines of $m_{hh}$ (solid) and $m_{H^\pm}$ (dashed), the gray region is unphysical because of $m^2_A<0$. Light coloured regions are excluded at 95$\%$C.L. by the Higgs fit, the red region is excluded by CMS direct searches for $A,H\to \tau^+\tau^-$.}
\end{center}
\end{figure}

As recalled in Section~\ref{sec:S_dec}, it is instructive to consider the MSSM using as much as possible the
same language, since the MSSM is the $\lambda=0$ limit of the NMSSM in the singlet-decoupled case.\footnote{Two recent papers \cite{Djouadi:2013vqa,Djouadi:2013uqa} analyze the Higgs system of the MSSM in a way similar to ours and give comments about the heavy Higgs searches in different channels (see also \cite{Maiani:2012ij,Maiani:2012qf}).}.

A first important difference of the MSSM versus the NMSSM is in a minimum value of $\Delta_t \gtrsim 85$ GeV that is needed to accommodate the 126 GeV Higgs boson as the lightest CP-even neutral scalar. Also for this reason, and because we have one parameter less than in the previous section, we let $\Delta_t$ vary. As a consequence, in analogy with Figure~\ref{fig:mSdecoupled-1}, we show in Figure~\ref{fig:MSSM-1} the allowed regions by current experimental data on the signal strengths of $h_1 = h_{\text{LHC}}$. From the point of view of the parameter space the main difference is that instead of $\lambda$ we use $\Delta_t$ as an effective parameter.

\begin{figure}[t]
\begin{center}
\includegraphics[width=0.48\textwidth]{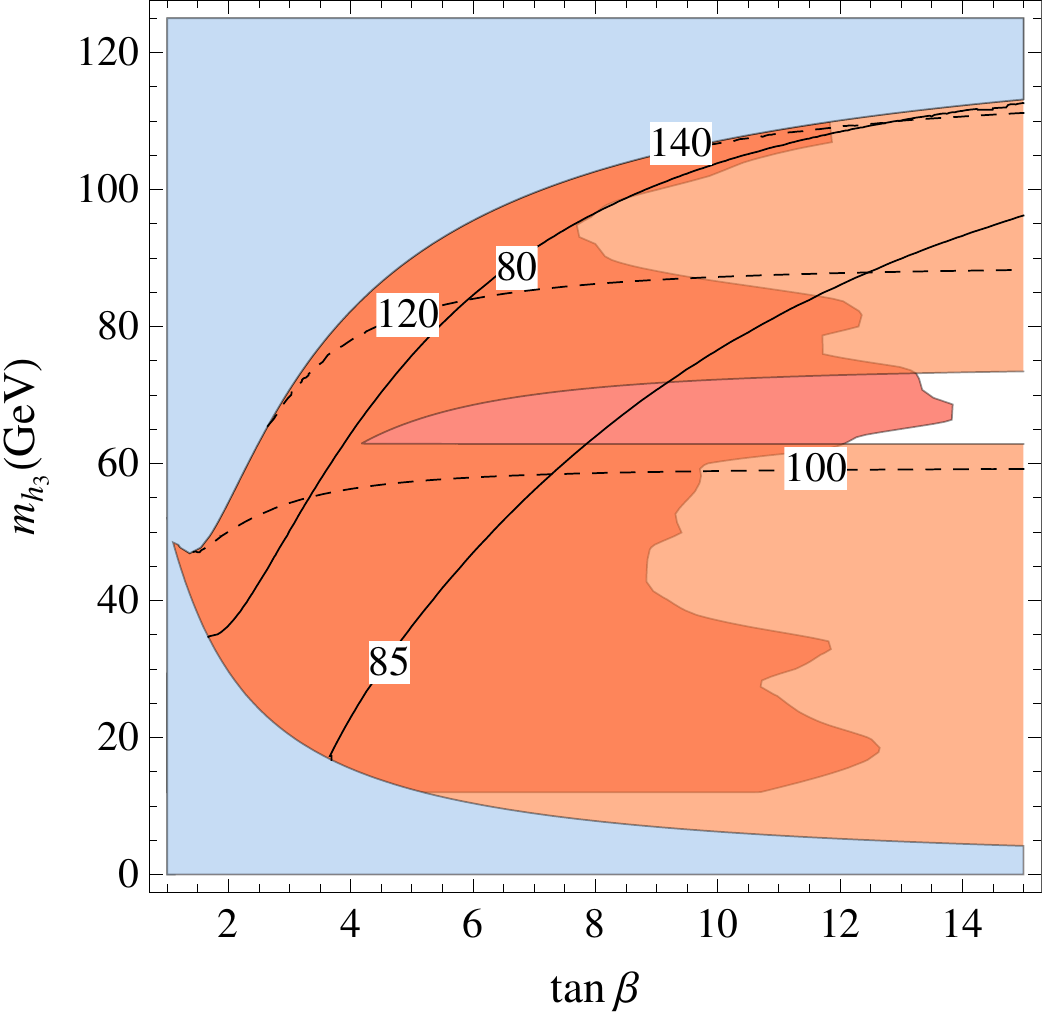}\hfill
\includegraphics[width=0.48\textwidth]{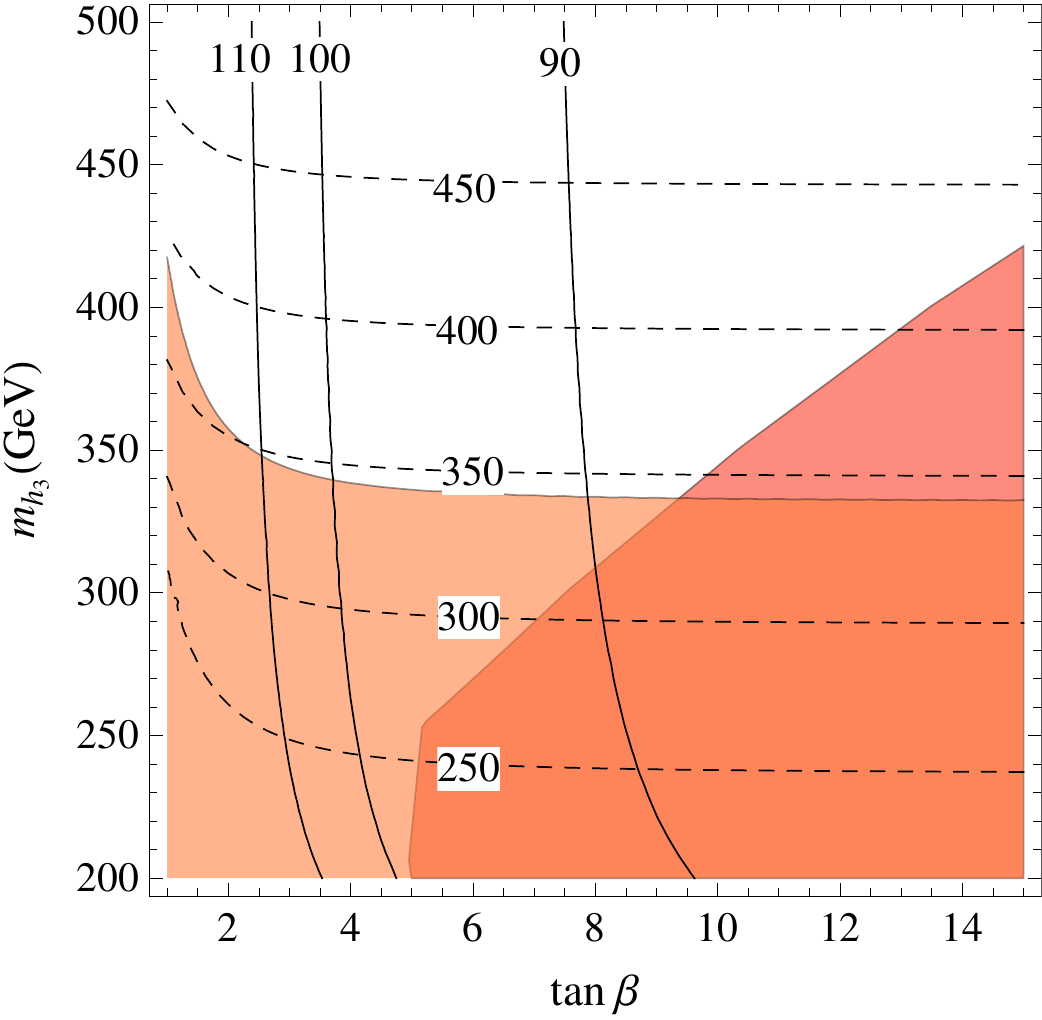}
\caption{\label{fig7} MSSM. Isolines of $\Delta_t$ (solid) and $m_{H^\pm}$ (dashed) at $(\mu A_t)/\langle m_{\tilde{t}}^2\rangle \ll 1$. Left: $h_{\rm LHC}>h_3$, red region is excluded by LEP direct searches for $h_3 \to b\bar b$. Right: $h_{\rm LHC}<h_3$, red region is excluded by CMS direct searches for $A, H\to \tau^+\tau^-$ \cite{CMS-PAS-HIG-12-050}. The orange region is excluded at 95\%C.L. by the current measurements of the signal strengths, the blue region is unphysical.}
\end{center}
\end{figure}
The analogue of Fig. \ref{fig1} are shown in Fig. \ref{fig7}. As expected, both Fig. \ref{fig:MSSM-1} and \ref{fig7} make clear that a large value of $\Delta_t$ is needed to make the MSSM consistent with a 125 GeV Higgs boson.

At the same time, and even more than in the NMSSM case, the projection of the measurements of the signal strengths of 
$h_{\text{LHC}}$ is expected to scrutinize most of the parameter space. We have checked that this is indeed the case with the indirect  sensitivity to $m_{h_3}$ in the right panel of Fig. \ref{fig7}, which will be excluded up to about 1 TeV, as well as with the closure of the white region in the left side of the same figure.
Notice that a similar exclusion will hold also for the CP-odd and charged Higgs bosons, whose masses are fixed in terms of the one of $h_3$.
A warning should be kept in mind, however, relevant to the case $h_3 < h_{\text{LHC}}$:  the one loop corrections to the mass matrix controlled by $(\mu A_t)/\langle m_{\tilde{t}}^2\rangle$ modify the left side of Fig. \ref{fig7} for $(\mu A_t)/\langle m_{\tilde{t}}^2\rangle \gtrsim 1$, changing in particular the currently and projected allowed regions.

\begin{figure}
\begin{center}
\includegraphics[width=.48\textwidth]{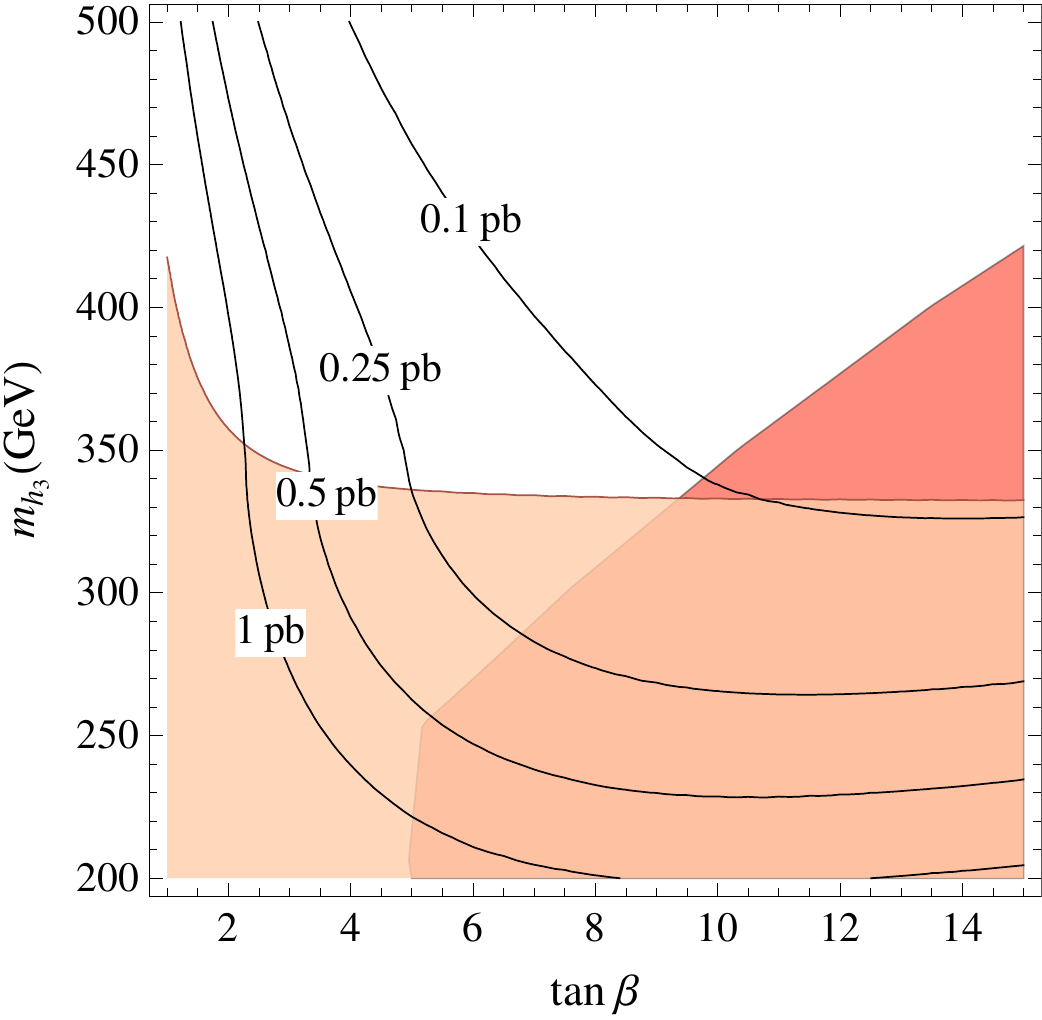}\hfill
\includegraphics[width=.48\textwidth]{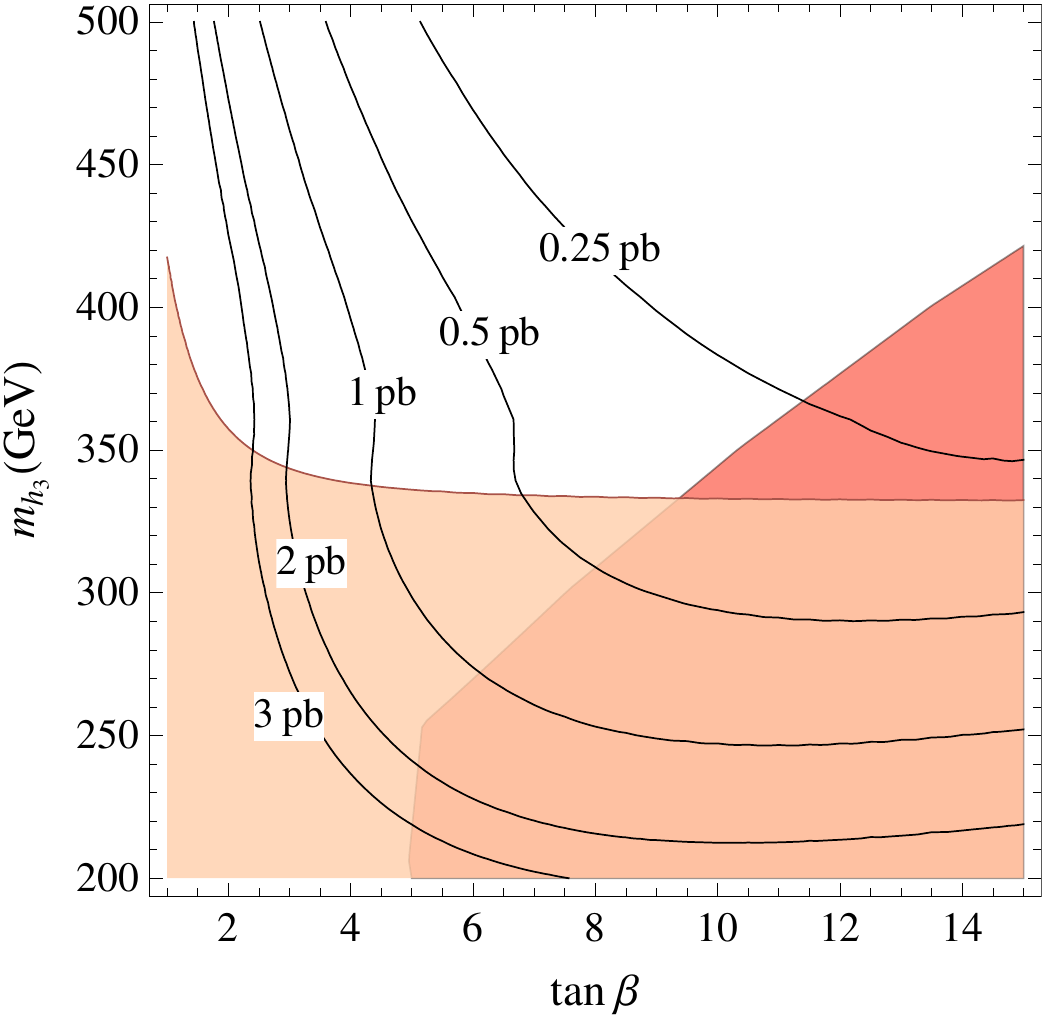}
\caption{\label{fig:MSSM-Xsec}\small MSSM. Isolines of gluon fusion production cross section $\sigma(gg\to h_3)$. Light coloured region is excluded at 95$\%$C.L., the red region is excluded by CMS direct searches for $A,H\to \tau^+\tau^-$. Left: LHC8. Right: LHC14.}
\end{center}
\end{figure}

\begin{figure}
\begin{center}
\includegraphics[width=.48\textwidth]{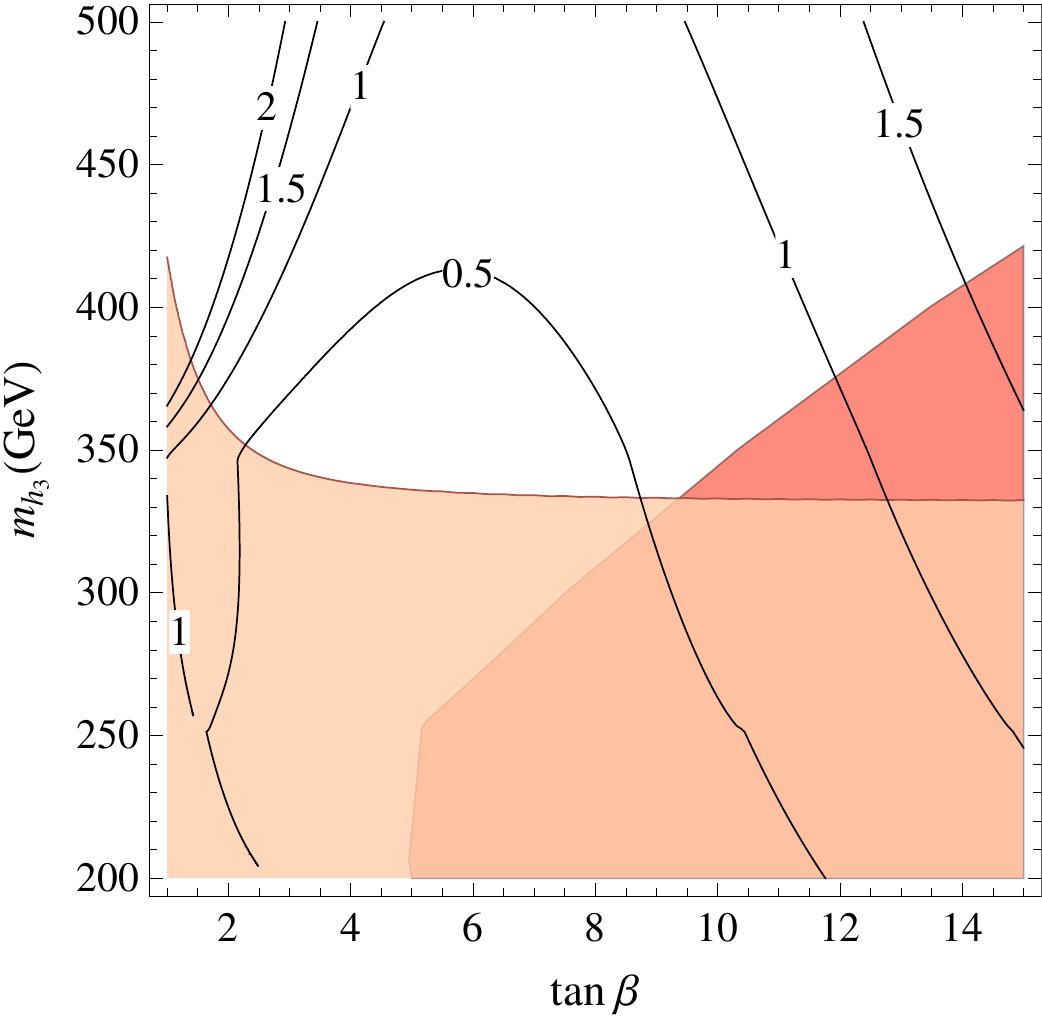}\hfill
\includegraphics[width=.48\textwidth]{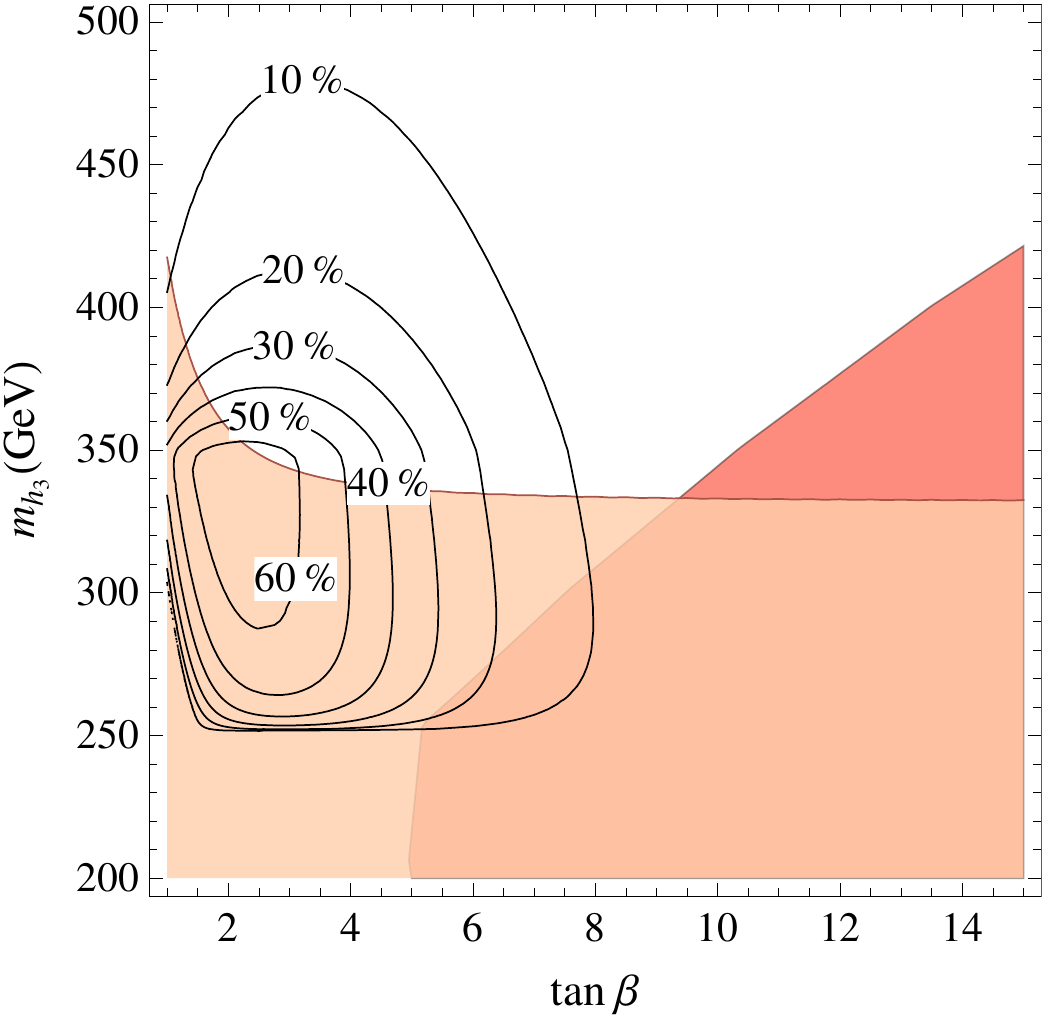}
\caption{\label{fig:MSSM-BRs}\small MSSM. Left: Isolines of the total width $\Gamma_{h_3}$ (GeV). Right: isolines of BR$(h_3\to h h)$. The light coloured region is excluded at 95\%C.L., the red region is excluded by CMS direct searches.}
\end{center}
\end{figure}

\begin{figure}
\begin{center}
\includegraphics[width=.48\textwidth]{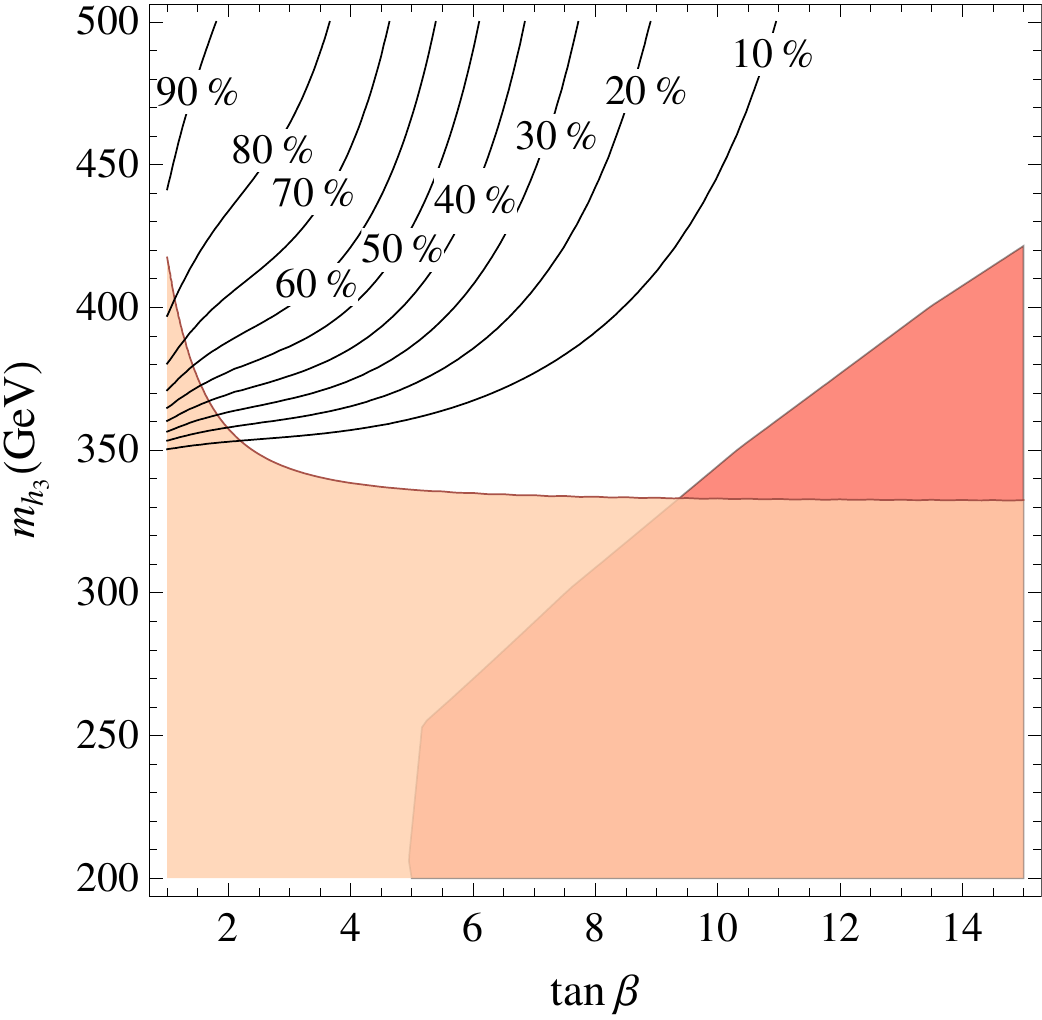}\hfill
\includegraphics[width=.48\textwidth]{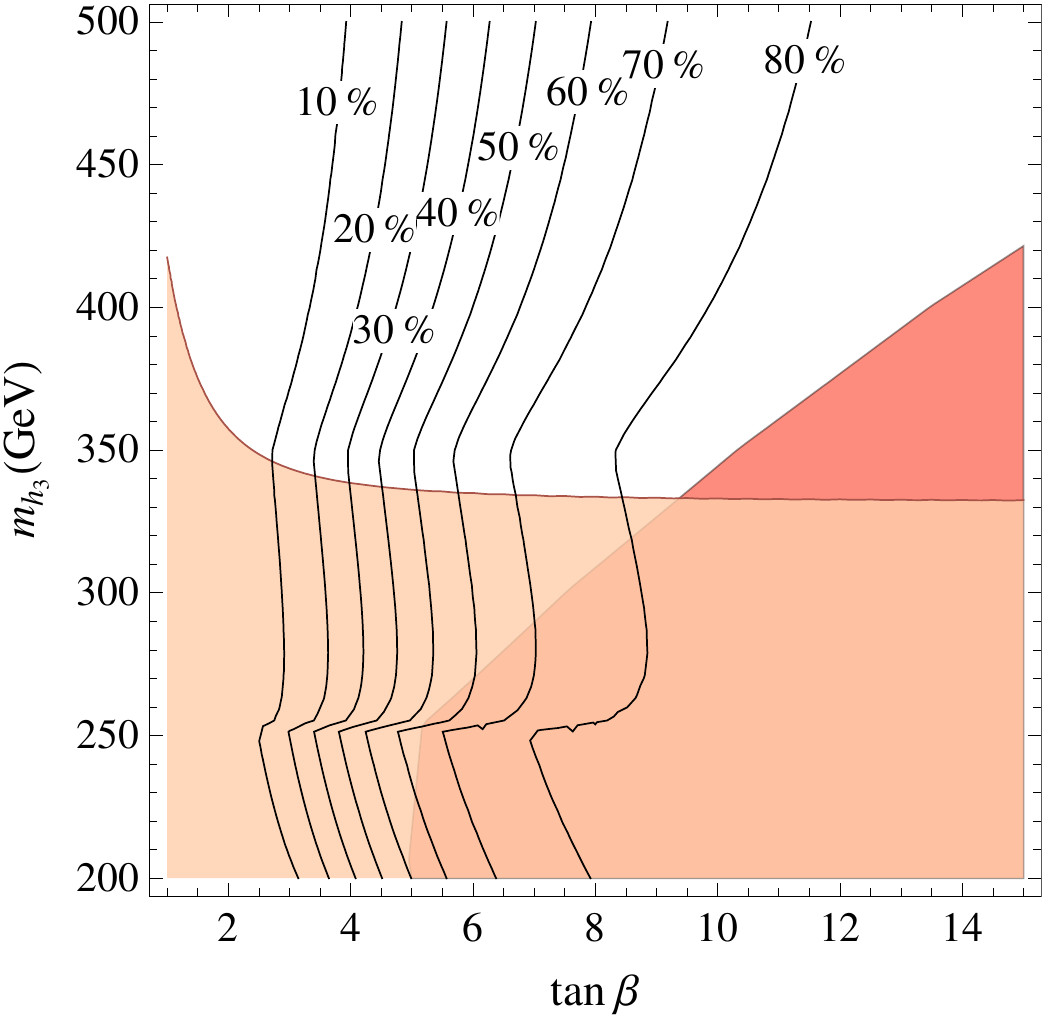}
\caption{\label{fig:MSSM-BRf}\small MSSM. Left: isolines of BR$(h_3\to t\bar t)$. Right: isolines of BR$(h_3\to b \bar{b})$. The light coloured region is excluded at 95$\%$C.L., the red region is excluded by CMS direct searches.}
\end{center}
\end{figure}

Finally, in analogy with Figures~\ref{fig:mSdecoupled-Xsec}-\ref{fig:mSdecoupled-BRf}, we show in Figures~\ref{fig:MSSM-Xsec}-\ref{fig:MSSM-BRf} the gluon-fusion production cross sections and the widths of $h_3$ for the MSSM case. For the production cross sections we have adopted the same procedure of the Singlet decoupled case, and performed a further check of our results with the ones recently presented in \cite{Arbey:2013jla} and \cite{Djouadi:2013vqa}, finding a very good agreement.


In the MSSM $m_A^2$ in (\ref{mA_mh}) at $\lambda=0$ is the squared mass of the neutral pseudoscalar $A$,  unlike the case of the general NMSSM, where $m_A^2$ in (\ref{mA_mh}) is only an auxiliary quantity.
In the same  $(\tan{\beta}, m_{h_3})$ plane $\sigma (gg\rightarrow A)$ is therefore also determined, which allows to delimit the currently excluded region by the direct searches for $A, h_3\rightarrow \tau^+ \tau^-$. Such a region is known to be significant, especially for growing $\tan{\beta}$. In Figures~\ref{fig:MSSM-1}-\ref{fig:MSSM-BRf} we draw the region excluded by such search, as inferred from \cite{CMS-PAS-HIG-12-050}.

\section{The NMSSM at $\lambda\gtrsim 1$ and gauge coupling unification}
\label{sec:Lambda_model}
As said in Section~\ref{sec:NMSSM_natural}, a $\lambda \gtrsim 0.7$ would run to higher values and become non perturbative before the unification scale\cite{Espinosa:1991gr}, thus ruining one of the phenomenological reasons that motivated supersymmetry. For values of $\lambda \gtrsim 2$ at the weak scale, this coupling would become non perturbative before $\sim 10$ TeV, calling for a strong sector at that scale that would generically affect EWPT, thus providing a phenomenological upper bound on $\lambda$.
To stay in the window $0.7 \lesssim \lambda \lesssim 2$ one can then either give up on the supersymmetric unification of gauge couplings, or find ways to change the above running. To pursue this latter option, the most immediate idea can be that of adding vector-like matter in complete $SU(5)$ multiplets, to slow down the running of $\lambda$ by increasing the gauge couplings at high energies. However, in this way one cannot go above weak scale values of $\lambda \simeq 0.8$, otherwise it is not possible to reproduce the measured value of the gauge couplings \cite{Masip:1998jc,Barbieri:2007tu}.

We propose here a solution that makes also use of two vector-like five-plets as follows. For ease of exposition let us call them  $F_{u,d} + \bar{F}_{u,d}$, where $F_u$ is a $5$ and $F_d$ a $\bar{5}$, thus containing one $SU(2)$-doublet each, $h_u$ and $h_d$, with the same quantum numbers of the standard $H_u, H_d$ used so far. Correspondingly $\bar{F}_{u,d}$ contain two doublets that we call $\bar{h}_{u,d}$. Needless to say all these are superfields. Let us further assume that the superpotential is such that:
\begin{itemize}
\item The five-plets interact with a singlet $S$ and pick up $SU(5)$-invariant masses consistently with a Peccei-Quinn symmetry;
\item The standard doublets $H_u, H_d$ mix by mass terms with $h_u$ and $h_d$, still maintaining the Peccei-Quinn symmetry, and do not interact directly with $S$.
\end{itemize}
The corresponding superpotential is 
\begin{equation}
f = \lambda_S S F_u F_d + M_u F_u \bar{F}_u + M_d F_d \bar{F}_d + m_u H_u \bar{h}_u + m_d H_d \bar{h}_d + \lambda_t H_u Q t,
\label{f}
\end{equation}
where we have also made explicit the  Yukawa coupling of the top to $H_u$.
Below these masses, all taken to be comparable, this $f$ term leaves three massless supermultiplets: 
\begin{equation}
S,~~~ \hat{H}_u = c_u H_u + s_u h_u,~~~\hat{H}_d = c_d H_d + s_d h_d,
\end{equation}
which interact through the superpotential
\begin{equation}
\hat{f}= \hat{\lambda} S \hat{H}_u \hat{H}_d + \hat{\lambda}_t \hat{H}_u Q t, ~~~\hat{\lambda} = \lambda_S s_u s_d,~~~~\hat{\lambda}_t=\lambda_t c_u .
\end{equation}
This superpotential, completed by Peccei-Quinn symmetry breaking terms at the Fermi scale, defines the effective NMSSM as discussed so far. To cure the growth of $\hat{\lambda}$ at increasing energies, the masses in (\ref{f}) will have to be crossed while $\hat{\lambda}$ is still semi-perturbative. For $\hat{\lambda} =1\div 1.5$ these masses are above 1000 TeV.

 At greater energies the running of the gauge couplings is affected, compared to the standard supersymmetric case,  by the supermultiplet $H_u$ with a top Yukawa coupling increased by a factor $1/c_u$ and by the  degenerate complete $SU(5)$-multiplets $F_{u,d} + \bar{F}_{u,d}$. To avoid a Landau pole before the GUT scale in the top Yukawa coupling  $c_u$ has to be bigger than about $1/\sqrt{2}$. As to the effect of $F_{u,d} + \bar{F}_{u,d}$, they do not alter the relative one loop running of the gauge couplings but might give rise to an exceeding growth of all of them before $M_{\text{GUT}}$ due to the presence of the coupling $\lambda_S$, which at some point will get strong. To avoid this a change of regime in the $SU(5)$-symmetric sector will have to intervene to keep under control the anomalous dimensions of the $F_{u,d}, \bar{F}_{u,d}$ superfields.

\section{Summary and partial conclusions}
\label{sec:NMSSM_concl}
Given the current experimental informations, the Higgs sector of the NMSSM appears to allow a minimally fine-tuned description of electroweak symmetry breaking, at least in the context of supersymmetric extensions of the SM.
Motivated by this fact we have outlined a possible overall strategy to search for signs of its $CP$-even states, taking into account the impact of the measured signal strengths of $h_{\rm LHC}$.

To have a simple characterization of the properties of the extended Higgs system we have focused on relations between physical parameters, and suggested a relatively simple analytic description of four different situations:
\begin{itemize}
\item Singlet-decoupled, $h_3  < h_{\rm LHC} < h_2 (= S)$;
\item Singlet-decoupled, $h_{\rm LHC} < h_3  < h_2 (= S)$;
\item $H$-decoupled, $h_2  < h_{\rm LHC} < h_3 (= H)$;
\item $H$-decoupled, $h_{\rm LHC} < h_2  < h_3 (= H)$.
\end{itemize}
To make this possible we have not included any radiative effects of superpartners other than the top-stop loop corrections to the quartic Higgs couplings, \eqref{delta-t}. 
While providing at least a useful reference case, we think that this is motivated by the consideration of superpartner masses at their ``naturalness limit''. 
We have not been sticking to a particular NMSSM, which might imply specific constraints on the physical parameters that we consider, but we have assumed to live in the case of negligible CP violation in the Higgs sector. 
In our view the advantages of having an overall coherent analytic picture justify the introduction of these assumptions.

Let us first summarize the impact of the measurements --current and foressen-- of the $h_{\text{LHC}}$ signal strengths on the above picture.
Even though they are close to those expected in the SM, in the NMSSM they still allow for a new heavier state nearby, unlike in the case of the MSSM, where a CP-even scalar heavier than $h_{\mathrm{LHC}}$ and below about 300 GeV is unlikely \cite{DAgnolo:2012mj}. This is true in both the limiting cases that we have considered, as visible in Figures~\ref{fig1} and \ref{fig:mAdecoupled-1}, to be contrasted with Figure~\ref{fig7}.
On the other hand, the same Figures show that the measured signal strengths of $h_{\rm LHC}$ do limit the possible values of $\lambda$, at least for moderate values of $\tan{\beta}$, which is the region mostly motivated by naturalness: $\lambda \approx 1$ is still largely allowed in the $H$-decoupled case, whereas it is borderline in the singlet decoupled situation.
As commented upon in Section~\ref{sec:Lambda_model}, we think that $\lambda \gtrsim 1$ can be compatible with gauge coupling unification.
The case of another CP-even state lighter than $h_{\mathrm{LHC}}$ is allowed in all cases, sometimes in a very small region, with the LEP direct searches playing an important role in the exclusions. Note however that in the $S$-decoupled and MSSM cases this would require a charged Higgs lighter than 100 GeV, which is generically difficult to reconcile with indirect constraints from $B \to X_s \gamma$ \cite{Misiak:2006zs}.
\\A quantitative estimate of the sensitivity of the foreseen measurements at LHC14 with 300 $\text{fb}^{-1}$ makes it likely that the singlet-decoupled case will be thoroughly explored, as evident from Figure~\ref{fig2}, while the singlet-mixing effects could remain hidden.
We also found that, in the MSSM with $(\mu A_t)/\langle m_{\tilde{t}}^2\rangle \lesssim 1$, the absence of deviations in the $h_{\text{LHC}}$ signal strengths would push the mass of the other Higgs bosons up to a TeV. 

Most importantly from the point of view of the direct searches,\footnote{A first attempt at studying heavier Higgs decays in the NMSSM with $\lambda > 1$ was made in \cite{Cavicchia:2007dp}.} these features reflect themselves in the behaviour of the new states, quite different in the two NMSSM cases, especially in their decay properties. 
\begin{itemize}
 \item The state $h_2$ of the $H$-decoupled case, when heavier than about 250 GeV, has a large BR into a pair of $h_{\rm LHC}$, with $VV$ as subdominant decay (Figures~\ref{fig:mAdecoupled-hh}-\ref{fig:mAdecoupled-WW}). With the production cross sections shown in Figure~\ref{fig:mAdecoupled-Xsec} its direct search at LHC8 and LHC14 may be challenging, although perhaps not impossible \cite{Gouzevitch:2013qca}.\\
 When the $h_{\rm LHC} h_{\rm LHC}$ decay channel is not allowed by phase space, and for both $h_2 > h_{\rm LHC}$ and $h_{\rm LHC} < h_2$ cases, the signal strengths of $h_2$ are simply those of a SM Higgs, suppressed by a $s_\gamma^2$ factor. With this in mind it is easy to see that e.g. the LHC searches of a Higgs boson into $ZZ$\cite{ATLAS-CONF-2013-013,CMS-PAS-HIG-13-014} and $WW$\cite{Chatrchyan:2013yoa} are starting to probe the allowed region, at least in a small mass window. On the contrary, those in $\tau \tau$\cite{CMS-PAS-HIG-13-004,ATLAS-CONF-2012-160}, $bb$\cite{CMS-PAS-HIG-13-012,ATLAS-CONF-2013-079} and $\gamma \gamma$\cite{CMShint} have not (yet) reached such a level of sensitivity.
 \item On the other hand, the reduced value of $\lambda$ allowed in the singlet decoupled case makes the $b\bar{b}$ channel, and so the $\tau\bar{\tau}$, most important, below the $t\bar{t}$ threshold (Figures~\ref{fig:mSdecoupled-BRs} and \ref{fig:mSdecoupled-BRf}). This makes the state $h_3$ relatively more similar to the CP-even $H$ state of the MSSM (Figures~\ref{fig:MSSM-BRs} and \ref{fig:MSSM-BRf}), which is being actively searched.
\end{itemize}
From the point of view of indirect searches it is also interesting that, in the $H$-decoupled case, large deviations from the SM value are possible in the triple Higgs coupling $g^3_{h_{\rm LHC}}$, contrary to the $S$-decoupled and MSSM cases.\\
Finally, in case of a positive signal, direct or indirect, it may be important to try to interpret it in a fully mixed scheme, involving all the three $CP$-even states. To this end the analytic relations of the mixing angles to the physical masses given in Eqs.~\eqref{eq:sin:gamma:general}-\eqref{eq:sin:2alpha:general} offer a useful tool, as illustrated in the examples of a $\gamma\gamma$ signal of Fig.~\ref{fig6}.

It will be interesting to follow the progression of the searches of the Higgs system of the NMSSM, directly or indirectly through the more precise measurements of the $h_1$ properties. We believe that the framework outlined here should allow to systematize these searches in a clear way. We also think that they should be pursued actively and independently from the searches of the superpartners.


\part{Composite Higgs models}
\label{part:CHM}

\chapter[Composite Higgs models vs flavour and EWPT]{Composite Higgs models facing flavour and electroweak precision tests}
\label{cha:CHM}

Here we are concerned with the implications of the Higgs boson discovery for the view that tries to explain a natural Fermi scale in terms of the Higgs particle as a composite pseudo-Goldstone boson \cite{Kaplan:1983fs,Georgi:1984af,Contino:2003ve, Agashe:2004rs}. More precisely, we shall concentrate our attention on the compatibility of such interpretation of the newly found particle with constraints from flavour and electroweak precision tests (EWPT). 

The common features emerging from the modelling of the strong dynamics responsible for the existence of the composite pseudo-Goldstone Higgs boson are:\footnote{For a review we refer the reader to \cite{Contino:2010rs}.} 
\begin{enumerate}[i)]
\item a breaking scale of the global symmetry group, $f$, somewhat larger than the EWSB scale $v \approx 246$~GeV;
\item a set of $\rho$-like vector resonances of typical mass $m_\rho = g_\rho f$;
\item a set of spin-$\frac{1}{2}$  resonances, vector-like under the Standard Model (SM) gauge group, of typical mass $m_\psi = Y f$;
\item bilinear mass-mixing terms between the composite and the elementary fermions, ultimately responsible for the masses of the elementary fermions themselves \cite{Kaplan:1991dc}.
\end{enumerate}
These same mass mixings are crucial in explicitly breaking the global symmetry of the strong dynamics, i.e. in triggering EWSB, with a resulting Higgs boson mass
\begin{equation}\label{mh}
m_h = C \frac{\sqrt{N_c}}{\pi} m_t Y,
\end{equation}
where $N_c=3$ is the number of colours, $m_t$ is the top mass and $C$ is a model dependent coefficient of ${O}(1)$, barring  unnatural fine-tunings \cite{Contino:2006qr, Pomarol:2012qf, Redi:2012ha, Matsedonskyi:2012ym, Marzocca:2012zn, Panico:2012uw}. This very equation makes manifest that the measured mass of 125 GeV calls for a semi-perturbative coupling $Y$ of the fermion resonances and, in turn, for their relative lightness, if one wants to insist on a breaking scale $f$ not too distant from $v$ itself.
For a reference value of $f = 500\text{--}700$ GeV, which in PNGB Higgs models is enough to bring all the Higgs signal strengths in agreement with the currently measured values \cite{Giardino:2013bma},  one expects fermion resonances with typical mass not exceeding about 1 TeV, of crucial importance for their direct searches at the LHC. These searches are currently sensitive to masses in the  500--700 GeV range, depending on the charge of the spin-$\frac{1}{2}$ resonance and on the decay channel~\cite{CMS:2012ab,Chatrchyan:2012vu,Chatrchyan:2012af,ATLAS-CONF-2012-130}.
In this Chapter we aim to investigate the compatibility of this feature with flavour and EWPT.

To address this question, we consider a number of  different options for the transformation properties of the spin-$\frac{1}{2}$ resonances under the global symmetries of the strong dynamics, motivated by the need to be consistent with the constraints from the EWPT, as well as different options for the flavour structure/symmetries, motivated by  the many significant flavour bounds.
To make the Chapter readable, after defining the setup for the various cases in Section~\ref{sec:setup}, we analyze in succession the different options for the flavour structures/symmetries: {\it Anarchy} in Section~\ref{sec:anarchy}, $U(3)^3$ in Section~\ref{sec:u3}, $U(2)^3$ in Section~\ref{sec:u2}. Section~\ref{sec:ewpt} describes the constraints from EWPT that apply generally to all flavour models. Orthogonal to the rest of the Chapter, Section~\ref{sec:CHM_leptons} discuss the implementation of a flavour symmetry in the lepton sector, and its consequences. A partial summary and the conclusions are contained in Section~\ref{sec:summary}.

\section{Setup}\label{sec:setup}

To keep the discussion simple and possibly not too model dependent, we follow the {\it partial compositeness} approach of ref.~\cite{Contino:2006nn}. The vector resonances transform in the adjoint representation of a global symmetry respected by the strong sector, which contains the SM gauge group. To protect the $T$ parameter from tree-level contributions, we take this symmetry to be $G_c=SU(3)_c\times SU(2)_L\times SU(2)_R\times U(1)_X$. We assume all vector resonances to have mass $m_\rho$ and coupling $g_\rho$. For the explicit form of their effective Lagrangian we refer to \cite{Contino:2006nn}.

The choice of the fermion representations has important implications for the electroweak precision constraints. We will consider three cases, as customary in the literature.
\begin{enumerate}
\item The elementary $SU(2)_L$ quark doublets, $q_L$, mix with composite vector-like $SU(2)_L$ doublets, $Q$, one per generation. The elementary quark singlets, $u_R$ and $d_R$, couple both to an $SU(2)_R$ doublet $R$. We will call this the {\bf doublet model}.
\item The elementary $SU(2)_L$ quark doublets mix with a  composite $L = (2,2)_{2/3}$ of $SU(2)_L\times SU(2)_R\times U(1)_X$, and the elementary quark singlets couple both to a composite triplet $R=(1,3)_{2/3}$. The model also contains a $(3,1)_{2/3}$ to preserve LR symmetry.  We will call this the {\bf triplet model}.
\item The elementary $SU(2)_L$ quark doublets mix with a $L_U = (2,2)_{2/3}$ and a $L_D = (2,2)_{-1/3}$ of $SU(2)_L\times SU(2)_R\times U(1)_X$, the former giving masses to up-type quarks, the latter to down-type quarks. The elementary up and down quark singlets couple to a $(1,1)_{2/3}$ and a $(1,1)_{-1/3}$ respectively. We will call this the {\bf bidoublet model}.
\end{enumerate}

For concreteness, the part of the Lagrangian involving fermions  reads
\begin{itemize}
\item In  the doublet model
\begin{gather}\label{doubletL}
\mathcal L_s^\text{doublet} =\
-\bar Q^i m_{Q}^i Q^i
-\bar R^i m_{R}^i R^i
+ \left( Y^{ij} \text{tr}[ \bar Q^i_L \mathcal H R_R^j]  + \text{h.c} \right)
\,,
\\
\mathcal L_\text{mix}^\text{doublet} =
m_{Q}^j\lambda_{L}^{ij}\bar q_L^i Q_{R}^j
+
m_{R}^i\lambda_{Ru}^{ij}\bar U_L^i u_{R}^j
+
m_{R}^i\lambda_{Rd}^{ij}\bar D_L^i d_{R}^j
\,.
\end{gather}
where $\mathcal H=(i\sigma_2H^*,H)$ and $R=(U~D)^T$ is an $SU(2)_R$ doublet;

\item  In the triplet model
\begin{gather}
\mathcal L_s^\text{triplet} =\
-\text{tr}[ \bar L^i m_{L}^i L^i ]
-\text{tr}[ \bar R^i m_{R}^i R^i ]
-\text{tr}[ \bar R^{\prime\, i} m_{R}^iR^{\prime\, i}]\notag\\
\qquad\qquad+ Y^{ij} \text{tr}[ \bar L_L^i \mathcal H R_R^j] + Y^{ij}\text{tr}[\mathcal H\,  \bar L_L^i R_R^{\prime\, j}]  + \text{h.c}
\,,
\label{tripletL} \\
\mathcal L_\text{mix}^\text{triplet} =
m_{L}^j\lambda_{L}^{ij}\bar q_L^i Q_{R}^j
+
m_{R}^i\lambda_{Ru}^{ij}\bar U_L^i u_{R}^j
+
m_{R}^i\lambda_{Rd}^{ij}\bar D_L^i d_{R}^j
\,.
\end{gather}
where $Q$ is the $T_{3R}=-\frac{1}{2}$ $SU(2)_L$ doublet contained in $L$ and $U, D$ are  the elements in the triplet $R$ with charge 2/3 and -1/3 respectively;
\item In the bidoublet model
\begin{gather}\label{bidoubletL}
\mathcal L_s^\text{bidoublet} =\
-\text{tr}[ \bar L_U^i m_{Q_u}^i L_U^i ]
-\bar U^i m_{U}^i U^i
+ \left( Y_U^{ij} \text{tr}[ \bar L_U^i \mathcal H ]_L U_R^j  + \text{h.c} \right)
+ (U\to D)
\,,
\\
\mathcal L_\text{mix}^\text{bidoublet} =
m_{Q_u}^j\lambda_{Lu}^{ij}\bar q_L^i Q_{Ru}^j
+
m_{U}^i\lambda_{Ru}^{ij}\bar U_L^i u_{R}^j
+ (U,u\to D,d)
\,,
\end{gather}
where again $Q_u$ and $Q_d$ are the doublets in $L_U$ and $L_D$ which have the same gauge quantum numbers of the SM left-handed quark doublet.
\end{itemize}
Everywhere $i,j$ are flavour indices.
The field content in all three cases is summarized in Table~\ref{tab:fields}.\footnote{Note that we have omitted ``wrong-chirality'' Yukawa couplings like $\tilde Y^{ij}{\rm tr}[\bar Q_R^i\mathcal{H}R_L^j]$ for simplicity. They are not relevant for the tree-level electroweak and flavour constraints and do not add qualitatively new effects to the loop contributions to the $T$ parameter.}

We avoid an explicit discussion of the relation between the above simple effective Lagrangians and more basic models which include the Higgs particle as a pseudo-Goldstone boson. Here it suffices to say that the above Lagrangians are suitable to catch the main phenomenological properties of more fundamental models. For this to be the case, the truly basic assumption is that the lowest elements of towers of resonances, either of spin-$\frac{1}{2}$ or of spin 1, normally occurring in more complete models, are enough to describe the main phenomenological consequences, at least in as much as tree-level effects are considered.
For simplicity we also assume the composite fermions to have all the same mass.
To set the correspondence between the {\it partial compositeness} Lagrangians that we use and models with the Higgs as a pseudo-Goldstone boson, one can take the composite Yukawa couplings $Y^{ij}$ in \eqref{doubletL},\eqref{tripletL} and \eqref{bidoubletL} to be proportional to the parameter $Y$ in \eqref{mh}, and identify the common fermion mass with $m_\psi =Yf$, up to a model dependent factor of ${O}(1)$.

\begin{table}
\renewcommand{\arraystretch}{1.15}
\centering
\begin{tabular}{lccccc}
\hline
model && $SU(3)_c$ & $SU(2)_L$ & $SU(2)_R$ & $U(1)_X$ \\
\hline
\multirow{2}{1.5cm}{doublet}
&$Q$ & $\mathbf{3}$ & $\mathbf{2}$ & $\mathbf{1}$ & $\frac{1}{6}$ \\
&$R$ & $\mathbf{3}$ & $\mathbf{1}$ & $\mathbf{2}$ & $\frac{1}{6}$ \\
\hline
\multirow{3}{1.5cm}{triplet}
&$L$ & $\mathbf{3}$ & $\mathbf{2}$ & $\mathbf{2}$ & $\frac{2}{3}$ \\
&$R$ & $\mathbf{3}$ & $\mathbf{1}$ & $\mathbf{3}$ & $\frac{2}{3}$ \\
&$R'$ & $\mathbf{3}$ & $\mathbf{3}$ & $\mathbf{1}$ & $\frac{2}{3}$ \\
\hline
\multirow{4}{1.5cm}{bidoublet}
&$L_U$ & $\mathbf{3}$ & $\mathbf{2}$ & $\mathbf{2}$ & $\frac{2}{3}$ \\
&$L_D$ & $\mathbf{3}$ & $\mathbf{2}$ & $\mathbf{2}$ & $-\frac{1}{3}$ \\
&$U$ & $\mathbf{3}$ & $\mathbf{1}$ & $\mathbf{1}$ & $\frac{2}{3}$ \\
&$D$ & $\mathbf{3}$ & $\mathbf{1}$ & $\mathbf{1}$ & $-\frac{1}{3}$ \\
\hline
\end{tabular}
\caption{Quantum numbers of the fermionic resonances in the three models considered. All composite fields come in vector-like pairs. The $X$ charge is related to the standard hypercharge as $Y=T_{3R}+X$.}
\label{tab:fields}
\end{table}

\subsection{Flavour structure}\label{sec:flasy}

Quark masses and mixings are generated after electroweak symmetry breaking from the composite-elementary mixing.  The states with vanishing mass at $v=0$ obtain the standard Yukawa couplings, in matrix notation,
\begin{align}\label{SMYuk}
\hat y_u &\approx s_{Lu} \cdot U_{Lu} \cdot Y_U \cdot U_{Ru}^\dagger \cdot s_{Ru}
\end{align}
where
\begin{gather}
\lambda_{Lu}=\text{diag}(\lambda_{Lu1},\lambda_{Lu2},\lambda_{Lu3}) \cdot U_{Lu}\,,
\\
\lambda_{Ru}=U_{Ru}^{\dag}\cdot \text{diag}(\lambda_{Ru1},\lambda_{Ru2},\lambda_{Ru3})\,,
\\
s_{X}^{ii}={\lambda_{X i}}/{\sqrt{1+(\lambda_{X i})^2}},
~ X = L,R~,
\end{gather}
and similarly for $\hat y_d$. Here and in the following, the left-handed mixings are different for $u$ and $d$ quarks, $s_{Lu}\neq s_{Ld}$, only in the bidoublet model.
At the same time, in the $v=0$ limit, the remaining states have mass $m_{\psi}$ or $m_{\psi}/\sqrt{1+(\lambda_X)^2}$, respectively if they mix or do not mix with the elementary fermions.

While the effective Yukawa couplings $\hat y_{u,d}$ must have the known hierarchical form, the Yukawa couplings in the strong sector, $Y_{U,D}$, could be structureless {\em anarchic} matrices (see e.g. \cite{Grossman:1999ra,Huber:2000ie,Gherghetta:2000qt,Agashe:2004cp,Blanke:2008zb,Bauer:2009cf,KerenZur:2012fr,Csaki:2008zd}).
However, to ameliorate flavour problems, one can also impose global flavour symmetries on the strong sector. We discuss three cases in the following.

\subsubsection*{Anarchy}

In the anarchic model, the $Y_{U,D}$ are anarchic matrices, with all entries of similar order, and the Yukawa hierarchies are generated by hierarchical mixings $\lambda$.
From a low energy effective theory point of view, the
requirement to reproduce the observed quark masses and mixings fixes the relative size of the mixing parameters up to -- a priori unknown -- functions of the elements in $Y_{U,D}$. We follow the common approach to replace functions of Yukawa couplings by appropriate powers of ``average'' Yukawas $Y_{U*,D*}$, keeping in mind that this introduces $O(1)$ uncertainties in all observables. In this convention, assuming $\lambda_{X3}\gg\lambda_{X2}\gg\lambda_{X1}$, the quark yukawas are given by
\begin{align}
y_{u}&=Y_{U*} s_{Lu 1}s_{Ru 1}~,&
y_{c}&=Y_{U*} s_{Lu 2}s_{Ru 2}~,&
y_{t}&=Y_{U*} s_{Lu 3}s_{Ru 3}~.
\label{eq:anarchy-yukawas}
\end{align}
and similarly for the $Q=-1/3$ quarks. In the doublet and triplet models, the entries of the CKM matrix are approximately given by
\begin{align}
V_{ij} &\sim \frac{s_{L i}}{s_{L j}} ~,
\label{eq:anarchy-ckm-doublet}
\end{align}
where $i<j$. Using Eqs.~(\ref{eq:anarchy-yukawas}) and (\ref{eq:anarchy-ckm-doublet}), one can trade all but one of the $s_{L,R}$ for known quark masses and mixings. We choose the free parameter as
\begin{equation}
x_t \equiv s_{L3}/s_{Ru3}.
\label{eq:x-doublet}
\end{equation}
In the bidoublet model, instead of (\ref{eq:anarchy-ckm-doublet}) one has
 in general two different contributions to $V_{ij}$%
\begin{align}
V_{ij} &\sim
\frac{s_{Ld i}}{s_{Ld j}}\pm \frac{s_{Lu i}}{s_{Lu j}} ~.
\label{eq:anarchy-ckm-bidoublet}
\end{align}
Given the values of all quark masses and mixings, the hierarchy $\lambda_{X3}\gg\lambda_{X2}\gg\lambda_{X1}$ is only compatible with $s_{Lu i}/s_{Lu j}$ being at most comparable to $s_{Ld i}/s_{Ld j}$. In view of this, the two important parameters are
\begin{align}
x_t &\equiv s_{Lt}/s_{Rt} ~,
&
z &\equiv s_{Lt}/s_{Lb} ~.
\label{eq:x-bidoublet}
\end{align}
The requirement to reproduce the large top quark yukawa ($m_t = \frac{y_t}{\sqrt{2}}v$)
\begin{equation}
y_t = s_{Lt} Y_{U*} s_{Rt},
\end{equation}
restricts $x_t$ to a limited range around one\footnote{\text{In our numerical analysis, we will take $y_t=0.78$, which is the running $\overline{\text{MS}}$ coupling at 3~TeV.}},
\begin{equation}
\frac{y_t}{Y_{U*}} < x_t < \frac{Y_{U*}}{y_t} ~,
\end{equation}
while we take $z$ throughout to be greater than or equal to 1.

From now on we identify $Y_{U*}$ and $Y_{D*}$ with the parameter $Y$ of \eqref{mh}.

\subsubsection*{\boldmath $U(3)^3$}

In the $U(3)^3$ models \cite{Cacciapaglia:2007fw,Barbieri:2008zt,Redi:2011zi} one tries to ameliorate the flavour problem of the anarchic model by imposing a global flavour symmetry, at the price of giving up the potential explanation of the generation of flavour hierarchies. Concretely,
one assumes the strong sector to be invariant under the diagonal group $U(3)_{Q+U+D}$ or $U(3)_{Q^u+U}\times U(3)_{Q^d+D}$. The composite-elementary mixings are the only sources of breaking of the flavour symmetry of the composite sector and of the $U(3)_{q}\times U(3)_{u}\times U(3)_{d}$ flavour symmetry of the elementary sector. We consider two choices.
\begin{enumerate}
\item In {\it left-compositeness}, to be called {\boldmath\UtreLC} in short, the left mixings are proportional to the identity, thus linking $q$ to $Q~(Q^u, Q^d)$ into $U(3)_{Q+U+D+q}$ (or $U(3)_{Q^u+Q^d+U+D+q}$),
and the right mixings $\lambda_{Ru}$, $\lambda_{Rd}$ are the only source of $U(3)^3$ breaking.

\item In {\it right-compositeness}, to be called {\boldmath\UtreRC} in short, the right mixings link $u$ to $U$ and $d$ to $D$ into $U(3)_{Q^u+U+u}\times U(3)_{Q^d+D+d}$, while
the left mixings $\lambda_{Lu}$, $\lambda_{Ld}$ are the only source of $U(3)^3$ breaking.

\end{enumerate}
All the composite-elementary mixings are then fixed by the known quark masses and CKM angles, up to the parameters $x_t$ (and, in the bidoublet model, $z$), which are defined as in (\ref{eq:x-doublet},\,\ref{eq:x-bidoublet}).
Compared to the anarchic case, one now expects the presence of resonances related to the global symmetry $U(3)_{Q+U+D}$ or $U(3)_{Q^u+U}\times U(3)_{Q^d+D}$, which in the following will be called flavour gauge bosons\footnote{We will only allow flavour gauge bosons related to the $SU(3)$ subgroups of the $U(3)$ factors.} and assumed to have the same masses $m_\rho$ and $g_\rho$ as the gauge resonances. Note that left-compositeness can be meaningfully defined for any of the three cases for the fermion representations, whereas right-compositeness allows to describe flavour violations only in the bidoublet model.

To fix the conventions for a later analysis, in $U(3)^3_\text{RC}$  the effective Yukawa couplings have the form 
\begin{equation}
\bar{q}_L \hat{s}_{Lu} Y_U s_{Ru} u_R
\label{Yuk_RC}
\end{equation}
 (and similarly for the down quarks) where $\hat{s}_{Lu}$ is a generic $3\times 3$ mixing matrix and $Y_U$, $s_{Ru}$ are both proportional to the unit matrix. In $U(3)^3_\text{LC}$ the role of the mixings is reversed and the Yukawa couplings take the form  
\begin{equation}
\bar{q}_L {s}_{Lu} Y_U \hat{s}_{Ru} u_R.
\label{Yuk_LC}
\end{equation}

\subsubsection*{\boldmath $U(2)^3$}

In $U(2)^3$ models one considers a $U(2)_q\times U(2)_u\times U(2)_d$ symmetry, under which the first two generations of quarks transform as doublets and the third generation as singlets, broken in specific directions dictated by minimality \cite{Barbieri:2011ci,Barbieri:2012uh}. Compared to \Utre, one has a larger number of free parameters, but can break the flavour symmetry {\em weakly}, since the large top yukawa is invariant under \Udue.
Analogously to the \Utre\ case:

\begin{enumerate}
\item In left-compositeness, to be called {\boldmath\UdueLC}, the left mixings are diagonal with the first two entries equal to each other and the only sources of $U(2)^3$ breaking reside in the right-handed mixings.
\item In right-compositenss, to be called {\boldmath\UdueRC}, the right mixings are diagonal with the first two entries equal to each other and the only sources of $U(2)^3$ breaking reside in the left-handed mixings.
\end{enumerate}
Again we expect the presence of flavour gauge bosons associated with the global symmetries of the strong sector. As before right-compositeness can be meaningfully defined only in the bidoublet model.

Let us now be more specific in the definition of the framework. The strong sector can be taken invariant under a $U(2)_{Q+U+D}$ flavour symmetry acting on the first two generations of composite quarks. In right-compositeness -- meaningful only in the bidoublet model -- in order to generate the CKM matrix one has to consider a larger $U(2)_{Q^u+U}\times U(2)_{Q^d+D}$ symmetry.
Let us define
\begin{align}
Q ^{u} &= \begin{pmatrix}\boldsymbol{Q^{u}}\\ Q_3^{u}\end{pmatrix}, & U &= \begin{pmatrix}\boldsymbol{U}\\T \end{pmatrix}, & q_L &= \begin{pmatrix}\ELqL\\ \EHqL\end{pmatrix}, & u_R &= \begin{pmatrix}\ELuR\\ \EHuR\end{pmatrix},
\end{align}
where the first two generation doublets are written in boldface, and the same for down-type quarks. The mixing Lagrangians in the cases of {\it left-compositeness} and {\it right-compositeness} are respectively\footnote{We write the Lagrangians for the bidoublet model. The doublet and triplet cases are analogous, with $Q^u$ and $Q^d$ replaced by a single $Q$.}
\begin{align}
\mathcal{L}_\text{mix}^{U(2)^3_\text{LC}} &=
m_{U3}\lambda_{Lu3}\, \EHqLbar \CHquR +
m_{U2}\lambda_{Lu2} \,\ELqLbar\CLquR +
m_{U3}\lambda_{Ru3}\, \CHuLbar \EHuR
\notag\\
&+m_{U2}\,d_u\, (\CLuLbar\V)\EHuR +
m_{U2}\,\CLuLbar \Delta_u \ELuR +
\text{h.c.}
+ (u,U,t,T\to d,D,b,B)
\label{mixing2L}
\end{align}
and
\begin{align}
\mathcal{L}_\text{mix}^{U(2)^3_\text{RC}} &=
m_{U3}\lambda_{Ru3}\, \CHuLbar\EHuR +
m_{U2}\lambda_{Ru2} \, \CLuLbar\ELuR +
m_{U3}\lambda_{L(u)3}\, \EHqLbar\CHquR
\notag\\
&+m_{U3}\,d_u\, (\ELqLbar\V)\CHquR +
m_{U2}\,\ELqLbar \Delta_u \CLquR +
\text{h.c.}
+ (u,U,t,T\to d,D,b,B).
\label{mixing2R}
\end{align}
The mixings in the first line of \eqref{mixing2L} and \eqref{mixing2R} break the symmetry of the strong sector down to $U(2)_q\times U(2)_u\times U(2)_d$. This symmetry is in turn broken minimally by the spurions of \eqref{eq:MU2_spurions}, where we have renamed $\Delta Y_{u,d}$ to $\Delta_{u,d}$ to avoid confusion with the strong Yukawa couplings.
Using $U(2)^3$ transformations of the quarks they can be put in the simple form
\begin{align}
\V &= \begin{pmatrix} 0 \\ \epsilon_L \end{pmatrix},&
\Delta_u &=
\begin{pmatrix}
c_u & s_u e^{i\alpha_u} \\
-s_u e^{-i\alpha_u} & c_u 
\end{pmatrix}
\begin{pmatrix}
\lambda_{Xu1} & 0 \\
0 & \lambda_{Xu2}
\end{pmatrix}, & (u\leftrightarrow d),
\end{align}
where $X=R,L$ in left- and right-compositeness, respectively.

The SM Yukawa couplings \eqref{SMYuk} can be written in terms of the spurions as
\begin{align}\label{SMYukLC}
\hat y_u &= \begin{pmatrix}a_u\, \Delta_u & b_t e^{i\phi_t}\V\\ 0 & y_t\end{pmatrix}, &
\hat y_d &= \begin{pmatrix}a_d\, \Delta_d & b_b e^{i\phi_b}\V\\ 0 & y_b\end{pmatrix},
\end{align}
where
\begin{align}
y_t &= Y_{U3}s_{Lu3}s_{Ru3},
\\
a_u &= Y_{U2}s_{Lu2},& b_t &= Y_{U2}s_{Lu2}\,d_u, & &\text{ in left-compositeness},\\ a_u &= Y_{U2}s_{Ru2},& b_t &= Y_{U3}s_{Ru3}\,d_u,& &\text{ in right-compositeness},
\end{align}
$s_{Xi} = \lambda_{Xi}/\sqrt{1 + (\lambda_{Xi})^2}$, and similarly for $a_d$, $b_b$ and $y_b$.
Here and in the following we consider all the parameters real, factoring out the phases everywhere as in \eqref{SMYukLC}. The $\hat y_{u,d}$ are diagonalized to a sufficient level of approximation by pure unitary transformations of the left-handed quarks \cite{Barbieri:2012uh}
\begin{align}\label{UL}
U_u &\simeq \begin{pmatrix}c_{u} & s_u e^{i\alpha_u} & -s_u s_t e^{i(\alpha_u + \phi_t)}\\
-s_u e^{-i\alpha_u} & c_u & -c_u s_t e^{i\phi_t}\\
0 & s_t e^{-i\phi_t} & 1
\end{pmatrix}, &
U_d &\simeq \begin{pmatrix}c_{d} & s_d e^{i\alpha_d} & -s_d s_b e^{i(\alpha_d + \phi_b)}\\
-s_d e^{-i\alpha_d} & c_d & -c_d s_b e^{i\phi_b}\\
0 & s_b e^{-i\phi_b} & 1
\end{pmatrix},\end{align}
where
\begin{align}
s_t &= Y_{U2}s_{Lu2}\frac{d_u\epsilon_L}{y_t}, & s_b &= Y_{D2}s_{Ld2}\frac{d_d\epsilon_L}{y_b},& &\text{in left-compositeness},
\label{eq:stbLC}
\\
s_t &= Y_{U3}s_{Ru3}\frac{d_u\epsilon_L}{y_t}, & s_b &= Y_{D3}s_{Rd3}\frac{d_d\epsilon_L}{y_b},& &\text{in right-compositeness}.
\end{align}

The CKM matrix is $V = U_uU_d^{\dag}$ and, after a suitable redefinition of quark phases, takes the form
\begin{equation}
V  \simeq
\left(\begin{array}{ccc}
 1- \lambda^2/2 &  \lambda & s_u s e^{-i \delta}  \\
-\lambda & 1- \lambda^2/2   & c_u s  \\
-s_d s \,e^{i (\phi+\delta)} & -s c_d & 1 \\
\end{array}\right),
\label{eq:CKMstand}
\end{equation}
where
\begin{align}
s_uc_d - c_us_d e^{-i\phi} &\equiv \lambda e^{i\delta}, & s_b e^{i\phi_b} - s_t e^{i\phi_t}\equiv s e^{i\chi}.
\end{align}

\section{General electroweak precision constraints}\label{sec:ewpt}

In this section we discuss electroweak precision constraints that hold independently of the flavour structure. Among the models considered, only \UtreLC\ is subject to additional electroweak constraints, to be discussed in Section~\ref{sec:U3EW}.

\subsection{Oblique corrections}

As well known, the $S$ parameter receives a tree-level contribution, which for degenerate composite vectors reads \cite{Contino:2006nn}
\begin{equation}
S = \frac{8 \pi v^2}{m_\rho^2} ~,
\label{S-corr}
\end{equation}
independently of the choice of fermion representations. It is also well known that $S$ and $T$ both get at one loop model-independent ``infrared-log'' contributions \cite{Barbieri:2007bh}
\begin{equation}
\hat{S} =  \left(\frac{v}{f}\right)^2 \frac{g^2}{96\pi^2} \log{\frac{m_\rho}{m_h}}~,
\qquad
 \hat{T} = -   \left(\frac{v}{f}\right)^2 \frac{3 g^2 t^2_w}{32\pi^2} \log{\frac{m_\rho}{m_h}}~.
 \label{inf-logs}
\end{equation}
where $\hat{S} = \alpha_\text{em}/(4 s^2_w)S$ and $\hat{T}=\alpha_\text{em}T$.

Experimentally, a recent global electroweak fit after the discovery of the Higgs boson \cite{Baak:2012kk} finds $S - S_{\text{SM}}=0.03\pm0.10$ and $T- T_{\text{SM}}=0.05\pm0.12$. Requiring $2\sigma$ consistency with these results of the tree level correction to $S$, Eq.~(\ref{S-corr}), which largely exceeds the infrared logarithmic contribution of (\ref{inf-logs}) and has the same sign, gives the bound
\begin{equation}
m_\rho>2.6\,\text{TeV} \,.
\label{eq:boundmrho}
\end{equation}

The one loop correction to the $T$ parameter instead strongly depends on the choice of the fermion representations. 
We present here simplified formulae valid in the three models for a common fermion resonance mass $m_\psi$ and developed to first nonvanishing order in $\lambda_{Lt}, \lambda_{Rt}$, as such only valid for small $s_{Lt}, s_{Rt}$. An explicit derivation of such formulae was not presented explicitely in \cite{Barbieri:2012tu}, the interested reader finds it in Appendix \ref{app:Tparameter}.

In the {doublet model} the leading contribution to $\hat{T}$, proportional to $\lambda_{Rt}^4$, reads
\begin{equation}
\hat{T} = \frac{71}{140} \frac{N_c}{16 \pi^2} \frac{m_t^2}{m_\psi^2} \frac{Y^2}{x_t^2}\,.
\label{eq:T-doublet}
\end{equation}

In the {bidoublet model} one obtains from a leading $\lambda_{Lt}^4$ term 
\begin{equation}
\hat{T} = - \frac{107}{420}\frac{N_c}{16 \pi^2} \frac{m_t^2}{m_\psi^2} x_t^2 Y_U^2\,.
\label{eq:T-bidoublet}
\end{equation}

In the {triplet model} the leading contributions are
\begin{equation}
\hat{T} = \Big(\log\frac{\Lambda^2}{m_\psi^2} - \frac{1}{2}\Big)\frac{N_c}{16 \pi^2} \frac{m_t^2}{m_\psi^2} \frac{Y^3}{y_t x_t}\,, \quad \text{and} \quad \hat{T} = \frac{197}{84}\frac{N_c}{16 \pi^2} \frac{m_t^2}{m_\psi^2} x_t^2 Y^2\,,
\label{eq:T-triplet}
\end{equation}
where the first comes from $\lambda_{Rt}^2$ and the second from $\lambda_{Lt}^4.$ Note the logarithmically divergent contribution to the $\lambda_{Rt}^2$ term that is related to the explicit breaking of the $SU(2)_R$ symmetry in the elementary-composite fermion mixing and would have to be cured in a more complete model.

Imposing the experimental bound at $2\sigma$, Eqs.~(\ref{eq:T-doublet}, \ref{eq:T-bidoublet}, \ref{eq:T-triplet}) give rise to the bounds on the first line in Table~\ref{tab:bounds-ew} (where we set $\log{(\Lambda/m_\psi)} = 1$). Here however there are two caveats. First, as mentioned, Eqs.~(\ref{eq:T-doublet}, \ref{eq:T-bidoublet}, \ref{eq:T-triplet})  are only valid for small mixing angles. Furthermore, for moderate values of $f$, a cancellation could take place between the fermionic contributions and the infrared logs of the bosonic contribution to $T$.
As we shall see, the bounds  from $S$ and $T$ are anyhow not the strongest ones that we will encounter: they are compatible with  $m_\psi\lesssim 1$ TeV for $Y = 1$ to $2$ and $g_\rho = 3$ to $5$. Note that here and in the following $m_{\psi}$ represents the mass of the composite fermions that mix with the elementary ones, whereas, as already noticed, the ``custodians'' have mass $m_{\psi}/\sqrt{1+(\lambda_X)^2}$.

\begin{table}[tbp]
\renewcommand{\arraystretch}{1.3}
\centering
\begin{tabular}{cccc}
\hline
Observable & \multicolumn{3}{c}{Bounds on $m_{\psi}$ [TeV]} \\
& doublet  & triplet & bidoublet \\
\hline
$T$ & $0.28 ~Y/x_t$ &$0.51 ~\sqrt{Y^3/x_t}$, \; $0.60  ~x_t Y$  & $0.25 ~x_tY_U$\\
$R_b$ ($g_{Zbb}^L$) & $5.6  ~\sqrt{x_tY}$ & & $6.5  ~Y_D \sqrt{x_t/Y_{U}}/z$ \\
$B\to X_s\gamma$ ($g_{Wtb}^R$) & $0.44  ~\sqrt{Y/x_t}$  & $0.44  ~\sqrt{Y/x_t}$ & 0.61\\
\hline
\end{tabular}
\caption{Lower bounds on the fermion resonance mass $m_\psi=Y f$ in TeV from electroweak precision observables. A blank space means no significant bound.}
\label{tab:bounds-ew}
\end{table}

\subsection{Modified $Z$ couplings}\label{sec:Zbb}

In all three models for the electroweak structure, fields with different $SU(2)_L$ quantum numbers mix after electroweak symmetry breaking, leading to modifications in $Z$ couplings which have been precisely measured at LEP. Independently of the flavour structure, an important constraint comes from the $Z$ partial width into $b$ quarks, which deviates by $2.5\sigma$ from its best-fit SM value \cite{Baak:2012kk}
\begin{align}
R_b^\text{exp} &= 0.21629(66)~,
&
R_b^\text{SM} &= 0.21474(3)~.
\label{eq:Rbexp}
\end{align}
Writing the left- and right-handed $Z$ couplings as
\begin{equation}
\frac{g}{c_w}\bar b \gamma^\mu \left[
(-\tfrac{1}{2}+\tfrac{1}{3}s_w^2+\delta g^L_{Zbb}) P_L
+(\tfrac{1}{3}s^2_w+\delta g^R_{Zbb})P_R
\right]bZ_\mu
\,,
\end{equation}
one gets
\begin{align}
\delta g_{Zbb}^L &=
\frac{v^2Y_{D}^2}{2m_{D}^2}\frac{xy_t}{Y_U} \,a+
\frac{g_\rho^2v^2}{4m_\rho^2}\frac{xy_t}{Y_U}\,b \,,
&
\delta g_{Zbb}^R &=
\frac{v^2Y_{D}^2}{2m_{D}^2}\frac{y_b^2 Y_U}{x_t y_t Y_D^2} \,c+
\frac{g_\rho^2v^2}{4m_\rho^2}\frac{y_b^2 Y_U}{x_t y_t Y_D^2}\,d \,,
\label{eq:Zbb1}
\end{align}
with the coefficients
\begin{center}
\begin{tabular}{c|ccc}
& doublet & triplet & bidoublet \\
\hline
$a$ & $1/2$   & $0$ & $1/(2 z^2)$\\
$b$ & $1/2$ & $0$ & $1/z^2$
\end{tabular}
\qquad
\begin{tabular}{c|ccc}
& doublet & triplet & bidoublet \\
\hline
$c$ & $-1/2$   & $-1/2$ & $0$\\
$d$ & $-1/2$ & $-1$ & $0$
\end{tabular}
\end{center}
The vanishing of some entries in (\ref{eq:Zbb1}) can be simply understood by the symmetry considerations of ref.~\cite{Agashe:2006at}. 
As manifest from their explicit expressions the contributions proportional to $a$ and $c$ come from mixings between elementary and composite fermions with different $SU(2)\times U(1)$ properties, whereas the contributions proportional to $b$ and $d$ come from $\rho$-$Z$ mixing. Taking $Y_U=Y_D=Y$, $m_D = Yf$ and $m_\rho = g_\rho f$, all these contributions scale however in the same way as $1/(f^2 Y)$.

It is important to note that $\delta g^L_{Zbb}$ is always positive or 0, while $\delta g^R_{Zbb}$ is always negative or 0, while the sign of the SM couplings is opposite. As a consequence, in all 3 models considered, the tension in Eq.~(\ref{eq:Rbexp}) is {\em always increased}. Allowing the discrepancy to be at most $3\sigma$, we obtain the numerical bounds in the second row of Table~\ref{tab:bounds-ew}. The bound on $m_\psi$ in the doublet model is highly significant since $x_t Y > 1$, whereas it is irrelevant in the triplet model and can be kept under control in the bidoublet model for large enough $z$ (but see below). In the triplet model, there is a bound from the modification of the right-handed coupling, which is however insignificant.

\subsection{Right-handed $W$ couplings}\label{sec:Wtb}

Analogously to the modified $Z$ couplings, also the $W$ couplings are modified after EWSB. Most importantly, a right-handed coupling of the $W$ to quarks is generated. The most relevant experimental constraint on such coupling is the branching ratio of $B\to X_s\gamma$, because a right-handed $Wtb$ coupling lifts the helicity suppression present in this loop-induced decay in the SM \cite{Vignaroli:2012si}. Writing this coupling as
\begin{equation}
\frac{g}{\sqrt{2}}\delta g^R_{Wtb} 
(\bar t \gamma^\mu
P_R
b)W_\mu^+
\,,
\end{equation}
one gets
\begin{align}
\delta g_{Wtb}^R &=
\frac{v^2Y_UY_D}{2m_{Q}m_{U}}\frac{y_b}{x_t Y_U} \,a+
\frac{g_\rho^2v^2}{4m_\rho^2}\frac{y_b}{x_t Y_U}\,b \,,
\label{eq:Wtb1}
\end{align}
with the coefficients
\begin{center}
\begin{tabular}{c|ccc}
& doublet & triplet & bidoublet \\
\hline
$a$ & $1$ & $1$ & $-2x_t y_t/Y$\\
$b$ & $1$ & $1$ & $0$
\end{tabular}
\end{center}
The coefficients in the bidoublet model vanish at quadratic order in the elementary-composite mixings as a consequence of a discrete symmetry \cite{Agashe:2006at}. The nonzero value for $a$ in the table is due to the violation of that symmetry at quartic order \cite{Vignaroli:2012si}.
The contribution to the Wilson coefficient $C_{7,8}$, defined as in \cite{Altmannshofer:2012az}, reads
\begin{equation}
C_{7,8} = \frac{m_t}{m_b} \frac{\delta g^R_{Wtb}}{V_{tb}} A_{7,8}(m_t^2/m_W^2)
\end{equation}
where $A_7(m_t^2/m_W^2)\approx -0.80$ and $A_8(m_t^2/m_W^2)\approx -0.36$.

Since the $B\to X_s\gamma$ decay receives also UV contributions involving composite dynamics, we impose the conservative bound that the SM plus the IR contributions above do not exceed the experimental branching ratio by more than $3\sigma$. In this way we find the bound in the last row of Table~\ref{tab:bounds-ew}.

\section{Constraints on the anarchic model}\label{sec:anarchy}

We now discuss constraints that are specific to the anarchic model, as defined above, and hold in addition to the bounds described in the previous section.

\subsection{Tree-level $\Delta F=2$ FCNCs}

In the anarchic model exchanges of gauge resonances give rise to $\Delta F=2$ operators at tree level. Up to corrections of order $v^2/f^2$, the Wilson coefficients of the operators
\begin{align}
Q_V^{dLL} &= (\bar d^i_L\gamma^\mu d^j_L)(\bar d^i_L\gamma^\mu d^j_L) \,,
&
Q_V^{dRR} &= (\bar d^i_R\gamma^\mu d^j_R)(\bar d^i_R\gamma^\mu d^j_R) \,,
\\
Q_V^{dLR} &= (\bar d^i_L\gamma^\mu d^j_L)(\bar d^i_R\gamma^\mu d^j_R) \,,
&
Q_S^{dLR} &= (\bar d^i_R d^j_L)(\bar d^i_L d^j_R) \,,
\end{align}
can be written as
\begin{align}
C_D^{dAB} &= \frac{g_\rho^2}{m_\rho^2}
g_{Ad}^{ij}g_{Bd}^{ij}
c_D^{dAB},& A,B &= L,R,\quad D = V,S,
\label{eq:DF2}
\end{align}
and with the obvious replacements for up-type quarks, relevant for $D$-$\bar D$ mixing.

The couplings $g_{qA}^{ij}$ with $i\neq j$ contain two powers of elementary-composite mixings. In the doublet and triplet models, one can use Eqs.~(\ref{eq:anarchy-yukawas})--(\ref{eq:x-doublet}) to write them as ($\xi_{ij} = V_{tj}V_{ti}^*$)
\begin{align}
g_L^{ij}&\sim s_{Ldi}s_{Ldj}\sim \xi_{ij} \frac{x_t y_t}{Y} \,,
\\
g_{Ru}^{ij}& \sim s_{Rui}s_{Ruj}\sim \frac{y_{u^i}y_{u^j}}{Y y_t x_t \xi_{ij}} \,,
\\
g_{Rd}^{ij}& \sim s_{Rdi}s_{Rdj}\sim \frac{y_{d^i}y_{d^j}}{Y y_t x_t \xi_{ij}} \,.
\end{align}
In the bidoublet model,
one has
\begin{align}
g_{Ld}^{ij}\sim g_{Lu}^{ij}&\sim \xi_{ij} \frac{x_t y_t}{Y_U} \,,
&
g_{Rd}^{ij}&\sim z^2\frac{Y_U}{Y_D^2}\frac{y_{d^i}y_{d^j}}{y_t x_t \xi_{ij}} \,.
&
g_{Ru}^{ij}&\sim \frac{y_{u^i}y_{u^j}}{Y_Uy_t x_t \xi_{ij}} \,.
\end{align}
The coefficients $c_D^{AB}$ are discussed in Appendix~\ref{sec:app-flavour}.

The experimental bounds on the real and imaginary parts of the Wilson coefficients have been given in \cite{Isidori:2011qw,Calibbi:2012at}.
Since the phases of the coefficients can be of order one and are uncorrelated, we derive the bounds assuming the phase to be maximal. We obtain the bounds in the first eight rows of Table~\ref{tab:bounds-anarchy}.
As is well known, by far the strongest bound, shown in the first row, comes from the scalar left-right operator in the kaon system which is enhanced by RG evolution and a chiral factor. Note in particular the growth with $z$ of the bound in the bidoublet case, which counteracts the $1/z$ behaviour of the bound from $R_b$. But also the left-left vector operators in the kaon, $B_d$ and $B_s$ systems lead to bounds which are relevant in some regions of parameter space. The bounds from the $D$ system are subleading.

\begin{table}[tbp]
\renewcommand{\arraystretch}{1.3}
\centering
\begin{tabular}{clll}
\hline
Observable & \multicolumn{3}{c}{Bounds on $m_{\psi}$ [TeV]} \\
& doublet & triplet & bidoublet \\
\hline
$\epsilon_K$ $(Q_S^{LR})$ & $14 $  & $14 $& $14  ~z $ \\
$\epsilon_K$ $(Q_V^{LL})$ & $2.7 ~x_t$& $3.9  ~x_t$  & $3.9  ~x_t$ \\
$B_d$-$\bar B_d$ $(Q_S^{LR})$ & $0.7 $ & $0.7  $ & $0.7 $ \\
$B_d$-$\bar B_d$ $(Q_V^{LL})$ & $2.3 ~x_t$ & $3.4  ~x_t$ & $3.4  ~x_t$ \\
$B_s$-$\bar B_s$ $(Q_S^{LR})$ & $0.6 $ & $0.6  $ & $0.6 $ \\
$B_s$-$\bar B_s$ $(Q_V^{LL})$ & $2.3 ~x_t$ & $3.4  ~x_t$ & $3.4  ~x_t$ \\
$D$-$\bar D$ $(Q_S^{LR})$ & $0.5 $ & $0.5  $ & $0.5 $ \\
$D$-$\bar D$ $(Q_V^{LL})$ & $0.4 ~x_t$ & $0.6  ~x_t$ & $0.6  ~x_t$ \\
$K_L\to\mu\mu$ ($f$--$\psi$) & $0.56  ~\sqrt{Y/x_t}$  & $0.56  ~\sqrt{Y/x_t}$ &\\
$K_L\to\mu\mu$ ($Z$--$\rho$) & $0.39  ~\sqrt{Y/x_t}$  & $0.56  ~\sqrt{Y/x_t}$ &\\
\hline
\end{tabular}
\caption{Flavour bounds on the fermion resonance mass $m_\psi$ in TeV in the anarchic model.}
\label{tab:bounds-anarchy}
\end{table}

\subsection{Flavour-changing $Z$ couplings}

Similarly to the modified flavour-conserving $Z$ couplings discussed in Section~\ref{sec:Zbb}, also {\em  flavour-changing} $Z$ couplings are generated in the anarchic model. In the triplet and doublet models, as well as in the bidoublet model, since the down-type contributions to the CKM matrix are not smaller than the up-type contributions in \eqref{eq:anarchy-ckm-bidoublet}, one has
\begin{gather}
\delta g_{Zd^id^j}^L \sim \frac{s_{Ldi}s_{Ldj}}{s_{Lb}^2} ~\delta g_{Zbb}^L \sim \xi_{ij} ~ \delta g_{Zbb}^L~, 
\label{eq:ZbsL}
\\
\delta g_{Zd^id^j}^R \sim \frac{s_{Rdi}s_{Rdj}}{s_{Rb}^2}~ \delta g_{Zbb}^R \sim \frac{y_{d^i}y_{d^j}}{y_b^2 \xi_{ij}} ~ \delta g_{Zbb}^R~.
\label{eq:ZbsR}
\end{gather}

In the $b\to s$ case, a global analysis of inclusive and exclusive $b\to s\ell^+\ell^-$ decays \cite{Altmannshofer:2012az} finds $|\delta g_{Zbs}^{L,R}|\lesssim 8\times 10^{-5}$, while in the $s\to d$ case, one finds $|\delta g_{Zsd}^{L,R}|\lesssim 6\times 10^{-7}$ from the $K_L\to\mu^+\mu^-$ decay \cite{Buras:2011ph}\footnote{The decay $K^+\to\pi^+\nu\bar\nu$ leads to a bound $|\delta g_{Zsd}^{L,R}|\lesssim 3 \times10^{-6}$ at 95\% C.L. and is thus currently weaker than $K_L\to\mu^+\mu^-$, even though it is theoretically much cleaner.}. Using (\ref{eq:ZbsL}) one finds that the resulting constraints on the left-handed coupling are comparable for $b\to s$ and $s \to d$. Since they are about a factor of 3 weaker than the corresponding bound from $Z\to b\bar b$, we refrain from listing them in Table~\ref{tab:bounds-anarchy}, but their presence shows that the strong bound from $R_b$ cannot simply be circumvented by a fortuitous cancellation.
In the case of the right-handed coupling, one finds that the constraint from $K_L\to\mu^+\mu^-$ is an order of magnitude stronger than the one from $b\to s\ell^+\ell^-$, and also much stronger than the bound on the right-handed coupling coming from $Z\to b\bar b$. The numerical bounds we obtain are shown in the last two rows of Table~\ref{tab:bounds-anarchy} from the contributions with fermion or gauge boson mixing separately since, in constrast to $Z\to b\bar b$, the two terms are multiplied by different $O(1)$ parameters in the flavour-violating case.

\subsection{Loop-induced chirality-breaking effects}

Every flavour changing effect discussed so far originates from tree-level chirality-conserving interactions of the vector bosons, either the elementary $W_\mu$ and $Z_\mu$ or the composite $\rho_\mu$. At loop level, chirality-breaking interactions occur as well, most notably with the photon and the gluon, which give rise  in general to significant  $\Delta F=1$ flavour-changing effects ($b\rightarrow s \gamma$, $\epsilon_K^\prime$, $\Delta A_{CP}(D\rightarrow PP)$), as well as to electric dipole moments of the light quarks. In the weak mixing limit between the elementary and the composite fermions, explicit calculations of some of the  $\Delta F=1$ effects have been made in \cite{Agashe:2008uz,Vignaroli:2012si,Gedalia:2009ws}, obtaining bounds in the range $m_\psi > (0.5\text{--}1.5)Y$\,TeV. For large CP-violating phases the generated EDMs for the light quarks can be estimated  consistent with the current limit on the neutron EDM only if $m_\psi > (3\text{--}5)Y$\,TeV, where the limit is obtained from the 
analysis of \cite{Barbieri:2012bh}. 

\subsection{Direct bounds on vector resonances}\label{sec:anarchy-jjres}

Direct production of vector resonances and subsequent decay to light quarks can lead to a peak in the invariant mass distribution of $pp\to jj$ events at the LHC. In the anarchic model, due to the small degree of compositeness of first generation quarks, the coupling of vector resonances to a first generation quark-antiquark pair is dominated by mixing with the SM gauge bosons and thus suppressed by $g_\text{el}^2/g_\rho$. For a 3~TeV gluon resonance at the LHC with $\sqrt{s}=8$~TeV, following the discussion in Appendix~\ref{sec:app-dijet} we expect
\begin{equation}
\sigma(pp\to G^*) = \frac{2\pi}{9s}\frac{g_3^4}{g_\rho^2}
\left[\mathcal L_{u\bar u}(s,m_\rho^2) +\mathcal L_{d\bar d}(s,m_\rho^2)\right]
\approx
\frac{5 ~\text{fb}}{g_\rho^2}\,.
\end{equation}
The ATLAS collaboration has set an upper bound of 7~fb on the cross section times branching ratio to two jets times the acceptance \cite{ATLAS-CONF-2012-088}, and a similar bound has been obtained by CMS \cite{CMS-PAS-EXO-12-016}. Given that the gluon resonance will decay dominantly to top quarks, we conclude that the bound is currently not relevant, even for small $g_\rho$.

A similar argument holds in the case of the dijet angular distribution, which can be used to constrain local four-quark operators mediated by vector resonances. Following the discussion in Appendix~\ref{sec:app-dijet-angular}, we obtain the bound
\begin{equation}
m_\rho > \frac{4.5~\text{TeV}}{g_\rho}
\end{equation}
which, in combination with the bound on $m_\rho$ from the $S$ parameter, is irrelevant for $g_\rho\gtrsim1.5$.

\subsection{Partial summary and prospects on anarchy}

If the bound coming from the $Q_S^{LR}$ contribution to $\epsilon_K$ is taken at face value, the fermion resonances should be far too heavy to be consistent with a naturally light Higgs boson and certainly unobservable, either directly or indirectly. Note in particular the growth of this bound with $z$ in the bidoublet model.

In view of the fact that this bound carries an $O(1)$ uncertainty, one might however speculate on what happens if this constraint is ignored.
As visible from Table~\ref{tab:bounds-anarchy}, with the exception of the first line, all the strongest bounds  on $m_\psi$ in the bidoublet or in the triplet models can be reduced down to about 1 TeV by taking $x_t = \frac{1}{3}$ to $\frac{1}{4}$. This however correspondingly requires $Y = 3$ to $4$ (and maximal right-handed mixing) which pushes up the bounds from $K_L\rightarrow \mu^+ \mu^-$ and is not consistent with $m_\psi = Yf$ and $f \gtrsim 0.5$~TeV. The loop-induced chirality-breaking effects on $\epsilon^\prime$ and $\Delta A_{CP}$ in $D\rightarrow P P$ decays would also come into play. Altogether, even neglecting the bound from $\epsilon_K(Q_S^{LR})$, fermion resonances below about 1.5 TeV seem hard to conceive.

\section{Constraints on $U(3)^3$}\label{sec:u3}

We now discuss the constraints specific to \Utre. In \UtreLC\ the sizable degree of compositeness of light left-handed quarks leads to additional contributions to electroweak precision observables; in \UtreRC\ FCNCs arise at the tree level. In both cases collider bounds on the compositeness of light quarks place important constraints. Our analysis follows and extends the analysis in \cite{Redi:2011zi}.

\subsection{Electroweak precision constraints specific to $U(3)^3$}\label{sec:U3EW}

The bounds from $R_b$ as well as the $S$ and $T$ parameters discussed in Section~\ref{sec:ewpt} are also valid in $U(3)^3$, with one modification: in  \UtreLC\ the contributions to the $\hat T$ parameter proportional to $s_{Lt}^4$ have to be multiplied by 3 since all three generations of left-handed up-type quarks contribute. The corresponding
bounds remain nevertheless relatively mild.

In addition, an important constraint arises from the partial width of the $Z$ into hadrons normalized to the partial width into leptons, which was measured precisely at LEP
\begin{align}
R_h^\text{exp} &= 20.767(25)~,
&
R_h^\text{SM} &= 20.740(17)~,
\end{align}
showing a $1.1\sigma$ tension with the best-fit SM prediction \cite{Baak:2012kk}.

In \UtreLC\ the modified left-handed $Z$ couplings of up and down quarks are equal to the ones of the $t$ and $b$ respectively, while the same is true in \UtreRC\ for the right-handed modified couplings. Analogously to the discussion in Section~\ref{sec:Zbb}, one can write the modified $Z$ coupling of the top as
\begin{equation}
\frac{g}{c_w}\bar t \gamma^\mu \left[
(\tfrac{1}{2}-\tfrac{2}{3}s_w^2+\delta g^L_{Ztt}) P_L
+(-\tfrac{2}{3}s^2_w+\delta g^R_{Ztt})P_R
\right]tZ_\mu
\,,
\end{equation}
and one has
\begin{align}
\delta g_{Ztt}^L &=
\frac{v^2Y_{U}^2}{2m_{U}^2}\frac{x_ty_t}{Y_U} \,a+
\frac{g_\rho^2v^2}{4m_\rho^2}\frac{x_ty_t}{Y_U}\,b \,,
&
\delta g_{Ztt}^R &=
\frac{v^2Y_{U}^2}{2m_{U}^2}\frac{y_t}{x_t Y_U} \,c+
\frac{g_\rho^2v^2}{4m_\rho^2}\frac{y_t}{x_t Y_U}\,d \,,
\label{eq:Zbb}
\end{align}
with
\begin{center}
\begin{tabular}{c|ccc}
& doublet & triplet & bidoublet \\
\hline
$a$ & $-1/2$   & $-1$ & $-1/2$\\
$b$ & $-1/2$ & $-1$ & $-1$
\end{tabular}
\qquad
\begin{tabular}{c|ccc}
& doublet & triplet & bidoublet \\
\hline
$c$ & $1/2$   & $0$ & $0$\\
$d$ & $1/2$ & $0$ & $0$
\end{tabular}
\end{center}
Since the right-handed $Z$ coupling to $b$ and $t$ receives no contribution in the bidoublet model, there is no additional bound from $R_h$ in \UtreRC.
In \UtreLC we find the numerical bounds shown in Table~\ref{tab:bounds-U3LC}.

In \UtreLC\ an additional bound arises from violations of quark-lepton universality.
Writing the $W$ couplings as
\begin{equation}
\frac{g}{\sqrt{2}}(1+\delta g^L_{W}) \bar u\, V_{ui}\gamma^\mu 
P_L
d_i W_\mu^+
\,,
\end{equation}
we find
\begin{align}
\delta g_{W}^L &=
\frac{Y_{U}^2v^2}{2m_{U}^2}\frac{x_ty_t}{Y_U} \,a_u +
\frac{Y_{D}^2v^2}{2m_{D}^2}\frac{x_ty_t}{Y_U} \,a_d +
\frac{g_\rho^2v^2}{4m_\rho^2}\frac{x_ty_t}{Y_U}\,b \,,
\end{align}
with
\begin{center}
\begin{tabular}{c|ccc}
& doublet & triplet & bidoublet \\
\hline
$a_u$ & $-1/2$ & $-1/2$ & $-1/2$
\\
$a_d$ & $-1/2$ & $-1/2$ & $-1/(2z^2)$
\\
$b$ & $-1$ & $-1$ & $-1$
\end{tabular}
\end{center}
The usual experimental constraint on the strength of the $W\bar{u}d_i$ couplings, normalized to the leptonic ones, is expressed by $(1+\delta g^L_{W})^2 \sum_i|V_{ui}|^2-1=(-1\pm6)\times10^{-4}$, which, from the unitarity of the $V_{ij}$ matrix, becomes $2 \delta g^L_{W}= (-1\pm6)\times10^{-4}$.  By requiring it to be fulfilled within $2\sigma$, we find the numerical bounds in Table~\ref{tab:bounds-U3LC}.

Finally we note that, in contrast to the anarchic case, there are no {\em flavour-changing} $Z$ couplings neither in \UtreLC\ nor in \UtreRC. In the former case this is a general property of chirality-conserving bilinears, while in the latter it is a consequence of the fact that only the down-type mixings $\lambda_{Ld}$ affect the $Z$ vertex, which thus becomes flavour-diagonal in the mass basis.

\subsection{Tree-level  $\Delta F=2$ FCNCs}

While in \UtreLC\ there are no tree-level FCNCs at all \cite{Redi:2011zi}, minimally flavour violating tree-level FCNCs are generated in \UtreRC\ \cite{Barbieri:2012uh,Redi:2012uj}.
This can be shown as follows. Before going to the physical basis, the relevant interactions with the composite resonances have the form in $U(3)^3_\text{RC}$ 
\begin{equation}
\rho_\mu (\bar{q}_L \hat{s}_{Lu} \gamma_\mu \hat{s}_{Lu}^{\dag} q_L)
\label{int_RC}
\end{equation}
and in $U(3)^3_\text{LC}$ 
\begin{equation}
\rho_\mu (\bar{q}_L {s}_{Lu} \gamma_\mu {s}_{Lu}^* q_L).
\label{int_LC}
\end{equation}
In $U(3)^3_\text{RC}$ the physical bases for up and down quarks are reached by proper $3\times 3$ unitary transformations that diagonalize $\hat{s}_{Lu}$ and $\hat{s}_{Ld}$
\begin{equation}
U_L^u\hat{s}_{Lu} U_R^{u\dag}=  \hat{s}_{Lu}^{\rm diag}~~~~ U_L^d\hat{s}_{Ld} U_R^{d\dag}=  \hat{s}_{Ld}^{\rm diag},
\end{equation} 
 so that the CKM matrix is $V = U_L^u U_L^{d\dag}$. In the same physical basis the interaction  (\ref{int_RC}) in the down sector becomes
\begin{equation}
\rho_\mu (\bar{d}_L V^{\dag} \hat{s}_{Lu}^{\rm diag} \gamma_\mu (\hat{s}_{Lu}^{\rm diag})^* V d_L) \approx 
\rho_\mu s_{Lt}^2 \xi_{ij} (\bar{d}_{Li}  \gamma_\mu  d_{Lj}),~~~~~\xi_{ij} = V_{ti}^* V_{tj}\,.
\label{int_dRC}
\end{equation}
Note that the ratio of the third to  the second entry in  $\hat{s}_{Lu}^{\rm diag}$ equals $y_t/y_c$. On the other hand a similar procedure for $U(3)^3_\text{LC}$ leaves (\ref{int_LC}) unaltered since ${s}_{Lu}$ is proportional to the identity matrix.

The Wilson coefficients of $\Delta F=2$ operators are given by (\ref{eq:DF2}), with the coefficients $c_D^{qAB}$ listed in Appendix~\ref{sec:app-flavour} and the couplings
\begin{align}
g_{Ld}^{ij} &= \xi_{ij}\frac{x_ty_t}{Y_U} \,,
&
g_{Rd}^{ij} &\approx 0 \,.
\label{U(3)RC-FC}
\end{align}
We obtain the bounds shown in Table~\ref{tab:bounds-U3RC}. The bound from $D$-$\bar D$ mixing turns out to be numerically irrelevant.

We stress that, in contrast to the anarchic case, there is no $O(1)$ uncertainty in these bounds since the composite Yukawas are proportional to the identity. Furthermore, since the model is minimally flavour violating, there is no contribution to the meson mixing phases and the new physics effects in the $K$, $B_d$ and $B_s$ systems are prefectly correlated.

\begin{table}[tbp]
\renewcommand{\arraystretch}{1.3}
\centering
\begin{tabular}{cccc}
\hline
Observable & \multicolumn{3}{c}{Bounds on $m_{\psi}$ [TeV]} \\
&  doublet & triplet  & bidoublet\\
\hline
$R_h$ & $7.2 ~\sqrt{x_tY}$  &$6.0 ~\sqrt{x_tY}$  &$4.9 ~\sqrt{x_tY_U}$\\
$V_\text{CKM}$ & $7.4 ~\sqrt{x_tY}$ & $7.4 ~\sqrt{x_tY}$ & $6.0 ~\sqrt{x_tY_U}$ \\
$pp\to jj$ ang. dist. & $3.4  ~x_t$ & $4.2  ~x_t$ & $4.2  ~x_t$  \\
\hline
\end{tabular}
\caption{Lower bounds on the fermion resonance mass $m_\psi$ in TeV in \UtreLC.}
\label{tab:bounds-U3LC}
\end{table}

\begin{table}[tbp]
\renewcommand{\arraystretch}{1.3}
\centering
\begin{tabular}{cc}
\hline
Observable & Bounds on $m_{\psi}$ [TeV] \\
\hline
$\epsilon_K(Q^{LL}_V)$ & $3.7 ~x_t$ \\
$B_d$-$\bar B_d$ & $3.2 ~x_t$ \\
$B_s$-$\bar B_s$  & $3.6 ~x_t$\\
$pp\to jj$ ang. dist. & $3.0/x_t$ \\
\hline
\end{tabular}
\caption{Lower bounds on the fermion resonance mass $m_\psi$ in TeV in \UtreRC\ (bidoublet model).}
\label{tab:bounds-U3RC}
\end{table}

\subsection{Loop-induced chirality-breaking effects}

Flavour-changing chirality-breaking effects in $U(3)^3$ occur when elementary-composite mixings are included inside the loops. At least for moderate mixings, the bounds are of the form $m_\psi > (0.5\text{--}1.5) \sqrt{Y/x_t} $ TeV in the \UtreLC\ case, or $m_\psi > (0.5\text{--}1.5)\sqrt{Y x_t}$ TeV in the \UtreRC\ case. The stronger bounds from  quark EDMs, similar to the ones of the anarchic case,  disappear if the strong sector conserves CP. This is automatically realized, in our effective Lagrangian description, if the ``wrong chirality'' Yukawas vanish or are aligned in phase with the $Y$'s. On the contrary, in the anarchic case this condition is in general not sufficient to avoid large EDMs.

\subsection{Compositeness constraints}

Since one chirality of first-generation quarks has a sizable degree of compositeness in the $U(3)^3$ models, a significant constraint arises from the angular distribution of dijet events at LHC, which is modified by local four-quark operators obtained after integrating out the heavy vector resonances related to the global $SU(3)_c\times SU(2)_L\times SU(2)_R\times U(1)_X$ as well as the flavour symmetry in the strong sector, $U(3)$ in the case of \UtreLC\ and $U(3)\times U(3)$ in the case of \UtreRC.

In general, there are ten four-quark operators relevant in the dijet angular distribution \cite{Domenech:2012ai}. Following the discussion in Appendix~\ref{sec:app-dijet-angular}, the relevant operators in \UtreLC\ are $\mathcal O_{qq}^{(1,8)}$. Their Wilson coefficients read
\begin{align}
C_{qq}^{(1)} = -\frac{a}{36}\frac{g_\rho^2}{m_\rho^2}\left(\frac{x_t y_t}{Y_U}\right)^2\,,
\qquad
C_{qq}^{(8)} = -\frac{g_\rho^2}{m_\rho^2}\left(\frac{x_t y_t}{Y_U}\right)^2\,,
\end{align}
where $a=5$ in the doublet model and $a=17$ in the triplet and bidoublet models.
Using the updated version of \cite{Domenech:2012ai},
we obtain the bound in the last row of Table~\ref{tab:bounds-U3LC}.
In \UtreRC\ the operators with right-handed quarks are relevant, i.e. $\mathcal O_{uu,dd,ud}^{(1)}$ and $\mathcal O_{ud}^{(8)}$.
Numerically, we find the bound on $\mathcal O_{uu}^{(1)}$ to give the most significant constraint on the model parameters. Its Wilson coefficient reads
\begin{align}
C_{uu}^{(1)} = -\frac{5}{9}\frac{g_\rho^2}{m_\rho^2}\left(\frac{y_t}{x_t Y_U}\right)^2\,.
\end{align}
and the resulting numerical constraint is shown in the last row of Table~\ref{tab:bounds-U3RC}.

\subsection{Direct bounds on vector resonances}\label{sec:u3-jjres}

As discussed in Section~\ref{sec:anarchy-jjres}, direct bounds on $m_\rho$ are obtained from searches for peaks in the invariant mass of dijets at LHC. In $U(3)^3$ the production cross sections can be larger than in the anarchic case due to the sizable degree of compositeness of first-generation quarks. Neglecting the contribution due to mixing of the vector resonances with the gauge bosons, the production cross section of a gluon resonance reads (see Appendix~\ref{sec:app-dijet})
\begin{align}
\sigma(pp\to G^*) &=  \frac{2\pi}{9s}g_\rho^2
\left[
s_{L,Ru}^4 \mathcal L_{u\bar u}(s,m_\rho^2)
+
s_{L,Rd}^4 \mathcal L_{d\bar d}(s,m_\rho^2)
\right],
\label{eq:sigmapprhoU3}
\end{align}
where the $L$ is valid in \UtreLC\ and the $R$ in \UtreRC.
In \UtreLC\ the branching ratio to two jets reads approximately
\begin{align}
\text{BR}(G^*\to jj) =
\frac{2 s_{Lu}^4+3 s_{Ld}^4+s_{Rb}^4}{3 s_{Lu}^4+s_{Rt}^4+3 s_{Ld}^4+s_{Rb}^4
}\,,
\end{align}
and is typically larger than in the anarchic case. Similarly, in \UtreRC\ one has
\begin{equation}
\text{BR}(G^*\to jj) =
\frac{
2 s_{Ru}^4
+
s_{Lb}^4
+
3 s_{Rd}^4
}{
s_{Lt}^4
+
3 s_{Ru}^4
+
s_{Lb}^4
+
3 s_{Rd}^4
}\,.
\end{equation}

To judge if the most recent experimental bounds by ATLAS \cite{ATLAS-CONF-2012-088} and CMS \cite{CMS-PAS-EXO-12-016} have already started to probe the $U(3)^3$ parameter space, we evaluate the cross section for maximal mixing, i.e. $x_t=Y/y_t$ in \UtreLC\ and $x_t=y_t/Y$ in \UtreRC, for a 3~TeV gluon resonance, i.e. only marginally heavier than allowed by the $S$ parameter (cf. Table~\ref{tab:bounds-ew}). For \UtreLC\ we obtain 
\begin{align}
\sigma(pp\to G^*) \approx 13g_\rho^2 ~\text{fb}~,
\qquad
\text{BR}(G^*\to jj) \approx 58\% ~(83\%) \text{ for } Y=1 ~(4\pi)~,
\end{align}
and for \UtreRC
\begin{align}
\sigma(pp\to G^*) \approx 30g_\rho^2 ~\text{fb}~,
\qquad
\text{BR}(G^*\to jj) \approx 69\% ~(67\%) \text{ for } Y=1 ~(4\pi)~.
\end{align}
This is to be compared to the ATLAS bound $\sigma\times\text{BR}\times A<7~\text{fb}$, where $A$ is the acceptance. We conclude that, assuming an acceptance of the order of 60\% \cite{ATLAS-CONF-2012-088}, maximal mixing is on the border of exclusion in \UtreLC\ and already excluded in \UtreRC\ for a 3~TeV gluon resonance.
We note however that maximal mixing is already disfavoured by the indirect bounds discussed above.

\subsection{Partial summary on $U(3)^3$}

As apparent from Tables~\ref{tab:bounds-U3LC} and \ref{tab:bounds-U3RC}, a fermion resonance at about 1 TeV is disfavoured.
In \UtreLC\ the crucial constrains come from the EWPT due to the large mixing of the first generations quarks in their left component. Note that $x_t Y$ cannot go below $y_t\sim1$. In  \UtreRC\ there is a clash between the tree-level $\Delta F=2$ FCNC effects, which decrease with $x_t$, and the bound from the $pp\to jj$  angular distributions due to the composite nature of the light quarks in their right component, which goes like $1/x_t$.
We stress again that these conclusions are more robust than in the anarchic case, since there is no uncertainty related to the composite Yukawas, which are flavour universal in the \Utre\ case.

\section{Constraints on $U(2)^3$}\label{sec:u2}

In \UdueLC\ and \UdueRC\ the first and second generation elementary-composite mixings are expected to be significantly smaller than the third generation ones, so the collider phenomenology is virtually identical to the anarchic case. The same is true for the electroweak precision constraints, although if to a lesser extent, and the most serious problems plaguing the \Utre\ models are absent. The most important difference concerns the flavour constraints.

\subsection{Tree-level  $\Delta F=2$ FCNCs}

Equations (\ref{int_RC}, \ref{int_LC}) remain formally true in $U(2)^3$  as well, with the following qualifications.  $Y_U, s_{Ru},  s_{Lu}$ are no longer proportional to the identity but are still  diagonal with only the first  two entries equal to each other. At the same time minimal breaking of $U(2)^3$ leads to a special form of the matrices $ \hat{s}_{Lu},  \hat{s}_{Ru}$ that allows to diagonalize approximately the Yukawa couplings  by pure left unitary transformations of the form \eqref{UL}.

In $U(2)^3_\text{RC}$ these transformations lead to exactly the same equation as (\ref{int_dRC}), whereas in the $U(2)^3_\text{LC}$ case equation (\ref{int_LC}) in the down sector goes into
\begin{equation}
\rho_\mu (\bar{d}_L U_d {s}_{Lu} \gamma_\mu {s}_{Lu}^* U_{d}^{\dag} d_L) \approx 
\rho_\mu s_{Lt}^2 \chi_{ij} (\bar{d}_{Li}  \gamma_\mu  d_{Lj}),~~~~~\chi_{ij} = U^d_{i3} U^{d*}_{j3},
\label{int_dLC}
\end{equation}
Remember that, contrary to the $U(3)^3_\text{RC}$ case, ${s}_{Lu}$, although still diagonal, is not proportional to the unit matrix. Hence a flavour violation survives with
\begin{equation}
\label{eq:rb}
r_b = \frac{s_b}{s} e^{i(\chi - \phi_b)}\,.
\end{equation}
In the bidoublet model, in addition to \eqref{int_dLC} there are also the terms coming from the mixing with the $\bar Q^d\gamma_{\mu}Q^d$ current, which are suppressed as $1/z^2$.
In the up-quark sector with right-compositeness only this suppressed contribution from $Q^d$ gives rise to flavour violation, while in left-compositeness the analog of \eqref{int_dLC} holds, with $U_d$ replaced by $U_u$.

The Wilson coefficients of $\Delta F=2$ operators generated in \UdueLC\ and \UdueRC\ are again given by (\ref{eq:DF2}). The flavour-changing couplings in \UdueLC\ read
\begin{align}
g_{Ld}^{i3} &= \xi_{i3} \, r_b\frac{x_t y_t}{Y_U} \,,
&
g_{Ld}^{12} &= \xi_{12} \, |r_b|^2 \frac{x_t y_t}{Y_U} \,,
&
g_{Rd}^{ij} &\approx 0 \,.
\label{FCU2LC}
\end{align}
As a consequence of the presence of $r_b$ there is a new, universal phase in $B_d$ and $B_s$ mixing, while the $K$-$\bar K$ amplitude is aligned in phase with the SM. We find the bounds in Table~\ref{tab:bounds-U2LC}. If the parameter $|r_b|$ is somewhat less than 1, these bounds can be in agreement with experiment even for light fermion resonances. We note that the contribution to the $\Delta C=2$ operator is proportional to $|1-r_b|^2$, so it cannot be reduced simultaneously. However, it turns out that it is numerically insignificant. Since furthermore the contribution is real -- a general prediction of the \Udue\ symmetry for $1\leftrightarrow 2$ transitions -- the expected improvement of the bound on CP violation in $D$-$\bar D$ mixing will have no impact.

In \UdueRC\ the flavour-changing couplings are the same as in \UtreRC,
\begin{align}
g_{Ld}^{i3} &= \xi_{i3} \,\frac{x_ty_t}{Y_U} \,,
&
g_{Ld}^{12} &= \xi_{12} \, \frac{x_t y_t}{Y_U} \,,
&
g_{Rd}^{ij} &\approx 0 \,.
\end{align}
Thus, as in \UtreRC, there is no new phase in meson-antimeson mixing and the NP effects in the $K$, $B_d$ and $B_s$ systems are perfectly correlated. The resulting bounds are shown in Table~\ref{tab:bounds-U2RC}.

\subsection{Loop-induced chirality-breaking effects}

One expects in general  flavour-changing chirality-breaking effects in $U(2)^3$ with bounds on the fermion resonances similar to the one of the anarchic case,  $m_\psi > (0.5\text{--}1.5)Y$\,TeV.
With CP conservation in the strong sector, however, the contributions to the quarks EDMs would arise only at higher orders in the $U(2)^3$ breaking terms, so that they would not be significant for the current limit on the neutron EDM.

\begin{table}[tbp]
\renewcommand{\arraystretch}{1.3}
\centering
\begin{tabular}{cccc}
\hline
Observable & \multicolumn{3}{c}{Bounds on $m_{\psi}$ [TeV]} \\
& doublet & triplet& bidoublet  \\
\hline
$\epsilon_K(Q^{LL}_V)$ & $2.3 ~x_t|r_b|^2$ & $3.3  ~x_t|r_b|^2$ & $3.3  ~x_t|r_b|^2$ \\
$B_d$-$\bar B_d$ & $2.3 ~x_t|r_b|$ & $3.4  ~x_t|r_b|$ & $3.4  ~x_t|r_b|$ \\
$B_s$-$\bar B_s$ & $2.3 ~x_t|r_b|$ & $3.4  ~x_t|r_b|$ & $3.4  ~x_t|r_b|$ \\
$K_L\to\mu\mu$ & $3.8 ~\sqrt{x_tY}|r_b|$ && $3.8  ~Y_D |r_b| \sqrt{x_t/Y_{U}}/z$ \\
$b\to s\ell\ell$ & $3.5 ~\sqrt{x_tY|r_b|}$ && $3.5  ~Y_D \sqrt{x_t|r_b|/Y_{U}}/z$ \\
\hline
\end{tabular}
\caption{Lower bounds on the fermion resonance mass $m_\psi$ in TeV in \UdueLC.}
\label{tab:bounds-U2LC}
\end{table}

\begin{table}[tbp]
\renewcommand{\arraystretch}{1.3}
\centering
\begin{tabular}{cc}
\hline
Observable & Bounds on $m_{\psi}$ [TeV] \\
\hline
$\epsilon_K(Q^{LL}_V)$ & $3.3 ~x_t$ \\
$B_d$-$\bar B_d$ & $2.8 ~x_t$ \\
$B_s$-$\bar B_s$  & $3.1 ~x_t$\\
\hline
\end{tabular}
\caption{Lower bounds on the fermion resonance mass $m_\psi$ in TeV in \UdueRC\ (bidoublet model).}
\label{tab:bounds-U2RC}
\end{table}

\subsection{Flavour-changing $Z$ couplings}

In \UdueRC\ flavour-changing $Z$ couplings are absent at tree level. In \UdueLC\ the left-handed couplings do arise, while the right-handed couplings are strongly suppressed. Similarly to the anarchic case, one can write them as
\begin{gather}
\delta g_{Zbd^i}^L \sim  \xi_{i3} ~ r_b ~ \delta g_{Zbb}^L~,
\qquad
\delta g_{Zsd}^L \sim  \xi_{12} ~ |r_b|^2 ~ \delta g_{Zbb}^L~.
\label{eq:ZbsL-U2}
\end{gather}
One obtains the bounds in the last two lines of Table~\ref{tab:bounds-U2LC}, which are weaker than the analogous bounds from $R_b$ unless $|r_b|>1$.
An important difference with respect to the anarchic case is the absence of sizable flavour-changing {\em right-handed} $Z$ couplings, which can be probed e.g. in certain angular observables in $B\to K^*\mu^+\mu^-$ decays \cite{Altmannshofer:2008dz}.

\subsection{Electroweak precision constraints}

Note that in \UdueLC, for $Y_{U2}\sim Y_{D2}\sim O(1)$ and $d_u,d_d \lesssim O(1)$, the expressions (\ref{eq:stbLC}) for $s_t, s_b$ lead to two possibilities:
\begin{enumerate}
 \item $s_t\ll s_b$, i.e. $|r_b|\approx1$;
 \item $s_t\sim s_b\sim|V_{cb}|$, which allows $|r_b|$ to deviate from 1 but requires at the same time $s_{Lu2}\epsilon_L\sim |V_{cb}|$.
\end{enumerate}
In the first case one would have $m_{\psi} \gtrsim 1\text{--}1.5$ TeV from the flavour bounds of Table~\ref{tab:bounds-U2LC}, while in the second case one can obtain a minimal value of $m_\psi \simeq 0.6$ TeV, for $|r_b| \sim 0.25$ and $Y \sim 1$. However, to avoid a too large $U(2)^3$-breaking -- i.e. a large $\epsilon_L$ -- the mixing angle of the first generations quarks $s_{Lu2}$ cannot be too small.
This in turn has to be confronted with the lower bounds on $m_\psi$ from $R_h$, $V_{\text{CKM}}$ and the dijet angular distribution shown in Table \ref{tab:bounds-U2LC-sL2}: to make them consistent with $m_\psi \simeq 0.6$ TeV, it must be $\epsilon_L \gtrsim 0.3$. Note anyhow that we are not treating $\epsilon_L$ as an expansion parameter.
\begin{table}[tbp]
\renewcommand{\arraystretch}{1.3}
\centering
\begin{tabular}{cccc}
\hline
Observable & \multicolumn{3}{c}{Bounds on $m_{\psi}$ [TeV]} \\
&  doublet & triplet  & bidoublet\\
\hline
$R_h$ & $7.2 ~s_{L2}Y_2$  &$6.8 ~s_{L2}Y_2$  &$5.6 ~s_{Lu2}Y_{U2}$\\
$V_\text{CKM}$ & $8.4 ~s_{L2}Y_2$  &$6.8 ~s_{L2}Y_2$  &$6.8 ~s_{Lu2}Y_{U2}$\\
$pp\to jj$ ang. dist. & $4.3  ~s_{L2}^2Y_{2}$ & $5.3  ~s_{L2}^2Y_{2}$ & $5.3  ~s_{Lu2}^2Y_{U2}$  \\
\hline
\end{tabular}
\caption{Lower bounds on the fermion resonance mass $m_\psi$ in TeV in \UdueLC\ from left-handed 1st and 2nd generation quarks mixed with the composite resonances by an angle $s_{Lu2}$.}
\label{tab:bounds-U2LC-sL2}
\end{table}

\subsection{Partial summary on $U(2)^3$}

Two important differences distinguish the $U(2)^3$ case from the $U(3)^3$ one: i) both for the bidoublet (at  large enough $z$) and for the triplet models, the  bounds from the EWPT or from compositeness become irrelevant; ii) a single complex parameter correlates the various observables, $r_b$ in the  \UdueLC\  case. As apparent from Table~\ref{tab:bounds-U2LC}, values of $x_t$ and $r_b$ somewhat smaller than one can reduce the bounds on the fermion resonance mass at or even below the 1 TeV level. This is also formally possible in \UdueRC, where $r_b=1$, but requires $x_t \lesssim 0.3$, i.e. $Y \gtrsim 3$, not consistent with $m_\psi = Y f$ and $f \gtrsim 0.5$~TeV.

\section{LFV in composite Higgs models and the $g - 2$ of the muon}
\label{sec:CHM_leptons}
To make a composite Higgs model fully realistic one must extend the discussion of Section~\ref{sec:setup} to the lepton sector as well. A rather unique way in which this can be done closely mimics the case of the quarks.  In the strong sector composite vector-like leptons, $L$, $E$ and $N$, are assumed to exist with the same quantum numbers of the elementary $l_L, e_R, \nu_R$ as well as Yukawa couplings and mass terms in analogy with \eqref{doubletL}\footnote{Although we do not think it to be phenomenologically necessary, to maximize the quark-lepton symmetry in the bidoublet model one could consider two $SU(2)_L$-doublets, $L^e$ and $L^\nu$.}. Similarly there will be mass mixing terms between the composite and the elementary fermions. The only asymmetry with the quark sector is in the presence of a  mass matrix for the elementary $\nu_R$, with elements much larger than any other scale. 
As a consequence the neutrino spectrum, per generation, consists of one light neutrino, which can be arranged to have standard left-handed weak interaction to a sufficient level of accuracy, two quasi-Dirac  neutrinos at the typical compositeness scale and one superheavy almost pure right-handed neutrino. Following the discussion of Section~\ref{sec:U2leptons} we assume that the right-handed neutrino mixing matrix, $\bar{N}_L \hat{m}_\nu \nu_R$, and the Majorana mass matrix of the right-handed neutrinos have a flavour structure such that the mass matrix of the light neutrinos  gives rise to the large mixing angles of the leptonic charge current.

What about the flavour properties of the light charged leptons to all orders in the strong interactions? Let us consider first the case in which the strong sector conserves  a diagonal leptonic $U(3)_L$ symmetry. As mixing terms we can consider:
\begin{equation}
\mathcal{L}^{R\text{-comp}}_\text{mix} = m_E \bar{E}_L e_R + \bar{l}_L \hat{m}_e L_R
\end{equation}
or
\begin{equation}
\mathcal{L}^{L\text{-comp}}_\text{mix} = m_E \bar{l}_L L_R + \bar{E}_L \hat{m}_e e_R,
\end{equation}
with $\hat{m}_e$ transforming as $(3, \bar{3})$ under $U(3)_l\times U(3)_{L+e}$ or $U(3)_{L+ l}\times U(3)_e$ respectively. 
As anticipated, in either case there is no leptonic flavour changing phenomenon at the Fermi scale other than the standard mixing in the leptonic charged-current interaction. This is because $\hat{m}_e$ can be set to diagonal form and $\hat{m}_\nu$ has no effect at the Fermi scale\footnote{For a recent comparison of the anarchic and flavour-symmetric scenarios for leptons in CHMs see \cite{Redi:2013pga}.}.
The discussion of lepton flavour violation in the case of a $U(2)$ symmetry proceeds along similar lines as in the quark case. The strong interaction Lagrangian respects a $(U(2)\times U(1))_L$ symmetry whereas the mixing Lagrangians are:
\begin{equation}
\mathcal{L}^{R\text{-comp}}_\text{mix}(U(2)) \approx  m_E(A_e \, \CHeLbar \EHeR + B_e\, \CLeLbar\ELeR) +
\tilde m_E(a_e \,\EHlLbar \CHlR + b_e \,(\ELlLbar\Ve)\CHlR + c_e\, \ELlLbar\Delta Y_e \CLlR) +
\text{h.c.}
\end{equation}
or
\begin{equation}
\mathcal{L}^{L\text{-comp}}_\text{mix}(U(2)) \approx m_E(A_e\, \EHlLbar \CHlR + B_e \,\ELlLbar\CLlR) +
\tilde m_E( a_e\, \CHeLbar \EHeR + b_e\, (\CLeLbar\Ve)\EHeR + c_e \,\CLeLbar\Delta Y_e \ELeR) +
\text{h.c.}
\end{equation}
Both in the case of Left- and of Right-compositeness there are flavour violating, chirality conserving transitions, whereas a difference exists for chirality breaking transitions, as in the quark case. This means that the leading order operators, $\tau \rightarrow \mu, e + \gamma$ as in (\ref{taumue}), are present only in the Left-compositeness case, with an amplitude proportional to $m_\tau$. On the contrary, as a subleading phenomenon in the expansion in $\Delta Y_e$ and  $
\Ve$, the  $\mu \rightarrow  e + \gamma$ operator as in (\ref{mue}) exists both in Left- and Right-compositeness.

As an aside remark we note that a magnetic moment operator of the composite charged leptons 
\begin{equation}
\dfrac{\lambda_L v}{\sqrt{2} M^2} (\bar{L}\sigma_{\mu\nu} E) e F_{\mu\nu}
\end{equation}
 gives rise, after mass mixing, to a magnetic moment operator for the standard charged leptons 
\begin{equation}
\frac{1}{M^2} (\bar{l}_i \sigma_{\mu\nu} m_i e_i) e F_{\mu\nu}, ~~~ i = e, \mu, 
\end{equation}
where $M$ is a typical compositeness scale and $m_i$ are the masses of the standard charged leptons. In turn this corrects the $g-2$ anomalies by an extra contribution:
\begin{equation}
\Delta a_i \equiv \dfrac{\Delta (g - 2)_i}{2} = \frac{4\, m_i^2}{M^2}.
\end{equation}
This could explain the putative discrepancy between theory and experiment of the muon anomaly, $\Delta a_\mu \approx 3 \cdot10^{-9}$\cite{Prades:2009qp}, with a mass $M\approx 4$ TeV, while being consistent with the current information on the electron $(g-2)$. Notice that $M$ includes a loop suppression factor, so that the actual mass of the composite lepton necessary to generate $\Delta a_\mu$ is parametrically lower than $M$ by a factor $\sim g_\psi^2/16\pi^2$. The exact correlation between the electron and the muon anomalies is a consequence of a $U(3)_L$   or a $(U(2)\times U(1))_L$ symmetries of the strong interaction Lagrangian.

\section{Summary and partial conclusions}\label{sec:summary}

One  way to implement a natural Fermi scale is to make the Higgs particle, one or more, a pseudo-Goldstone boson of a new strong interaction in the few TeV range. 
A meaningful question is if and how a Higgs boson of 125 GeV mass fits into this picture, which requires spin-$\frac{1}{2}$ resonances, partners of the top, with a semi-perturbative coupling to the strong sector and a mass not exceeding about 1 TeV. 

Not the least difficulty in addressing this question is the variety of  possible specific implementations of the Higgs-as-pseudo-Goldstone-boson picture, especially with regard to the different representations of the spin-$\frac{1}{2}$ resonances and the various ways to describe flavour. A further problem is represented by the limited calculability of key observables in potentially complete models, due to their strongly interacting nature. 

\begin{table}[tbp]
\renewcommand{\arraystretch}{1.3}
\centering
\begin{tabular}{cccc}
\hline
& doublet & triplet & bidoublet \\
\hline
\CircledA & $4.9$ & $1.7$ & $1.2*$ \\
\UtreLC & $6.5$ & $6.5$ & $5.3$ \\
\UtreRC &-&-& $3.3$ \\
\UdueLC & $4.9$ & $0.6$ & $0.6$ \\
\UdueRC &-&-& $1.1*$ \\
\hline
\end{tabular}
\caption{Minimal fermion resonance mass $m_\psi$ in TeV  compatible with all the bounds (except for the $Q_S^{LR}$ contribution to $\epsilon_K$ in the anarchic model), fixing $O(1)$ parameters in anarchy to 1 and assuming the parameter $|r_b|$ in \UdueLC\ to be $\sim 0.2$. The bounds with a $*$ are obtained for a  value of  $Y \approx 2.5$, that minimizes the flavour and EWPT constraints consistently with $m_\psi = Y f$ and $f \gtrsim 0.5$ TeV.}
\label{tab:mmin}
\end{table}

To circumvent these difficulties, we have adopted some simple {\it partial-compositeness} Lagrangians and assumed that they catch the basic phenomenological properties of the theories under consideration. This allows us to consider a grid of various possibilities, represented, although at the risk of being too simplistic, in Table~\ref{tab:mmin}, which tries to summarize all in one go the content of the more detailed tables \ref{tab:bounds-ew} to \ref{tab:bounds-U2RC} discussed throughout the paper, taking into account all constraints from flavour and EWPT.
For any given case, this table estimates a lowest possible value for the mass of the composite fermions that mix with the elementary ones and which are heavier than the ``custodians'' by a factor of $\sqrt{1 + (\lambda_X)^2}$. In the case of {\it anarchy} we are neglecting the constraint coming from $\epsilon_K$ (first line of Table~\ref{tab:bounds-anarchy}, particularly problematic for the bidoublet model, maybe accidentally suppressed) and the various 
$O(1)$ factors that plague most of the other flavour observables in Table~\ref{tab:bounds-anarchy}.
In every case we also neglect the constraint coming from  one-loop chirality-breaking operators, relevant to direct CP violation both in the $K$ and in the $D$ systems, as well as to the quark electric dipole moments. This is a subject that  deserves further detailed study.

We also note that measurements of Higgs boson properties, which have not been considered here, amount to lower bounds on the decay constant $f$ in the case of PNGB Higgs models, and are currently probing values of $500\text{--}700$ GeV. Once these bounds improve, Tables~\ref{tab:bounds-ew} to \ref{tab:bounds-U2RC} allow a straightforward qualitative understanding of their impact on flavour and electroweak observables. Since our predictions are based on a simple partial compositeness Lagrangian, they are in fact independent of the details of the Higgs sector and can even be applied to other theories, like 4D duals of Randall-Sundrum models.

\definecolor{green}{cmyk}{0.5,0,1,0.2}
\definecolor{lightgray}{rgb}{0.7,0.7,0.7}
\newcommand{\si}{{\color{green}\footnotesize$\bigstar$}}
\newcommand{\no}{{\color{lightgray}$\circ$}}

\begin{table}
\renewcommand{\arraystretch}{1.3}
\centering
\begin{tabular}{cccccc}
\hline
& ~\CircledA~ & \UtreLC & \UtreRC & \UdueLC & \UdueRC \\
\hline
$\epsilon_K$, $\Delta M_{d,s}$ & \si & \no & \si & \si & \si \\
$\Delta M_{s}/\Delta M_{d}$ & \si & \no & \no & \no & \no \\
$\phi_{d,s}$ & \si & \no & \no & \si & \no \\
$\phi_s-\phi_d$ & \si & \no & \no & \no & \no \\
$C_{10}$ & \si & \no & \no & \si & \no \\
$C_{10}'$ & \si & \no & \no & \no & \no \\
\hline
$pp\to jj$ & \no & \si & \si & \no & \no \\
$pp\to q'q'$ & \si & \no & \no & \si & \si \\
\hline
\end{tabular}
\caption{Observables where NP effects could show up with realistic experimental and/or lattice improvements in the most favourable cases.}
\label{tab:dna}
\end{table}

The general message that emerges from Table~\ref{tab:mmin}, taken at face value,  is pretty clear. To accommodate top partners at or below 1 TeV is often not  possible and requires a judicious choice of the underlying model: an approximate $U(2)^3$ flavour symmetry appears favorite, if not necessary. Note that the bounds with a $*$ (bidoublet model with anarchic or  \UdueRC\  flavour structure) are obtained for a value of $Y \approx 2.5$, that minimizes the flavour and EWPT constraints consistently with $m_\psi = Y f$ and $f \gtrsim 0.5$ TeV. There are two simple reasons for the emergence of $U(2)^3$: i) in common with $U(3)^3$, the suppression of flavour changing effects in four-fermion operators with both left- and right-handed currents, present in the anarchic case; ii) contrary to $U(3)^3$ but as in anarchy, the disentanglement of the properties (their degree of compositeness) of the first and third generation of quarks.

The source of the constraint that plays the dominant role in the various cases is diverse. Sometimes more than one observable gives comparable constraints. This is reflected in Table~\ref{tab:dna}, which summarizes 
where possible new physics
effects could show up\footnote{The observables are, from top to bottom: the direct CP violating parameter in $K$-$\bar K$ mixing and the $B_d$ and $B_s$ mass differences (as well as their ratio), the mixing phases $\phi_d,\phi_s$ in the $B_d$ and $B_s$ systems (as well as their difference), the Wilson coefficient of the axial vector semi-leptonic operator relevant for $b\to s\ell^+\ell^-$ transitions $C_{10}$ and its chirality-flipped counterpart $C_{10}'$, the angular distribution of dijet events at LHC as discussed above and the direct production of fermion resonances at LHC.}(for some observables with more experimental
data, for others if  lattice parameters and/or other theoretical inputs are improved). We keep in this table every possible case even though  some of them, according to Table~\ref{tab:mmin},   would have to live with a fine tuned Higgs boson squared mass and, as such, appear less motivated.

The attempt to include many different possibilities, though motivated,  is also a limit of the analysis presented in this work. A next step might consist in selecting a few emerging cases to analyze them in more detail, perhaps going beyond the {\it partial-compositeness}  effective description. For this we think that Table~\ref{tab:mmin} offers a useful criterion.
It is in any event important and a priori non trivial that some models with a suitable structure emerge that look  capable of accommodating a 125 GeV Higgs boson without too much fine tuning, i.e. with top partners in an interesting mass range for discovery at the LHC.

\makeatletter
\def\toclevel@chapter{-1}
\makeatother

\setcounter{secnumdepth}{-1}

\chapter{Conclusion}

The idea that the hierarchy problem of the Fermi scale is solved in a natural way is currently under challenge. Both direct and indirect experimental searches for such a kind of new physics have so far given negative results. At the same time, the reach of these searches will be soon extended by the next generation of experiments. This situation motivates a careful study both of the compatibility of natural theories with current bounds, and of the signals where these theories are more likely to show up in the near future.
I have contributed to pursue this program with the work presented in this thesis. I have done so from a general point of view as well as in the specific contexts of Supersymmetry and composite Higgs models. Below a general summary of such a work is given, for more specific considerations see the concluding sections of Chapters \ref{cha:EFT}, \ref{cha:U2_SUSY}, \ref{cha:SUSY_Higgs} and \ref{cha:CHM}.

After a critical discussion of the hierarchy problem in Chapter \ref{cha:chapter1}, in the first part of the thesis we have concentrated our attention on flavour physics. In Chapter \ref{cha:SMCKM} we have presented the CKM picture in some detail. In particular we have discussed the logical steps that establish it as the dominant source of the observed flavour and CP violation.
Then in Chapter \ref{cha:EFT} we have considered the approximate $U(2)^3$ flavour symmetry of the quark sector to be a more fundamental symmetry of Nature, and explored the phenomenological consequences of this assumption from an effective field theory point of view. One can identify a minimal set of breaking parameters appearing in the Yukawa matrices, and assign them fictitious transformation properties under $U(2)^3$, in such a way that the Lagrangian be invariant under the flavour symmetry. The framework is then defined by assuming that any new source of flavour violation is controlled by their size, which is in turn fixed by the form of the CKM matrix.
The outcome is that a generic new physics at the scale of a TeV is compatible with all current experimental bounds, and that a rich phenomenology is potentially behind the corner. In particular, the processes that deserve more attention are CP violating $B$ decays, where $d$ and $s$ final states are expected to be correlated as in the SM, and where new phases coming from $B$ meson mixing could potentially appear, e.g. with deviations in $\beta_s$. Contrary to the SM and to Minimal Flavour Violation, in general these processes will not be correlated with the corresponding flavour violating $K$ decays. Other observables deserving attention are $|V_{ub}|$ and the CKM angle $\gamma_{\text{CKM}}$, a more precise determination of them could provide a crucial test of this framework.
These conclusions stay true if all the possible $U(2)^3$ breaking terms in the Yukawa matrices are taken into account. Contrary to the previous case (Minimal $U(2)^3$), in this one (Generic $U(2)^3$) the size of the extra breaking parameters is not determined a priori, but bounded by various experimental observations. We have also tried an extension of this framework with a $U(2)^2$ acting on the charged lepton sector, finding that it yields to lepton flavour violation at a level compatible, again, with new physics at the TeV scale.
The $U(2)^3$ framework is also appealing from the point of view of natural theories, since it allows to split the NP associated with the third generation of quarks from the one associated with the first two. This statement has been made more explicit in some of the following Chapters of the thesis, with the discussion of its realization in specific models.

The second part of the thesis has dealt with Supersymmetry. We have briefly introduced it in Chapter \ref{cha:SUSYintro}, focusing on the SUSY flavour and CP problems and on Supersymmetry as a solution to the hierarchy problem. In this respect we have presented the NMSSM as a most natural scenario.
Then, in Chapter \ref{cha:U2_SUSY}, we have discussed the embedding of the Minimal $U(2)^3$ framework within SUSY. This framework naturally allows to take sfermions of the first two generations to be heavier than the third generation ones. In this way the supersymmetric solution to the hierarchy problem is not spoiled by direct collider constraints, and a dynamical suppression of EDMs is obtained. We have then recalled the features of $\Delta F = 2$ amplitudes. As in the EFT case, a new phase appears in $B_d$ and in $B_s$ mixings, which are correlated. In addition, the requirement to solve the $\epsilon_K$--$S_{\psi K}$ tension in the CKM unitarity fit results in an upper bound on the sbottom and gluino masses, of roughly 1.5 TeV.
The main body of the Chapter is constituted by a study of the $\Delta B = 1$ amplitudes. Compatibly with all existing bounds, we have found potentially sizeable CP violating contributions even in the absence of flavour blind phases, as opposite to the MFV case. Peculiar correlations among these observables emerge, that could provide a ground to test this model in case some new signal is observed.
After flavour, in Chapter~\ref{cha:SUSY_Higgs} we have concentrated on the supersymmetric CP-even scalars. Despite perhaps for the LSP, they could well constitute the lightest new particles around. In a generic NMSSM we have derived analytical relations among the mixing angles of the three scalar states and the other physical parameters. We have then studied the impact of the measurements --current and foreseen-- of the Higgs signal strengths on the above parameter space. Electroweak precision tests turn out no to significantly affect the regions of interest.
In these same regions, we have studied the phenomenology of the Higgs bosons, identifying a possible strategy for their search at the LHC.
All the above analyisis is carried out with some motivated simplifying assumption on the CP-even scalars spectrum, specifically we decouple one of the three Higgs states at a time. This allows to retain an analytical control over the whole picture, in terms of a small number of parameters. If the singlet-like state is decoupled, the parameter space will be likely explored by the future Higgs coupling measurements. On the contrary, when the doublet-like state is decoupled, such measurements will have a much weaker impact on the parameter space.
In any case, we have always controlled the effect of lowering the mass of the third scalar, via the explicit relations we derived. In case some new signal is observed, an analysis including all the three states will be needed, as we have shown with the example of a $\gamma \gamma$ signal of a lighter singlet-like scalar.

Finally the third part of the thesis, consisting of Chapter \ref{cha:CHM}, has been dedicated to composite Higgs models with partial compositeness. The aim has been to determine to which level fermion resonances below roughly a TeV and a semiperturbative coupling, as required to yield naturally the observed Higgs mass, are compatible with the many indirect constraints. To do so, we have considered one level of resonances, for which we have written explicit Lagrangians in the cases of three possible motivated representations of the composite spin-1/2 states. For simplicity we have assumed all the fermion resonances to possess the same mass before EWSB. Three possibilities for the flavour structure of the strong sector have also been taken into account: anarchic, $U(2)^3$-symmetric, and $U(3)^3$-symmetric.
In the rest of the Chapter we have coherently analyzed indirect experimental constraints on each of the above combinations. An anarchic flavour structure is unlikely to be compatible with fermions lighter than $1\div1.5$ TeV, irrespectively of their representation. This is mainly caused by flavour constraints, and does not even take into account the much more severe bound coming from $\epsilon_K$, which might be accidentally suppressed.
In the case of a $U(3)^3$-symmetric strong sector, where flavour bounds are substantially alleviated, the most important constraints are due to the sizeable degree of compositeness of light quarks (either left- or right-handed ones). In particular, new contributions to electroweak precision observables and to dijet angular distributions make composite fermion lighter than about 3 TeV hard to conceive within this structure.
An $U(2)^3$ flavour structure possesses the virtues of both the previous ones: the suppression of flavour violating processes is accompained by the possibility to chose a low degree of compositeness for the quarks of the first two generations. Indeed this has turned out to be the only case where resonances lighter than a TeV can be accomodated, at least for two out of the three fermion representations examinated.
The above analysis has also allowed to determine the observables where each model will be more likely to show up, with realistic experimental and/or lattice improvements.







\chapter{Acknowledgements}
I am deeply grateful to prof. Riccardo Barbieri for the effort he put in educate me to critical judgement and thinking, and for his passionate and inspiring way of mentoring. I am indebted with him also for his expert guidance to approach physics problems, and for his constant support.
In these years I have had the lucky opportunity to work in close contact with David, from which I learnt a lot, Dario, and Andrea: thank you all, I really enjoyed it. Recently it has been a pleasure to work again with Alessandro, which I thank also for his support. For the same reason I am glad to thank Gino. Many thanks to Paolo, also for the many discussions, to Kristjan, and to the rest of the people I have collaborated with: Pier Paolo, Alberto, Giuseppe, and Gian Francesco.
Thanks to all the others that made the high energy physics group at the Scuola both formative and enjoyable: Enrico and Caterina and, in my first year, Enrico, Marco and Paolo.
I would like to express my gratitude to Scuola Normale Superiore for its support, and for the unique human environment that I found there. Thanks also to the Lawrence Berkeley National Laboratory for the very kind hospitality, and to all the people that made my time with the theoretical physics group of Berkeley so nice and stimulating.




%
%
%

\setcounter{secnumdepth}{1}

\appendix

\chapter{Quark bilinears and effective operators in $U(2)^3$}\label{app:bilinears}

\section*{Diagonalization of the quark masses}
\label{sec:Diagonalization}

The effective operators of \eqref{eq:genL} are constructed from the most generic quark bilinears which contain the spurions and are formally invariant under $U(2)^3$.

To a sufficient approximation, the chirality conserving bilinears take the form
\begin{align}\label{cc}
\bar q_{Li}\gamma^{\mu}(X_L^{\alpha})_{ij} q_{Lj} &= a_L^{\alpha}\bar q_{3L}\gamma^{\mu}q_{3L} + b_L^{\alpha}\qLbar\gamma^{\mu}\qL + c_L^{\alpha}\bar q_{3L}\gamma^{\mu}\V^{\dag}\qL\notag\\
&+ d_L^{\alpha}(\qLbar \V)\gamma^{\mu}(\V^{\dag}\qL) + {\rm h.c.},\\
\bar u_{Ri}\gamma^{\mu}(X_{uR}^{\alpha})_{ij} u_{Rj} &= a_{uR}^{\alpha}\bar t_R\gamma^{\mu}t_R + b_{uR}^{\alpha}\uRbar\gamma^{\mu}\uR + c_{uR}^{\alpha}\bar t_R\gamma^{\mu}\Vu^{\dag}\uR\notag\\
&+ d_{uR}^{\alpha}(\uRbar \Vu)\gamma^{\mu}(\Vu^{\dag}\uR) + {\rm h.c.}\label{ccR},
\end{align}
where an analogous expression holds for the right-handed down quarks, we denote by uppercase letters the light generation doublets $\qL, \uR, \dR$, and all the parameters except the c's are real by hermiticity. These bilinears give rise to the four-fermion operators $\Delta\mathcal{L}^{4f}_{L,R}$, to $\Delta\mathcal{L}_{LR}^{4f}$, as well as to the kinetic terms.

Analogously, the chirality breaking bilinears are, to lowest order in the spurions,
\begin{align}
\bar q_{Li}(M_u^{\beta})_{ij} u_{Rj} &= \lambda_t\Big(a_u^{\beta}\bar q_{3L}t_R + b_u^{\beta}(\qLbar\V)t_R + c_u^{\beta}\qLbar\Delta Y_u \uR + d_u^{\beta}\bar q_{3L}(\Vu^{\dag}\uR)\nonumber\\
&+ e_u^{\beta}(\qLbar\V)(\Vu^{\dag}\uR)\Big) + {\rm h.c.}\, , \label{cb_u}
\end{align}
with an analogous expression for the down-quark sector and where all the parameters are complex. They generate the interaction terms $\Delta\mathcal{L}^{mag}$, $\Delta\mathcal{L}_{LR}^{4f}$, as well as the Yukawa couplings $Y_u, Y_d$.

Consider now the basis where the spurions take the form \eqref{genericV}, \eqref{genericY}. Moreover, notice that all the parameters in the kinetic and Yukawa terms, except one, can be made real through rephasings of the fields. With these redefinitions, the previous operators can be written in the form
\begin{align}\label{ccmatrix}
X_{Lu}^{\rm kin} &= A_{uL}\mathbbm{1} + B_{uL} L_{23}^u\I_{32}^L (L_{23}^u)^{T}, & X_{Lu}^{\rm int,\alpha} &= A_{uL}^{\alpha}\mathbbm{1} + B_{uL}^{\alpha}U_{23}^{u,\alpha}\I_{32}^L (U_{23}^{u,\alpha})^{\dag},\\
X_{Ru}^{\rm kin} &= A_{uR}\mathbbm{1} + B_{uR} (R_{23}^u)^T\I_{32}^{Ru} R_{23}^u, & X_{Ru}^{\rm int,\alpha} &= A_{uR}^{\alpha}\mathbbm{1} + B_{uR}^{\alpha}(U_{23}^{u,\alpha})^{\dag}\I_{32}^{Ru} U_{23}^{u,\alpha},
\end{align}
plus analogous expressions for the down sector, where $\I_{32}^{I} = {\rm diag}(0, O(\epsilon^2_{I}), 1)$, the $A$'s and $B$'s are real functions of the parameters of \eqref{cc}, \eqref{ccR};
\begin{align}
Y_u &= \lambda_t (L_{23}^u\I_3 R_{23}^u + L_{12}^u\Delta \tilde Y^{\rm diag}_u V_{12}^u), & M_u^{\beta} &= \lambda_t(a_u^{\beta}{U}_{23}^{u,\beta}\I_3{V}_{23}^{u,\beta} + d_u^{\beta} L_{12}^u\Delta \tilde Y^{\rm diag}_u V_{12}^u),\\
Y_d &= \lambda_b (U_{23}^d\I_3 R_{23}^d + U_{12}^d\Delta \tilde Y^{\rm diag}_d V_{12}^d), & M_d^{\beta} &= \lambda_b(a_d^{\beta}{U}_{23}^{d,\beta}\I_3{V}_{23}^{d,\beta} + d_d^{\beta} U_{12}^d\Delta \tilde Y^{\rm diag}_d V_{12}^d),
\end{align}
where $\I_3 = {\rm diag}(0,0,1)$, $\Delta \tilde Y_{u,d}^{\rm diag} = {\rm diag} (y_{u,d}, y_{c,s},0)$, and $y_{u,d,c,s}$ are the diagonal entries of $\Delta Y_{u,d}^{\rm diag}$. Here and in the following $U_{ij}$ ($V_{ij}$) stand always for unitary left (right) matrices in the $(i,j)$ sector, while $L_{ij}$ ($R_{ij}$) indicate orthogonal left (right) matrices. In particular, in the notations of Section~\ref{sec:spurions}, $U_{12}^d = \Phi_L L_{12}^d$ and $V_{12}^{u,d} = \Phi_R^{u,d} R_{12}^{u,d}$.

We want to derive the expressions for these operators in the physical basis where the quark masses are diagonal, and the kinetic terms are canonical.
The kinetic terms are put in the canonical form by real rotations in the $(2,3)$ sector plus wavefunction renormalizations of the fields. One can check that these transformations do not alter, to a sufficient accuracy, the structure of the other operators, but cause only $O(1)$ redefinitions of the parameters.

The mass terms are diagonalized approximately by the transformation
\begin{align}
Y_u\mapsto Y_u^{\text{diag}} &= (L_{12}^u)^T(L_{23}^u)^T Y_u R_{23}^u R_{12}^u \equiv (L^u)^T Y_u R^u,\\
Y_d\mapsto Y_d^{\text{diag}} &=(U_{12}^d)^{\dag}(U_{23}^d)^{\dag} Y_d R_{23}^d V_{12}^d \equiv (U^d)^{\dag} Y_d V^d,
\end{align}
up to transformations of order $\epsilon_L y_{u,d,c,s}$, $\epsilon^u_R y_{u,c}$ and $\epsilon^d_R y_{d,s}$. Therefore one goes to the physical basis for the quarks by
\begin{align}
u_L&\mapsto L^u u_L, & d_L&\mapsto U^d d_L & u_R&\mapsto R^u u_R & d_R&\mapsto V^d d_R,
\end{align}
and the Cabibbo-Kobayashi-Maskawa matrix is
\begin{equation}
V_{CKM} \simeq (R_{12}^u)^T(R_{23}^u)^TU_{23}^d U_{12}^d\equiv (R_{12}^u)^T U_{23}^{\epsilon} U_{12}^d,
\end{equation}
where $U_{23}^{\epsilon}$ is a unitary transformation of order $\epsilon_L$.

In the physical mass basis the chirality conserving operators become
\begin{align}
X_{dL, \rm int}^{\alpha}&\mapsto A_{dL}^{\alpha}\mathbbm{1} + B_{dL}^{\alpha}(U_{12}^d)^{\dag} U_{23}^{d,\alpha}\I_{32}^L(U_{23}^{d,\alpha})^{\dag}U_{12}^d,\\
X_{dR, \rm int}^{\alpha}&\mapsto A^{\alpha}_{dR}\mathbbm{1} + B^{\alpha}_{dR}(V_{12}^d)^{\dag} V_{23}^{d,\alpha}\I_{32}^{Rd}(V_{23}^{d,\alpha})^{\dag}V_{12}^d,
\end{align}
and the $\sigma_{\mu\nu}$-terms are
\begin{equation}
M_d^{\beta}\mapsto \lambda_b\big(a_d^{\beta}(U_{12}^d)^{\dag} U_{23}^{\beta}\I_3 V_{23}^{\beta} V_{12}^d + c_d^{\beta} \Delta\tilde Y_d^{\rm diag}\big),
\end{equation}
plus analogous expressions for the up sector.

\section*{List of interaction bilinears}
\label{sec:AllBilinears}
The following results are obtained after rotation to the mass basis, and factorizing out explicitly all the phases, CKM matrix elements and quark masses. We define $\xi_{ij} = V_{ti}^*V_{tj}$, and $\zeta_{ij} = V_{ib}V_{jb}^*$. In chirality breaking bilinears $\alpha = \gamma(g)$ for (chromo)electric dipole operators, while $\alpha = cb$ for other generic interaction bilinears. All the parameters are real.

\subsection*{Up quark sector}
Chirality conserving LL, RR currents:
\begin{align}
X_{12}^{uL} &= c_D\zeta_{uc}, & X_{12}^{uR} &= \tilde c_D e^{i(\phi_1^u - \phi_2^u)}\zeta_{uc}\frac{s^u_R}{s^u_L}\left(\frac{\epsilon^u_R}{\epsilon_L}\right)^2,\\
X_{13}^{uL} &= c_t e^{i\phi_t}\zeta_{ut}, & X_{13}^{uR} &= \tilde c_t e^{i(\tilde\phi_t + \phi_1^u)}\zeta_{ut}\frac{s^u_R}{s^u_L}\frac{\epsilon^u_R}{\epsilon_L},\\
X_{23}^{uL} &= c_t e^{i\phi_t}\zeta_{ct}, & X_{23}^{uR} &= \tilde c_t e^{i(\tilde\phi_t + \phi_2^u)}\zeta_{ct}\frac{\epsilon^u_R}{\epsilon_L}.
\end{align}
Flavour conserving dipole operators:
\begin{align}
M_{11}^u &= c_D^{\rm \alpha}e^{i(\phi_D^{\rm \alpha} - \phi_1^u)}\zeta_{uu}\frac{s^u_R}{s^u_L}\frac{\epsilon^u_R}{\epsilon_L}, & M_{22}^u &= c_D^{\rm \alpha}e^{i(\phi_D^{\rm \alpha} - \phi_2^u)}\zeta_{cc}\frac{\epsilon^u_R}{\epsilon_L}, & M_{33}^u &= a_t e^{i\alpha_t}.
\end{align}
Flavour changing, chirality breaking operators:
\begin{align}
M_{12}^u &= c_D^{\rm \alpha}e^{i(\phi_D^{\rm \alpha} - \phi_2^u)}\zeta_{uc}\frac{\epsilon^u_R}{\epsilon_L}, & M_{21}^u  &= c_D^{\rm \alpha}e^{i(\phi_D^{\rm \alpha} - \phi_1^u)}\zeta_{uc}^*\frac{s^u_R}{s^u_L}\frac{\epsilon^u_R}{\epsilon_L},\\
M_{13}^u &= c_t^{\rm \alpha}e^{i\phi_t^{\rm \alpha}}\zeta_{ut}, & M_{31}^u &= \tilde c_t^{\rm \alpha}e^{i(\tilde\phi_t^{\rm \alpha} - \phi_1^u)}\zeta_{ut}^*\frac{s^u_R}{s^u_L}\frac{\epsilon^u_R}{\epsilon_L},\\
M_{23}^u &= c_t^{\rm \alpha}e^{i\phi_t^{\rm \alpha}}\zeta_{ct}, & M_{32}^u &= \tilde c_t^{\rm \alpha}e^{i(\tilde\phi_t^{\rm \alpha} - \phi_2^u)}\zeta_{ct}^*\frac{\epsilon^u_R}{\epsilon_L}.
\end{align}

\subsection*{Down quark sector}
Chirality conserving LL, RR currents:
\begin{align}
X_{12}^{dL} &= c_K\xi_{ds}, & X_{12}^{dR} &= \tilde c_K e^{i(\phi_1^d - \phi_2^d)}\xi_{ds}\frac{s^d_R}{s^d_L}\left(\frac{\epsilon^d_R}{\epsilon_L}\right)^2,\\
X_{13}^{dL} &= c_B e^{i\phi_B}\xi_{db}, & X_{13}^{dR} &= \tilde c_B e^{i(\tilde\phi_B + \phi_1^d)}\xi_{db}\frac{s^d_R}{s^d_L}\frac{\epsilon^d_R}{\epsilon_L},\\
X_{23}^{dL} &= c_B e^{i\phi_B}\xi_{sb}, & X_{23}^{dR} &= \tilde c_B e^{i(\tilde\phi_B + \phi_2^d)}\xi_{sb}\frac{\epsilon^d_R}{\epsilon_L}.
\end{align}
Flavour conserving dipole operators:
\begin{align}
M_{11}^d &= \lambda_b\,c_K^{\rm \alpha}e^{i(\phi_K^{\rm \alpha} - \phi_1^d)}\xi_{dd}\frac{s^d_R}{s^d_L}\frac{\epsilon^d_R}{\epsilon_L}, & M_{22}^d &= \lambda_b\,c_K^{\rm \alpha}e^{i(\phi_K^{\rm \alpha} - \phi_2^d)}\xi_{ss}\frac{\epsilon^d_R}{\epsilon_L}, & M_{33}^d &= \lambda_b\,a_b e^{i\alpha_b}.
\end{align}
Flavour changing, chirality breaking operators:
\begin{align}
M_{12}^d &= \lambda_b\,c_K^{\rm \alpha}e^{i(\phi_K^{\rm \alpha} - \phi_2^d)}\xi_{ds}\frac{\epsilon^d_R}{\epsilon_L}, & M_{21}^d  &= \lambda_b\,c_K^{\rm \alpha}e^{i(\phi_K^{\rm \alpha} - \phi_1^d)}\xi_{ds}^*\frac{s^d_R}{s^d_L}\frac{\epsilon^d_R}{\epsilon_L},\\
M_{13}^d &= \lambda_b\,c_B^{\rm \alpha}e^{i\phi_B^{\rm \alpha}}\xi_{db}, & M_{31}^d &= \lambda_b\,\tilde c_B^{\rm \alpha}e^{i(\tilde\phi_B^{\rm \alpha} - \phi_1^d)}\xi_{db}^*\frac{s^d_R}{s^d_L}\frac{\epsilon^d_R}{\epsilon_L},\\
M_{23}^d &= \lambda_b\,c_B^{\rm \alpha}e^{i\phi_B^{\rm \alpha}}\xi_{sb}, & M_{32}^d &= \lambda_b\,\tilde c_B^{\rm \alpha}e^{i(\tilde\phi_B^{\rm \alpha} - \phi_2^d)}\xi_{sb}^*\frac{\epsilon^d_R}{\epsilon_L}.
\end{align}

\chapter{Coefficients of four fermion operators in CHMs}
\label{app:4fermion}

\section*{Tree-level $\Delta F=2$ FCNCs}\label{sec:app-flavour}

In a model with flavour anarchy in the strong sector, the coefficients defined in Eq.~(\ref{eq:DF2}) can be written as
\begin{align}
c_V^{dLL} &= -\frac{1}{6} -\frac{1}{2}\left[X(Q)^2+T_{3L}(Q)^2+T_{3R}(Q)^2\right]
\,,\\
c_V^{dRR} &= -\frac{1}{6} -\frac{1}{2}\left[X(D)^2+T_{3L}(D)^2+T_{3R}(D)^2\right]
\,,\\
c_V^{dLR} &= \frac{1}{6}-\left[X(Q)X(D)+T_{3L}(Q)T_{3L}(D)+T_{3R}(Q)T_{3R}(D)\right]
\,,\\
c_S^{dLR} &= 1
\,,
\end{align}
where the first terms come from heavy gluon exchange and the terms in brackets from neutral heavy gauge boson exchange. $Q$ refers to the charge $-1/3$ fermion mixing with $q$ and $D$ to the charge $-1/3$ fermion mixing with $d_R$. In the bidoublet model, we consider only the contribution from $Q_u$ for simplicity, which is enhanced if $z>1$.
The numerical coefficients relevant for the models discussed above are collected in Table~\ref{tab:df2coeffs}.

In $U(3)^3$ there is an additional contribution from flavour gauge bosons. However the only relevant $\Delta F = 2$ operator is $Q_V^{dLL}$ in \UtreRC, for which one obtains $c_V^{dLL} = -{29}/{36}$ instead of the value reported in the table.

In $U(2)^3$, since all the flavour effects are generated by mixing with third generation partners, which are not charged under any of the $U(2)$ flavour groups, there is no relevant additional effect coming from flavour gauge bosons, and the coefficients of Table \ref{tab:df2coeffs} are valid.

\begin{table}[t]
\renewcommand{\arraystretch}{1.5}
\centering%
\begin{tabular}{cccc}
\hline
 & doublet & triplet & bidoublet\\
\hline
$c_V^{dLL}$ & $-\frac{11}{36}$ & $-\frac{23}{36}$ & $-\frac{23}{36}$\\
$c_V^{dRR}$ & $-\frac{11}{36}$ & $-\frac{8}{9}$ & $-\frac{2}{9}$\\
$c_V^{dLR}$ & $\frac{5}{36}$ & $-\frac{7}{9}$ & $\frac{7}{18}$\\
$c_S^{dLR}$ & 1 & 1 & 1\\
\hline
\end{tabular}
\caption{Coefficients relevant for $\Delta F = 2$ operators in anarchy and $U(2)^3$.}
\label{tab:df2coeffs}
\end{table}

\section*{Compositeness constraints from the dijet angular distribution}\label{sec:app-dijet-angular}

Exchanges of gauge resonances and flavour gauge bosons give rise to four-fermion operators involving only the first generation which contribute to the angular distribution of high-mass dijets at LHC. As shown in \cite{Domenech:2012ai}, only ten operators have to be considered, which we list here for convenience
\begin{align}
\mathcal O^{(1)}_{uu}&=(\bar{u}_R\gamma^\mu u_R)(\bar{u}_R\gamma_\mu u_R)~,
&
\mathcal O^{(1)}_{dd}&=(\bar{d}_R\gamma^\mu d_R)(\bar{d}_R\gamma_\mu d_R)~,
\nonumber\\
\mathcal O^{(1)}_{ud}&=(\bar{u}_R\gamma^\mu u_R)(\bar{d}_R\gamma_\mu d_R)~,
&
\mathcal O^{(8)}_{ud}&=(\bar{u}_R\gamma^\mu T^A u_R)(\bar{d}_R\gamma_\mu T^A d_R)~,
\nonumber\\
\mathcal O^{(1)}_{qq}&=(\bar{q}_L\gamma^\mu q_L)(\bar{q}_L\gamma_\mu q_L)~,
&
\mathcal O^{(8)}_{qq}&=(\bar{q}_L\gamma^\mu T^A q_L)(\bar{q}_L\gamma_\mu T^A q_L)~,
\nonumber\\
\mathcal O^{(1)}_{qu} &= (\bar{q}_{L} \gamma^\mu q_{L}) (\bar{u}_{R} \gamma_{\mu}  u_R)~,
&
\mathcal O^{(8)}_{qu} &= (\bar{q}_{L} \gamma^\mu T^A q_{L}) (\bar{u}_{R} \gamma_{\mu}  T^A u_R)~,
\nonumber\\
\mathcal O^{(1)}_{qd} &= (\bar{q}_{L} \gamma^\mu q_{L}) (\bar{d}_{R} \gamma_{\mu}  d_R)~,
&
\mathcal O^{(8)}_{qd} &= (\bar{q}_{L} \gamma^\mu T^A q_{L}) (\bar{d}_{R} \gamma_{\mu}  T^A d_R)~.
\label{eq:jj-operators}
\end{align}
The coupling of a first generation quark mass eigenstate to a heavy vector resonance receives contributions from fermion composite-elementary mixing as well as vector boson composite-elementary mixing. For example, the coupling of the up quark to the gluon resonance reads
\begin{equation}
\bar u \gamma^\mu T^a\left(
g_\rho s_{Lu}^2 P_L +g_\rho s_{Ru}^2 P_R + \frac{g_3^2}{g_\rho}
\right)u G^*_\mu ~.
\label{eq:gG}
\end{equation}
Neglecting electroweak gauge couplings, one can then write the Wilson coefficients of the above operators as
\begin{equation}
C_{ab}^{(1,8)} = \frac{g_\rho^2}{m_\rho^2}\left[
s_{a}^2
s_{b}^2
c_{ab}^{(1,8)}
+
\left(
\frac{g_3^4}{g_\rho^4}
-(s_{a}^2+s_{b}^2)
\frac{g_3^2}{g_\rho^2}
\right)
d_{ab}^{(1,8)}
\right],
\end{equation}
where $(a,b)=(q,u,d)$ and $s_{u,d}^2\equiv s_{Ru,d}^2$, $s_q^2\equiv s_{L}^2$ (in the bidoublet model, for simplicity we will neglect terms with $s_{Ld}^2$ over terms with $s_{Lu}^2$).
The numerical coefficients $c_{ab}^{(1,8)}$ depend on the electroweak structure and on the flavour group and are collected in Table~\ref{tab:dijet} together with the $d_{ab}^{(1,8)}$.

\begin{table}[tbp]
\renewcommand{\arraystretch}{1.5}
\centering
\begin{tabular}{lcccccccccc}
\hline
&$c_{uu}^{(1)}$ & $c_{dd}^{(1)}$ & $c_{ud}^{(1)}$ & $c_{ud}^{(8)}$ &
$c_{qq}^{(1)}$ & $c_{qq}^{(8)}$ & $c_{qu}^{(1)}$ & $c_{qu}^{(8)}$ & $c_{qd}^{(1)}$ & $c_{qd}^{(8)}$ \\
\hline
doublet $U(3)^3_\text{LC}$ & $-\frac{17}{36}$ & $-\frac{17}{36}$ & $-\frac{1}{9}$ & $-1$ & $-\frac{5}{36}$ & $-1$ & $-\frac{13}{36}$ & $-1$ & $-\frac{13}{36}$ & $-1$ \\
triplet $U(3)^3_\text{LC}$& $-\frac{5}{9}$ & $-\frac{19}{18}$ & $-\frac{7}{9}$ & $-1$ & $-\frac{17}{36}$ & $-1$ & $-\frac{7}{9}$ & $-1$ & $-\frac{7}{9}$ & $-1$ \\
bidoublet $U(3)^3_\text{LC}$ &  $-\frac{5}{9}$ & $-\frac{7}{18}$ & $-\frac{1}{9}$ & $-1$ & $-\frac{17}{36}$ & $-1$ & $-\frac{7}{9}$ & $-1$ & $-\frac{1}{9}$ & $-1$ \\
bidoublet $U(3)^3_\text{RC}$ & $-\frac{5}{9}$ & $-\frac{7}{18}$ & $\frac{2}{9}$ & $-1$ & $-\frac{17}{36}$ & $-1$ & $-\frac{7}{9}$ & $-1$ & $\frac{2}{9}$ & $-1$ \\
\hline
\hline
&$d_{uu}^{(1)}$ & $d_{dd}^{(1)}$ & $d_{ud}^{(1)}$ & $d_{ud}^{(8)}$ &
$d_{qq}^{(1)}$ & $d_{qq}^{(8)}$ & $d_{qu}^{(1)}$ & $d_{qu}^{(8)}$ & $d_{qd}^{(1)}$ & $d_{qd}^{(8)}$ \\
\hline
all models & $-\frac{1}{6}$ & $-\frac{1}{6}$ & $0$ & $-1$ & $0$ & $-\frac{1}{2}$ & $0$ & $-1$ & $0$ & $-1$\\
\hline
\end{tabular}
\caption{Coefficients $c_{ab}^{(1,8)}$ relevant for dijet bounds in the doublet, triplet and bidoublet models as well as the coefficients $d_{ab}^{(1,8)}$, which are independent of the flavour and electroweak groups.}
\label{tab:dijet}
\end{table}

\section*{Production and decay of vector resonances}\label{sec:app-dijet}

The production cross section of a gluon resonance in $pp$ collisions reads
\begin{align}
\sigma(pp\to G^*) &=  \frac{2\pi}{9s}
\left[
(|g_L^u|^2+|g_R^u|^2) \mathcal L_{u\bar u}(s,m_\rho^2)
+
(|g_L^d|^2+|g_R^d|^2) \mathcal L_{d\bar d}(s,m_\rho^2)
\right],
\label{eq:sigmapprho}
\end{align}
where
\begin{equation}
\mathcal L_{q\bar q}(s,\hat s)=
\int_{\hat s/s}^1 \frac{dx}{x}
f_q(x,\mu)
f_{\bar q}\!\left(\frac{\hat s}{xs},\mu\right)
\end{equation}
is the parton-parton luminosity function at partonic (hadronic) center of mass energy $\sqrt{\hat s}$ ($\sqrt{s}$)
and the couplings $g_{L,R}^{u,d}$ are defined as $\mathcal L \supset \bar u_L \gamma^\mu T^a g_L^u
u_L G^*_\mu$ and can be read off Eq.~\ref{eq:gG}. Again, there is a contribution due to fermion mixing, which is only relevant in $U(3)^3$ models due to the potentially sizable compositeness of the first generation, while the contribution due to vector mixing is always present.
The total width reads
\begin{equation}
\Gamma(G^*\to q \bar q) = \sum_{q=u,d}\sum_{i=1}^3\frac{m_\rho}{48\pi}\left(|g^{q^i}_L|^2+|g^{q^i}_R|^2\right)
\end{equation}
while the branching ratio to dijets is simply the width without the top contribution divided by the total width\footnote{Neglecting the top quark mass in the kinematics, which is a good approximation for multi-TeV resonances still allowed by the constraints}.

\chapter{The T parameter in CHM}
\label{app:Tparameter}
We compute the one-loop contributions to $\Delta \rho = \hat T$ arising from top partner loops in models with partial compositeness. We do the calculation in the elementary-composite basis for fermions and vector bosons.

We consider a composite sector which has a custodial $SU(2)_L\times SU(2)_R$ symmetry, and the corresponding $\rho_L$ and $\rho_R$ vector resonances. After electroweak symmetry breaking, the elementary $W$ boson mixing insertions with $\rho_L$ and $\rho_R$ are given by, respectively ($v = 174$~GeV):
\begin{equation}\label{mixings}
  i\frac{g}{g_{\rho}}m_{\rho}^2\equiv -im_{\rho}^2\Delta_L \qquad \text{and} \qquad
  i\frac{g}{g_{\rho}}\frac{v^2 g_{\rho}^2}{2}\equiv -im_{\rho}^2\Delta_R\,.
\end{equation}
We'll consider degenerate fermion masses $M_f\simeq Y f$ and degenerate vector masses $m_{\rho}\simeq g_{\rho} f$.

Since the composite sector is symmetric under the custodial $SU(2)_L\times SU(2)_R$, elementary fermion insertions are needed in the loop in order to have a non-vanishing contribution. Moreover at least four Higgs insertions are needed overall to generate the operator $\mathcal{O}_H = (H^{\dag}D_{\mu}H)^2$.

The contributions from the following two point functions contribute to $\hat T$ after elementary-composite mixings on the external legs:
\begin{itemize}
\item $\hat T^{LL}$ from $W$-$\rho_L$ mixing on both external legs of $\langle\rho_L\rho_L\rangle$: four Higgs insertions inside the fermion loop are needed;
\item $\hat T^{WL}$ from $W$-$\rho_L$ mixing on the $\rho_L$ leg of $\langle W\rho_L\rangle$ and $\langle\rho_L W\rangle$: four Higgs insertions inside the fermion loop;
\item $\hat T^{RR}$ from $W$-$\rho_R$ mixing on both external legs of $\langle\rho_R\rho_R\rangle$: all the Higgs insertions are in the $L$-$R$ mixings;
\item $\hat T^{LR}$ from $W$-$\rho_L$ and $W$-$\rho_R$ mixing on the external legs of $\langle\rho_L\rho_R\rangle$ and $\langle\rho_R\rho_L\rangle$: two Higgs insertions are in the $L$-$R$ mixing and two are in the fermion loop;
\item $\hat T^{WR}$ from $W$-$\rho_R$ mixing on the $\rho_R$ leg of $\langle W\rho_R\rangle$ and $\langle\rho_R W\rangle$: two Higgs insertions are in the $L$-$R$ mixing and two are in the fermion loop.
\end{itemize}

We calculate the leading terms in $v/f$ and in the elementary/composite quark mixings in the one-doublet (1D), two-bidoublets (2B) and triplet (T) models. We consider only third-generation quarks, since all the other mixings are negligible.

\section*{One-doublet model}
The composite quarks are one doublet $Q$ under $SU(2)_L$ and one doublet $\mathcal{Q}$ under $SU(2)_R$ which couple respectively to the elementary quarks $q_L$ and $u_R,d_R$.
\begin{align}
Q &= \begin{pmatrix}T\\ B\end{pmatrix},& D_{\mu}Q = \partial_{\mu}Q - ig_{\rho}(\rho_L)_{\mu}^a T_L^a Q,\\
\mathcal{Q} &= \begin{pmatrix}U\\ D\end{pmatrix},& D_{\mu}\mathcal{Q} = \partial_{\mu}\mathcal{Q} - ig_{\rho}(\rho_R)_{\mu}^a T^a_R \mathcal{Q}.
\end{align}
The lagrangian reads
\begin{equation}\begin{aligned}
\L &= \frac{g_{\rho}}{2}\rho_L^3(\bar T T - \bar B B) + \frac{g_{\rho}}{\sqrt{2}}\rho_L^- \bar T B + \frac{g_{\rho}}{2}\rho_R^3(\bar U U - \bar D D) + \frac{g_{\rho}}{\sqrt{2}}\rho_R^- \bar U D\\
& + Yv(\bar T U + \bar B D) + M_f\lambda_L (\bar u_L T + \bar d_L B) + M_f\lambda_R \bar u_R U + {\rm h.c.}\,.
\end{aligned}\end{equation}
In the limit $\tilde Y\to 0$ (only standard Yukawas) one gets
\begin{align}
\hat T^{LL}_{1D} &= \frac{13}{105}\frac{N_c}{(4 \pi)^2}\frac{m_t^2 Y^2}{M_f^2 x_t^2}, & \hat T^{WL}_{1D} &= 0,\\
\hat T^{LR}_{1D} &= \frac{13}{60}\frac{N_c}{(4 \pi)^2}\frac{m_t^2 Y^2}{M_f^2 x_t^2} & \hat T^{WR}_{1D} &= 0,\\
\hat T^{RR}_{1D} &= \frac{1}{6}\frac{N_c}{(4 \pi)^2}\frac{m_t^2 Y^2}{M_f^2 x_t^2},
\end{align}
and the total contribution is thus
\begin{empheq}[box={\fboxsep=10pt\ovalbox}]{align}
\hat T_{1D} = \frac{71}{140}\frac{N_c}{(4 \pi)^2}\frac{m_t^2 Y^2}{M_f^2 x_t^2}.
\end{empheq}
Including also the $\tilde Y$ terms one has
\begin{equation}
\hat T_{1D} = \frac{N_c}{(4 \pi)^2}\frac{m_t^2 Y^2}{M_f^2 x_t^2}\left(\frac{1}{210}(26 + 6 X + 9 X^2 - X^3 + 5 X^4) + R^2(13 + X + 8 X^2) + \frac{R^4}{6}\right),
\end{equation}
where $X\equiv \tilde Y/Y$ and $R\equiv M_f / (Y f)$.

\section*{Two-bidoublets model}
The composite quarks are two $SU(2)_L\times SU(2)_R$ bidoublets $Q^u$ and $Q^d$ which mix with $q_L$, and two singlets $U$ and $D$ which mix with $u_R$ and $d_R$ respectively. Here we neglect all terms with $Q^d$, since they are suppressed by factors $z = \lambda_L^u/\lambda_L^d$.
\begin{align}
Q^u &= \begin{pmatrix}T & X\\ B & T'\end{pmatrix}, & D_{\mu}Q^u &= \partial_{\mu}Q^u - i g_{\rho}(\rho_L)_{\mu}^a T_L^a Q^u + ig_{\rho}(\rho_R)_{\mu}^a Q^u T_R^a.
\end{align}
The lagrangian reads
\begin{equation}\begin{aligned}
\L &= \frac{g_{\rho}}{2}\rho_L^3(\bar T T - \bar B B + \bar X X - \bar T' T') + \frac{g_{\rho}}{\sqrt{2}}\rho_L^-(\bar T B + \bar X T')\\
& +\frac{g_{\rho}}{2}\rho_R^3(\bar X X + \bar T' T' - \bar T T - \bar B B) - \frac{g_{\rho}}{\sqrt{2}}\rho_R^-(\bar X T + \bar T' B)\\
& +Yv(\bar T + \bar T')U + M_f\lambda_L (\bar u_L T + \bar d_L B) + \lambda_R \bar u_R U + {\rm h.c.}\,.
\end{aligned}\end{equation}
In the limit $\tilde Y\to 0$ one gets
\begin{align}
\hat T^{LL}_{2B} &= -\frac{4}{105}\frac{N_c}{(4 \pi)^2}\frac{m_t^2 x_t^2 Y^2}{M_f^2}, & \hat T^{WL}_{2B} &= \frac{1}{15}\frac{N_c}{(4 \pi)^2}\frac{m_t^2 x_t^2 Y^2}{M_f^2},\\
\hat T^{LR}_{2B} &= -\frac{1}{5}\frac{N_c}{(4 \pi)^2}\frac{m_t^2 x_t^2 Y^2}{M_f^2}, & \hat T^{WR}_{2B} &= -\frac{5}{12}\frac{N_c}{(4 \pi)^2}\frac{m_t^2 x_t^2 Y^2}{M_f^2},\\
\hat T^{RR}_{2B} &= \frac{1}{3}\frac{N_c}{(4 \pi)^2}\frac{m_t^2 x_t^2 Y^2}{M_f^2},
\end{align}
and the total contribution is
\begin{empheq}[box={\fboxsep=10pt\ovalbox}]{align}
\hat T_{2B} = -\frac{107}{420}\frac{N_c}{(4 \pi)^2}\frac{m_t^2 x_t^2 Y^2}{M_f^2}.
\end{empheq}
Including also the $\tilde Y$ terms one has
\begin{equation}
\hat T_{2B} = \frac{N_c}{(4 \pi)^2}\frac{m_t^2 Y^2 x_t^2}{M_f^2}\left(\frac{12 + 272 X + 485 X^2 - 302 X^3 + 82 X^4}{420} - R^2\frac{37 + 98 X - 24 X^2}{60} + \frac{R^4}{3}\right),
\end{equation}
where $X\equiv \tilde Y/Y$ and $R\equiv M_f / (Y f)$.

\section*{Triplet model}
The composite quarks are one bidoublet $Q\sim ({\bf 2},{\bf 2})$ that mixes with $q_L$ and two triplets $\T^l\sim ({\bf 3},{\bf 1})$ and $\T^r\sim({\bf 1},{\bf 3})$; the $SU(2)_R$-triplet $\T^r$ mixes with $u_R$ and $d_R$.
\begin{align}
Q^u &= \begin{pmatrix}T & X\\ B & T'\end{pmatrix}, & D_{\mu}Q^u &= \partial_{\mu}Q^u - i g_{\rho}(\rho_L)_{\mu}^a T_L^a Q^u + ig_{\rho}(\rho_R)_{\mu}^a Q^u T_R^a,\\
\T^l &= \begin{pmatrix}U^l/\sqrt{2} & -X^l\\D^l & -U^l/\sqrt{2}\end{pmatrix}, & D_{\mu}\T^l &= \partial_{\mu}\T^l - ig_{\rho}(\rho_L)_{\mu}^a [T_L^a\T^l],\\
\T^r &= \begin{pmatrix}U^r/\sqrt{2} & -X^r\\D^r & -U^r/\sqrt{2}\end{pmatrix}, & D_{\mu}\T^r &= \partial_{\mu}\T^r - ig_{\rho}(\rho_R)_{\mu}^a [T_R^a\T^r].
\end{align}
The lagrangian reads
\begin{equation}\begin{aligned}\label{triplet}
\L &=g_{\rho}\rho_L^3\left[\frac{1}{2}(\bar T T - \bar B B + \bar X X - \bar T' T') + \bar X^l X^l - \bar D^l D^l\right]\\
& + g_{\rho}\rho_R^3\left[\frac{1}{2}(\bar X X + \bar T' T' - \bar T T - \bar B B) + \bar X^r X^r - \bar D^r D^r\right]\\
& +g_{\rho}\rho_L^-\left[\bar X^l U^l + \bar U^l D^l + \frac{\bar T B + \bar X T'}{\sqrt{2}}\right] + g_{\rho}\rho_R^-\left[\bar X^r U^r + \bar U^r D^r - \frac{\bar X T + \bar T' B}{\sqrt{2}}\right]\\
& + Yv\left[\frac{\bar T - \bar T'}{\sqrt{2}}(U^l + U^r) + \bar B(D^l + D^r) - \bar X (X^l + X^r)\right]\\
& + M_f\lambda_L(\bar u_L T + \bar d_L B) + M_f\lambda_R \bar u_R U^r  + {\rm h.c.}
\end{aligned}\end{equation}

\subsection*{$\lambda_R^2$ contributions}
The leading contributions are proportional to $\lambda_R^2$. In the limit of vanishing $\tilde Y$ one gets:
\begin{align}
\hat T^{LL}_{T,R} &= -\frac{1}{2}\frac{N_c}{(4 \pi)^2}\frac{m_t^2 Y^3}{M_f^2 y_t x_t}, & \hat T^{WL}_{T,R} &= 0,\\
\hat T^{LR}_{T,R} &= 0, & \hat T^{WR}_{T,R} &= 0,\\
\hat T^{RR}_{T,R} &= \frac{N_c}{(4 \pi)^2}\frac{m_t^2 Y^3}{M_f^2 y_t x_t}\log\Big(\frac{\Lambda^2}{M_f^2} + 1\Big),
\end{align}
and the total $\lambda_R^2$ contribution is
\begin{empheq}[box={\fboxsep=10pt\ovalbox}]{align}\label{lambda2}
\hat T_{T,R} = \left(\log\Big(\frac{\Lambda^2}{M_f^2} + 1\Big) - \frac{1}{2}\right)\frac{N_c}{(4 \pi)^2}\frac{m_t^2 Y^3}{M_f^2 y_t x_t}.
\end{empheq}
Including also the $\tilde Y$ terms one has
\begin{equation}\begin{aligned}
\hat T_{T,R} = \frac{N_c}{(4 \pi)^2}\frac{m_t^2 Y^3}{M_f^2 y_t x_t}&\left(-\frac{15 - 2 X^2 - 26 X^2 + 24 X^3 - 17 X^4}{480}\right.\\
&\left.\quad+ \frac{R^2}{48}X(5X - 1) + R^4\log\Big(\frac{\Lambda^2}{M_f^2} + 1\Big)\right).
\end{aligned}\end{equation}

\subsection*{$\lambda_L^4$ contributions}
There are many diagrams that give contributions to $\hat T$ proportional to $\lambda_L^4$. In the limit $\tilde Y\to 0$ one gets
\begin{align}
\hat T^{LL}_{T,L} &= \frac{3}{7}\frac{N_c}{(4 \pi)^2}\frac{m_t^2 Y^2 x_t^2}{M_f^2}, & \hat T^{WL}_{T,L} &= \frac{8}{15}\frac{N_c}{(4 \pi)^2}\frac{m_t^2 x_t^2 Y^2}{M_f^2},\\
\hat T^{LR}_{T,L} &= \frac{19}{30}\frac{N_c}{(4 \pi)^2}\frac{m_t^2 x_t^2 Y^2}{M_f^2}, & \hat T^{WR}_{T,L} &= \frac{5}{12}\frac{N_c}{(4 \pi)^2}\frac{m_t^2 x_t^2 Y^2}{M_f^2},\\
\hat T^{RR}_{T,L} &= \frac{1}{3}\frac{N_c}{(4 \pi)^2}\frac{m_t^2 x_t^2 Y^2}{M_f^2},
\end{align}
and the total $\lambda_L^4$ contribution is
\begin{empheq}[box={\fboxsep=10pt\ovalbox}]{align}
\hat T_{T,L} = \frac{197}{84}\frac{N_c}{(4 \pi)^2}\frac{m_t^2 x_t^2 Y^2}{M_f^2}.
\end{empheq}
Including also the $\tilde Y$ terms one has
\begin{equation}\begin{aligned}
\hat T_{T,L} = \frac{N_c}{(4 \pi)^2}\frac{m_t^2 Y^2 x_t^2}{M_f^2}&\left(\frac{404 + 1168 X + 737 X^2 - 78 X^3 - 2 X^4}{420}\right.\\
&\left.\quad + R^2\frac{63 + 101 X - 22 X^2}{60} + \frac{R^4}{3}\right).
\end{aligned}\end{equation}

\subsection*{Interpretation of the divergence}
We can treat the $SU(2)_R$ symmetry as a gauge symmetry associated with the $\rho_R$ vector bosons. The divergence appearing in \eqref{lambda2}, which comes from $\Pi_{RR} = \Pi_{\rho^3_R\rho^3_R} - \Pi_{\rho^+_R\rho^-_R}$, can then be understood as a consequence of the explicit breaking of $SU(2)_R$ by $\lambda_R$.

We want to determine the scale $\Lambda$ that appears in the logarithm in \eqref{lambda2}. A way to do that is to construct a renormalizable completion of the theory where $\Pi_{RR}$ is finite, and match the divergences obtained integrating out the new degrees of freedom.

Let's introduce a right-handed ``Higgs'' doublet $\lambda\sim ({\bf 1}, {\bf 2})$ which generates $\lambda_R$ and the $\rho_R$ masses when getting a vev. In the three cases 1D, 2B ,T one has
\begin{align}\label{UV}
\L_T &= \bar\T^r (\slashed{D} - M_f)\T^r + |D_{\mu}\lambda|^2 - V(\lambda) + \frac{1}{M_F}(\lambda^{\dag}\sigma^a\lambda)\bar\T^r_a u_R,\\
\L_{1D} &= \bar\Q(\slashed{D} - M_f)\Q + |D_{\mu}\lambda|^2 - V(\lambda) + (\bar \Q\lambda)u_R,\\
\L_{2B} &= \bar Q(\slashed{D} - M_f)Q + |D_{\mu}\lambda|^2 - V(\lambda) + (\bar Q\lambda)q_L.
\end{align}
In the 1D and 2B cases, where we have a renormalizable $SU(2)_R$-invariant lagrangian, $\Pi_{RR}$ has to be finite since there is no mass splitting $\rho_R^3-\rho_R^+$ at tree-level. In the triplet case, insertions of the dimension 5 operator generate the divergent terms.

The model becomes renormalizable introducing a new heavy fermion $F = (U,D)\sim (\bf{1},\bf{2})$
\begin{equation}
\L = \bar\T^r (\slashed{D} - M_f)\T^r + |D_{\mu}\lambda|^2 - V(\lambda) + \bar F(\slashed{D} - M_F)F + Y_F(\bar F\sigma^a\lambda)\T_a^r + Y'_F (\bar F\lambda)u_R.
\end{equation}
After the breaking of $SU(2)_R$, $\langle \lambda\rangle = (f,0)$ and
\begin{equation}\label{broken}
\L_{\slashed{R}} = \bar\T^r (\slashed{D} - M_f)\T^r + |D_{\mu}\lambda|^2 - V(\lambda) + \bar F(\slashed{D} - M_F)F + Y_F f(\bar U U^r + \bar D D^r) + Y'_F f\bar Uu_R.
\end{equation}
Now, integrating out the field $F$ one recovers the original mass mixing
\begin{equation}\label{lambdaR}
\lambda_R = \frac{Y_F Y'_F f^2}{M_f M_F}.
\end{equation}

We can calculate the contribution to $\hat T$ coming from $\Pi_{RR}$ in the model \eqref{broken}. In the limit of large $M_F$ we obtain
\begin{equation}\label{TF}
\hat T_F = \frac{N_c}{(4\pi)^2}\frac{m_t^2 Y^3}{M_f^2 y_t x_t}\left(\frac{1+S^2}{2}\log\Big(\frac{M_F^2}{M_f^2}\Big) + \frac{1}{48}(8 - 22 S^2 - 47 S^4)\right),
\end{equation}
where $S = Y_F'/Y_F$ and we used \eqref{lambdaR} to express the result in terms of $m_t$, $y_t$ and $x_t$. If now we identify $M_F\cong \alpha\Lambda$, with $\alpha$ arbitrary, and impose the matching conditions between \eqref{TF} and \eqref{lambda2} we get $\alpha = \exp(61/96)$ and $Y_F = Y_F'$, and thus
\begin{equation}
\Lambda = M_F e^{-61/96}\simeq 0.53 M_F.
\end{equation}
Making the assumption $M_F\sim Y_F f$ one has $\lambda_R \simeq Y_F/Y.$



\phantomsection 
\addcontentsline{toc}{chapter}{Bibliography}

\setcounter{secnumdepth}{-1}
\small

\bibliographystyle{David}
\bibliography{ThesisFilippo}

\end{document}